\documentclass[%
aps,
rmp,
reprint,
superscriptaddress,
nofootinbib,
amsmath,
amssymb
]{revtex4-2}
\usepackage{
    physics,
    graphicx,
    multirow,
    mathtools, 
    placeins, 
    nicefrac, 
    hyperref, 
    threeparttable,
    enumitem,
    booktabs,
    makecell,
    longtable
}

\newcolumntype{C}[1]{>{\Centering\arraybackslash}m{#1}}
\newcolumntype{L}[1]{>{\RaggedRight\arraybackslash}m{#1}}
\newcommand\cell[2]{\multicolumn{1}{#1}{#2}} 

\usepackage[capitalize]{cleveref}
\Crefname{section}{Sec.}{Secs.}

\usepackage[dvipsnames]{xcolor}
\hypersetup{
    colorlinks,
    linkcolor={blue!50!black},
    citecolor={blue!50!black},
    urlcolor={blue!80!black}
}

\usepackage[caption=false]{subfig} 
\graphicspath{{Figures/}}

\newcommand{\fiteqn}[1]{\resizebox{\hsize}{!}{$#1$}} 

\renewcommand{\var}[1]{{\mathrm{Var}[#1]}}

\newcommand{\mse}[1]{{\mathrm{MSE}[#1]}}
\newcommand{\bias}[1]{{\mathrm{Bias}[#1]}}
\newcommand{\expect}[1]{{\mathrm{E}[#1]}}

\newcommand{\range}[1]{{\mathrm{R}[#1]}}
\newcommand{\mean}[1]{{\overline{#1}}}
\newcommand{\est}[1]{{\hat{#1}}}

\DeclarePairedDelimiter\pbra{\langle\!\langle}{\rvert}
\DeclarePairedDelimiter\pket{\lvert}{\rangle\!\rangle}
\DeclarePairedDelimiterX\pbraket[2]{\langle\!\langle}{\rangle\!\rangle}{#1 \delimsize\vert #2}
\DeclareMathOperator*{\sign}{sgn}

\DeclareMathOperator*{\argmin}{arg\,min}

\newcommand\RMPtext{%
  \begingroup
  \renewcommand\thefootnote{}\footnote{Submitted to \emph{Reviews of Modern Physics}.\\ Please address all correspondence to Zhenyu Cai.}%
  \addtocounter{footnote}{-1}%
  \endgroup
}

\begin{document}
    
\RMPtext

\title{Quantum Error Mitigation}

\newcommand{\addressOxMat}{\affiliation{Department of Materials, University of Oxford, Oxford, OX1 3PH, United Kingdom}}
\newcommand{\addressQM}{\affiliation{Quantum Motion, 9 Sterling Way, London N7 9HJ, United Kingdom}}
\newcommand{\addressGoogle}{\affiliation{Google Quantum AI, Venice, California 90291, USA}}
\newcommand{\addressNTT}{\affiliation{NTT Computer and Data Science Laboratories, NTT Corporation, Musashino 180-8585, Japan}}
\newcommand{\addressGSCA}{\affiliation{Graduate School of China Academy of Engineering Physics, Beijing 100193, China}}

\author{Zhenyu Cai}\email{cai.zhenyu.physics@gmail.com} \addressOxMat\addressQM
\author{Ryan Babbush}\addressGoogle
\author{Simon C. Benjamin}\addressOxMat\addressQM
\author{Suguru Endo}\addressNTT
\author{William J. Huggins}\addressGoogle
\author{Ying Li}\addressGSCA
\author{Jarrod R. McClean}\addressGoogle
\author{Thomas E. O’Brien}\addressGoogle

\date{\today}

\begin{abstract}
    For quantum computers to successfully solve real-world problems, it is necessary to tackle the challenge of \emph{noise}: the errors which occur in elementary physical components due to unwanted or imperfect interactions. The theory of quantum fault tolerance can provide an answer in the long term, but in the coming era of `NISQ' machines we must seek to mitigate errors rather than completely remove them. This review surveys the diverse methods that have been proposed for quantum error mitigation, assesses their in-principle efficacy, and then describes the hardware demonstrations achieved to date. We identify the commonalities and limitations among the methods, noting how mitigation methods can be chosen according to the primary type of noise present, including algorithmic errors. Open problems in the field are identified and we discuss the prospects for realising mitigation-based devices that can deliver quantum advantage with an impact on science and business. 
\end{abstract}

\maketitle

\tableofcontents

\section{Introduction}\label{sec:intro}
The central promise of quantum computing is to enable algorithms that have been shown to provide both polynomial and super-polynomial speed-ups over the best-known classical algorithms for a special set of problems. These problems range from simulating quantum mechanics~\cite{feynmanSimulatingPhysicsComputers1982} to purely algebraic problems such as factoring integers~\cite{shorPolynomialTimeAlgorithmsPrime1999}. The strongest challenge to the viability of practical quantum computing has always been its sensitivity to errors and noise. It was realised early on that the coupling of quantum systems to their environment sets an ultimate time limit and size limit for any quantum computation~\cite{unruhMaintainingCoherenceQuantum1995}. This constraint poses a formidable challenge to the ambitions of realising a quantum computer, since it set bounds on the scalability of any algorithm.  With the advent of quantum error correction (QEC)~\cite{shorSchemeReducingDecoherence1995,steaneErrorCorrectingCodes1996,calderbankGoodQuantumErrorcorrecting1996}, this challenge has been solved at least in theory. The celebrated threshold theorem~\cite{aharonovFaulttolerantQuantumComputation1997,kitaevQuantumComputationsAlgorithms1997} showed that if errors in the quantum hardware could be reduced below a finite rate, known as the threshold, a fault-tolerant quantum computation could be carried out for arbitrary length even on noisy hardware. However, besides the technical challenge of building hardware that achieves the threshold, the implementation of a fault-tolerant universal gate set with current codes, such as the surface code~\cite{fowlerSurfaceCodesPractical2012}, generates a qubit overhead that seems daunting at the moment. For example, recent optimised approaches show that scientific applications that are classically intractable may require hundreds of thousands of qubits~\cite{kivlichanImprovedFaultTolerantQuantum2020}, while industrial applications will require millions of qubits~\cite{leeEvenMoreEfficient2021}. There is ongoing theoretical research to find alternative codes with a more favourable overhead, and recent progress gives reasons to be optimistic~\cite{gottesmanFaultTolerantQuantumComputation2014,breuckmannQuantumLowDensityParityCheck2021,panteleevAsymptoticallyGoodQuantum2022,dinurLocallyTestableCodes2022}. Nevertheless the challenge of realising full-scale fault-tolerant quantum computing is a considerable one. 

This of course motivates the question of whether other approaches, prior to the era of fully fault-tolerant systems, might achieve quantum advantage with significant practical impacts. One might hope so, given the continual and remarkable progress that has been made in quantum computational hardware.
In recent years, it has become routine to see reports of experiments demonstrating high-quality control over multiple qubits [see e.g. \textcite{xueQuantumLogicSpin2022,madjarovHighfidelityEntanglementDetection2020,jurcevicDemonstrationQuantumVolume2021,ebadiQuantumPhasesMatter2021,asavanantGenerationTimedomainmultiplexedTwodimensional2019}], some even reaching beyond 50 qubits [see e.g. \textcite{aruteQuantumSupremacyUsing2019,wuStrongQuantumComputational2021}]. Meanwhile other experiments have indeed demonstrated early-stage fault-tolerant potentials~[see e.g. \textcite{ryan-andersonImplementingFaulttolerantEntangling2022,googlequantumaiSuppressingQuantumErrors2023,krinnerRealizingRepeatedQuantum2022,takedaQuantumErrorCorrection2022,abobeihFaulttolerantOperationLogical2022,postlerDemonstrationFaulttolerantUniversal2022,eganFaulttolerantControlErrorcorrected2021}]. Of course, the works mentioned here are far from exhaustive as it is impossible to capture the all the breakthroughs on different fronts across the diverse range of platforms, we refer the reader to \textcite{acinQuantumTechnologiesRoadmap2018,altmanQuantumSimulatorsArchitectures2021} and the references therein for the key milestones in different platforms. 

The primary goal of \emph{quantum error mitigation} (QEM) is to translate this continuous progress in quantum hardware into immediate improvements for quantum information processing. While accepting that the hardware imperfections will limit the complexity of quantum algorithms, nevertheless we can expect that every advance should enable this boundary to be pushed further. As this review will demonstrate, the mitigation approach indeed proves to be both practically effective and quite fascinating as an intellectual challenge. 

When exploring the prospects for achieving quantum advantage through error mitigation, it is crucial to consider suitable forms of circuits. It is understood that in the era of noisy, intermediate-scale quantum (NISQ) devices, only certain approaches may be able to achieve meaningful and useful results. Due to the limited coherence times and the noise floor present in quantum hardware, one typically resorts to the idea of quantum computation with short-depth circuits. Motivating examples include variational quantum circuits in physics simulations~\cite{peruzzoVariationalEigenvalueSolver2014,mccleanTheoryVariationalHybrid2016,weckerProgressPracticalQuantum2015}, approximate optimisation algorithms~\cite{farhiQuantumApproximateOptimization2014}, and even heuristic algorithms for quantum machine learning~\cite{biamonteQuantumMachineLearning2017}.  Typically in applications of these kinds, the algorithm can be understood as applying a short-depth quantum circuit to a simple initial state and then estimating the expectation value of a relevant observable. 
Such expectation values ultimately lead to the output of the algorithm, which must be accurate enough to be useful in some context (for example, for estimating the energies of molecular states, a useful level of \emph{chemical accuracy} corresponds to $1 \mathrm{kcal}/\mathrm{mol}$~\cite{helgakerMolecularElectronicStructureTheory2000}). This leads to the most essential feature of QEM: the ability to minimise the noise-induced bias in expectation values on noisy hardware. However, this can also be achieved by QEC and many other long-established tools like decoherence-free subspaces and dynamical decoupling sequences (derived from optimal quantum control)~\cite{lidarReviewDecoherenceFreeSubspaces2014,suterColloquiumProtectingQuantum2016}. Therefore this feature alone is not sufficient to capture the QEM techniques that we wish to cover in this review.

It is challenging to find a universally acceptable definition of quantum error mitigation. For the purposes of this review, we will define the term `\emph{quantum error mitigation}' as algorithmic schemes that reduce the noise-induced bias in the expectation value by post-processing outputs from an ensemble of circuit runs, using circuits at the \emph{same noise level} as the original unmitigated circuit \emph{or above}. That is to say QEM will only reduce the effective damage due to noise for \emph{the whole ensemble of circuit runs} (with the help of post-processing), but when we zoom into each individual circuit run, the circuit noise level remains unchanged or even increases. This is in contrast to the other techniques like QEC which aim to reduce the effect of noise on the output in \emph{every single circuit run}. 
 
Since QEM performs post-processing using data directly from noisy hardware, it will become impractical if the amount of noise in the whole circuit is so large that it completely damages the output. In practice, this usually means that for a given hardware set-up, there is a maximum circuit size (circuit depth times qubit number) beyond which QEM will become impractical, usually due to an infeasible number of circuit repetitions. In contrast to QEC, there is no specific error threshold that one must surpass before QEM can be useful; different qubit operation error rates simply lead to different circuit sizes for which QEM will be practical. In other words, as quantum hardware continues to advance, we will be able to apply QEM to ever larger quantum circuits for more challenging applications, without requiring big jumps in the technologies.
 
There are certain desirable features for error mitigation; the methods reviewed here meet the following criteria to differing extents. First, the mitigation method should ideally only require a very modest qubit overhead to remain practical on current and near-term quantum hardware. Nevertheless, error mitigation techniques should provide an accuracy guarantee for the method. Such a guarantee should ideally provide a formal error bound on the mitigated expectation values that indicate how well the method works at varying levels of noise. The bounds would then indicate which concrete hardware improvements would lead to improved estimates.
Needless to say, methods that are conceptually simple and easy to implement experimentally lead to practically feasible approaches. Lastly, a reliable error mitigation method should require few assumptions (or, no assumptions) about the final state that is prepared by the computation. Making strong assumptions about the final state, for example that the state is a product state, may restrict the method to scenarios where a computational advantage over classical approaches may not be given.   

We will start by introducing the basic notion of QEM in \cref{sec:qem_concepts}, with the details of different QEM techniques presented in \cref{sec:methods}. The comparison and combinations of these individual techniques will then be discussed in \cref{sec:compare_and_comb}. In \cref{sec:applications} we will explore the application of QEM in different noise scenarios. Finally, we will discuss the open problems in the field in \cref{sec:open_prob} and offer a conclusion in \cref{sec:concl}

\section{Concepts}\label{sec:qem_concepts}
\subsection{Narrative introduction to concepts and terminology}
In this section we will introduce certain key concepts and terminology that are common to all the QEM methods. However, do note that the approach used in this section might not be the native way for introducing individual techniques in \cref{sec:methods}. In those cases, we will keep terminology that is unique to the given technique self-contained in the respective section.

Near-term quantum devices have imperfections that degrade the desired output information. A QEM protocol will aim to minimise this degradation. We will use the term \emph{primary circuit} to describe the process that, ideally, would produce the \emph{perfect output state} $\rho_0$ whose properties we are interested in. In practice, due to the noise present in the primary circuit, the actual output state is some \emph{noisy state} $\rho$ instead. 

Typically the ideal output information we seek is the expectation value of some \emph{observable of interest} $O$ of the ideal output $\rho_0$. Commonly we would obtain this information simply by averaging the measurement results of repeated execution, as opposed to, say, some phase estimation techniques that can obtain the result through single-shot measurements but require deeper circuits that are more relevant to the fault-tolerant computing era. Therefore, even if we had ideal hardware, we would still need to perform repeated executions to determine the average. We will use $N_{\mathrm{cir}}$ to denote the \emph{number of circuit executions}, or `shots', that we employ -- this will include any executions of variant circuits called for in the QEM protocol. Even in the noiseless limit, the finite $N_{\mathrm{cir}}$ will usually imply a finite inaccuracy in our estimated average, often called \emph{shot noise}. However, with perfect noiseless hardware, there would be zero \emph{bias}. In other words, there would be no systematic shift to the estimated mean versus the true value (the infinite sampling limit). Given that our hardware is not perfect, there will generally be finite bias. QEM protocols aim to reduce this bias, but this often means an increase in the variance (for a fixed number of circuit executions $N_{\mathrm{cir}}$). One could increase $N_{\mathrm{cir}}$ to compensate but this cost should be acknowledged; the cost is the \emph{sampling overhead} of that error-mitigation method versus the ideal noiseless case. In the following subsections we will make these terms and concepts more precise.

\subsection{Error-mitigated estimators}\label{sec:bias_variance}

Our goal is to estimate the expectation value $\Tr[O\rho_0]$ of some observable of interest $O$. Using the outputs of the primary circuit and its variants, we can construct an estimator $\est{O}$ for our target parameter $\Tr[O\rho_0]$. The quality of a given estimator can be assessed in different ways. One way is to use prediction intervals, which calculate the interval within which the outcome of the estimator will fall with a given probability, offering a rigorous bound on the worst-case deviation of the estimator. Here however, in order to see the different factors that contribute to the deviation of the estimator more clearly, we will instead focus on the expected (average-case) square deviation of our estimator $\est{O}$ from the true value $\Tr[O\rho_0]$, which is called the \emph{mean square error}:
\begin{align}\label{eqn:mse}
    \mse{\est{O}} = \expect{(\est{O} - \Tr[O\rho_0])^2}.
\end{align}
The ultimate goal of error mitigation is to reduce $\mse{\est{O}}$ as much as possible, but this needs to be achieved using only finite resources.
To quantify this, it is useful to decompose the mean square error of an estimator into two components, the bias and the variance of the estimator:
\begin{align*}
    \mse{\est{O}} = \bias{\est{O}}^2 + \var{\est{O}}
\end{align*}
with the bias and the variance defined as:
\begin{align*}
    \bias{\est{O}} &= \expect{\est{O}} - \Tr[O\rho_0]\\
    \var{\est{O}} &= \expect{\est{O}^2} - \expect{\est{O}}^2.
\end{align*}
In this review when we say `bias', sometimes we are referring to the magnitude of the bias $\abs*{\bias{\est{O}}}$, the exact meaning should be obvious from the context.

The simplest way to construct the estimator $\est{O}$ is by directly measuring $O$ on the \emph{noisy} output state of the primary circuit $\rho$, and the measurement output is simply denoted using the random variable $\est{O}_{\rho}$. After running the noisy primary circuit $N_{\mathrm{cir}}$ times, we can take the average of these noisy outputs (obtaining the noisy sample mean) to estimate the ideal expectation value $\Tr[O\rho_0]$. This noisy sample mean estimator is denoted as $\mean{O}_{\rho}$ and its mean square error is given by
\begin{align}\label{eqn:mean_mse}
    \mse{\mean{O}_{\rho}} = \big(\underbrace{\Tr[O\rho] - \Tr[O\rho_0]}_{\bias{\mean{O}_{\rho}} = \bias{\est{O}_{\rho}}}\big)^2 + \underbrace{\frac{\Tr[O^2\rho] -  \Tr[O\rho]^2}{N_{\mathrm{cir}}}}_{\var{\mean{O}_{\rho}}=\var{\est{O}_{\rho}}/N_{\mathrm{cir}}}.
\end{align} 
We can see that the error contribution due to the variance, which is often called \emph{shot noise}, will reduce as we increase the number of circuit runs $N_{\mathrm{cir}}$. In the limit of a large number of circuit executions, the mean square error $\mse{\mean{O}_{\rho}}$ will be mainly limited by the bias of the estimator  $\abs{\bias{\mean{O}_{\rho}}}$, which is a systematic error that cannot be reduced by increasing the number of circuit runs. 

\begin{figure}[ht!]
    \centering
    \includegraphics[width = 0.48\textwidth]{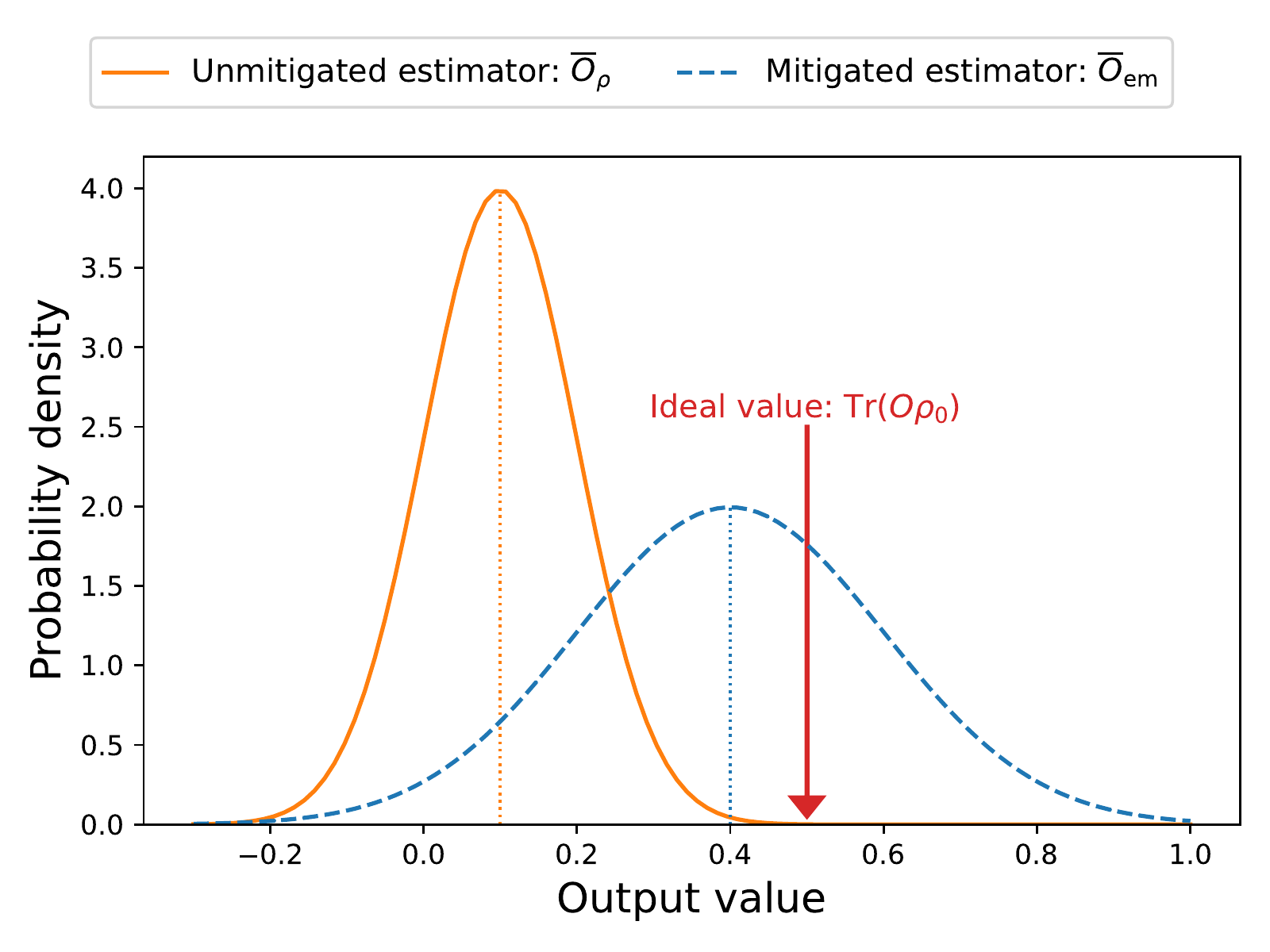}
    \caption{The probability density distributions of the unmitigated estimator and the error-mitigated estimator. We see a decrease of bias and an increase of variance after performing error mitigation.}
    \label{fig:bias_variance}
\end{figure}

In order to reduce the bias, we can apply QEM using data obtained from the noisy primary circuit and its variants, as will be discussed in \cref{sec:methods}. We will construct an error-mitigated estimator $\mean{O}_{\mathrm{em}}$ using the \emph{same number of circuit runs} $N_{\mathrm{cir}}$. We would want to construct the error-mitigated estimator $\mean{O}_{\mathrm{em}}$ in such a way that it can achieve a smaller bias than the naive noisy estimator $\mean{O}_{\rho}$,
\begin{align*}
    \abs*{\bias{\mean{O}_{\mathrm{em}}}} \leq \abs*{\bias{\mean{O}_{\rho}}}.
\end{align*}
This reduction in the bias is usually achieved by constructing a more complex estimator that extracts and amplifies the useful information buried within the noise. As a result, the error-mitigated estimator is also more sensitive to the variation in the sampled data, and thus its variance will usually increase,
\begin{align*}
    \var{\mean{O}_{\mathrm{em}}} \geq \var{\mean{O}_{\rho}}.
\end{align*}
Such a bias-variance trade-off is illustrated in \cref{fig:bias_variance} and can be found in almost all areas of parameter estimation. As we will see later, different ways of performing error mitigation often lead to different trade-offs between bias and variance, giving the user a choice between a quickly-converging QEM method with large residual error, and one that is more costly but more accurate.

We can define an `one-shot' error-mitigated estimator $\est{O}_{\mathrm{em}}$ that will satisfy $\expect{\est{O}_{\mathrm{em}}} = \expect{\mean{O}_{\mathrm{em}}}$ and $\var{\est{O}_{\mathrm{em}}} = N_{\mathrm{cir}}\var{\mean{O}_{\mathrm{em}}}$. The number of circuit runs needed for given estimator $\est{X}$ to achieve the shot noise level $\epsilon$ is given by $N_{\mathrm{shot}}^{\epsilon}(\est{X}) = \var{\est{X}}/\epsilon^2$. To reach the same shot noise level as the original noisy estimator, the error-mitigated estimator will require more circuit runs. This factor of increase in the number of circuit runs is called the \emph{sampling overhead}, which is given by:
\begin{align}\label{eqn:samp_cost}
    C_{\mathrm{em}} =  \frac{N_{\mathrm{shot}}^\epsilon(\est{O}_{\mathrm{em}})}{N_{\mathrm{shot}}^\epsilon(\est{O}_{\rho})} = \frac{\var{\est{O}_{\mathrm{em}}}}{\var{\est{O}_{\rho}}}.
\end{align}

The sampling overhead can also be estimated using the range of the estimator through Hoeffding's inequality. The \emph{range} of a random variable $\est{X}$, denoted as $\range{\est{X}}$, is the difference between the maximum and minimum possible values taken by $\est{X}$. Using Hoeffding's inequality, the number of samples that is sufficient to guarantee an estimation of $\expect{\est{X}}$ to $\epsilon$-precision with $1- \delta$ probability is given by
\begin{align}\label{eqn:samp_number_hoeff}
    N_{\mathrm{hff}}^{\epsilon,\delta}(\est{X}) = \frac{\ln(2/\delta)}{2\epsilon^2} \range{\est{X}}^2
\end{align}
which can be used to estimate the sampling overhead required for the error-mitigated estimator $\est{O}_{\mathrm{em}}$ to achieve the same $\epsilon$ and $\delta$ as the unmitigated estimator $\est{O}_{\rho}$:
\begin{align}\label{eqn:samp_cost_hoeff}
    C_{\mathrm{em}} =  \frac{N_{\mathrm{shot}}^\epsilon(\est{O}_{\mathrm{em}})}{N_{\mathrm{shot}}^\epsilon(\est{O}_{\rho})} \sim \frac{N_{\mathrm{hff}}^{\epsilon,\delta}(\est{O}_{\mathrm{em}})}{N_{\mathrm{hff}}^{\epsilon,\delta}(\est{O}_{\rho})} = \frac{\range{\est{O}_{\mathrm{em}}}^2}{\range{\est{O}_{\rho}}^2}.
\end{align} 

\subsection{Faults in the circuit}\label{sec:circuit_faults}
To gain intuition about the performance and costs of QEM, we need a way to quantify the damages due to noise in the circuit. Doing this by modelling noise in complete generality on a quantum system can be very challenging~\cite{lidarIntroductionDecoherenceNoise2013}. One useful approximation, which is widely employed in the field of quantum error correction, is to model noise as discrete, probabilistic events named \emph{faults} that can occur at the various \emph{locations} in the circuit including gates, idling steps and measurements~\cite{gottesmanIntroductionQuantumError2009,terhalQuantumErrorCorrection2015}. For simplicity, we will assume all locations are afflicted with Pauli noise with the error probability of location $f$ given by $p_f$ \footnote{All of the arguments will still apply if we define $1-p_f$ as the coefficient of the error-free part of the Kraus representation of some general noise (we will select the Kraus representation that gives the largest $1-p_f$). In this case, $1-p_f$ is \emph{not} the average gate fidelity. For example, for any non-identity unitary channel, $1-p_f$ will always be $0$, which is not the case for average gate fidelity. More generally, we can always apply Pauli twirling to transform all noise to Pauli noise such that our arguments become valid.}. If we further assume the error events in all locations are \emph{independent}, then the total probability that there is no fault in the circuit is given by 
\begin{equation}\label{eqn:no_error_rate_1}
    P_0=\prod_f(1-p_f),
\end{equation}
which we will simply call the \emph{fault-free probability} of the circuit. It decays exponentially with the increase of the number of fault locations in the circuit (and thus the number of qubits and the circuit depth). Do note that the fault-free probability is \emph{not} the fidelity of the output state, but it is a lower bound for the fidelity under our noise assumption.

We can also quantify the amount of noise in the circuit using the average number of faults in each circuit run, which is given by:
\begin{equation}\label{eqn:circ_fault_rate}
    \lambda=\sum_f p_f,
\end{equation}
and will be called the \emph{circuit fault rate}. In the simple case that all $M$ fault locations in the circuit have the same error rate $p$, we then simply have $\lambda = Mp$. If the circuit contains a large number (more than dozens) of fault locations, and the circuit fault rate is of the order of unity $\lambda \sim 1$, then the number of faults occurring in a given circuit run can be modelled using a \emph{Poisson distribution} with mean $\lambda$ using the Le Cam's theorem~\cite{lecamApproximationTheoremPoisson1960,endoPracticalQuantumError2018,caiMultiexponentialErrorExtrapolation2021}, i.e. the probability that $\ell$ faults occur in the circuit is given by $P_\ell = e^{-\lambda} \lambda^\ell/\ell!$. In this way, the fault-free probability is given by 
\begin{equation}\label{eqn:no_error_rate}
    P_0=e^{-\lambda},
\end{equation}
which decay exponentially with the circuit fault rate.
Note that for the intuitive arguments made in the next subsection, and indeed for general estimation about the feasibility of error mitigation, approximate estimates of $P_0$ and $\lambda$ are often good enough to be useful.

\subsection{Exponential scaling of the sampling overhead}\label{sec:exp_samp_cost}
We can perform \emph{bias-free} QEM if we are able to post-select for the fault-free circuit runs without needing any additional circuit components. The fraction of circuit runs that are selected is simply given by the fault-free probability $e^{-\lambda}$ in \cref{eqn:no_error_rate}, which means that we will still require $ e^{\lambda}$ times more circuit runs to obtain the same number of ``effective'' circuit runs as a noise-free machine and achieve the same level of shot noise. Hence, even allowing for the ``magical'' post-selection of fault-free circuit runs, the sampling overhead $C_{\mathrm{em}} = e^{\lambda}$ will still increase exponentially with the circuit fault rate (and thus the circuit size). This implies a sampling overhead of $C_{\mathrm{em}} \sim 150$ when $\lambda = 5$ and $C_{\mathrm{em}} \sim 10^4$ when $\lambda = 9$, which provides intuition on why QEM is unlikely to be efficient when the circuit fault rate is beyond $\order{1}$. Of course, this does not constitute a rigorous bound on the overhead of QEM.

Now let us move beyond trying to extract the error-free state and instead focus on obtaining the right expectation value for the observable of interest. \textcite{caiMultiexponentialErrorExtrapolation2021,wangNoiseinducedBarrenPlateaus2021} have shown that the expectation value of Pauli observables under Pauli gate noise is bounded by an exponential decay curve against the increase of the circuit fault rate $\lambda$. In order to resolve such an exponentially small quantity at large $\lambda$, we will need an exponential number of samples ($C_{\mathrm{em}} = \order{e^{\beta\lambda}}$ for some positive $\beta$). This exponential scaling of sample overhead still applies when we consider error-mitigated estimators that are linear combinations of the output of such noisy circuits (see \cref{sec:monte_carlo_sampling_exp}). 

For a given noisy circuit with the circuit fault rate $\lambda$, rather than performing active correction to reduce $\lambda$ in each circuit run like in quantum error correction, QEM relies on post-processing the outputs from an ensemble of circuit runs with the same circuit fault rate $\lambda$ or above. Hence, through the simple examples above, we see that QEM cannot efficiently tackle noisy circuits with large $\lambda$ \emph{on its own} due to the exponential sampling overhead. However, as we will see later, due to the much lower implementation cost for QEM in terms of additional circuit components and qubits, it has become an effective means to stretch the application potential of near-term noisy devices and will be a useful tool to help alongside quantum error correction in the longer term.

\section{Methods}\label{sec:methods}
After introducing the overall concept of QEM, we will now look at how the various error-mitigated estimators are actually constructed through performing different QEM methods.

\subsection{Zero-noise extrapolation}\label{sec:zne}

In this section, we will make use of noisy states obtained at different circuit fault rates. The state obtained at the circuit fault rate $\lambda$ will be denoted as $\rho_{\lambda}$. The noisy expectation value $\Tr[O\rho_{\lambda}]$ can be viewed a function of $\lambda$. In this way, the ideal expectation value we want is just the value of the function at $\lambda = 0$. Trying to obtain this zero-noise value using data points at different circuit fault rates brings us to the concept of \emph{zero-noise extrapolation} (also called \emph{error extrapolation}), which was first introduced by \textcite{liEfficientVariationalQuantum2017,temmeErrorMitigationShortDepth2017}.

\begin{figure}[ht!]
    \centering
    \includegraphics[width = 0.48\textwidth]{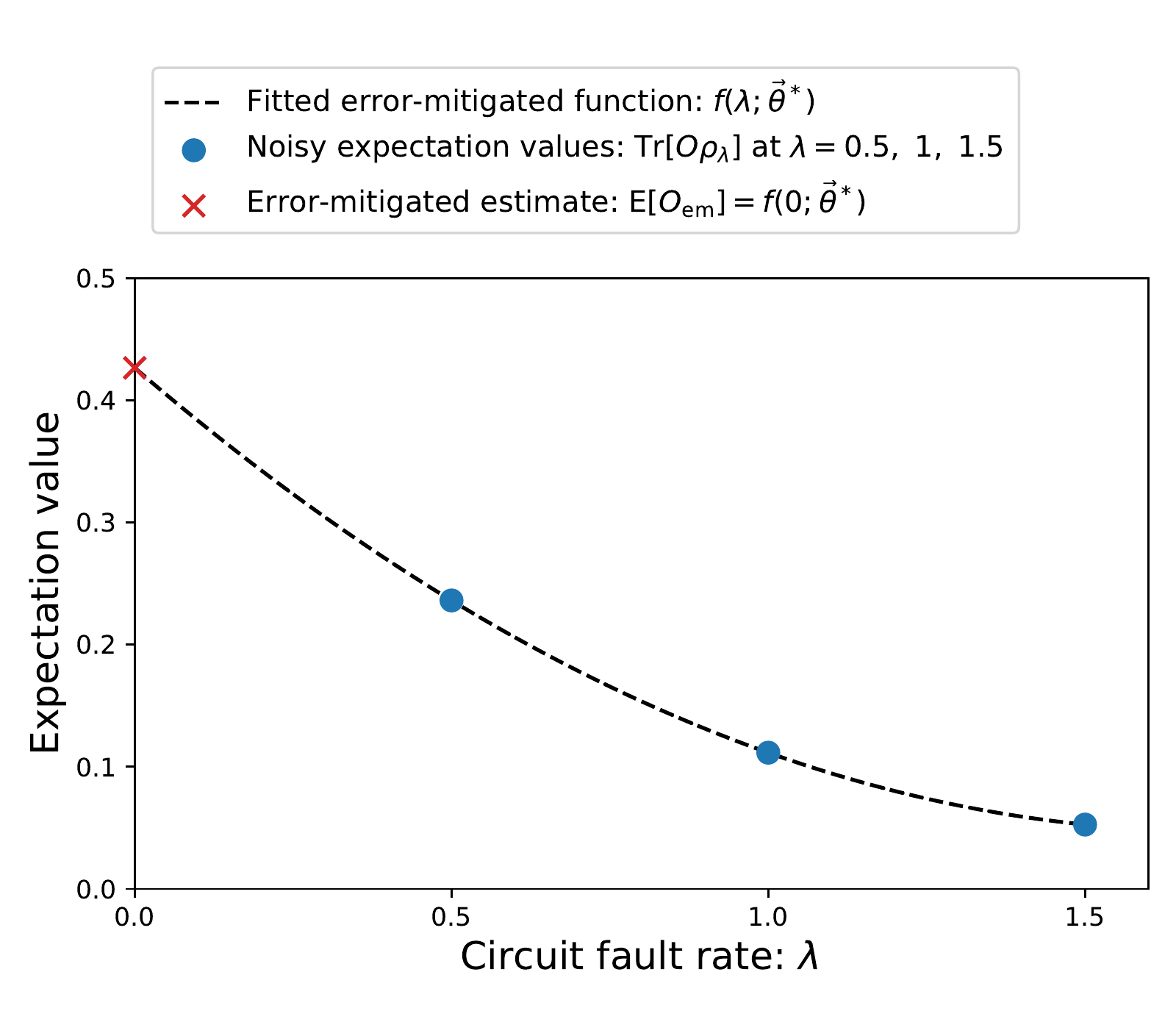}
    \caption{Obtaining the error-mitigated estimate using zero-noise extrapolation. Here we are performing extrapolation using three noisy expectation values at the circuit fault rate of $\lambda = 0.5,\ 1,\  1.5$, where $0.5$ is the lowest circuit fault rate we can achieve and the other two are obtained through boosting the noise in the device.}
    \label{fig:zne}
\end{figure}

Using $\lambda_1$ to denote the smallest circuit fault rate we can achieve, we can probe $\Tr[O\rho_{\lambda}]$ at a range of \emph{boosted} error rate $\{\lambda_{m}\}$ with $\lambda_{m} < \lambda_{m+1}$ to obtain a set of data points $\{(\lambda_{m}, \Tr[O\rho_{\lambda_{m}}])\}$. We can model $\Tr[O\rho_{\lambda}]$ using a parametrised function $f(\lambda;\vec{\theta})$ and fit it to the data points to obtain a set of optimal parameters $\vec{\theta}^*$. The error-mitigated estimate of the zero-noise output $\Tr[O\rho_0]$ is then given by
\begin{align*}
    \expect{\est{O}_{\mathrm{em}}} = f(0;\vec{\theta}^*).
\end{align*}
A simple illustration of zero-noise extrapolation is shown in \cref{fig:zne}.

If \emph{$\lambda$ is small}, \textcite{temmeErrorMitigationShortDepth2017} showed that $\Tr[O\rho_{\lambda}]$ can be approximated using a polynomial function in the same spirit as truncated Taylor expansion:
\begin{align}\label{eqn:poly_approx}
    \Tr[O\rho_{\lambda}] \approx f(\lambda; \vec{\theta})  = \sum_{\ell = 0}^{M-1} \theta_\ell \frac{\lambda^\ell}{\ell!}.
\end{align}
Here we have a polynomial of degree $M-1$, which has $M$ different free parameters and the zero-noise estimate we want is $\theta_0^* = f(0;\vec{\theta}^*)$. The simplest case is linear extrapolation with $M = 2$~\cite{liEfficientVariationalQuantum2017}. If we try to fit \cref{eqn:poly_approx} with $M$ data points, which is the minimal number of data points needed, we can perform Richardson extrapolation as discussed in \textcite{temmeErrorMitigationShortDepth2017}. The corresponding error-mitigated estimate obtained using the set of data points $\{(\lambda_{m}, \Tr[O\rho_{\lambda_{m}}])\}$ is~\cite{giurgica-tironDigitalZeroNoise2020}
\begin{align}\label{eqn:expect_richardson}
    \expect{\est{O}_{\mathrm{em}}} = \theta_0^* = \sum_{m=1}^{M} \Tr[O\rho_{\lambda_{m}}] \prod_{k \neq m}\frac{\lambda_k}{\lambda_k - \lambda_{m}}.
\end{align}
The bias in our estimate is mostly due to our omission of the higher degree terms in the polynomial approximation, and thus we should expect $\bias{\est{O}_{\mathrm{em}}} = \order{\lambda^{M}}$.

The error-mitigated expectation value in \cref{eqn:expect_richardson} is a linear combination of the set of noisy expectation values $\{\Tr[O\rho_{\lambda_{m}}]\}$, thus it can be estimated using the Monte Carlo sampling method(see \cref{sec:monte_carlo_sampling}) and the corresponding sampling overhead for Richardson extrapolation is given by:
\begin{align}\label{eqn:ric_samp_cost}
    C_{\mathrm{em}} \sim  \left(\sum_{m=1}^M \abs{\prod_{k \neq m}\frac{\lambda_k}{\lambda_k - \lambda_{m}}}\right)^2.
\end{align}

We see that if any of the probed circuit fault rates $\lambda_{m}$ is too big or the gap between any two data points $\abs{\lambda_{m} - \lambda_k}$ is too small, then $C_{\mathrm{em}}$ will blow up and the extrapolation would become infeasible. For the simple case of equal-gap Richardson extrapolation ($\lambda_{m} = m \lambda_1$), the sampling overhead in \cref{eqn:ric_samp_cost} becomes $C_{\mathrm{em}} \sim (2^M - 1)^2$, which grow \emph{exponentially} with the number of data points $M$. The example here is mainly for illustrating the scaling behaviour of the sampling overhead. In practice, Richardson extrapolation usually takes more general data gap $\Delta$ with data points at $\lambda_{m} = \lambda_1 + (m - 1) \Delta$. Even more sophisticated Richardson extrapolation beyond the equal-gap variant is possible, which can reduce the sampling overhead~\cite{krebsbachOptimizationRichardsonExtrapolation2022}.

As mentioned, theoretically Richardson extrapolation will only be valid for small $\lambda$ due to the approximation we made in \cref{eqn:poly_approx}. However, a recent experiment by \textcite{kimScalableErrorMitigation2023} showed that Richardson extrapolation can be effective at large $\lambda$ in practice. There they performed a 26-qubit simulation of the 2D transverse field Ising model in the limit of strong entanglement, and showed that the error-mitigated evolution of the magnetisation is competitive in comparison to standard tensor network methods. To look for an extrapolation method that is naturally compatible with large $\lambda$, we can consider a large circuit with Pauli noise. Such a circuit in the limit of $\lambda \rightarrow \infty$ will become a random circuit with \emph{zero} expectation value for bounded traceless observables as mentioned in \cref{sec:exp_samp_cost}. A polynomial extrapolation function diverges at $\lambda \rightarrow \infty$ and thus does not fit the intuition above, which motivates the extrapolation schemes that use an exponential decay curve or even a multi-exponential decay curve instead. Exponential extrapolation has proven to be able to achieve smaller biases than Richardson and more generally polynomial extrapolation in some numerical simulations and cloud experiments~\cite{endoPracticalQuantumError2018,giurgica-tironDigitalZeroNoise2020}, especially for Pauli noise~\cite{caiMultiexponentialErrorExtrapolation2021}. It is worth noting that going beyond Richardson extrapolation, the error-mitigated estimator obtained through least-square fitting often does not have a closed-form representation, and thus there are not yet analytical expressions for their biases and variance in most of the cases. It is possible to combine exponential and Richardson extrapolation by performing Richardson extrapolation on the function $e^\lambda\Tr[O\rho_\lambda]$ instead of $\Tr[O\rho_\lambda]$, which can give explicit bounds on the sampling overheads and biases~\cite{caiPracticalFrameworkQuantum2021}. 

When we try to boost the noise in the circuit, we must know the exact factor of increase in the circuit fault rate. Ideally, we would want to increase the error strength in the various fault locations without changing their error models, which can be challenging to implement in experiments. \textcite{temmeErrorMitigationShortDepth2017} showed that under the assumption that the noise was time-invariant, the noise could be effectively amplified by stretching and recalibrating the gate pulses, which was demonstrated on a superconducting platform~\cite{kandalaErrorMitigationExtends2019}. When targeting only single-qubit gate errors, these experiments can successfully perform Richardson extrapolation up to the fourth order, highlighting the accuracy of the noise amplification. Rescaling noise for two-qubit cross-resonance gates in these architectures can be more challenging due to reasons like more complicated drive Hamiltonian and drive-dependent coherence times~\cite{chowSimpleAllMicrowaveEntangling2011}. Nevertheless, pulse-stretching experiments with slowed-down two-qubit cross-resonance gates have been shown to be effective with linear extrapolation~\cite{kandalaErrorMitigationExtends2019}, most recently demonstrated in experiments with up to 26 qubits and a depth of 120~\cite{kimScalableErrorMitigation2023}. 

Alternatively, \textcite{dumitrescuCloudQuantumComputing2018,giurgica-tironDigitalZeroNoise2020,heZeronoiseExtrapolationQuantumgate2020} tried to boost the circuit fault rate by inserting a sequence of abundant gates that are equivalent to the identity if operated noiselessly. This is easier to implement and calibrate compared to the pulse-stretching method. However, while this will work well with depolarising gate noise, it may change the gate error model for more general gate noise such that the factor of increase in the error strength is no longer well-defined~\cite{kimScalableErrorMitigation2023}. To address this, \textcite{henaoAdaptiveQuantumError2023} replaced the gate-based inverse with a pulse-based inverse that is applicable beyond depolarising noise. They also adapted the weights of the noise-amplified circuits to the strength of the noise in order to extend beyond the weak noise assumption in Richardson Extrapolation. Note that for both pulse stretching and gate insertion, an $M$-time increase in the noise strength might lead to up to an $M$-time increase in the circuit runtime. If the error models at the various fault locations are completely known, \textcite{liEfficientVariationalQuantum2017}  suggested that it is possible to controllably amplify the noise by probabilistically inserting gates to simulate the faults. In fact, as first noted in \textcite{caiMultiexponentialErrorExtrapolation2021} and later in \textcite{mariExtendingQuantumProbabilistic2021}, it would be more efficient to use these probabilistically inserted gates to perform probabilistic error cancellation (\cref{sec:pec}) instead, which yields new data points at \emph{reduced} error strength. Of course, learning a representative noise model can be challenging for large-scale devices, especially for correlated noise.

Due to the simplicity of zero-noise extrapolation, it is one of the most widely implemented QEM methods. Very recently, \textcite{kimEvidenceUtilityQuantum2023} have simulated the time dynamics of the Ising model up to a circuit size of 127 qubits and 60 layers of 2-qubit gates, and with the help of ZNE, they are able to produce results in agreement with state-of-the-art classical simulations for that circuit size~\cite{tindallEfficientTensorNetwork2023,begusicFastClassicalSimulation2023,kechedzhiEffectiveQuantumVolume2023}. In addition to the experiments mentioned above, it was also successfully demonstrated in a wide range of other experiments, especially through cloud platforms~[see e.g. \textcite{klcoQuantumclassicalComputationSchwinger2018,tacchinoQuantumImplementationArtificial2020,keenQuantumclassicalSimulationTwosite2020,yeter-aydenizPracticalQuantumComputation2020,garmonBenchmarkingNoiseExtrapolation2020}].

\subsection{Probabilistic error cancellation}\label{sec:pec}
An alternative quantum error mitigation method referred to as \emph{probabilistic error cancellation} was introduced by \textcite{temmeErrorMitigationShortDepth2017}. A particular feature of this method is that it can fully remove the bias of expectation values of generic quantum circuits ($\bias{\est{O}_{em}} = 0$). This comes at the expense of a sampling overhead $C_{em}$ that grows exponentially with the circuit fault rate $\lambda$. The key idea is noting that the noise-free expectation value can be written as a linear combination of expectation values from a set of noisy quantum circuits. This real-valued, linear combination can be interpreted as a \emph{quasi-probability decomposition} which can be sampled, c.f. \cref{sec:monte_carlo_sampling}, according to a Monte Carlo procedure. In this review we formalise the method using the superoperator representation~\cite{gilchristVectorizationQuantumOperations2011}, in which density matrices $\rho$ are vectorised into $\pket{\rho}$ and quantum channels are written as matrices acting on $\pket{\rho}$, denoted using the curly font like $\mathcal{U}$. Taking the trace with the observable $O$ is then written as the inner product with the vectorised observable $\Tr[O\rho] = \pbraket{O}{\rho}$. In this section, we employ this representation over the standard density matrix formalism for clearer representations of the linear combination and decomposition of a set of noisy quantum circuits.

To construct an estimator for an ideal channel $\mathcal{U}$ from noisy operations, we need to choose a set of \emph{noisy} basis operations $\{\mathcal{B}_n\}$ that we can implement in the physical hardware. These operations are for example noisy gates, state preparation operations and measurements. These operations are assumed to be learned from the noisy hardware in experiments through some form of tomography. An example of a complete set of basis operations is discussed in \textcite{endoPracticalQuantumError2018}. With a sufficiently large basis, we can decompose the ideal operation into 
\begin{align}\label{eqn:ideal_gate_decomp}
    \mathcal{U} = \sum_{n} \alpha_n \mathcal{B}_n.
\end{align}
with real coefficients $\alpha_n$. Some of the coefficients $\alpha_n$ in this expansion can be negative, which means that this decomposition does not necessarily correspond to a probabilistic mixture of physical maps. The expansion therefore is often not a physical map that can be implemented directly. However, if we are applying $\mathcal{U}$ on some input state $\rho_{\mathrm{in}}$ in order to measure some observable $O$ and obtain the 
the ideal expectation value $\Tr[O\rho_0] = \pbraket{O}{\rho_0}$, then the ideal expectation value can be decomposed into:
\begin{align}\label{eqn:expect_pec}
    \pbraket{O}{\rho_0} = \pbra{O}\mathcal{U}\pket{\rho_{\mathrm{in}}} = \sum_{n} \alpha_n \pbra{O} \mathcal{B}_n \pket{\rho_{\mathrm{in}}},
\end{align}
i.e. the ideal expectation value $\pbra{O}\mathcal{U}\pket{\rho_{\mathrm{in}}}$ can be decomposed into a linear combination of noisy expectation values $\{\pbra{O} \mathcal{B}_n\pket{\rho_{\mathrm{in}}}\}$ that we can obtain individually.

When the expansion in \cref{eqn:expect_pec} has many terms, the linear combination of noisy expectation values in \cref{eqn:expect_pec} can be estimated using the Monte Carlo sampling method (\cref{sec:monte_carlo_sampling}). The method samples different basis operations $\mathcal{B}_n$ according to their weight in the expansion, i.e. the noisy circuit corresponding to $\mathcal{B}_n$ is chosen with the probability $|\alpha_n| Q^{-1}$, where $Q = \sum_{n} \abs{\alpha_n}$, and the circuit output is multiplied with $\mbox{sign}(\alpha_n) Q$ before being used to estimate the error-mitigated expectation value $\expect{\est{O}_{\mathrm{em}}}$. This multiplicative factor leads to an increase in the variance and correspondingly to a sampling overhead (given by \cref{eqn:monte_cost}):
\begin{align}\label{eqn:pec_samp_cost}
    C_{\mathrm{em}} &\sim  Q^2 = \left(\sum\nolimits_{n} \abs{\alpha_n}\right)^2.
\end{align}

In practice, we can often implement the noisy version of the target operation, denoted as $\mathcal{U}_{p}$, which can be written as $\mathcal{U}_{p} = (1-p) \mathcal{U} + p \mathcal{N}$ where $p$ is the operation error rate and $\mathcal{N}$ is the noise element. It can be rewritten as:
\begin{align}\label{eqn:ideal_gate_decomp_2}
    \mathcal{U}& = \frac{1}{1-p} \mathcal{U}_{p}  - \frac{p}{1-p} \mathcal{N}. 
\end{align}
Compared to \cref{eqn:ideal_gate_decomp}, we see that $\mathcal{U}_{p}$ is one of the basis operations $\mathcal{U}_{p} = \mathcal{B}_{1}$. In the simple case that $\mathcal{N}$ is also one of the basis operations that we can implement: $\mathcal{N} = \mathcal{B}_{2}$, the sampling overhead for removing the noise in $\mathcal{U}_{p}$ is given by:
\begin{align}\label{eqn:pec_samp_cost_gate}
    C_{\mathrm{em}} \sim Q^2 = \left(\frac{1}{1-p} + \frac{p}{1-p}\right)^2 = \left(\frac{1 + p}{1-p}\right)^2.
\end{align}
More generally, $\mathcal{N}$ must be decomposed into the rest of the basis $\mathcal{N} = \sum_{n \neq 1} w_n \mathcal{B}_{n}$. For the common scenario in which $\mathcal{U}_{p}$ suffers from Pauli noise and we can perform high fidelity Pauli gates to be used as our basis $\{\mathcal{B}_{n}\}$, the sampling overhead is still given by \cref{eqn:pec_samp_cost_gate}.

Up to now we are using the basis set to cancel out the noisy component $\mathcal{N}$ in $\mathcal{U}_{p} = (1-p) \mathcal{U} + p \mathcal{N}$. Alternatively, there is another means of decomposition to effectively ``invert'' the noise channel $\mathcal{E}_p$ in $\mathcal{U}_{p} = \mathcal{E}_p\mathcal{U}$ as discussed in \textcite{temmeErrorMitigationShortDepth2017} and \textcite{endoPracticalQuantumError2018} and its resultant sampling overhead is similar. However, this noise-inversion implementation requires inserting additional operations after every noisy operation and thus we might need up to twice the original circuit runtime.

To perform the error cancellation described above, we need to have the full description of the noisy operation $\mathcal{U}_{p}$. This can only be efficiently characterised for individual local gates. Any circuit we want to implement can be decomposed into a sequence of $M$ ideal gates $\{\mathcal{U}_m\}$, and the ideal expectation value we want to obtain will be
\begin{align*}
    \pbraket{O}{\rho_0} =  \pbra{O}\prod_{m=1}^M\mathcal{U}_m\pket{\rho_{\mathrm{in}}}.
\end{align*}
Similar to \cref{eqn:expect_pec}, we decompose the individual gates into the noisy basis using \cref{eqn:ideal_gate_decomp}
\begin{align}\label{eqn:pec_linear_sum}
    \pbraket{O}{\rho_0} = \expect{\est{O}_{\mathrm{em}}} &= \pbra{O} \prod_{m = 1}^{M} \left(\sum\nolimits_{n_m} \alpha_{mn_m} \mathcal{B}_{n_m}\right)\pket{\rho_{\mathrm{in}}}\nonumber\\
    & =  \sum_{\vec{n}} \alpha_{\vec{n}} \pbra{O} \mathcal{B}_{\vec{n}}\pket{\rho_{\mathrm{in}}}
\end{align}
where $\vec{n} = \{n_1, n_2, \cdots, n_M\}$ is the set of labels for a sequence of basis elements, and we have defined $\mathcal{B}_{\vec{n}} = \prod_{m = 1}^{M}\mathcal{B}_{n_m}$, $\alpha_{\vec{n}} = \prod_{m = 1}^{M} \alpha_{mn_m}$ and $Q = \sum_{\vec{n}} \abs{\alpha_{\vec{n}}}$. Note that the set of noisy basis elements $\{\mathcal{B}_n\}$ here includes the basis for all the gates in the circuit, it is thus over-complete and contains the noisy version of all the target gates. Again we can obtain samples of $\est{O}_{\mathrm{em}}$ using Monte Carlo sampling as discussed above with the label of a single basis $n$ replaced by the label of a sequence of basis $\vec{n}$.

The overall sampling overhead of mitigating the errors in the whole circuit is simply the product of the sampling overhead of each gate. As explored in \textcite{temmeErrorMitigationShortDepth2017} and subsequently \textcite{endoPracticalQuantumError2018}, if we assume all the gates suffer from Pauli noise with the same error rate $p$, the circuit size $M$ is large and the circuit fault rate $\lambda = Mp$ is finite, then taking the product of the gate-level sampling overhead in \cref{eqn:pec_samp_cost_gate} we have:
\begin{align}\label{eqn:pec_samp_cost_circ}
    C_{\mathrm{em}} = \prod_{m = 1}^{M}\left(\frac{1+p}{1-p}\right)^2 \approx e^{4Mp} = e^{4\lambda}. 
\end{align}
In fact, the overhead here is still valid even if the gates in the circuit have different error rates, as long as the circuit faults follow a Poisson distribution (see \cref{sec:circuit_faults}). More generally, as first discussed by \textcite{caiMultiexponentialErrorExtrapolation2021} and later by \textcite{mariExtendingQuantumProbabilistic2021}, it is possible to only apply probabilistic error cancellation partially. Denoting the resultant circuit fault rate as $\lambda_{\mathrm{em}}$, the corresponding sampling cost is simply $C_{\mathrm{em}} = e^{4\left(\lambda - \lambda_{\mathrm{em}}\right)}$, which grows exponentially with the reduction in the circuit fault rate. 

As mentioned, to be able to perform the decomposition in \cref{eqn:ideal_gate_decomp}, we need to have a set of basis elements that span the ideal operations, and full characterisation of this basis. \textcite{endoPracticalQuantumError2018} have shown that such basis can be constructed using single-qubit Clifford gates and $Z$-measurement with reasonable fidelity, while the characterisation can be carried out efficiently using gate set tomography. In practice, we usually supplement this set of basis elements with the noisy version of the target operation, such that it is over-complete as discussed. Only noisy operations with local noise can be efficiently characterised using the protocol above. 

A recent experimental implementation of the probabilistic error cancellation method used a correlated Pauli-noise model that is supported over the full gate layer on the device~\cite{vandenbergProbabilisticErrorCancellation2023}. The noise model is given in terms of a sparse Pauli-Lindbladian ${\cal L}(\rho) = \sum_{k} \lambda_k \left(P_k \rho P_k - \rho \right)$ for a set of Pauli matrices $P_k$,  which can be efficiently learned for a polynomial number of Pauli terms (usually low-weight Pauli terms). Although the noise model is correlated across the full circuit, it can be inverted efficiently and its quasi-probability distribution can be sampled exactly. A similar approach was also adopted by \textcite{ferracinEfficientlyImprovingPerformance2022} which used cycle benchmarking to characterise the low-weight correlated Pauli noise components for performing probabilistic error cancellation. It is also possible to mitigate these correlated noise components by performing noise characterisation using matrix product operators~\cite{guoQuantumErrorMitigation2022}, or with the help of learning-based methods~\cite{strikisLearningBasedQuantumError2021} as will be discussed in~\cref{sec:learning-based}. 

From \cref{eqn:ideal_gate_decomp,eqn:pec_samp_cost}, we see that the sampling overhead of probabilistic error cancellation is highly dependent on the basis we choose. For the standard basis proposed in \textcite{endoPracticalQuantumError2018}, it will perform well when the gates in the circuit suffer from Pauli noise (which can be achieved via Pauli twirling). Going beyond that, \textcite{takagiOptimalResourceCost2021} has derived a lower bound for the sampling overhead under general noise models. Such a lower bound on the sampling overhead has proven to be a good measure for many properties of the noise channel that we try to mitigate~\cite{jiangPhysicalImplementabilityLinear2021,regulaOperationalApplicationsDiamond2021,guoNoiseEffectsPurity2023}, but the basis required to reach this lower bound may not be implementable by the given hardware. To find a better practical basis beyond the standard basis, \textcite{piveteauQuasiprobabilityDecompositionsReduced2022} proposed to use variational circuits to construct the basis and the authors numerically tested this using real hardware noise models. For actual experiments, probabilistic error cancellation has been successfully demonstrated by \textcite{songQuantumComputationUniversal2019,zhangErrormitigatedQuantumGates2020} on superconducting and trapped-ion platforms. \textcite{sunMitigatingRealisticNoise2021} has shown that probabilistic error cancellation can also be applied to continuous noise processes, for which partial mitigation is possible by expanding the noise process into a perturbation series~\cite{hamaQuantumErrorMitigation2022}. It is also possible to use the non-Markovianity in noise to reduce the sampling cost~\cite{hakoshimaRelationshipCostsQuantum2021}.

\subsection{Measurement error mitigation}\label{sec:meas_miti}

Depending on the error mitigation procedure it is necessary to make a distinction between the types of errors that occur during a calculation. Errors that occur during the state preparation and measurement stage are referred to as SPAM errors~\cite{merkelSelfconsistentQuantumProcess2013,linIndependentStateMeasurement2021}. The error that occurs at the final measurement stage introduces an additional bias in the expectation value of interest. To put this into a more concrete form, let us continue to use the super-operator representation introduced in \cref{sec:pec}. When we perform measurements and obtain the binary string $x\in \{0,1\}^N$ as the output, ideally we want to perform projective measurements in the computational basis $\{\pbra{x}\}$. However, some measurement noise $\mathcal{A}$ might occur and transform the projective measurements into some positive operator-value measures (POVM) $\{\pbra{E_{x}}\} = \{\pbra{x}\mathcal{A}\}$, leading to a different output statistic.

The origins of the measurement errors are as diverse as the hardware that is used to implement quantum processors. For example, a dominant error in the measurement of superconducting qubits is due to thermal excitations and $T_1$-decay~\cite{wallraffStrongCouplingSingle2004,blaisCavityQuantumElectrodynamics2004}, while for ion traps a major source of uncertainty arises from the difficulty of detecting an ion's dark state and collisions in the trap~\cite{nagourneyShelvedOpticalElectron1986,bergquistObservationQuantumJumps1986,sauterObservationQuantumJumps1986}. Other architectures experience noise in the measurement stage from different sources~\cite{harocheExploringQuantumAtoms2006}. From the perspective of the measurement error protocols most frequently considered in the literature, it is sufficient to consider a simplified model that is agnostic regarding the actual origin of the noise. This model makes the assumption that the noise channel $\mathcal{A}$ has the computational subspace $\{\pbra{x}\ |\ x\in \{0,1\}^N\}$ as its invariant subspace, i.e. the resultant POVM basis $\{\pbra{E_{x}}\}$ lives within the computational subspace. This is not the most general measurement error model, but it is nonetheless the most frequently considered model as other coherent errors are usually assumed to be part of the computational stage instead of the measurement stage. Under this assumption, the POVM $\pbra{E_{x}}$ can be decomposed into the computational basis:
\begin{align}\label{eqn:noisy_meas_decomp}
    \pbra{E_{x}} = \sum_{y}\pbraket{E_{x}}{y} \pbra{y}  = \sum_{y}\pbra{x}\mathcal{A}\pket{y} \pbra{y} = \sum_{y} A_{xy} \pbra{y}.
\end{align}
Here the \emph{assignment matrix} $A$ is a \emph{transition matrix} (stochastic matrix) whose entries $\pbra{x}\mathcal{A}\pket{y}$ represent the transition probability from the measurement result $y$ to $x$ due to the noise channel $\mathcal{A}$~\cite{chowDetectingHighlyEntangled2010}. The entry $A_{xy} = \pbra{x}\mathcal{A}\pket{y} = \pbraket{E_{x}}{y}$ can be obtained by estimating the probability of the $x$ outcome when we prepare the computational state $\pket{y}$ (assumed to be almost perfect) and perform the set of noisy measurement $\{\pbra{E_{x}}\}$. If $A$ is full rank, then we can invert \cref{eqn:noisy_meas_decomp} and obtain:
\begin{align}\label{eqn:ideal_meas_decomp}
    \pbra{y} = \sum_{x} A_{yx}^{-1}\pbra{E_{x}}
\end{align}
i.e. we can simulate the behaviour of the ideal measurement $\{\pbra{y}\}$ using a linear combination of the noisy measurements $\{\pbra{E_{x}}\}$, just as we have done in probabilistic error cancellation in \cref{sec:pec}. Hence, the associated sampling overhead will increase exponentially with the measurement faults rate, similar to \cref{eqn:pec_samp_cost_circ} for probabilistic error cancellation.

For an incoming state $\pket{\rho}$, the output distribution using the ideal measurements is given by the vector $\vec{p}_0 = \{\pbraket{y}{\rho}\}$ while the output distribution using the noisy measurements is $\vec{p}_{\mathrm{noi}} = \{\pbraket{E_{x}}{\rho}\}$. Applying \cref{eqn:noisy_meas_decomp,eqn:ideal_meas_decomp} on $\pket{\rho}$, we then have:
\begin{align}\label{eqn:inverse_prob}
    \vec{p}_{\mathrm{noi}} = A \vec{p}_{0}\quad \Rightarrow\quad \vec{p}_0 = A^{-1} \vec{p}_{\mathrm{noi}}.
\end{align}
For a given observable $O$ with the spectrum $\vec{O} = \{O_x\}$, i.e. $\pbra{O} = \sum_{x} O_x \pbra{x}$, its ideal expectation value is:
\begin{align}\label{eqn:inverse_matrix_expect}
    \pbraket{O}{\rho} = \vec{O}^{T} \vec{p}_0 = \vec{O}^{T} A^{-1} \vec{p}_{\mathrm{noi}}.
\end{align}
Hence, the ideal expectation value can be obtained once we know the assignment matrix $A$ and the noisy output distribution $\vec{p}_{\mathrm{noi}}$. 

In early experiments with only a few qubits, \textcite{kandalaHardwareefficientVariationalQuantum2017} have performed a full readout tomography of the noisy output distribution $\vec{p}_{\mathrm{noi}}$ in the computational basis. Then, all entries of the assignment matrix $A$ were estimated, which can be used to estimate the ideal expectation value using \cref{eqn:inverse_matrix_expect}. This approach is not efficiently scalable as the size of $A$ scales exponentially with the number of qubits $N$. 

The simplest way to tackle the problem is to assume that the measurement errors of different qubits are not correlated, which implies that the assignment matrix is just the tensor product of the assignment matrices of the individual qubits $A = \bigotimes_{n = 1}^N A_n$. However, realistic noise encountered in experiments may not be captured accurately by this simplified model and it can be observed that correlations between individual bit-flips are in fact present~\cite{heinsooRapidHighfidelityMultiplexed2018}. 

\textcite{bravyiMitigatingMeasurementErrors2021} try to construct the assignment matrix using continuous-time Markov processes with the generators (transition rate matrices) $\{G_i\}$ being single- and two-qubit operators:
\begin{equation}
    A = e^{G} \quad \text{with}  \quad G = \sum_{i=1}^{2 N^2} r_i G_i.
\end{equation}
The assignment matrix $A$ is now determined by $2N^2$ positive coefficients $\{r_i\}$ that can be learned by only considering a polynomial number of input bit strings. Once the coefficients are learned, we can easily construct the inverse matrix $A^{-1} = e^{-G}$ for error mitigation. 

As mentioned in \cref{sec:twirling}, we can perform Pauli twirling on the noise channel $\mathcal{A}$ by conjugating it with random Pauli operators, which will remove all the off-diagonal elements of $\mathcal{A}$ in the Pauli basis and produce a Pauli channel $\mathcal{D}$. This can be used to simplify measurement error mitigation as discussed by \textcite{vandenbergModelfreeReadouterrorMitigation2022, chenRobustShadowEstimation2021}. Since we are only interested in the action of $\mathcal{A}$ on the computational subspace, we only need to consider Pauli basis elements that are the tensor products of $Z$ denoted as $\{\pbra{Z^{x}}\}$ with $x$ being the bit string that marks the qubits that are acted by $Z$. Suppose ideally we would like to perform the Pauli measurement $\pbra{Z^{x}}$, however we can only perform the noisy measurement $\pbra{Z^{x}}\mathcal{A}$. Now using Pauli twirling, we can transform the noisy measurement into
\begin{align}\label{eqn:noisy_Z_decomp}
    \pbra{Z^{x}}\mathcal{D} = D_{x}  \pbra{Z^{x}}
\end{align}
using the fact that the twirled channel $\mathcal{D}$ is diagonal in the Pauli basis with the entries being $D_{x}  =  \pbra{Z^{x}}\mathcal{D} \pket{Z^{x}} = \pbra{Z^{x}}\mathcal{A} \pket{Z^{x}}$. Thus the noisy measurement $\pbra{Z^{x}}\mathcal{D}$ is simply the ideal measurement $\pbra{Z^{x}}$ rescaled by the factor $D_{x}$. For a given input state $\pket{\rho}$, we have:
\begin{align*}
    \underbrace{\pbraket{Z^{x}}{\rho}}_{\text{ideal}} & = D_{x}^{-1} \underbrace{\pbra{Z^{x}}\mathcal{D}\pket{\rho}}_{\text{twirled noisy}}.
\end{align*}
Hence, by transforming the observable into $Z^{x}$ and performing Pauli twirling, we only need to rescale the noisy expectation value by the factor $D_{x}^{-1}$ to obtain the ideal expectation value. Note that since conjugation with $Z$ will have trivial effects within the computational subspace, we only need to conjugate the noise channel with a random operator in $\{I, X\}^{\otimes N}$ (i.e. random bit-flips) to achieve the effect of Pauli twirling mentioned above. 

In practice, the ideal output distribution $\vec{p}_{0}$ is often sparse. With weak measurement noise, we would expect the corresponding noisy output distribution $\vec{p}_{\mathrm{noi}}$ to also be sparse and have non-zero probability at all positions that are non-zero in $\vec{p}_{0}$. Using this fact, \textcite{nationScalableMitigationMeasurement2021} proposed to focus on the action of measurement noise $\mathcal{A}$ within the subspace spanned by the basis of $\vec{p}_{\mathrm{noi}}$ with non-zero probability, which gives an assignment matrix with a much smaller dimension. The ideal output distribution $\vec{p}_{0}$ can then be obtained by inverting the assignment matrix within this subspace. The inversion can be sped up by considering a matrix-free preconditioned iteration algorithm.

Due to sampling noise, the estimation of ideal distribution $\vec{p}_{0}$ we obtained using matrix inversion in \cref{eqn:inverse_prob} may contain negative values are thus is not a valid probability distribution. However, it is still able to provide an unbiased expectation value estimate using \cref{eqn:inverse_matrix_expect}~\cite{bravyiMitigatingMeasurementErrors2021}. On the other hand, instead of using matrix inversion, one might try to solve a constrained optimisation problem with the cost function $\norm{A\vec{p}_0 - \vec{p}_{\mathrm{noi}}}_2^2$ such that $\vec{p}_0$ is a valid probability distribution. This can be solved using maximum likelihood~\cite{chenDetectorTomographyIBM2019,gellerRigorousMeasurementError2020,maciejewskiMitigationReadoutNoise2020} or iterative Bayesian unfolding~\cite{nachmanUnfoldingQuantumComputer2020}. 

There is a wide range of other techniques for combating measurement errors. In \textcite{hamiltonScalableQuantumProcessor2020}, a representation based on cumulant expansion is proposed to capture correlations between observables. \textcite{kwonHybridQuantumClassicalApproach2021} look into the use of Clifford twirling in the context of measurement error mitigation.
Measurement error protocols that are directly tailored to VQE type calculations are possible~\cite{barronMeasurementErrorMitigation2020}.
Other approaches propose the use of final pre-measurement entangling circuits to combat noise~\cite{hicksActiveReadouterrorMitigation2022}. \textcite{tannuMitigatingMeasurementErrors2019} propose to exploit a potential asymmetry in the noise strength of the assignment matrix $A$ by flipping bit-values into a configuration less likely to be affected by noise. The experimental observations have been used to train classical neural networks to infer predictions of the correct expectation values~\cite{palmieriExperimentalNeuralNetwork2020}. As measurement noise is a major obstacle in almost all experimental set-ups, implementation of measurement error mitigation is almost \emph{ubiquitous in all near-term experiments}.

\subsection{Symmetry constraints}\label{sec:sym}

A simple but effective scheme for suppressing errors is to identify errors that break the symmetries of the ideal quantum state, and remove them via post-selection.
This notion originates from quantum error correction~\cite{gottesmanStabilizerCodesQuantum1997, terhalQuantumErrorCorrection2015}, in which we explicitly define a set of measurements to detect and correct all local errors at the cost of additional qubit overhead.
Though explicitly correcting errors is required for scalability~\cite{shorFaulttolerantQuantumComputation1996}, \emph{quantum error detection} of artificially-added symmetries has been widely recognised as an important milestone towards this end goal~\cite{niggQuantumComputationsTopologically2014,kellyStatePreservationRepetitive2015,corcolesDemonstrationQuantumError2015,gottesmanQuantumFaultTolerance2016,linkeFaulttolerantQuantumError2017}. In practical applications, quantum circuits often possess inherent symmetries that can be used for error mitigation without the need to execute a quantum circuit on an error-detection code.
Measuring these \emph{inherent symmetries} and discarding circuit runs that produce the wrong results produces a (post-selected) state $\rho_{\mathrm{sym}}$. The broad class of schemes that directly or indirectly measure $\Tr[O\rho_{\mathrm{sym}}]$ are known collectively as \emph{symmetry verification}~\cite{mcardleErrorMitigatedDigitalQuantum2019,bonet-monroigLowcostErrorMitigation2018}. 
If the symmetry measurements are perfect, $\rho_{\mathrm{sym}}$ must have non-decreasing overlap on the ideal state $\rho_0$ compared to the noisy state $\rho$ as we have thrown away states with zero overlaps.
In practice, symmetry measurements may themselves introduce errors into the state, thus choosing the right symmetries and optimising their measurements are at the core of symmetry verification.

\textcite{mcardleErrorMitigatedDigitalQuantum2019,bonet-monroigLowcostErrorMitigation2018} proposed various easily-accessible symmetry operators $S$ for symmetry verification, drawing from those that naturally emerge in physical systems.
In this context, given a physical system, its symmetry operators $S$ are operators that commute with the system Hamiltonian $H$: $[H,S]=HS-SH=0$. When this is the case, $H$ and $S$ may be simultaneously diagonalized, i.e. energy eigenstates $\ket{\Psi_j}$ can be chosen in such a way that $S\ket{\Psi_j} = s\ket{\Psi_j}$,
where $s$ is an eigenvalue of $S$.
Thus, measuring $S$ on a prepared quantum state, and post-selecting on the correct symmetry eigenvalue $s$ for the target energy eigenstate, should project one closer to the said energy eigenstate.
Furthermore, time evolution by $e^{iHt}$ leaves the eigenspaces of $S$ invariant, which means that dynamic properties of the physical system may be studied entirely within these eigenspaces as well. Common examples of symmetries are the parity ($\prod_iZ_i$) and the $Z$ component of the total spin ($\sum_iZ_i$) of a spin system, or the particle number $\sum_in_i$ of a fermionic system. Note that the Jordan-Wigner transformation maps $\sum_in_i$ to $\sum_iZ_i$, modulo a constant shift. Given an $N$-qubit simulation, the eigensubspaces of these symmetries have dimension $2^{N-1}$, $\binom{N}{\nicefrac{(N-s)}{2}}$ and $\binom{N}{s}$, respectively, with $s$ being the eigenvalue of the given eigensubspace. With the right $s$, these subspaces are large enough for executing classically intractable quantum algorithms.
Note though that it is not necessary to enforce symmetries during an entire circuit in order to verify them at the end~\cite{dallaire-demersLowdepthCircuitAnsatz2019}. Even when using a circuit that does not conserve the symmetry of the target physical problem, symmetry verification can still be used for projecting back into the appropriate spin or number sector~\cite{yenExactApproximateSymmetry2019,khamoshiCorrelatingAGPQuantum2020,tsuchimochiSpinprojectionQuantumComputation2020}, which can be viewed as mitigating algorithmic errors (\cref{sec:algo_error}).

As mentioned, the process of \emph{direct} symmetry verification is carried out by measuring both the symmetry operators $S$ and the target observable $O$ in every circuit run and discarding runs that produce the wrong output for $S$ (i.e. fail the symmetry check). Since symmetries $S$ are typically global observables, measuring them alongside the target observable $O$ is non-trivial. The additional circuit components required for their measurements can introduce additional errors, reducing or even nullifying the effect of error mitigation. Various ways to measure multiple operators (e.g. the symmetry operator and the target observable) in the same circuit run are discussed in \cref{sec:meas_tech}. One way is to use a Hadamard test to measure a Pauli symmetry like the number parity~\cite{mcardleErrorMitigatedDigitalQuantum2019,bonet-monroigLowcostErrorMitigation2018}. Other practical schemes involve measuring qubit-wise commuting operators~\cite{izmaylovRevisingMeasurementProcess2019}, i.e. when both $S$ and $O$ can be obtained through post-processing the same set of single-qubit Pauli measurements across all qubits (see \cref{sec:meas_tech}). This implies that if we are using only single-qubit rotations and readout to perform the measurements, symmetries such as the $Z$-component of the total spin $\sum_iZ_i$ or parity $\prod_iZ_i$ cannot be measured simultaneously with any operator that is not diagonal in the computational basis.

The issue of qubit-wise commutativity presents a specific problem in chemistry where the fermion hopping operator $c_i^{\dag}c_j+c_j^{\dag}c_i$ (our target observable) commutes with the particle number $\sum_i n_i$ (the symmetry operator), but not qubit-wise. An identical problem presents in spin physics between the operators $X_iX_j+Y_iY_j$ and $\sum_iZ_i$. This may be solved by noting that the rotation
\begin{equation}\label{eqn:fsim_rotation}
    \exp\Big[i\frac{\pi}{8}\left(X_iY_j-Y_jX_i\right)\Big]
\end{equation}
maps $X_iX_j+Y_iY_j$ to $Z_i-Z_j$ while leaving $Z_i+Z_j$ invariant~\cite{hugginsEfficientNoiseResilient2021}.
In terms of fermionic systems, this is the operator $e^{i\frac{\pi}{2}\mathrm{FSWAP}_{i,j}}$, where $\mathrm{FSWAP}=c^{\dag}_ic_j+c^{\dag}_jc_i-n_i-n_j$~\cite{googlequantumaiandcollaboratorsHartreeFockSuperconductingQubit2020}.
In a spin system, this allows for the joint measurement of hopping terms and the $Z$-component of the total spin using only a constant depth circuit (assuming all-to-all coupling).
However, this measurement does not parallelise efficiently; one can only measure hopping terms simultaneously between disjoint pairs of qubits, requiring $O(N)$ distinct measurements to estimate all spin-spin hopping terms simultaneously with the $Z$-component of the total spin.
In contrast, without the requirement to simultaneously measure the total spin-$Z$ component, all spin-spin hopping terms may be estimated by just two distinct choices of single-qubit rotation and readout.
In a fermionic system, we need $O(N)$ distinct measurements in the first place for estimating all fermionic hopping terms due to the lack of mutually commuting terms~\cite{bonet-monroigNearlyOptimalMeasurement2020}, and so simultaneous measurements of the particle number do not present a significant overhead.

Instead of constructing circuits to simultaneously diagonalise and measure the symmetry operators $S$ and the target observable $O$, \textcite{bonet-monroigLowcostErrorMitigation2018} showed that it is possible to perform effective post-selection via \emph{post-processing} in symmetry verification,  which turns out to be closely related to subspace expansion~\cite{mccleanHybridQuantumclassicalHierarchy2017} (see \cref{sec:sub_expand}). Let us consider the simple example in which there is only one single Pauli symmetry operator $S$, and the ideal state lives within the $+1$-eigenspace of $S$ defined by the projector $\Pi=\frac{1}{2}(1+S)$. In this way, if the state prepared prior to the post-selection is $\rho$, the post-selected state is then
\begin{equation}\label{eqn:sym_state}
    \rho_{\mathrm{sym}} = \frac{\Pi\rho \Pi}{\Tr[\Pi\rho \Pi]}.
\end{equation}
The symmetry-verified expectation value for the target observable $O$ is given by
\begin{equation}
     \Tr[O\rho_{\mathrm{sym}}] = \frac{\Tr[O\Pi\rho \Pi]}{\Tr[\Pi\rho \Pi]}= \frac{\Tr[O_{\mathrm{sym}} \rho ]}{\Tr[\Pi \rho]},
\end{equation}
where $O_{\mathrm{sym}}=\Pi O\Pi$ is the symmetrised observable, i.e. the verified expectation value $\Tr[O\rho_{\mathrm{sym}}]$ is simply the quotient between the noisy expectation value of $O_{\mathrm{sym}}$ and $\Pi$. The symmetrised observable $O_{\mathrm{sym}}$ and the symmetry projector $\Pi$ can be further decomposed into the Pauli basis, e.g. if $O$ is Pauli, the Pauli decomposition of $O_{\mathrm{sym}}$ is simply $O_{\mathrm{sym}} = \Pi O\Pi = \frac{1}{4}\left(O + SO + OS + SOS\right)$. In this way, the expectation values of $O_{\mathrm{sym}}$ and $\Pi$ can be obtained by measuring the expectation values of these Pauli basis operators (or via Monte Carlo sampling in \cref{sec:monte_carlo_sampling}). In the method described above, we only need to measure one Pauli observable in a given circuit run, which can be carried out using only single-qubit Pauli measurements without needing additional circuit components. The simple examples above can be further generalised into the cases of multiple symmetries and non-Pauli symmetries with a change in the definition of the projector $\Pi$, reflecting a change in the symmetry subspace.
We can also decompose the observables into bases beyond the Pauli basis, as long as the components are easy to measure. Instead of combining the measurement results of the basis to reconstruct the projected observable $O_{\mathrm{sym}}$ and the projector $\Pi$, \textcite{caiQuantumErrorMitigation2021} showed that it is possible to achieve smaller biases by combining these measurement results with different weights at the cost of larger sampling overhead.

Note that as we try to describe different ways of performing symmetry verification above, we clearly separate post-selection away from post-processing (for a clear distinction between different methods) even though technically post-selection is a specific type of post-processing in the most general context of QEM.

For direct in-circuit symmetry verification, the fraction of circuit runs that are ``useful'' is simply the ``pass rate'' of the symmetry checks given by $\Tr[\Pi \rho ]$.
The corresponding sampling overhead is simply the \emph{inverse} of this ``pass rate'': $C_{\mathrm{em}} \sim \Tr[\Pi \rho ]^{-1}$. The fault-free post-selection discussed in \cref{sec:exp_samp_cost} can be viewed as the ideal symmetry verification that can detect all faults. It can achieve zero bias, but correspondingly its sampling overhead also sets the upper bound for all possible direct symmetry verification. As discussed above, we can greatly simplify the measurement circuit through post-processing, but this will come at a higher sampling overhead, which scales as $C_{\mathrm{em}} \sim \Tr[\Pi \rho ]^{-2}$~\cite{hugginsEfficientNoiseResilient2021,caiQuantumErrorMitigation2021}.

Though it is simplest to rely on the native symmetries of a system for the purposes of error mitigation, one can consider adding more symmetries artificially, or unitarily transforming a system to improve the error mitigation power of a set of symmetries.
This is important as symmetry-based methods cannot mitigate against errors that commute with all symmetries of the system.
It was shown in \textcite{bonet-monroigLowcostErrorMitigation2018} that one can unitarily transform chemistry Hamiltonians so that no single-qubit operator commutes with all symmetries, and that this can be preferable even to removing qubits from the system. They also described a basic scheme to add artificial symmetries to a system.
However, this scheme makes local system operators highly non-local, and thus is relatively unscalable.

A solution to the above was found in the Bravyi-Kitaev superfast (BKSF) transformation, where artificial symmetries are used to transform local fermionic Hamiltonians to local qubit Hamiltonians~\cite{bravyiFermionicQuantumComputation2002, setiaBravyiKitaevSuperfastSimulation2018}.
These symmetries are necessary for implementing a geometrically-local fermion-to-qubit transformation in more than one dimension, and at the same time they also provide a natural boon for symmetry-based QEM.
The list of fermion-to-qubit transformations has seen significant development and optimisation in recent years~\cite{setiaSuperfastEncodingsFermionic2019, steudtnerFermiontoqubitMappingsVarying2018, steudtnerQuantumCodesQuantum2019}, with attempts to optimise the number of local errors that can be mitigated~\cite{jiangMajoranaLoopStabilizer2019, derbyCompactFermionQubit2021a}.
It was pointed out in \textcite{jiangMajoranaLoopStabilizer2019} that this need not be contained to fermionic lattice models; as time evolution on a fermionic system may be mapped to a series of operations that are local on a lattice (without any asymptotic growth in the circuit depth), one can implement this using a mapping intended for a local fermionic lattice without paying the price of some system observables being extensively large. \textcite{jiangMajoranaLoopStabilizer2019} further found an encoding (the `Majorana loop stabiliser code', or MLSC) that can mitigate or even correct all single-qubit errors, making it in effect an error correcting code of distance $3$. Unfortunately, it appears that these methods cannot be extended to construct codes of arbitrarily large distances due to the Eastin-Knill theorem~\cite{eastinRestrictionsTransversalEncoded2009}.

Encoding the qubits into QEC codes is also a way to add artificial symmetry. Performing direct in-circuit verification in this case is simply quantum error detection. However, if the symmetries (stabilisers) of the code are hard to be directly measured due to high weight or connectivity constraints, then as discussed above, one may instead perform post-processing verification for QEC codes~\cite{mccleanDecodingQuantumErrors2020}. This was later extended to include mid-circuit stabiliser checks~\cite{tsubouchiVirtualQuantumError2023} and bosonic codes~\cite{endoQuantumErrorMitigation2022a}.

A large number of experiments have demonstrated stabiliser measurements in error correction codes throughout the 2010's;~\cite{niggQuantumComputationsTopologically2014,kellyStatePreservationRepetitive2015,corcolesDemonstrationQuantumError2015,linkeFaulttolerantQuantumError2017,vuillotErrorDetectionHelpful2018}.
However, to the best of our knowledge the first experimental demonstration of symmetry verification using \emph{natural} symmetries ($\prod_iZ_i$) in a quantum algorithm was by \textcite{sagastizabalExperimentalErrorMitigation2019} through post-processing, as low-cost techniques to measure natural symmetries were not known prior to this. Later, \textcite{googlequantumaiandcollaboratorsHartreeFockSuperconductingQubit2020} made direct simultaneous measurement of the number operator and the fermionic 1-RDM using the FSWAP rotation (\cref{eqn:fsim_rotation}) and combined this with McWeeny purification (\cref{sec:pur}). \textcite{stanisicObservingGroundstateProperties2022} has successfully demonstrated the verification of multiple symmetries (number, particle-hole symmetry, and total spin) and also combine it with learning-based methods (\cref{sec:learning-based}). Due to the simplicity and effectiveness of symmetry verification, it has been employed in a wide range of other experiments [see e.g. \textcite{googlequantumaiandcollaboratorsObservationSeparatedDynamics2020,neillAccuratelyComputingElectronic2021,stanisicObservingGroundstateProperties2022,dborinSimulatingGroundstateDynamical2022,googlequantumaiandcollaboratorsFormationRobustBound2022}].

\subsection{Purity constraints}\label{sec:pur}

Many quantum algorithms target the preparation of an ideal state $\rho_0$ that is pure: $\rho_0 =
\ketbra{\psi_0}$.
Many common noise channels are stochastic, which will turn our ideal pure state $\rho_0$ into some noisy mixed state $\rho$. At a high level, error mitigation techniques based on purity constraints attempt to reduce the bias in the expectation value by trying to approximate the pure state closest to $\rho$. If we look at the spectral decomposition of $\rho$,
\begin{equation}
  \label{eqn:spectral_decomp}
  \rho = \sum_{i=1}^{2^N} p_i \ketbra{\phi_i},
\end{equation}
where the eigenvalues are ordered $p_i \geq p_j$ for $i < j$ and we assume $p_1 > p_2$ for simplicity, then the closest pure state to $\rho$ in the trace distance is simply the dominant eigenvector $\ketbra{\phi_1}$. In principle, one could use the quantum principle component analysis to sample from the eigenbasis of $\rho$~\cite{lloydQuantumPrincipalComponent2014}. 
Combined with post-selection this would allow for the efficient preparation of the dominant eigenvector. However, the additional circuit required is too deep for mitigating errors in near-term devices and thus simpler strategies are needed. 

One such strategy, referred to as \emph{virtual distillation} (VD)~\cite{hugginsVirtualDistillationQuantum2021} or \emph{error suppression by derangement} (ESD)~\cite{koczorExponentialErrorSuppression2021}, uses collective measurements of $M$ copies of $\rho$ in order to access expectation values with respect to the $M^{\mathrm{th}}$ degree purified state 
\begin{equation}
    \label{eqn:powers_of_rho}
    \rho^{(M)}_{\mathrm{pur}} = \frac{\rho^M}{\Tr[\rho^M]} = \frac{1}{\sum_{i=1}^{2^N} p_i^M} \sum_{i=1}^{2^N} p_i^M \ketbra{\phi_i}. 
\end{equation}
We see that under the assumption $p_1$ is strictly greater than $p_2$, we have $\lim_{M\rightarrow\infty}\rho^{(M)}_{\mathrm{pur}}=\ketbra{\phi_1}$.
The rate at which this is achieved is exponential in the number of copies used for purification $M$.
The remaining bias in the VD/ESD estimator at $M\rightarrow\infty$ comes from the deviation between $\ketbra{\phi_1}$ and the target state $\rho_0$, which is sometimes known as the \emph{coherent mismatch} or \emph{noise floor}~\cite{koczorExponentialErrorSuppression2021,hugginsVirtualDistillationQuantum2021}.
As the largest sources of noise in state-of-the-art quantum devices are typically incoherent, this coherent mismatch can be expected to be significantly smaller than the error in the unmitigated state.
Indeed, numerical and analytic studies have confirmed that the error suppression from VD/ESD can be of multiple orders of magnitude for large systems, even using as little as $M=2$ copies of the state~\cite{koczorDominantEigenvectorNoisy2021,koczorExponentialErrorSuppression2021,hugginsVirtualDistillationQuantum2021}.

\begin{figure}[ht!]
    \centering
    \includegraphics[width = 0.48\textwidth]{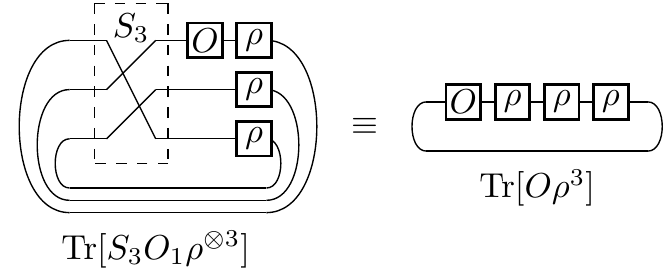}
    \caption{A diagrammatic proof of \cref{eqn:shift} for 3 copies ($M=3$) with the observable acting on the first copy ($m=1$). The proof here uses tensor network notations~\cite{bridgemanHandwavingInterpretiveDance2017} and it can be easily extended to more copies and/or with $O$ acting on the $m^{\text{th}}$ copy rather than the first. }
    \label{fig:VD_proof}
\end{figure}

It was shown by \textcite{koczorExponentialErrorSuppression2021,hugginsVirtualDistillationQuantum2021} that one can estimate expectation values of the purified states in \cref{eqn:powers_of_rho} without ever having to prepare them on a quantum device.
The expectation value of this purified state with respect to the observable of interest $O$ is $\Tr[O\smash{\rho^{(M)}_{\mathrm{pur}}}] = \Tr[O \rho^M]/\Tr[\rho^M]$.
In VD/ESD, this is obtained by estimating $\Tr[O \rho^M]$ and $\Tr[\rho^M]$ in separate measurements.
Let $S_M$ be the cyclic permutation operator between $M$ copies of $\rho$, the quantity $\Tr[O \rho^M]$ can be estimated using
\begin{equation}\label{eqn:shift}
  \Tr[O \rho^M] = \Tr[S_M O_m \rho^{\otimes M}] = \Tr[S_M \overline{O} \rho^{\otimes M}]
\end{equation}
where $O_m$ is the operator $O$ acting on the $m^\mathrm{th}$ copy and $\overline{O}=\frac{1}{M}\sum_m O_m$ is the observable symmetrised under copy permutation. Thus $\Tr[O \rho^M]$ can be obtained by measuring $S_M O_m$ or $S_M \overline{O}$ on $M$ noisy copies of $\rho$. A diagrammatic proof of \cref{eqn:shift} is shown in \cref{fig:VD_proof}. One can extend this to estimate $\Tr[\rho^M]$ by putting $I$ in place of $O$. 

\begin{figure}[ht!]
    \centering
    \includegraphics[width = 0.37\textwidth]{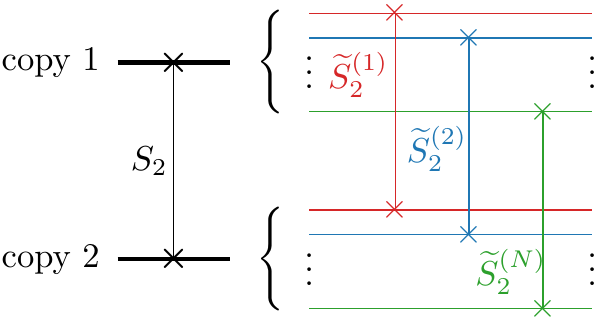}
    \caption{Decomposition of the copy-swap operator $S_2$ into transversal qubit-swap operators $\widetilde{S}_2$. Similar decomposition also applies to other cyclic copy-permutation operators $S_M$ with $M > 2$.}
    \label{fig:copy_swap}
\end{figure}

Measuring a global operator like $S_M$ can be challenging, but it can be decomposed into transversal operations among different copies $S_M = \bigotimes_{n = 1}^N \widetilde{S}_M^{(n)}$ where $\widetilde{S}_M^{(n)}$ cyclically permute the $n^\mathrm{th}$ qubits of different copies as shown in \cref{fig:copy_swap}. If the observable $O$ acts on a single qubit, then the symmetrised observable $\overline{O}$ will commute with all $\widetilde{S}_M^{(n)}$, and thus $S_M \overline{O}$ can be obtained by measuring low-weight operators $\widetilde{S}_M^{(n)}$ and $O_m$ for all $n$ and $m$ and then post-process. This requires only transversal operations among the identically-labelled qubits of each copy of $\rho$, avoiding global measurement.
Explicit circuits for $O = Z$ and $M=2,3$ without ancilla qubits are given by \textcite{hugginsVirtualDistillationQuantum2021}; more general measurement may be achieved by a Hadamard test using ancilla qubits~\cite{hugginsVirtualDistillationQuantum2021,koczorExponentialErrorSuppression2021}. If the observable $O$ is not single-qubit, but it is a tensor product of single-qubit operators: $O= \bigotimes_{n=1}^N G^{(n)}$ (e.g. $O$ is Pauli), then the observable $S_M O_m$ in \cref{eqn:shift} can be decomposed into a tensor product of low-weight operators $\bigotimes_{n=1}^N \widetilde{S}_M^{(n)}G_m^{(n)}$ which can be measured in a transversal manner~\cite{caiLoopedPipelinesEnabling2023}. We can use Hadamard tests to measure each $\widetilde{S}_M^{(n)}G_m^{(n)}$, which require $N$ ancilla qubits in total. To efficiently carry out any of the low-weight measurement schemes above, we need transversal operations among different copies, which can be challenging to implement in practice and may involve long-range interactions. A hardware architecture with native transversal operations among different copies has been proposed~\cite{caiLoopedPipelinesEnabling2023}, in which an example implementation of VD/ESD is shown with almost no space-time overhead. 

In the more general case, one can imagine either unitarily transforming $O$ to a single-qubit observable (if possible) or linearly decomposing it into a sum of simpler terms (which is always possible). One could additionally imagine decomposing $S_MO_m$ or $S_M\overline{O}$ into a linear combination of Pauli operators and measuring the whole set of operators using shadow tomography~\cite{huangPredictingManyProperties2020,seifShadowDistillationQuantum2023,huLogicalShadowTomography2022}.
This has the advantage that one may reuse a single copy of $\rho$ rather than using multiple physical copies.
However, the sampling cost in this case scale exponentially with the number of qubits and thus may be difficult to scale beyond small system sizes~\cite{seifShadowDistillationQuantum2023}.

Instead of estimating $\Tr[O\rho^2]$ using two copies of $\rho$ separated in space, it is possible to do the estimation using two copies separated in time.
This technique was given multiple names since its initial incarnations~\cite{obrienErrorMitigationVerified2021,huoDualstatePurificationPractical2022,caiResourceefficientPurificationbasedQuantum2021}, but here we refer to it as \emph{echo verification} (EV). To perform EV, ideally we would want to measure the ideal state projector $\rho_0 = \ketbra{\psi_0}$ at the end of the circuit for post-selection, which will return $1$ if the output state is the ideal state and $0$ otherwise, just like a symmetry check. The measurement of the projector $\rho_0$ is carried out by applying the inverse of the primary circuit~\cite{obrienErrorMitigationVerified2021} and the noise in the inverse circuit means that we are effectively measuring the noisy operator $\overline{\rho}$ instead of $\rho_0$. If we measure the observable of interest $O$ along with the noisy projector $\overline{\rho}$ in the same circuit run and post-select according to the outcome of $\overline{\rho}$, we are effectively performing measurement of $O$ on the state $(\overline{\rho}\rho+\rho\overline{\rho})/(2\Tr[\overline{\rho}\rho])$~\cite{huoDualstatePurificationPractical2022}. Compared to \cref{eqn:powers_of_rho}, we see that this is simply the $2^{\text{nd}}$ degree purified state with $\overline{\rho}$ in place of $\rho$. The measurement of $O$ alongside $\overline{\rho}$ is usually carried out using a Hadamard test~\cite{obrienErrorMitigationVerified2021}. Alternatively, it is possible to achieve a similar degree of error suppression without the ancilla by preparing a superposition of $\rho_0$ with a known eigenstate of $O$ and applying the gate $O$ on the quantum state (assuming $O$ is unitary)~\cite{luAlgorithmsQuantumSimulation2021,obrienErrorMitigationVerified2021}, though this is not formally equivalent to $2^{\text{nd}}$ degree purification. The error mitigation power of EV has been demonstrated in numerical simulation~\cite{obrienErrorMitigationVerified2021} and a four-qubit experiment~\cite{huoDualstatePurificationPractical2022}. The results differ depending on the exact circuit implementation of the methods, even in simplified noise models~\cite{obrienErrorMitigationVerified2021}, indicating that further optimising the circuit implementation can be an interesting direction to investigate. \textcite{guNoiseResilientPhaseEstimation2023} recently proved that applying EV to control-free phase estimation~\cite{luAlgorithmsQuantumSimulation2021,russoEvaluatingEnergyDifferences2021,obrienErrorMitigationVerified2021} corrects any noise source with Hermitian Kraus operators to first-order.

Due to their similarity, EV and VD/ESD are directly comparable in terms of performance and resource requirements. If one wishes to suppress the incoherent contributions of $\rho$ beyond $2^{\text{nd}}$ degree purification, then one can only use VD/ESD instead of EV. However, EV has a smaller qubit footprint and requires less circuit overhead for measurement (especially as it removes the need to perform operations across multiple copies of the same state). In fact, EV can be combined with VD/ESD to achieve a high degree of purification with a lower qubit and circuit overhead~\cite{caiResourceefficientPurificationbasedQuantum2021}. There is also a difference between these two approaches in terms of their sampling overhead, similar to the difference between the sampling overhead of direct and post-processing symmetry verification (\cref{sec:sym}).
For EV, the error-mitigation is done through direct post-selection, and thus the sampling overhead is simply the inverse of the success probability $C_{\mathrm{em}} \sim \Tr[\overline{\rho}\rho]^{-1} \sim \Tr[\rho^2]^{-1}$.
For the two-copy version of VD/ESD, the error-mitigation is done through effective post-selection (post-processing), and thus the sampling overhead increases more steeply, scaling as $C_{\mathrm{em}} \sim \Tr[\rho^2]^{-2}$.
More generally, the sampling cost of VD/ESD scales exponentially in the number of copies $M$ (as $\Tr[\rho^M]$ is exponentially small in $M$ unless $\rho$ is pure). If the faults in the circuit follow a Poisson distribution (\cref{sec:circuit_faults}), we then have  $\Tr[\rho^2]\lesssim P_0^2 =  e^{-2\lambda}$ where $P_0$ is the fault-free probability of the circuit and $\lambda$ is the circuit fault rate. This suggests that both EV and VD/ESD can incur a sampling cost growing exponentially with the circuit fault rate, just as we have discussed for general bias-free QEM in \cref{sec:exp_samp_cost}.
Furthermore, both methods (in their standard form) lack the parallelisability of unmitigated expectation value estimation. 
As mentioned above VD/ESD typically estimates expectation values of only $N$ operators (rather than their products).
EV is restricted further; it was shown in~\cite{pollaOptimizingInformationExtracted2023} that EV cannot be efficiently parallelised for even commuting observables, reducing the information extracted by the method to a single bit per state preparation.
Though this can be significantly optimised over a simple Pauli decomposition of a complex $O$~\cite{pollaOptimizingInformationExtracted2023}, it presents roughly an $O(N)$ overhead in sampling compared to VD/ESD.

The original formulation of EV in terms of projection on the initial state gives a clear connection between purification- and (symmetry) projection-based techniques (\cref{sec:sym}).
Though they are clearly not equivalent, this raises the question of whether a similar connection can be made to VD/ESD.
Some connection was originally made in \textcite{hugginsVirtualDistillationQuantum2021}, where it was pointed out that in the eigenbasis $\{|\phi_j\rangle\}$ of $\rho$, the swap operator measurement has zero expectation value on terms in $\rho\otimes\rho$ except for those of the form $|\phi_j\rangle\langle\phi_j|\otimes|\phi_j\rangle\langle\phi_j|$. \textcite{caiQuantumErrorMitigation2021} show that VD/ESD can naturally arise from the verification of copy-permutation symmetry by replacing the symmetry projector with a general linear combination of permutation operators, which is also connected to subspace expansion using symmetry operators (will be discussed in \cref{sec:sub_expand}). \textcite{yoshiokaGeneralizedQuantumSubspace2022} further exploited the connection between VD/ESD and subspace expansion and looked at the effect of performing error mitigation using a linear combination of states with different degrees of purification. Both EV and VD/ESD have been successfully implemented experimentally in \textcite{obrienPurificationbasedQuantumError2023}, where they have performed variational ground state energy estimation up to 20 qubits (2 copies of 10-qubit states for VD/ESD) and achieved 1-2 orders of magnitude error reduction by applying these QEM methods.

\subsection{Subspace expansions}\label{sec:sub_expand}

In some quantum tasks, one has knowledge of not only the ideal circuit and potential noise sources, but also the structure of the task at hand.  One such class of methods that can take advantage of the knowledge of the problem are \emph{quantum subspace expansion} techniques~\cite{mccleanHybridQuantumclassicalHierarchy2017,mccleanDecodingQuantumErrors2020,takeshitaIncreasingRepresentationAccuracy2020}. Many tasks in quantum computing, such as optimisation or state preparation, can be phrased as the desire to minimise the objective function $\bra{\psi} H \ket{\psi}$ with respect to the state $\ket{\psi}$ for some known Hermitian operator $H$.  Very often we are unable to directly prepare the optimal state, either due to a coherent error in our implementation of the ideal circuit, or a simple lack of knowledge of the ideal circuit. However, there is usually a set of $M$ different states $\{\ket{\phi_i}\}$ (linearly independent but not necessarily orthogonal to each other) that we can easily prepare, which can be used to construct our target state through linear combination $\ket{\psi_{\vec{w}}} = \sum_{i=0}^{M-1} w_i \ket{\phi_i}$. In this way, the problem of finding the optimal state becomes finding the optimal set of coefficients $\vec{w}^*$:
\begin{align}\label{eqn:pure_state_expansion}
    \vec{w}^* = \argmin_{\vec{w}} \bra{\psi_{\vec{w}}} H \ket{\psi_{\vec{w}}} \quad\quad \text{s.t. \ } \braket{\psi_{\vec{w}}}  = 1.
\end{align}
This has the well-known exact solution in the form of a generalised linear eigenvalue problem
\begin{align}\label{eqn:gen_eigen}
    \overline{H} W = \overline{S} W E,
\end{align}
where 
\begin{align}
    \overline{H}_{ij} = \bra{\phi_i} H \ket{\phi_j}, \quad\quad \overline{S}_{ij} = \braket{\phi_i}{\phi_j},
\end{align}
i.e. $\overline{H}$ is the $M \times M$ matrix representations of $H$ in the chosen basis set $\{\ket{\phi_i}\}$ and $\overline{S}$ is the overlap matrix for the basis set. In \cref{eqn:gen_eigen}, $W$ and $E$ are the matrices of eigenvectors and eigenvalues, respectively, of the solved problem.  The eigenvector in $W$ with the lowest eigenvalue is precisely the optimal combination of coefficients $\vec{w}^*$ for the state $\ket{\psi_{\vec{w}^*}}$. 

If we want to obtain the improved expectation value of some observable $O$ with respect to our new found state $\ket{\psi_{\vec{w}^*}}$, it is simply given as:
\begin{align*}
    \bra{\psi_{\vec{w}^*}}O\ket{\psi_{\vec{w}^*}} = \sum_{i,j=0}^{M-1} w_i^* w_j^* \bra{\phi_i}O\ket{\phi_j}
\end{align*}
i.e. we can construct it using the optimal weight $\vec{w}^*$ and the measurement results of $\bra{\phi_i}O\ket{\phi_j}$, without needing to explicitly prepare $\ket{\psi_{\vec{w}^*}}$. For the special case of $O = H$, the improved expectation value is simply given by the smallest eigenvalue in $E$ when we solve \cref{eqn:gen_eigen}. If desired however, one could prepare the state $\ket{\psi_{\vec{w}^*}}$ via LCU methods~\cite{childsHamiltonianSimulationUsing2012} in order to use it as the input state for a subsequent quantum routine.

If one takes the limit of choosing $\ket{\phi_i}$ to be a complete basis for the whole Hilbert space, then solving the optimisation problem will return the ideal state, however choosing an exponentially large space to perform classical optimisation obviously defeats the purpose of using a quantum computer to begin with. Hence, how to choose the right set of basis state $\{\ket{\phi_i}\}$ is the key to the success of quantum subspace expansion. The first basis state $\ket{\phi_0}$ we select is usually the best state that we can prepare before performing quantum subspace expansion. In this way, in the worst case we will simply obtain back $\ket{\phi_0}$ through subspace expansion.  

For the other basis states, the original work of \textcite{mccleanHybridQuantumclassicalHierarchy2017} suggests that we can draw inspiration from the configuration interaction (CI) expansions~\cite{helgakerMolecularElectronicStructureTheory2000} in quantum chemistry, which is commonly used for improving energy and properties of mean-field states as well as determining excited states for response properties. 
There each of the other basis states is generated by applying an \emph{expansion basis} operator $G_i$ on the original state such that $G_i\ket{\phi_0} = \ket{\phi_i}$. Knowledge of a good set of expansion basis operators $\{G_i\}$ can come from symmetry considerations, excitation operators, or simply knowing that correcting (as opposed to replacing) the state $\ket{\phi_0}$ requires that the additional states are connected directly or indirectly through $H$. In this way, the expanded state is now $\ket{\psi_{\vec{w}}} = \Gamma_{\vec{w}} \ket{\phi_0}$ where $\Gamma_{\vec{w}} = \sum_{i=0}^{M-1} w_i G_i$ is the \emph{expansion operator}, which is a weighted sum of expansion basis operators. Once the expansion basis operators are determined, the matrix elements required for solving the optimisation equation in \cref{eqn:gen_eigen} can be measured on a quantum computer through
\begin{equation}\label{eqn:expand_matrix}
    \begin{aligned}
        \overline{H}_{ij} &= \bra{\phi_0} G_i^\dagger H G_j \ket{\phi_0} = \Tr[G_i^\dagger H G_j\rho ] \\
        \overline{S}_{ij} &= \bra{\phi_0} G_i^\dagger G_j \ket{\phi_0} = \Tr[G_i^\dagger G_j\rho ]
    \end{aligned}
\end{equation}
without needing detailed knowledge of the original state $\rho = \ketbra{\phi_0}$.  

So far both the starting state $\ket{\phi_0}$ and the expanded state $\ket{\psi_{\vec{w}^*}}$ are pure states and thus any errors that we have removed are coherent errors. This is in stark contrast to our focus on incoherent errors in the last section (\cref{sec:pur}). The right-hand side of \cref{eqn:expand_matrix} is suggestive of the fact that we can apply expansion around some mixed state $\rho$ for removing incoherent errors. This is indeed first conjectured in the original works of \textcite{mccleanHybridQuantumclassicalHierarchy2017} and later confirmed by several experimental implementations of the method~\cite{collessComputationMolecularSpectra2018,sagastizabalExperimentalErrorMitigation2019,urbanekChemistryQuantumComputers2020}. These observations are put on more solid theoretical footing by \textcite{bonet-monroigLowcostErrorMitigation2018, mccleanDecodingQuantumErrors2020}. The effective state after performing subspace expansion on a noisy state $\rho$ is shown to be~\cite{mccleanDecodingQuantumErrors2020}:
\begin{align}
    \rho_{\mathrm{sub}} = \frac{\Gamma_{\vec{w}}\rho \Gamma_{\vec{w}}^\dagger}{\Tr[\Gamma_{\vec{w}}\rho \Gamma_{\vec{w}}^\dagger]}.
\end{align}
We see that this is very similar to the symmetry-verified state in \cref{eqn:sym_state} with the symmetry projector $\Pi$ replaced by the expansion operator $\Gamma_{\vec{w}}$. This implies that using the symmetry operators as our expansion basis operators, we can recover the symmetry subspace by performing subspace expansion. \textcite{caiQuantumErrorMitigation2021} try to further generalise this by searching for expanded ``states'' of the form $\Gamma_{\vec{w}}\rho/\Tr[\Gamma_{\vec{w}}\rho]$ instead, which allows us to also incorporate purification-based QEM in \cref{sec:pur} under this formalism. 

A number of recent works have also looked into other possible sets of expansion basis operators $\{G_k\}$.  For example, when the operators $\{G_k\}$ are chosen to be powers of the Hamiltonian $\{H^k\}$, we see that these methods coincide with quantum Krylov subspace methods like qLanczos~\cite{mottaDeterminingEigenstatesThermal2020} or other methods based on filtering the eigenspectrum via functions of the Hamiltonian~\cite{suchslandAlgorithmicErrorMitigation2021}.  Another recent work that was mentioned before~\cite{yoshiokaGeneralizedQuantumSubspace2022} has included the operators that are powers of the density matrix, making close ties to the purification-based QEM methods.  By doing so, optimal combinations of states can now exploit problem-specific knowledge related to purity in addition to general knowledge.

\subsection{$N$-representability}\label{sec:nrep}

Often in the context of quantum simulation (and sometimes more broadly), the goal is ultimately to measure a set of observables corresponding to a marginal of the total density matrix known as a reduced density matrix (RDM). Given a general quantum state $\rho$ on $N$ qubits, the set of $p$-qubit RDMs are obtained by integrating out $q$-qubits (such that $N - q = p$) of the joint distribution,
\begin{equation}
^{p}\rho_{m_{1}, ..., m_{p}} = \Tr_{n_{1}, n_{2}, ..., n_{q}}\left[\rho\right]
\end{equation}
resulting in $\binom{n}{p}$ different RDMs each of dimension $2^{p} \times 2^{p}$. The coefficients $n_{1}, ..., n_{q}$ on the trace operator indicate which qubits are integrated out of $\rho$ and coefficients $m_{1}, ..., m_{p}$ label the subsystem marginal. The result of this marginalization is a distribution over $p$-qubits.

The connection to error-mitigation is that these RDMs are known to have special geometric structures; not all marginals that one can write down are consistent with having come from a valid (or in the parlance of this field, ``representable'') wave function. In principle there exists a set of conditions that constrain the space of representable RDMs. Articulating and evaluating these equality and inequality constraints is known as the ``$N$-representability problem'' \cite{mazziottiPureRepresentabilityConditions2016}, and the problem is formally QMA-Complete \cite{liuQuantumComputationalComplexity2007}. However, for most RDMs of interest, one can write down and evaluate at least some of the $N$-representability conditions. That knowledge can often be used to mitigate errors, in a spirit similar to the application of symmetry constraints, but generally using different methods.

The focus on measuring RDMs is especially common when simulating many-body systems of identical particles. For example, because real fermions interact pairwise, most properties of interest can be obtained using just the 1-particle and 2-particle reduced density matrices (and because the particles are identical, there is only a single 1-RDM and a single 2-RDM). Using the second quantized fermionic creation and annihilation operators acting on site $p$, $a^\dagger_p$ and $a_p$, the fermionic 1-RDM and 2-RDM can be expressed as
\begin{align}
^{1}D_{j}^{i} &= \Tr[a_{i}^{\dagger}a_{j}  \;^{n}D ] = \bra{\psi} a_{i}^{\dagger}a_{j} \ket{\psi} \\
^{2}D_{rs}^{pq} &= \Tr[a_{p}^{\dagger}a_{q}^{\dagger}a_{s}a_{r} \;^{n}D ] = \bra{\psi}  a_{p}^{\dagger}a_{q}^{\dagger}a_{s}a_{r} \ket{\psi}
\end{align}
where ${}^k D$ is the $k$-particle RDM and the equalities on the right correspond to the case of pure states. For a system of $N$ sites this 2-RDM is only of dimensions $N^2 \times N^2$ (in contrast to $2^N \times 2^N$ for the full density matrix), and yet completely determines the energy of a fermionic system with pairwise interactions.

Some of the simpler constraints we can express on the 1- and 2-RDMs are as follows:
\begin{enumerate}
\item 
Hermiticity of the density matrices
\begin{align}
^{1}D_{i}^{j} &= \left(\;^{1}D_{j}^{i}\right)^{*}\\
^{2}D_{rs}^{pq} &= \left(\;^{2}D_{pq}^{rs}\right)^{*}
\end{align}

\item 
Antisymmetry of the $2$-particle marginal
\begin{align}
^{2}D_{rs}^{pq} = - \;^{2}D_{sr}^{pq} = -\; ^{2}D_{rs}^{qp} = \;^{2}D_{sr}^{qp}
\end{align}

\item 
The $(p-1)$-marginal is related to the $p$-marginal by contraction--e.g. the $2$-marginal can be contracted to the $1$-marginal
\begin{align}
^{1}D_{j}^{i} = \frac{1}{n-1}\sum_{k}\;^{2}D_{jk}^{ik}
\end{align}

\item 
The trace of each marginal is fixed by the number of particles in the system
\begin{align}
\Tr[\;^{1}D] & = n\\
\Tr[\;^{2}D] & = n ( n -1 )
\end{align}

\item 
The marginals are proportional to density matrices and are thus positive semidefinite
\begin{align}
\{{}^{1}D, {}^{2}D\} \succeq 0 \, .
\end{align}
\end{enumerate}
This is not an exhaustive list of all $N$-representability constraints.

The work of \textcite{rubinApplicationFermionicMarginal2018} first pointed out that knowledge of these constraints could be used for mitigating errors when measuring RDMs. The essential idea is that in the course of a NISQ simulation, one might measure an RDM that violates $N$-representability conditions as a consequence of errors corrupting the estimation of the tomography elements composing the RDM. However, one can use even a partial list of RDM conditions in order to project the noisy and unrepresentable RDM estimate back to the nearest RDM consistent with a list of RDM constraints. Because the RDM constraints all take the form of equality and inequality constraints, this can be performed using semidefinite programming. Such an approach was demonstrated numerically in \textcite{rubinApplicationFermionicMarginal2018} and experimentally in \textcite{smartEfficientTwoelectronAnsatz2020}.

Of special interest for error-mitigation, there is a subdiscipline of the $N$-representability field known as ``pure state $N$-representability'' which is concerned with describing the geometry of pure density matrices (i.e. $N$-representability that must hold for pure states). In principle, using pure state representability would potentially allow one to measure the 2-RDM of a partially decohered state and then project that estimate back to the nearest RDM consistent with having come from a pure state. While this idea was first discussed in \textcite{rubinApplicationFermionicMarginal2018}, it has been difficult to realise in practice since pure $N$-representability conditions are extremely difficult to compute. The state-of-the-art in the field is that specialised computer algebra systems are needed to generate the pure state conditions \cite{klyachkoQuantumMarginalProblem2006,deprinceVariationalOptimizationTwoelectron2016}. Nevertheless, some work has succeeded in using some of these conditions in quantum simulations
\cite{smartQuantumclassicalHybridAlgorithm2019}.

In the case when the 1-RDM of a fermionic system is expected to be idempotent ($D^2=D$), a special type of purification known as McWeeny purification~\cite{mcweenyRecentAdvancesDensity1960} is possible.
This purification scheme is achieved by iterating on a non-idempotent 1-RDM estimate as
\begin{equation}
    {}^1 D_{i+1} = 3 \left({}^1 D_i\right)^2 - 2 \left({}^1 D_i\right)^3 
\end{equation}
until idempotency is restored. The only fermionic states with idempotent 1-RDMs are Slater determinants, which limits the applicability of this scheme in strongly-correlated systems.
Despite the lack of theoretical justification, \textcite{mccaskeyQuantumChemistryBenchmark2019} demonstrated moderate mitigation success from McWeeny purification in a correlated four-qubit chemistry simulation.
The most notable success however, came from the application in \textcite{googlequantumaiandcollaboratorsHartreeFockSuperconductingQubit2020} to a Hartree-Fock state, for which the idempotency assumption is justified.
In this work McWeeny purification demonstrated between one and two orders of magnitude of error suppression, on top of the other mitigation methods used.
The hope remains that similar results can be demonstrated in other experiments by enforcing the pure-state representability constraints discussed in the previous paragraph, or by applying purification-based QEM methods discussed in \cref{sec:pur}.

\subsection{Learning-based}\label{sec:learning-based}

\begin{figure*}[ht!]
    \centering
    \includegraphics[width = 0.9\textwidth]{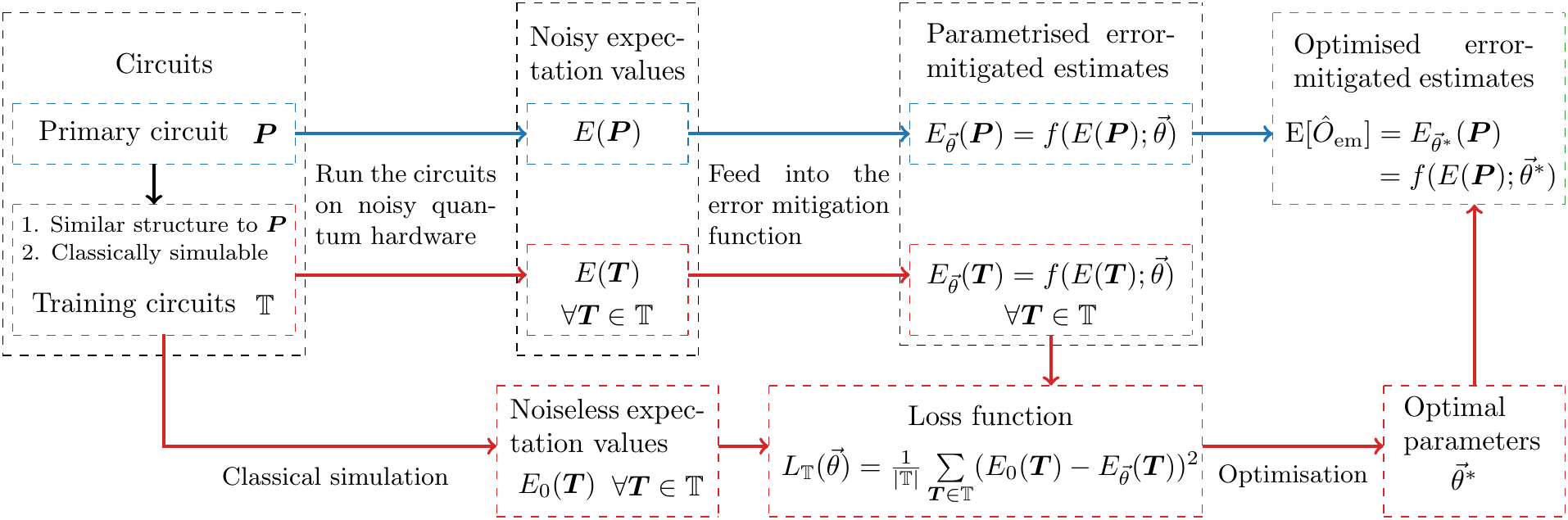}
    \caption{A diagram showing the process of learning-based quantum error mitigation. For the cases in which we need to construct variants of the input circuit for the error mitigation function like in \cref{eqn:em_rsp}, the whole process is similar, but instead of only running the input circuits on the quantum hardware, we need to run all the different variants of the input circuits.}
    \label{fig:learning_based}
\end{figure*}

Given the primary circuit $\boldsymbol{P}$, its noisy expectation value is denoted as $E(\boldsymbol{P})$ and we are trying to estimate its noiseless expectation value $E_{0}(\boldsymbol{P})$. To achieve this, we obtain the error-mitigated expectation value $E_{\vec{\theta}}(\boldsymbol{P})$ as a function of the primary circuit and a set of parameters $\vec{\theta}$. In previous sections, we see that the function parameters $\vec{\theta}$ in different QEM methods are obtained through different noise calibration processes.
However, we can also obtain $\vec{\theta}$ through \emph{learning-based methods}~\cite{strikisLearningBasedQuantumError2021,czarnikErrorMitigationClifford2021} using training circuits. We will construct a \emph{training circuit} $\boldsymbol{T}$ that is
\begin{enumerate}
    \item similar to $\boldsymbol{P}$, usually in terms of circuit structures, so that it contains similar circuit faults as $\boldsymbol{P}$;
    \item classically simulable, i.e. its ideal expectation value $E_0(\boldsymbol{T})$ can be obtained via classical simulation.
\end{enumerate}
Knowing the exact value of $E_0(\boldsymbol{T})$ enables us to find a good good set of parameters $\vec{\theta}$ for the error mitigation function $E_{\vec{\theta}}$ by minimising the difference between $E_0(\boldsymbol{T})$ and $E_{\vec{\theta}}(\boldsymbol{T})$. Then we will assume the error mitigation protocol $E_{\vec{\theta}}$ obtained from the training circuit $\boldsymbol{T}$ also works well for the primary circuit $\boldsymbol{P}$ due to their similarity, which give us the error-mitigated result $E_{\vec{\theta}}(\boldsymbol{P})$ as an estimate of the ideal result $E_{0}(\boldsymbol{P})$. More generally we can have more than one training circuit, which is denoted using the training set $\mathbb{T}$. The simplest loss function we can construct to obtain the optimal $\vec{\theta}$ is
\begin{align}\label{eqn:loss_func}
    L_{\mathbb{T}}(\vec{\theta}) = \frac{1}{\abs{\mathbb{T}}}\sum_{\boldsymbol{T} \in \mathbb{T}} (E_0(\boldsymbol{T})-E_{\vec{\theta}}(\boldsymbol{T}))^2.
\end{align}
The whole process of learning-based QEM is summarised in \cref{fig:learning_based}. If the error-mitigated estimate $E_{\vec{\theta}}(\boldsymbol{T})$ is linear in $\vec{\theta}$, which is the case for many QEM schemes, then the optimal $\vec{\theta}$ can be obtained using linear least squares.

The simplest error mitigation function simply rescales and shifts the noisy expectation value to approximate the ideal expectation value: 
\begin{align}\label{eqn:rescale_and_shift}
    E_{0}(\boldsymbol{A}) \approx E_{\vec{\theta}}(\boldsymbol{A}) = \theta_0 + \theta_1 E(\boldsymbol{A}).
\end{align}
Here the input circuit $\boldsymbol{A}$ can be the primary circuit $\boldsymbol{P}$ or the training circuits $\boldsymbol{T} \in \mathbb{T}$. \textcite{czarnikErrorMitigationClifford2021} first proposed to use such a linear function to mitigate errors, whose coefficients can be obtained by training using the Clifford variants of the primary circuit. This has been shown to be effective in experiments~\cite{miInformationScramblingQuantum2021,urbanekMitigatingDepolarizingNoise2021}. When performing fermion simulations, it is also possible to train using the closest free-fermion model as proposed and demonstrated by~\textcite{googlequantumaiandcollaboratorsObservationSeparatedDynamics2020}, or using fermionic linear optics~\cite{montanaroErrorMitigationTraining2021} as demonstrated experimentally by \textcite{stanisicObservingGroundstateProperties2022}. In some particular use cases like measuring the out-of-time-order correlator~\cite{miInformationScramblingQuantum2021}, by simply removing/replacing a small number of gates of the primary circuit, one can analytically derive the output of the resultant circuits, which can also be used as the training circuits. \textcite{rosenbergExperimentalErrorMitigation2022} have experimentally compared the different ways to perform such rescaling methods. 

The linear error mitigation function in \cref{eqn:rescale_and_shift} can naturally arise when we assume all noise sources in the circuit are globally depolarising. In such a case, the resultant noisy state is just a mixture of the ideal state and the completely mixed state: $\rho = P_0\rho_0 + (1 - P_0)I/2^N$~\cite{vovroshSimpleMitigationGlobal2021,miInformationScramblingQuantum2021} where $P_0$ is the fault-free probability of the circuit (\cref{sec:circuit_faults}). Hence, the ideal expectation value takes the simple form of
\begin{align*}
    \underbrace{\Tr[O\rho_0]}_{E_{0}(\boldsymbol{P})} = \frac{1}{P_0} \underbrace{\Tr[O\rho]}_{E(\boldsymbol{P})} - \frac{1-P_0}{P_02^N}\Tr[O],
\end{align*}
which is in the same form as \cref{eqn:rescale_and_shift}. The assumption of global depolarising noise motivates the linear error mitigation function, but is \emph{not} a necessary assumption for applying this error mitigation function using training circuits. There is evidence that global depolarising noise is an effective phenomenological error model emerging from gate-wise error models when the gate number is large~\cite{qinErrorStatisticsScalability2023}. On the other hand, if indeed there is only global depolarising noise in the circuit, we can actually estimate the rescaling factor $P_0^{-1}$ using  \cref{eqn:no_error_rate} if we know the circuit fault rate, or obtain $P_0^{-1}$ by measuring
\begin{align*}
\Tr[\rho^2] = P_0^2 + P_0(1-P_0)/2^{N-1} + (1-P_0)^2/2^{2N}
\end{align*}
and solving the quadratic equation~\cite{vovroshSimpleMitigationGlobal2021}. Multiple ways to measure $\Tr[\rho^2]$ have been discussed in purification-based QEM (\cref{sec:pur}). These circuits for obtaining $P_0^{-1}$ do \emph{not} have the same circuit structure as the primary circuit $\boldsymbol{P}$, so instead of viewing them as the training circuits for applying learning-based rescaling and shifting, it may be more appropriate to view them as the noise calibration circuits for performing probabilistic error cancellation (\cref{sec:pec}) against global depolarising channels or for a special case of linear error extrapolation (\cref{sec:zne})~\cite{caiMultiexponentialErrorExtrapolation2021}.

In \cref{eqn:rescale_and_shift}, $E_{\vec{\theta}}(\boldsymbol{A})$ is simply a function of the noisy expectation value $E(\boldsymbol{A})$ of the input circuit $\boldsymbol{A}$, thus only the input circuit $\boldsymbol{A}$ needs to be run on the quantum hardware. This is not the case for many of the QEM methods we discussed before. In general for a given input circuit $\boldsymbol{A}$, we need to construct a set of \emph{response measurement circuits} $\{\boldsymbol{A}_{\mathrm{rsp},i}\}$ that are variants of the input circuit $\boldsymbol{A}$ by adding/replacing gates and/or adding measurements. These circuits can be, e.g. circuits of different noise levels for zero-noise extrapolation, circuits with different added gates for probabilistic error cancellation and circuits with different added measurements for symmetry verification. The error-mitigated expectation value will be a function of the outputs of all of these response measurement circuits instead of just the input circuit:
\begin{align}\label{eqn:em_rsp}
    E_{\vec{\theta}}(\boldsymbol{A}) = f(\{E(\boldsymbol{A}_{\mathrm{rsp},i})\};\vec{\theta}).
\end{align}

One such example would be the error mitigation estimate for probabilistic error cancellation in \cref{eqn:pec_linear_sum}:
\begin{align*}
    \fiteqn{\expect{\est{O}_{\mathrm{em}}} 
        = \sum_{\vec{n}} \alpha_{\vec{n}} \pbra{O} \mathcal{B}_{\vec{n}}\pket{\rho_{\mathrm{in}}}
        \mapsto E_{\vec{\theta}}(\boldsymbol{A}) = \sum_{\vec{n}}\theta_{\vec{n}} E(\boldsymbol{A}_{\mathrm{rsp},\vec{n}}).}
\end{align*}
Here the response measurement circuit $\boldsymbol{A}_{\mathrm{rsp},\vec{n}}$ corresponds to preparing $\rho_{\mathrm{in}}$, applying the sequence of operation $\mathcal{B}_{\vec{n}}$ and measuring $O$, which as mentioned differs from $\boldsymbol{A}$ by additions or replacements of some subset of gates. The parameters $\theta_{\vec{n}}$ can be obtained as $\alpha_{\vec{n}}$ through device calibration as discussed in \cref{sec:pec}. On the other hand, we can also obtain $\theta_{\vec{n}}$ via learning-based methods~\cite{strikisLearningBasedQuantumError2021}. Applying learning-based methods to probabilistic error cancellation implies that we do not have enough information about the gate errors in the primary circuit (otherwise we will apply probabilistic error cancellation directly). Hence, we would need to assume Pauli gate errors in the primary circuit or apply Pauli twirling, such that the set of response measurement circuits can simply be constructed by adding Pauli gates. Other than optimising over $\theta_{\vec{n}}$ directly, we see that the response circuit coefficient $\alpha_{\vec{n}}$ is actually the product of the coefficients for individual gates in \cref{sec:pec}. In a similar way, we can write $\theta_{\vec{n}}$ as the product of the coefficients for individual gates and optimise over the gate coefficients instead. This would greatly simplify the optimisation problem if the number of gate types in the circuit is small. By incorporating the appropriate response measurement circuit $\boldsymbol{A}_{\mathrm{rsp},\vec{n}}$, it is also possible to mitigate spatially and temporally correlated using learning-based methods.

Let us continue to use probabilistic error cancellation with Clifford training~\cite{strikisLearningBasedQuantumError2021} as the example to illustrate how the training circuits can be constructed. In probabilistic error cancellation, we want to remove all faults in the circuit. To find a way to mitigate these faults using the training circuits, the same faults must exist in these training circuits. Such training circuits can be constructed by replacing gates in the primary circuit with gates that have the same error channels. If we compile the primary circuit such that all the multi-qubit gates in the primary circuit are Clifford, we then only need to replace the single-qubit gates to construct Clifford training circuits that are classically simulable. If we further assume that all the single-qubit gates have the same error channel \emph{or} they have negligible error rates compared to the multi-qubit gates, then the fault distribution of these Clifford training circuits will be the same as the primary circuit. We will use $\mathbb{C}$ and $\mathbb{U}$ to denote the set of all possible circuits generated by replacing the single-qubit gates in the primary circuit with random single-qubit Clifford and random single-qubit unitary, respectively (note that $\mathbb{C} \subset \mathbb{U}$ and $\boldsymbol{P} \in \mathbb{U}$). By constructing the training circuits in the way outlined above, the training loss function $L_{\mathbb{T}}(\vec{\theta})$ in \cref{eqn:loss_func} is a homogeneous polynomial of degree $2$ in matrix elements of the single-qubit gates in the circuits. Since the Clifford group is a unitary $2$-design, the loss function satisfies $ L_{\mathbb{C}}(\vec{\theta}) = L_{\mathbb{U}}(\vec{\theta})$~\cite{wangScalableEvaluationQuantumCircuit2021}. Therefore, by training over the Clifford circuits and minimising $L_{\mathbb{C}}(\vec{\theta})$, we are minimising the errors in the more general unitary circuits in $\mathbb{U}$. When $L_{\mathbb{C}}(\vec{\theta})$ goes to zero, we have $L_{\mathbb{U}}(\vec{\theta}) = 0$, which implies $E_{\vec{\theta}}(\boldsymbol{A}) = E_0(\boldsymbol{A})$, i.e. all errors are perfectly mitigated, for all $\boldsymbol{A}\in \mathbb{U}$ including the primary circuit $\boldsymbol{P}$.

Since the size of the Clifford set $\mathbb{C}$ grows exponentially with the number of qubits, it is impractical to evaluate $E_{0}(\boldsymbol{T})$ and the corresponding noisy response measurements $\{E(\boldsymbol{T}_{\mathrm{rsp},i})\}$ for all Clifford training circuits $\boldsymbol{T} \in \mathbb{C}$. Furthermore, the majority of noisy Clifford training circuits have near-zero expectation values which are costly to evaluate in terms of sampling overhead. One way to circumvent this is to truncate the Clifford training set by using only circuits with large noiseless expectation values $\abs{E_{0}(\boldsymbol{T})}$ due to their more significant contribution to the loss function, which was shown to be effective in numerical simulations~\cite{strikisLearningBasedQuantumError2021,czarnikImprovingEfficiencyLearningbased2022}. It is also possible to use Monte-Carlo sampling and variational update of the parameters $\vec{\theta}$ to overcome the large size of $\mathbb{C}$~\cite{strikisLearningBasedQuantumError2021}. Since it is possible to classically simulate circuits with a small number of non-Clifford gates, we can keep a few single-qubit gates in the primary circuit untouched when we define the training set. Alternative ways to truncate the Clifford set may be explored, e.g. based on their similarity to the primary circuits. 

If the training set is large enough such that the training can target a wide range of application circuits (computational tasks), then the training can be carried out at the device calibration stage and the training sampling overhead may be omitted when considering a particular computational task. On the other hand, when the training set is small such that the training is targeting a small set of application circuits (computational tasks), then the training is best viewed as a part of a task itself and the training overhead needs to be included in any resource audit for that computational task. Note that a smaller training set usually also means a smaller sampling overhead for training. The relation between the size of the training set $\abs{\mathbb{T}}$ and the performance of applying the trained result onto the target circuit $\boldsymbol{P}$ is only rigorously developed for the case of training using the full Clifford set $\mathbb{C}$ for probabilistic error cancellation as discussed above. More general studies into the trade-off between them would be essential for different learning-based methods.

So far we have only explicitly talked about learning-based methods applied to rescaling and shifting the noisy result~(\cref{eqn:rescale_and_shift}) and to probabilistic error cancellation. However, learning-based methods are in general compatible with almost all QEM methods that we have discussed, e.g. error extrapolation~\cite{loweUnifiedApproachDatadriven2021} and purification-based method~\cite{bultriniUnifyingBenchmarkingStateoftheart2023}. There are also suggestions that it can be applied to symmetry-based methods~\cite{caiQuantumErrorMitigation2021}. Some of the possible roles of learning-based methods in other QEM methods will be further discussed in the next section.

\section{Comparisons and Combinations}\label{sec:compare_and_comb}

\subsection{Comparison among QEM methods}\label{sec:qem_compare}
In the last section, we have provided a detailed account of the individual QEM methods. In order to see the connections and differences among them more clearly and provide a discussion of their respective costs in a more coherent framework, \textcite{caiPracticalFrameworkQuantum2021} divides the process of QEM into two stages:
\begin{enumerate}
    \item \textbf{Noise calibration:} measures the strength of some given noise components;
    \item \textbf{Response measurement:} measures how the observable of interest responds to changes in the noise components that were calibrated in the last step.
\end{enumerate}
Combining both components will inform us about how the observable of interest changes due to the presence of the calibrated noise components, and thus will enable us to construct an error-mitigated estimator protected from the calibrated noise components.
We have already discussed such a structure for QEM methods when introducing the learning-based methods (\cref{sec:learning-based}), there the noise calibration process is simply the training process. Of course the division between noise calibration and response measurement is not always clear-cut. For most of the QEM methods that we have discussed, the error-mitigated estimator will be a linear combination of the results obtained from the response measurement, and the way to combine them (the weightings of each term) will be determined by the noise calibration. Treating QEM as a two-stage process enables us to discuss the cost of noise calibration and response measurement separately.

\subsubsection{Noise calibration overhead}\label{sec:QEM_two_stage}
In the previous sections when talking about sampling overhead, we are mostly referring to the response measurement sampling overhead. We often assumed certain knowledge about the noise without discussing the cost of obtaining it.  Now let us look more closely into the cost of the noise calibration for various types of QEM methods.

\paragraph{Gate error mitigation} These QEM methods target the \emph{gate errors} in the primary circuit. Their noise calibration can simply be carried out using standard gate noise benchmarking/characterisation techniques~\cite{klieschTheoryQuantumSystem2021,eisertQuantumCertificationBenchmarking2020}, e.g. gate infidelity estimation for \emph{zero-noise extrapolation} (\cref{sec:zne}), full gate error characterisation for \emph{probabilistic error cancellation} (\cref{sec:pec}) and detector error characterisation for \emph{measurement error mitigation} (\cref{sec:meas_miti}). These can all be done in the device calibration stage and if this is the case, then ideally noise calibration is effectively free at the stage of applying QEM. However, fully correlated noise models are exponentially expensive (in terms of qubit number) to characterise and various ways to circumvent this have been mentioned in \cref{sec:pec,sec:meas_miti}. Moreover, device parameters can drift in time and thus routine re-calibration might be needed. In such a case, the noise calibration cost is no longer negligible at the QEM application stage and low-cost device calibration techniques would be essential~ [see e.g. \textcite{googlequantumaiandcollaboratorsObservationSeparatedDynamics2020}].

\paragraph{State error mitigation} These QEM methods target the \emph{errors on the output state} of the primary circuit. Since we do not know the exact form of the ideal state, we will try to probe errors that violate known constraints on the output state like \emph{symmetry constraints} (\cref{sec:sym}) and \emph{purity constraints} (\cref{sec:pur}). Their noise calibration will measure the strength of the noise that violates these constraints, e.g. the fraction of the circuit runs that fail the symmetry verification. In these cases, both noise calibration and response measurement involve measuring additional operators on the unmitigated noisy state. Hence, in some settings, the noise calibration can be performed alongside response measurement by simply measuring additional operators without additional circuit runs~\cite{bonet-monroigLowcostErrorMitigation2018,caiQuantumErrorMitigation2021,caiPracticalFrameworkQuantum2021}, which means the noise calibration is essentially free.

\paragraph{Observable error mitigation} Not all errors in the primary circuit will affect our observable of interest, and this class of QEM methods only target the \emph{error components that are damaging to our observable}.  Methods like \emph{subspace expansion} (\cref{sec:sub_expand}) and \emph{$N$-representability} (\cref{sec:nrep}) target error components that violate some given constraints on the noiseless observables (which can be the observable of interest or its components). For observable error mitigation, the noise calibration will be dependent on the observable of interest.  Hence, the noise calibration accuracy required is highly dependent on the problems we try to solve and thus the associated cost has not been analytically derived. 

As mentioned, \emph{learning-based methods}(\cref{sec:learning-based}) are a means to perform the noise calibration process using training circuits. Compared to the circuits used in the original noise calibration process in different QEM methods, the training circuits are usually more closely related to the primary circuit, so that we can target faults that are more specific to the primary circuit and hopefully can reduce the calibration cost. For example, when trying to construct the training circuit, we can make use of the structure of the primary circuit for gate error mitigation, or we can make use of the known observable of interest for state error mitigation. The training cost for learning-based methods is thus highly problem-specific and usually hard to analytically quantify, just like the cost for noise calibration in observable error mitigation.

The categorisation of QEM methods above is mainly for providing better intuitions rather than a definitive guide. For example, the RDM measured in the $N$-representability method can be viewed as an observable and then the method is seen as a type of observable error mitigation; alternatively it can be viewed as a state in a reduced subspace, so that the method is a form of state error mitigation. After our discussion of the noise calibration requirements above, we can now move onto the more general comparisons between QEM methods beyond the noise calibration stage.

\subsubsection{Mean square errors}
As discussed in \cref{sec:bias_variance}, the most straightforward metric for comparing two different error-mitigated estimators is just to compare their mean square errors (\cref{eqn:mse}). However, general comparisons between two QEM methods cannot be made in a similar manner because for each QEM method, there is a wide range of error-mitigated estimators we can construct using different hyper-parameters (e.g. the degree of purification for purification-based QEM, the number of symmetries in symmetry-based QEM, the number of data points in zero-noise extrapolation, etc). Each of these estimators has a different trade-off between bias and variance and thus a different mean square error. Comparisons between two QEM methods are further complicated by their different sets of assumptions (e.g. whether there are known constraints to states, knowledge about the noise) and different hardware requirements (e.g. qubit number and connectivity). Hence, in order to compare two QEM methods, very often we need to know the exact primary circuit whose noise we try to mitigate and the set of experimental constraints we need to adhere to (on things like qubit numbers, runtime and hardware error rate), such that we can specify exact implementations of the two given QEM methods and deduce the corresponding mean square errors. Even in such cases, the bias and variance often can only be calculated analytically for some canonical implementations. Hence, in most of the QEM literature mentioned in this review, when trying to assess or compare the performance of QEM methods, their bias and variance are often obtained using numerical simulations or physical experiments by applying them to some specific use case. This is not a good indicator of the general performance of a given QEM method, but this can usually demonstrate key characteristics of the QEM methods like their target noise types, their scaling behaviour, etc. 

While recognising that it is hard to make general comparisons between different QEM methods, nevertheless we endeavour to summarise the costs and performance of some \emph{canonical implementations} of typical QEM methods in \cref{tab:qem_summary}. The intention is to capture the essential distinguishing features of these QEM approaches.

\subsection{Benchmarking QEM from other perspectives}\label{sec:qem_other_persp}
Besides computing the mean square error for some specific implementation of a given QEM method, we can also look into other general characteristics of the QEM method by viewing the process of QEM from another perspective. 

\subsubsection{State discrimination}\label{sec:state_distinct_perf}
QEM allows us to better estimate the expectation values of the various noisy states output from a noisy system. However, the data-processing inequality never allows the distinguishability between these noisy states to increase~\cite{nielsenQuantumComputationQuantum2010}. This fact was used by \textcite{takagiFundamentalLimitsQuantum2022} to obtain explicit bounds on the range that the error-mitigated estimator may take, which in turn determines the number of samples needed to achieve a given level of performance.

Their work considered an error-mitigation protocol in which the first step is obtaining $M$ noisy copies of the error-free state $\rho_0$ through a process denoted as $\mathcal{L}$:
\begin{align*}
    \mathcal{L}(\rho_0) = \bigotimes_{m = 1}^M \mathcal{E}_m(\rho_0)
\end{align*}
where $\{\mathcal{E}_m\}$ are the different effective noise channels acting on different copies. The second step is constructing an error-mitigated observable $O_{\mathrm{em}}$ based on the QEM technique and the observable of interest $O$, such that measuring $O_{\mathrm{em}}$ on these noisy copies will output the error-mitigated estimator $\est{O}_{\mathrm{em}}$:
\begin{align*}
    \expect{\est{O}_{\mathrm{em}}} = \Tr[O_{\mathrm{em}}\mathcal{L}(\rho_0)].
\end{align*}
The \emph{maximum bias} that can be achieved by the QEM method for all possible observables of interest $O$ and target states $\rho_0$ is given as:
\begin{align*}
    B_{\mathrm{max}}  = \max_{O, \rho_0} \bias{\est{O}_{\mathrm{em}}} = \max_{O, \rho_0} \left[\Tr[O_{\mathrm{em}}\mathcal{L}(\rho_0)] - \Tr[O\rho_0]\right].
\end{align*}
Using this, \textcite{takagiFundamentalLimitsQuantum2022} managed to prove a lower bound for the range of the error-mitigated estimator:
\begin{align}\label{eqn:samp_cost_trace_dist}
    \range{\est{O}_{\mathrm{em}}} \geq  \max_{\rho_0, \sigma_0} \frac{D_{\mathrm{tr}}(\rho_0, \sigma_0) - 2 B_{\mathrm{max}}}{D_{\mathrm{tr}}(\mathcal{L}(\rho_0), \mathcal{L}(\sigma_0))} 
\end{align}
where $D_{\mathrm{tr}}$ denotes the trace distance. The range $\range{\est{O}_{\mathrm{em}}}$ can be used to obtain the \emph{sufficient} number of samples required via \cref{eqn:samp_number_hoeff}. Similar scaling relationship of $\range{\est{O}_{\mathrm{em}}}$ for the case of depolarising noise was also proven by \textcite{wangCanErrorMitigation2021}. 

Building on the framework above, \textcite{takagiUniversalSamplingLower2023} have further obtained the explicit lower bound for the \emph{necessary} number of samples $M$ for achieving the target accuracy $\delta$ (this deviation includes both the bias and shot noise) with success probability $1-\epsilon$:
\begin{align}\label{eqn:sample_lower_bound}
    M\geq \max_{\substack{\rho_0,\sigma_0\\ D_{\mathrm{tr}}(\rho_0,\sigma_0)\geq 2\delta}} \min_{\mathcal{E}\in\{\mathcal{E}_m\}} \frac{2(1-2\epsilon)^2}{\ln 2\  S(\mathcal{E}(\rho_0)||\mathcal{E}(\sigma_0))}.
\end{align}
Here $S(\rho || \sigma) = \Tr(\rho \log \rho) - \Tr(\rho \log \sigma)$ is the relative entropy, which is related to the trace distance through the quantum Pinsker's inequality $ D_{\mathrm{tr}}(\rho, \sigma) \leq \sqrt{\ln 2\ S(\rho || \sigma) / 2}$~\cite{hiaiSufficiencyKmsCondition2008}. It has been shown that the relative entropy of between two output states from noisy circuits under local depolarizing noise decreases exponentially with the circuit depth~\cite{muller-hermesEntropyProductionDoubly2016,kastoryanoQuantumLogarithmicSobolev2013}, which implies that the sampling cost in \cref{eqn:sample_lower_bound} will grow exponentially with the circuit depth in this case. Note that by using fidelity instead of relative entropy, a bound tigher than \cref{eqn:sample_lower_bound} can also be obtained as discussed in \textcite{takagiUniversalSamplingLower2023}.

\textcite{quekExponentiallyTighterBounds2022} managed to construct a circuit structure for which the relative entropy of two different output states decreases exponentially in \emph{both circuit depth and the number of qubits}, which allows them to prove that the worst-case sampling lower-bound for QEM scale exponentially with the circuit size (instead of just depth), confirming our intuition obtained in earlier sections like \cref{sec:exp_samp_cost}. They further showed that for geometrically local circuit, this exponential scaling is determined by the number of gates in the light-cone of observables rather than the circuit size. Note that the bounds obtained in this section are more for the purpose of demonstrating the fundamental limits on the performance of a given QEM setup, rather than to be used as a metric for comparing the practical performance of different QEM methods. 

\subsubsection{Quantum estimation theory}\label{sec:quantum_estimation_theory}
In estimation theory, the variance of any unbiased estimator can be lower-bounded by the inverse of the Fisher information through the Cram\'er-Rao inequality. Without loss of generality, we will assume that our goal is to estimate $\Tr(O\rho_0)$ for some traceless observable $O$, but the state is suffering from the noise channel $\mathcal{E}$. We can still obtain the ideal expectation value if we have a way to measure the operator $Y = \mathcal{E}^\dagger(O)$ on the noisy state $\mathcal{E}(\rho_0)$. \textcite{watanabeOptimalMeasurementNoisy2010} showed that the quantum Fisher information for estimating the observable $O$ using the noisy state $\mathcal{E}(\rho_0)$ is given by $J_O=(\Tr(\mathcal{E}(\rho_0) Y^2) - \Tr(\mathcal{E}(\rho_0)Y)^2)^{-1}$. Using the quantum Cram\'er-Rao inequality, the number of samples $M$ needed for evaluating the noiseless estimator with additive error $\epsilon$ can then be bounded as:
\begin{equation}
    M \geq \frac{1}{\epsilon^2 J_O}. 
\end{equation}
Thus, they concluded that the cost-optimal strategy for obtaining the unbiased estimator of $\Tr (O\rho_0)$ from the noisy state $\mathcal{E}(\rho_0)$ is simply to measure $\mathcal{E}(\rho_0)$ with the observable $Y$.

\textcite{tsubouchiUniversalCostBound2023} applied these arguments to the NISQ scenario with the estimator being an error-mitigated estimator and obtained two lower bounds on the sampling cost of QEM. Their first lower bound is for generic layered quantum circuits under a wide class of Markovian noise. By showing that the quantum Fisher information decays exponentially with the depth of the layered quantum circuit, they demonstrated the exponential growth of the sampling overhead as the depth of the circuit increases. In the special case of global depolarizing errors, the required number of samples $M$ for achieving an additive error $\epsilon$ in the standard deviation can be lower-bounded as $M\gtrsim \frac{\Tr(O^2)}{2^N\epsilon^2}(1-p)^{-2L}$, which can be saturated by simply rescaling the measurement result by a factor of $(1-p)^{-L}$ (see \cref{sec:learning-based}). For random quantum circuits under local noise, they further showed that the sampling cost grows exponentially with both the circuit depth and the number of qubits through both analytical arguments and numerical simulations. The scaling behaviour found here is consistent with \cref{sec:state_distinct_perf}.

\subsubsection{State extraction}\label{sec:state_extract_perf}
In symmetry-based or purification-based QEM, it is natural to think of QEM as a process of extracting the symmetry-verified or purified state out of the noisy state~\cite{mcardleErrorMitigatedDigitalQuantum2019,bonet-monroigLowcostErrorMitigation2018,koczorExponentialErrorSuppression2021,hugginsVirtualDistillationQuantum2021}. More generally  
in most of the QEM methods, the error-mitigated expectation value $\expect{\est{O}_{\mathrm{em}}}$ is a linear function of the observable of interest $O$, which can be written as 
\begin{align*}
    \expect{\est{O}_{\mathrm{em}}} = \Tr[O\rho_{\mathrm{em}}].
\end{align*}
Here $\rho_{\mathrm{em}}$ can be viewed as the error-mitigated component that we try to extract out of the noisy state using the given QEM method~\cite{caiPracticalFrameworkQuantum2021}. 

A higher fidelity of the error-mitigated state $\rho_{\mathrm{em}}$ against the ideal state $\rho_0$, denoted as $F(\rho_0,\rho_{\mathrm{em}})$, is shown to correspond to lower biases in various numerical simulations~\cite{caiQuantumErrorMitigation2021,koczorExponentialErrorSuppression2021,hugginsVirtualDistillationQuantum2021}. More exactly, using results in \textcite{koczorDominantEigenvectorNoisy2021} and the Fuchs-van de Graaf inequality~\cite{fuchsCryptographicDistinguishabilityMeasures1999}, the bias after error mitigation can be bounded by:
\begin{equation}
    \begin{aligned}\label{eqn:bias_fid_bound}
        \bias{\est{O}_{\mathrm{em}}} &\leq 2 \norm{O}_{\infty} D_{\mathrm{tr}}(\rho_0, \rho_{\mathrm{em}}) \\
        &\leq 2 \norm{O}_{\infty} \sqrt{1-F(\rho_0,\rho_{\mathrm{em}})},
    \end{aligned}
\end{equation}
where $\norm{O}_{\infty}$ is the largest absolute eigenvalue of $O$.

As discussed in \textcite{caiPracticalFrameworkQuantum2021}, by assuming the ideal state $\rho_0$ to be a \emph{pure} state, we can decompose the noisy state $\rho$ into the error-mitigated component $\rho_{\mathrm{em}}$ and an erroneous component $\rho_{\mathrm{err}}$ (not necessarily a valid density matrix) that is orthogonal to the ideal state $\rho_0$ ($\Tr[\rho_{0}\rho_{\mathrm{err}}] = 0$):
\begin{align}\label{eqn:noisy_state_decomp}
    \rho = p_{\mathrm{em}}\rho_{\mathrm{em}} + (1-p_{\mathrm{em}}) \rho_{\mathrm{err}}.
\end{align}
Here $p_{\mathrm{em}}$ is the amount of error-mitigated component contained in the noisy state. Only a partial amount of this error-mitigated component can be successfully extracted using the given QEM method and this amount is denoted as $q_{\mathrm{em}}$ with $q_{\mathrm{em}} \leq p_{\mathrm{em}}$.  

In this picture of extracting error-mitigated states, the factor of improvement in the fidelity, which we will call the \emph{fidelity boost}, is given by: 
\begin{align*}
    B_{\mathrm{em}} = \frac{\Tr[\rho_0\rho_{\mathrm{em}}]}{\Tr[\rho_0\rho]} = p_{\mathrm{em}}^{-1}
\end{align*}
(where $\rho_0$ is assumed to be pure) and the corresponding sampling overhead is given as:
\begin{align}\label{eqn:cost_extract}
    C_{\mathrm{em}} \sim  q_{\mathrm{em}}^{-2}. 
\end{align}
\textcite{caiPracticalFrameworkQuantum2021} has given a list of examples of how $p_{\mathrm{em}}$ and $q_{\mathrm{em}}$ can be calculated to obtain $B_{\mathrm{em}}$ and $C_{\mathrm{em}}$, with some of the results shown in \cref{tab:qem_summary}.

The fraction of the error-mitigated state that is successfully extracted using the QEM method is called the \emph{extraction rate}:
\begin{align*}
    r_{\mathrm{em}} = \frac{q_{\mathrm{em}}}{p_{\mathrm{em}}} = \frac{B_{\mathrm{em}}}{\sqrt{C_{\mathrm{em}}}}.
\end{align*}
This is simply the ratio between the fidelity boost and the square root of the sampling overhead, and thus is a natural indicator for the \emph{cost-effectiveness} of a given QEM method. Successfully extraction of all of the error-mitigated components contained in the noisy state corresponds to the maximum extraction rate $r_{\mathrm{em}} = 1$, which can be achieved by symmetry verification~\cite{caiPracticalFrameworkQuantum2021}, indicating its cost-effectiveness. Note that if we are able to perform direct post-selection instead post-processing to extract the error-mitigated state (like in direct symmetry verification and echo verification), then the sampling overhead will be $ C_{\mathrm{em}} \sim  q_{\mathrm{em}}^{-1}$ instead and the extraction rate will be $r_{\mathrm{em}} =  \nicefrac{B_{\mathrm{em}}}{C_{\mathrm{em}}}$. Similar to \cref{sec:state_distinct_perf}, the arguments in this section also do not apply to QEM techniques that non-linearly combine the results from response measurement.

\subsection{Combinations of QEM methods}
Rather than trying to pick a better QEM method, one might try to combine different QEM methods instead, in the hope of being able to target more noise components and/or achieving a better trade-off between bias and variance. We have mentioned some possible connections and combinations of QEM methods when discussing individual techniques in \cref{sec:methods}, and we will try to provide additional insights in this section. 

The simplest way of combining different QEM methods is applying them in parallel such that each of them targets a different noise source in the circuit. For example using zero-noise extrapolation or probabilistic error cancellation to target noise in the main computation while using measurement error mitigation to tackle noise at the measurement stage. For a circuit consisting of Pauli-symmetry-preserving components affected by Pauli noise, we can detect and partially mitigate the circuit faults that anti-commute with the symmetry using symmetry verification, while the rest of the circuit faults that commute with the symmetry and thus are immune to symmetry verification can be mitigated using zero-noise extrapolation (through only scaling these `commuting' noise) or probabilistic error cancellation~\cite{caiMultiexponentialErrorExtrapolation2021}. As will be discussed in \cref{sec:algo_error}, QEM can be used to mitigate errors in the circuit compilation. Hence, it is possible to use one QEM method to mitigate these compilation errors, while using another QEM method to mitigate noise in the circuit. Note that we can also apply a given QEM method multiple times, each time targetting a different noise source, which has been demonstrated for zero-noise extrapolations~\cite{ottenAccountingErrorsQuantum2019}.

When the different QEM methods applied interfere with each other rather than working entirely independently, their order of application becomes important. The QEM method that is first applied will be called the \emph{base QEM method}. In a sense, applying one QEM method on top of another can be viewed as ``concatenating'' the QEM methods, and the overall sampling overhead is simply the product of the sampling overheads of both stages of QEM. If we want to apply symmetry- or purification-based QEM on top of another QEM method, we can simply view the base QEM method as a process of extracting error-mitigated states out of the noisy state as discussed in \cref{sec:state_extract_perf}, and then apply symmetry constraints or purity constraints to this error-mitigated state~\cite{caiPracticalFrameworkQuantum2021}. In this way, we are able to remove the remaining symmetry- or purity-violating noise that was not targeted by the base QEM method. In a similar way, for a given error-mitigated state, we can also perform subspace expansion around it or we can measure its RDM, which means that we can directly perform subspace expansion or $N$-representability on top of other QEM methods. \textcite{googlequantumaiandcollaboratorsHartreeFockSuperconductingQubit2020} has experimentally demonstrated the application of $N$-representability on top of symmetry verification.

One may apply zero-noise extrapolation on top of another QEM method by performing extrapolation using error-mitigated expectation values~\cite{mcardleErrorMitigatedDigitalQuantum2019}. However, the error models of the circuit can be altered by the base QEM method, such that the noise scaling factor might not be well-defined anymore. Furthermore, the shape of the extrapolation curves may be altered by the base QEM method. It is possible to circumvent this by using probabilistic error cancellation as the base QEM method since it can be applied partially without changing the error models, yielding data points of reduced noise strength for extrapolation as mentioned before in \cref{sec:zne}~\cite{caiMultiexponentialErrorExtrapolation2021}. For other base QEM methods, as long as we know the effective noise scaling factor and the circuit fault rate is small after the base QEM, Richardson extrapolation will still be applicable, e.g. it has been applied on top of purification-based methods in numerical simulations~\cite{koczorExponentialErrorSuppression2021} and on top of instances of probabilistic error cancellation that changes the error models~\cite{sunMitigatingRealisticNoise2021}. In some specific use cases, the new shape of the extrapolation curve can be analytically deduced, one such example is given in \textcite{caiMultiexponentialErrorExtrapolation2021} with symmetry verification being the base QEM method. More generally, we can try to use learning-based methods to predict the shape of the extrapolation curves after performing the base QEM, especially for the special case of rescaling and shifting the noisy expectation value (see \cref{sec:learning-based}). Applying rescaling on top of symmetry verification has been demonstrated in numerical simulations~\cite{stanisicObservingGroundstateProperties2022} and experiments~\cite{googlequantumaiandcollaboratorsObservationSeparatedDynamics2020}. As mentioned in \cref{sec:learning-based}, rescaling and shifting can be viewed as performing probabilistic error cancellation for mitigating global depolarising noise. However, probabilistic error cancellation in a more general context is difficult to perform on top of other QEM methods because the effective error model in the circuit will change due to the base QEM method.

Learning-based methods can be viewed as a way to perform the noise calibration process rather than a stand-alone QEM method as mentioned in \cref{sec:QEM_two_stage}. Hence, it can be easily combined with the other methods by replacing their noise calibration process~\cite{loweUnifiedApproachDatadriven2021,caiPracticalFrameworkQuantum2021}, which we have mentioned above in several examples. We can also use learning-based methods to obtain the optimal hyper-parameters for various QEM methods~\cite{bultriniUnifyingBenchmarkingStateoftheart2023}.

There were also attempts to construct a unified framework containing hyper-parameters that can lead to different existing QEM methods when taking different values. Such frameworks give us access to a wide range of other possible QEM implementations by taking hyper-parameter values sitting in between existing QEM methods. We can think of this as ``interpolating'' between different QEM methods instead of ``concatenating'' different QEM methods. These new ``interpolated'' QEM implementations can achieve different bias-variance trade-offs beyond the existing QEM methods and often also target a different set of noise components. This allows us to choose the best implementation that fits our target problem and experimental constraints rather than being restricted to the existing QEM methods. We have mentioned some of these examples in the previous section (\cref{sec:methods}), e.g. the combination of purification-based methods with subspace expansion~\cite{yoshiokaGeneralizedQuantumSubspace2022}, the generalisation of symmetry verification and subspace expansion~\cite{caiQuantumErrorMitigation2021} or the combination of zero-noise extrapolation and probabilistic error cancellation~\cite{mariExtendingQuantumProbabilistic2021}. Some QEM combinations mentioned earlier in this section can also be viewed as attempts to construct a unified framework~\cite{loweUnifiedApproachDatadriven2021,bultriniUnifyingBenchmarkingStateoftheart2023}. Note that many QEM methods can also be natively combined with shadow tomography~\cite{jnaneQuantumErrorMitigated2023}, which may provide performance advantages when we want to estimate a large number of observables.

\subsection{Comparison to the other error suppression methods}\label{sec:comp_qem_qec}
Quantum error mitigation is one family of strategies for suppressing errors.
It is natural to consider how it relates to other well-known error-suppression methods, such as decoherence-free subspace/subsystem (DFS), dynamical decoupling (DD) and quantum error correction (QEC). Symmetry is at the heart of all the methods mentioned here, thus we will first try to compare QEC against symmetry verification to gain some intuition.

In QEC, we perform regular symmetry (stabiliser) measurements to detect and correct errors accumulated in the quantum system. If we discard the quantum states that violate these symmetry checks instead of correcting them, we will have quantum error detection instead. Symmetry verification (\cref{sec:sym}) can usually be seen as an error-\emph{detection} scheme whose symmetries are given by the physical problem of interest and the detection can be carried out via post-processing. Such native symmetries and the possible use of post-processing detection typically mean that much fewer qubits and a much lower gate fidelity are required for symmetry verification compared to QEC to achieve the ``break-even'' point of performing better than the uncorrected/unmitigated circuit. Furthermore, performing logical operations in QEC code usually come with higher space-time overhead than the physical operations performed in QEM, leading to longer circuit run time and even more qubit overhead for QEC. 

Even though we are talking about symmetry verification in the above comparison, many of the intuitions provided are applicable to general QEM methods. As mentioned in \cref{sec:intro}, general QEM methods aim to correct the output \emph{distribution} in an \emph{ensemble} of circuit runs, while QEC aims to correct the output in every single circuit run. If we are trying to remove (almost) all bias like in fault-tolerant QEC, the sampling overhead of general QEM will grow \emph{exponentially} with the circuit size as discussed in \cref{sec:exp_samp_cost}. Using all the arguments above, we can compile a rough guide for the differences between QEC and QEM as outlined in \cref{tab:qec_and_qem}. Note that there are intersections between QEC and symmetry-based QEM like post-processing quantum error detection and adding more symmetries to QEM (at the cost of adding more qubits) as discussed in \cref{sec:sym}. Adding onto what we have discussed, \textcite{caoNISQErrorCorrection2022} also looked at the difference between QEM and QEC in the setting of communication instead of computation.

\begin{table}[htb]
    \centering
    \resizebox{0.47\textwidth}{!}{
        \begin{threeparttable}
            \begingroup
            \renewcommand{\arraystretch}{1.7}
            \setlength{\tabcolsep}{0.5em}
            \aboverulesep=0ex
            \belowrulesep=0ex
            \begin{ruledtabular}
                \begin{tabular}{@{}p{5.8em} p{7em} p{9em}@{}}
                    &QEC&  QEM\\\hline
                    Qubit\quad\  overhead& High  & Low \\
                    Circuit\quad\  runtime& Often scale with code distance & Similar to the unmitigated circuit\\
                    Sampling overhead & Const. & Exponential in the circuit size \tnote{1}\\
                    Error rate& Must be below code threshold  & Important to keep it low;
                    no threshold \tnote{2}\\
                   Mid-circuit measurement& Essential and frequent  & Infrequent or not required
                \end{tabular}
            \end{ruledtabular}
            \endgroup
            \begin{tablenotes}\footnotesize
                \item[1] Assuming we try to construct an (nearly) unbiased estimator like in QEC and with fixed gate error rates.
                \item[2] Lower error rate means smaller sampling overheads. No threshold for the gate noise in the unmitigated circuit, but there might be requirements on the additional gates needed for QEM.
            \end{tablenotes}
            \caption{A rough guide on the differences between QEC and QEM. The exact difference will be dependent on the QEC codes, the QEM method and the exact application scenario.}
            \label{tab:qec_and_qem}
        \end{threeparttable}
    }
\end{table}

Without knowing the specific application task, it will be difficult to compare the amount of resources required for QEC and QEM. There have been attempts to estimate the resource required to obtain \emph{classically intractable} Fermi-Hubbard ground state energy using variational eigensolver (VQE) with the help of symmetry verification and error extrapolation~\cite{caiResourceEstimationQuantum2020}. To overcome the sampling overhead, which is partly due to QEM, it was estimated that more than $100$ independent quantum processors are needed for parallelisation, each containing $50$ qubits and with a gate error rate of $10^{-4}$. The gate fidelity and the total number of qubits required here are actually similar to those required for solving the problem using a fault-tolerant algorithm~\cite{kivlichanImprovedFaultTolerantQuantum2020}. However, given the same total number of qubits, building hundreds of identical quantum processors (cores) using commercial fabrication techniques is usually easier than building the single big integrated processor required for the fault-tolerant algorithm. Note that what we have discussed here is not a sheer comparison between QEC and QEM, but more of a comparison between a NISQ algorithm with the help of QEM and a fault-tolerant algorithm for a specific application.

In practice, QEC and QEM are more likely to be complementary rather than competing methods since in many practical applications we can apply QEM on top of QEC as will be further discussed in \cref{sec:fault_tol_error}. It is also possible to use QEM to mitigate compilation errors, which is not possible using QEC as will be discussed in \cref{sec:algo_error}.

Unlike QEC and QEM, both dynamical decoupling (DD) and decoherence-free subspace/subsystem (DFS) methods are \emph{open-loop} quantum control techniques whose action does not depend on the status of the quantum state and thus no measurements are needed for their implementation~\cite{lidarReviewDecoherenceFreeSubspaces2014,suterColloquiumProtectingQuantum2016}. In fact, DFS is a passive technique that requires no action at all, we only need to choose the right symmetry subspace that is immune to the target set of noise. QEC and symmetry-based QEM all have some element of DFS in them in the sense that they are all immune to noise generated by the symmetry operators. DD on the other hand is achieved by applying a sequence of decoupling pulses to ``average'' out the harmful interaction between the quantum system and the environment. In the limit of many rounds of instantaneous decoupling pulse, group-based DD is shown to be equivalent to making continuous symmetry measurements to achieve the quantum Zeno effect~\cite{facchiUnificationDynamicalDecoupling2004,burgarthGeneralizedProductFormulas2019}.  Such a DD sequence can be applied using the native symmetry of the physical problem just like in symmetry-based QEM, which is shown to be effective in quantum simulation~\cite{tranFasterDigitalQuantum2021}.

\section{Applications}\label{sec:applications}

So far we have discussed QEM mostly in the general context of expectation value estimation without over-specifying the application scenarios. There are a range of software packages for carrying out the various QEM techniques that we have mentioned. For example, Qermit, Mitiq and Qiskit each provide their own implementation of zero-noise extrapolation, probabilistic error cancellation and measurement error mitigation~\cite{laroseMitiqSoftwarePackage2021,cirstoiuVolumetricBenchmarkingError2023,qiskitcontributorsQiskitOpensourceFramework2023}. In additional, Qermit and Mitiq also contain the Clifford variants of the learning-based method. There are also a wide range of other packages providing measurement error mitigation like qCompute, Pennylane, QREM, etc~\cite{baiduQCompute2023,bergholmPennyLaneAutomaticDifferentiation2018,maciejewskiQuantumReadoutErrors2020}. 

Through numerical experiments, QEM has been proven to be effective in a wide range of applications like linear equation solvers~\cite{vazquezEnhancingQuantumLinear2022}, quantum metrology~\cite{yamamotoErrormitigatedQuantumMetrology2022,conlonApproachingOptimalEntangling2023}, Monte-carlo simulations~\cite{yangAcceleratedQuantumMonte2021} and in numerous other examples that we mentioned before in \cref{sec:methods}. Moreover, many physical experiments have successfully employed QEM for noise suppression, with some pioneering and state-of-the-art examples summarised in \cref{tab:application_examples}, spanning across different applications scenarios like Fermi-Hubbard models~\cite{stanisicObservingGroundstateProperties2022} and multi-particle bound states~\cite{googlequantumaiandcollaboratorsFormationRobustBound2022}, with a circuit size up to 127 qubits and 60 layers of 2-qubit gates~\cite{kimEvidenceUtilityQuantum2023}. Different application scenarios and hardware platforms will give rise to different types of noise. In this section, we will look into how to adapt the implementation of QEM to different noise types and the expected results.

\begin{table*}[htb]
    \centering
        \begin{threeparttable}
            \begingroup
            \renewcommand{\arraystretch}{1.5}
            \setlength{\tabcolsep}{0em}
            \aboverulesep=0ex
            \belowrulesep=0ex
            \begin{ruledtabular}
                \begin{tabular}{@{}p{7em}  p{3em}  p{23em}   p{12em}@{}}
QEM Methods&Platform& Applications& Reference\\
ZNE & SC & Variational eigensolver & \textcite{kandalaErrorMitigationExtends2019,dumitrescuCloudQuantumComputing2018}\\
ZNE & SC & Real-time dynamic simulation&\textcite{kimScalableErrorMitigation2023,kimEvidenceUtilityQuantum2023}\\
ZNE & Ion& Entanglement entropy measurement& \textcite{foss-feigEntanglementTensorNetworks2022}\\
PEC & SC & Deterministic quantum computation with pure states &\textcite{songQuantumComputationUniversal2019}\\
PEC & SC & Real-time dynamic simulation&\textcite{vandenbergProbabilisticErrorCancellation2023}\\
PEC & Ion& Gate fidelity estimation&\textcite{zhangErrormitigatedQuantumGates2020}\\
PEC & Ion& Real-time dynamic simulation&\textcite{chenErrorMitigatedQuantumSimulation2023}\\
SYM & SC & Variational eigensolver & \textcite{sagastizabalExperimentalErrorMitigation2019}\\
SYM & Ion& Variational state preparation& \textcite{zhuGenerationThermofieldDouble2020}\\
PUR & SC & Variational eigensolver &\textcite{obrienPurificationbasedQuantumError2023}\\
SUB & SC & Variational eigensolver &\textcite{collessComputationMolecularSpectra2018}\\
LEA & SC & Variational eigensolver &  \textcite{dborinSimulatingGroundstateDynamical2022}\\
LEA & SC & Quantum information scrambling &  \textcite{miInformationScramblingQuantum2021}\\
SYM \& NRP& SC & Variational state preparation&  \textcite{googlequantumaiandcollaboratorsHartreeFockSuperconductingQubit2020}\\
SYM \& LEA & SC & Variational eigensolver & \textcite{stanisicObservingGroundstateProperties2022}\\
SYM \& LEA & SC & Real-time dynamic simulation & \textcite{googlequantumaiandcollaboratorsObservationSeparatedDynamics2020}
                \end{tabular}
            \end{ruledtabular}
            \endgroup
            \begin{tablenotes}\footnotesize
                \item[] SC, Ion: superconducting qubits and trapped-ion qubits.
                \item[] ZNE, PEC, SYM, PUR, SUB, NRP, LEA: zero-noise extrapolation, probabilistic error cancellation, symmetry constraints, purity constraints, subspace expansion, $N$-representability and learning-based methods. Note that measurement error mitigation is present in almost all experimental works and thus is not explicitly included here.
            \end{tablenotes}
            \caption{A list of examples of pioneering and state-of-the-art experimental applications of QEM.}
            \label{tab:application_examples}
        \end{threeparttable}
\end{table*}

\subsection{Coherent errors}\label{sec:coherent_noise}
The degree of coherence in an error channel can be quantified by how well it preserves the purity of an average incoming state~\cite{wallmanEstimatingCoherenceNoise2015}. In this sense, the ``most coherent'' errors are just unitary errors, while incoherent errors usually refer to stochastic Pauli channels. More generally, the term \emph{coherent errors} refers to a broad spectrum of noise that sits closer to the unitary errors than the Pauli errors.
There are cases in which coherent errors can be mitigated using coherent control solutions like dynamical decoupling~\cite{lidarReviewDecoherenceFreeSubspaces2014, suterColloquiumProtectingQuantum2016}. The remnant coherent errors can be mitigated using most of the QEM techniques mentioned before (with the exception of purity-based methods). In fact, subspace expansion (\cref{sec:sub_expand}) was originally constructed to target coherent errors~\cite{mccleanHybridQuantumclassicalHierarchy2017}. However for many other QEM methods, Pauli errors can be far easier to mitigate compared to coherent errors as discussed in the previous sections. The advantages of mitigating Pauli noise are summarised below:
\begin{itemize}
    \item Zero-noise extrapolation: arguments can be made that the expectation value of Pauli observable should decay following a multi-exponential curve under Pauli noise. Knowing the form of the extrapolation curve can lead to lower sampling costs and smaller biases. 
    \item Probabilistic error cancellation: Pauli noise requires less overhead to be characterised and removed using the standard basis, similarly for measurement error mitigation. 
    \item Symmetry constraints: when dealing with symmetry-preserving circuit components and Pauli symmetry, analytical arguments can be made about the proportion of noise removed when considering Pauli noise. This also enables us to select a good set of symmetries to use. 
    \item Purity constraints: errors that are less coherent can improve the ultimate accuracy that can be reached using purity constraints.
    \item Learning-based: standardising error channels by transforming them into Pauli noises will result in a more similar error model between the training circuits and target circuits. The effects of Pauli noise on the observable are usually easier to be characterised and thus require less training. 
\end{itemize}
Even for subspace expansion, though Pauli errors might not be easier to mitigate, they are certainly easier to analyse and thus enable us to select a better set of expansion operators.
Furthermore, coherent errors usually accumulate at a faster rate than their Pauli counterpart~\cite{sandersBoundingQuantumGate2015,gutierrezComparisonQuantumErrorcorrection2015,kuengComparingExperimentsFaultTolerance2016}. Hence, \emph{one of the most effective ways to deal with coherent errors in the context of QEM is simply transforming them into Pauli errors.} 

Any arbitrary quantum channel can be transformed into a Pauli channel using \emph{Pauli twirling}, which  originated from entanglement purification~\cite{bennettPurificationNoisyEntanglement1996,bennettMixedstateEntanglementQuantum1996,knillFaultTolerantPostselectedQuantum2004} and is widely used in areas like quantum benchmarking~\cite{klieschTheoryQuantumSystem2021} for transforming quantum states or noise channels into a standardised form. When applying Pauli twirling to environmental noise or noise from Clifford gates, we simply insert random Pauli gates during the circuit compilation (as discussed in \cref{sec:twirling}). Most of these random Pauli gates can in fact be absorbed into the existing Pauli gates~\cite{wallmanNoiseTailoringScalable2016} and thus it can often be implemented with a relatively small gate cost. 
It is also possible to twirl over the Clifford group, which will further homogenise the probability of different Pauli errors, resulting in a depolarising channel, which is even easier to analyse and mitigate. However, Clifford gates, especially multi-qubit ones, can be much harder to implement than Pauli gates.

\subsection{Logical errors in fault-tolerant quantum computation}\label{sec:fault_tol_error}
The ultimate goal of quantum computation is to have a fully scalable fault-tolerant quantum system in which we can suppress the logical errors to an arbitrary small level by increasing the size of the system. However, before we achieve that, it is likely that there will be an extended period of time in which quantum error correction is successfully implemented, but the minimum logical error rate achievable is still substantial due to reasons like hardware limitations in the system size, challenges in large-scale decoding, and/or insufficient resources for magic state distillation. In this `early fault-tolerant' era, many tasks of interest requiring an effectively zero error rate will be impossible to perform.
\emph{As long as the results of the fault-tolerant algorithm are obtained through expectation value estimation, the residual logical errors can be mitigated using all of the QEM methods mentioned before}. We can follow the exact same arguments outlined in the previous sections and simply replace the qubit registers, the operations and the error channels with their logical equivalents.

In particular, the role of probabilistic error cancellation in the Clifford$+T$ paradigm of universal fault-tolerant quantum computation was studied extensively in several works~\cite{suzukiQuantumErrorMitigation2022,piveteauErrorMitigationUniversal2021,lostaglioErrorMitigationQuantumAssisted2021}. The logical noise associated with Clifford gates can be transformed into logical Pauli channels using Pauli twirling by inserting random Pauli gates as mentioned in \cref{sec:coherent_noise}. To mitigate the resultant logical Pauli noise using probabilistic error cancellation, we only need to insert Pauli gates into the unmitigated circuit. These additional logical Pauli gates required for error mitigation can be applied virtually and noiselessly through updating the Pauli frame~\cite{suzukiQuantumErrorMitigation2022}, boosting the effectiveness of probabilistic error cancellation in the fault-tolerant setting. The errors in the noisy logical $T$ gate can be mitigated in a similar way, but to twirl them we need to apply additional logical Clifford gates instead of just Pauli gates (see \cref{sec:twirling}). \textcite{suzukiQuantumErrorMitigation2022} show that to achieve a logical circuit fault rate of $10^{-3}$ with a sampling overhead of $100$, probabilistic error cancellation can reduce the physical qubit overhead by $80\%$ for some classically intractable problems and more than $45\%$ in many practical fault-tolerant applications. 

The magic state distillation process needed for implementing fault-tolerant $T$ gates accounts for significant costs in fault-tolerant computation even with recent advances~\cite{ogormanQuantumComputationRealistic2017,litinskiMagicStateDistillation2019}. Therefore, it makes sense to study the details of error mitigation for imperfect logical $T$ gates. \textcite{piveteauErrorMitigationUniversal2021} consider logical $T$ gates implemented through either gate teleportation or code switching, and found that it is sufficient to consider the effective logical error as a dephasing channel (with the help of logical twirling), which is much easier to characterise and mitigate using probabilistic error cancellation. Assuming perfect Clifford gates, they showed that one can use probabilistic error cancellation to remove errors in a circuit that has $2000$ $T$ gates and a physical error rate of $10^{-3}$ at a sampling overhead of $1000$. This is well beyond the classically tractable regime of $\sim 50$ $T$ gates~\cite{bravyiImprovedClassicalSimulation2016}. \textcite{lostaglioErrorMitigationQuantumAssisted2021} also discuss applying QEM to noisy logical $T$ gates but with more emphasis on the resource theoretic aspect. There a resource measure called quantum-assisted robustness of magic is introduced, which indicates the speed up of quantum circuit simulation with noisy non-Clifford gates using probabilistic error cancellation. 

Besides gate errors, there will also be compilation errors when we try to approximate an arbitrary unitary gate using the Clifford$+T$ gate set. The form of the resultant compilation errors can be analytically computed for local gates, which can then be mitigated using probabilistic error cancellation~\cite{suzukiQuantumErrorMitigation2022}. 
The resultant compilation error will decrease exponentially with the increase of the number of $T$ gates used~\cite{kitaevQuantumComputationsAlgorithms1997,dawsonSolovayKitaevAlgorithm2006,kliuchnikovAsymptoticallyOptimalApproximation2013,rossOptimalAncillafreeClifford2016}. Hence, there is a trade-off between the number of $T$ gates used and the cost of the QEM we need to apply. We will discuss more instances of compilation errors beyond the context of gate synthesis in the next section.

\subsection{Algorithmic (compilation) errors}\label{sec:algo_error}
To execute a unitary operation, we need to first compile it into a sequence of gates that the quantum computer can efficiently execute. The compiled circuit does not always exactly represent the target unitary (often due to a limited circuit depth), and any resultant mismatches will be called \emph{compilation errors}. This will lead to errors in the output expectation value that we will call \emph{algorithmic errors}. We have discussed one instance of compilation errors in the context of gate synthesis in \cref{sec:fault_tol_error}. In applications like variational eigensolver, the target unitary is represented using some parametrised ansatz circuit, which can lead to algorithmic errors due to wrong values of parameters or limited representation power of ansatz circuits. Algorithmic errors will be present even if all gates can be executed perfectly without any noise, and thus they \emph{cannot} be removed through QEC. On the other hand, algorithmic errors can be removed through some of the QEM techniques that we have mentioned and this is one of the areas where QEM is non-trivially different from QEC. 

\textcite{endoMitigatingAlgorithmicErrors2019} made the first attempt to mitigate such algorithmic errors via zero-noise extrapolation, which essentially operates in the same way as what we described in \cref{sec:zne}, but with algorithmic errors in place of physical errors. For a given application, there will be different ways to compile the target unitary using different circuit structures (e.g. different circuit depths) or different circuit parameters, which will output data points with different algorithmic errors. If we know the relative scaling of the algorithmic errors between these data points, we can then apply zero-noise extrapolation on them to remove the algorithmic errors.
\textcite{endoMitigatingAlgorithmicErrors2019} looked at the specific problem of Hamiltonian simulation using Trotterisation~\cite{suzukiFractalDecompositionExponential1990,suzukiGeneralTheoryFractal1991}. Higher-order Trotterisation will have fewer algorithmic errors but deeper circuits. Balancing these two aspects means there is an optimal order of Trotterisation in a given practical setting. Data points of different algorithmic errors can be obtained using different orders of Trotterisation up to the optimal order, using which we can apply Richardson extrapolation to estimate the result of infinite-order Trotterisation with zero algorithmic errors. Zero-noise extrapolation can also be used to mitigate algorithmic errors in energy minimisation algorithms (quantum optimisation algorithms) like quantum annealing and variational eigensolver by extrapolating to the infinite-annealing-time limit or the zero-energy-variance limit~\cite{caoMitigatingAlgorithmicErrors2022}.

Mitigating algorithmic errors is even more natural for QEM methods that are based on the known constraints of the ideal state or the ideal observables since algorithmic errors can also violate these constraints just like gate errors. One such example was discussed by \textcite{hugginsVirtualDistillationQuantum2021}. There they look at the randomised Trotterisation~\cite{childsFasterQuantumSimulation2019,campbellRandomCompilerFast2019,ouyangCompilationStochasticHamiltonian2020} in which the compiled circuit is inherently probabilistic while the ideal state is known to be pure. Hence, it is natural to apply purification-based QEM here to remove the stochastic errors due to the randomised compilation process. Another example is subspace expansion applied in the context of variational eigensolver~\cite{mccleanHybridQuantumclassicalHierarchy2017} to overcome the limited representation power of the ansatz circuit.

\section{Open Problems}\label{sec:open_prob}
\subsection{Overarching problems}
Having thus surveyed the diverse concepts, implementations and applications of QEM in the previous sections, it is appropriate to now reflect on the key unanswered questions in the field. We identify several such questions below. 
\begin{enumerate}[leftmargin=1em, font=\sffamily]
    \item {\sffamily What is the full landscape of the zoo of QEM? }
    
    In \cref{sec:methods}, we have discussed a set of QEM methods and their variants, which are grouped into three categories in \cref{sec:QEM_two_stage}. However, this is by no means the full landscape of QEM. Are there better classifications of the different QEM methods? Moreover, are there new methods waiting to be discovered? 
    
    \item {\sffamily What are some good performance metrics for QEM?}
    
    In \cref{sec:bias_variance}, we identify the bias and variance of the error-mitigated estimator as its key performance metrics and in \cref{sec:compare_and_comb}, we further define some performance metrics by viewing QEM from other perspectives. However, these metrics are greatly dependent on the exact problem we try to solve and many of them can only be obtained analytically for some canonical cases. Are there other performance metrics for QEM that can better reflect their practical performance and/or can be analytically calculated for a wider range of methods? Can these metrics take into account the various additional information/resources required by different QEM methods? Instead of performance metrics, should we identify a set of well-defined use cases which can then be used to benchmark QEM methods through numerics or experiments, much like the MNIST or ImageNet database in computer vision research? 
    
    \item {\sffamily What is the optimal QEM strategy for some practical use case?}
    
    This is a natural question that follows from the first two. After we have a full view of all possible QEM methods out there and good performance metrics for their practical performance, we can then identify the optimal QEM strategy for some practical use cases. This is most likely to be a hybrid of different QEM methods. 
    
    \item {\sffamily What is the connection between QEM and QEC?}
    
    We have attempted to connect QEM to QEC in \cref{sec:comp_qem_qec}. There, most of the connection is made specifically for symmetry-based QEM. It would be interesting to see a more systemic approach to connect general QEM methods to QEC, or even to merge them under a larger common framework of error suppression. For this to happen, of course we would need a good description of the framework for QEM first, which is the first question we presented above.
    
    \item {\sffamily Can we extend QEM beyond expectation value estimation?}
    
    While expectation value estimation is at the heart of many useful algorithms to date, there are still many algorithms fall outside this paradigm, e.g. a repeat-until-success algorithm like Shor's algorithm, or algorithms that output a single-shot measurement, such as quantum phase estimation using the quantum Fourier transform. Can we identify an equivalent of QEM for these algorithms and usefully apply concepts such as those described in the present review? For example, it should be possible to improve single-shot algorithms via post-selection, in the same spirit of quantum error detection. 
    
\end{enumerate}

\subsection{Technical questions}
There are also many other more technical questions about the implementations and applications of different QEM techniques, many of which we have mentioned in this review when talking about the individual techniques. We will further list some example questions here.

\begin{enumerate}[leftmargin=1em]
    \item  Are there general protocols to scale the noise without changing the error model? If yes, can we deduce the shape of the extrapolation curve from the error model? (\cref{sec:zne})
    \item How can we efficiently handle drifts in the error model when applying probabilistic error cancellation in practice? (\cref{sec:pec})
    \item What is the training cost for learning-based QEM? (\cref{sec:learning-based})
    \item  Various quantum computing hardware architectures have been realised, at least at the prototype level, with differing connectivities and native gate sets. What is the interplay between the hardware feature set and the suitable choices of QEM strategy?
\end{enumerate}

\section{Conclusion}\label{sec:concl}
In this review, we have provided a comprehensive survey of QEM which has ranged from the basic concepts and motivations, through to the implementation details of specific techniques. Due to the low hardware requirements of QEM compared to QEC, as well as the broad range of mitigation techniques available to target diverse application scenarios, QEM has already become an integral part of many recent experimental demonstrations of quantum hardware. Even though we cannot rely on QEM alone to suppress \emph{all} errors in all cases, especially when the circuit fault rate is large, we can always aim for a sweet spot between the amount of noise removed and the resource required. Consequently we can expect that QEM techniques will continue to establish themselves as indispensable enablers, vital to maximising the reach of each generation of hardware. We anticipate that this will remain true even into the fault-tolerant era, since recent works have shown that it is possible to reduce the hardware requirements of fault-tolerance by applying QEM alongside QEC. Indeed there are applications of QEM concepts entirely beyond handling physical device imperfections and instead mitigating compilation (algorithmic) imperfections, i.e. honing the performance of algorithms which of necessity produce only approximate answers.

Despite already playing a significant role in practical applications, the current landscape of QEM is still dynamic and complex, with many unexplored territories. While we have made every effort to present the rather tangled threads of this topic in a clear way in this review, it is evident that a more systematic and unified structure for QEM is desirable as the field matures further. There are still many open problems left unanswered as summarised in \cref{sec:open_prob}. Solving these problems may be the key to a clearer and more structured view of QEM and its role in the grand scheme of error suppression. We hope that the community of theoretical and experimental researchers who will drive the field forwards, may find that the survey of ideas and methods presented in this review provides useful guidance along the way.

\begin{acknowledgments}
    We thank Kristan Temme, Abhinav Kandala, and Jay Gambetta for their valuable insights and useful discussions. ZC and SE are grateful to B\'alint Koczor, Ryuji Takagi, and Nobuyuki Yoshioka for helpful discussions. SE is also grateful for useful discussions with Yasunari Suzuki, Kento Tsubouchi, and Raam Uzdin. RB thanks Nicholas Rubin for helpful discussions about pure state $N$-representability. 
    ZC is supported by the Junior Research Fellowship from St John’s College, Oxford. 
    SCB acknowledges financial support from EPSRC Hub grants under agreement No. EP/T001062/1, and from the IARPA funded LogiQ project. ZC and SCB also acknowledge support from EPSRC’s Robust and Reliable Quantum Computing (RoaRQ) project (EP/W032635/1).
    SE is supported by Moonshot R\&D, JST, Grant No.\,JPMJMS2061; MEXT Q-LEAP Grant No.\,JPMXS0120319794, and PRESTO, JST, Grant No.\, JPMJPR2114.
    YL acknowledges the support of the National Natural Science Foundation of China (Grants No. 11875050 and No. 12088101) and NSAF (Grant No. U1930403).
\end{acknowledgments}

\section*{List of Symbols}
\vspace*{-\baselineskip}
\begingroup
\renewcommand{\arraystretch}{2}
\setlength{\tabcolsep}{0em}
{\makeatletter
\renewcommand\table@hook{\normalsize}
\makeatother
\newlength{\symbolwidth}
\setlength{\symbolwidth}{4em}
\begin{longtable}{@{}p{\symbolwidth}p{\dimexpr 24em-\symbolwidth\relax}@{}} 
    $C_{\mathrm{em}}$ & error mitigation sampling overhead\\
    $N$&number of qubits\\
    $N_{\mathrm{cir}}$&{number of circuit runs}\\
    $O$ & {observable of interest}\\
    $\est{O}_{\mathrm{em}}$&{error-mitigated estimator}\\
    $\est{O}_{\rho}$&{random variable from measuring $O$ on state $\rho$}\\
    $p$&{physical gate error rate}\\
    $P_0$&{circuit fault-free probability}\\
    $\lambda$&{circuit fault rate, the average number of faults per circuit run}\\
    $\rho$&{unmitigated noisy state}\\
    $\rho_0$&{ideal noiseless state}
\end{longtable}}
\endgroup

\appendix

\section{Practical Techniques in Implementations}\label{sec:prac_tech}

\subsection{Monte carlo sampling}\label{sec:monte_carlo_sampling}
In many QEM techniques, very often the error-mitigated expectation value $\expect{\est{O}_{\mathrm{em}}}$ is a linear sum of the expectation values of a set of $K$ random variables $\{\est{O}_{n}\}$ (that are the outputs of the set of response measurement circuits for the QEM technique):
\begin{align}\label{eqn:em_est_decomp}
    \expect{\est{O}_{\mathrm{em}}} =  \sum_{n = 1}^{K} \alpha_{n} \expect{\est{O}_{n}},
\end{align}
where $\{\alpha_{n}\}$ are real coefficients. A naive way for estimating $\expect{\est{O}_{\mathrm{em}}}$ would be performing estimation for the individual terms $\expect{\est{O}_{n}}$ up to a certain precision, then combine the results. In such a way, the variance of the error-mitigated estimator $\est{O}_{\mathrm{em}}$ is given as:
\begin{align*}
    \var{\est{O}_{\mathrm{em}}} =  \sum_{n = 1}^{K} \abs{\alpha_{n}}^2 \var{\est{O}_{n}},
\end{align*}
The component random variables $\est{O}_{n}$, being generated from circuits that are variants of the primary circuit, can be expected to have a similar variance as the unmitigated estimator $\est{O}_{\rho}$ generated from the noisy primary circuit: $\var{\est{O}_{n}} \sim \var{\est{O}_{\rho}}$. Hence, we have:
\begin{align*}
    \var{\est{O}_{\mathrm{em}}} =  \left(\sum_{n = 1}^{K} \abs{\alpha_{n}}^2\right) \var{\est{O}_{\rho}},
\end{align*}
Therefore, each component observable $\est{O}_{n}$ is associated with a sampling overhead of $\sum_{n = 1}^{K} \abs{\alpha_{n}}^2$. Since there are $K$ of them, the total sampling overhead is
\begin{align}\label{eqn:naive_cost}
    C_{\mathrm{em}}^{\mathrm{naive}} = K \sum_{n = 1}^{K} \abs{\alpha_{n}}^2.   
\end{align}
Such a method is not scalable if the number of terms $K$ is large, which is the case for many QEM methods. Instead, we can construct the estimator $\est{O}_{\mathrm{em}}$ using Monte Carlo methods. We can rewrite \cref{eqn:em_est_decomp} into
\begin{align}\label{eqn:lin_comb_err_miti}
    \expect{\est{O}_{\mathrm{em}}} = A\sum_{n=1}^K \frac{\abs{\alpha_{n}}}{A} \sign(\alpha_{n}) \expect{\est{O}_{n}} = A \expect{\est{O}_{\mathrm{mix}}},
\end{align}
where $A = \sum_{n=1}^{K} \abs{\alpha_{n}}$. Here we have define a new random variable $\est{O}_{\mathrm{mix}}$ which is a probabilistic mixture of the set of random variables $\{\sign(\alpha_{n}) \est{O}_{n}\}$ with $\sign(\alpha_{n}) \est{O}_{n}$ being chosen with the probability $\frac{\abs{\alpha_{n}}}{A}$. Each sample of $\est{O}_{\mathrm{em}}$ is just a sample of $\est{O}_{\mathrm{mix}}$ scaled by the factor $A$
\begin{align*}
    \est{O}_{\mathrm{em}} = A \est{O}_{\mathrm{mix}}.
\end{align*} 
Using the same $\var{\est{O}_{n}} \sim \var{\est{O}_{\rho}}$ assumption, we then have $\var{\est{O}_{\mathrm{mix}}} \sim \var{\est{O}_{\rho}} $ and thus
\begin{align*}
    \var{\est{O}_{\mathrm{em}}}  = A^2 \var{\est{O}_{\mathrm{mix}}} \sim A^2 \var{\est{O}_{\rho}}.
\end{align*}
Hence, the sampling overhead (\cref{eqn:samp_cost}) of performing error-mitigated estimation using $\est{O}_{\mathrm{em}}$ instead of $\est{O}_{\rho}$ is then
\begin{align}\label{eqn:monte_cost}
    C_{\mathrm{em}}^{\mathrm{MC}} = A^2 = \big(\sum\nolimits_{n=1}^{K} \abs{\alpha_{n}}\big)^2.
\end{align}
Using the Cauchy–Schwarz inequality to compare the sampling cost in \cref{eqn:naive_cost} and \cref{eqn:monte_cost}, we have
\begin{align*}
    \big(\sum\nolimits_{n=1}^K \abs{\alpha_{n}}\big)^2 &\leq K \sum\nolimits_{n=1}^K \abs{\alpha_{n}}^2\\
    \implies C_{\mathrm{em}}^{\mathrm{MC}} &\leq  C_{\mathrm{em}}^{\mathrm{naive}}.
\end{align*}
i.e. the Monte Carlo method is always more sample efficient under our assumptions. 

Instead of making assumptions about the variance of the component random variables $\est{O}_{n}$, we can also obtain a similar sampling overhead using Hoeffding's inequality as shown in \cref{eqn:samp_cost_hoeff}. This uses the fact that the component random variables $\est{O}_{n}$ usually have the same range as $\est{O}_{\rho}$ since they are usually obtained from the measurement of the same observable, or even if the observables are different, they are often all Pauli observables which have the same range. 

Here we have talked about using Monte Carlo sampling for estimating the error-mitigated expectation value. Similar arguments can also be applied to the estimation of loss functions in learning-based methods and other similar situations.

\subsubsection{Exponential sampling overhead}\label{sec:monte_carlo_sampling_exp}
As mentioned in \cref{sec:exp_samp_cost}, if the component random variables $\{\est{O}_n\}$ are obtained from measuring Pauli observables on circuits suffering from Pauli gate noise with a circuit fault rate of $\lambda$ or more, then its expectation value will decay exponentially with $\lambda$:
\begin{align*}
    \expect{\est{O}_n} = \order{e^{-\beta_n \lambda}}
\end{align*}
for some positive $\beta_n$. We also have:
\begin{align*}
    \var{\est{O}_n} = \expect{\est{O}_n^2} - \expect{\est{O}_n}^2 = 1 - \order{e^{-2\beta_n \lambda}}
\end{align*}
where we have used $\est{O}_n^2 = 1$ for Pauli observable.

As mentioned above, $\est{O}_{\mathrm{mix}}$ is the probabilistic mixture of the set of random variable $\{\sign(\alpha_{n})\est{O}_n\}$ with the probability distribution $\{p_{n} = \frac{\abs{\alpha_{n}}}{A}\}$. Using the properties of random variables from mixture distribution, we have:
\begin{align*}
     \expect{\est{O}_{\mathrm{mix}}}  = \sum_{n} p_n \sign(\alpha_{n})\expect{\est{O}_n} = \order{e^{-\beta \lambda}}
\end{align*}
with $\beta = \min_{n} \beta_n$ and we also have:
\begin{align*}
    \var{\est{O}_{\mathrm{mix}}}  &= \expect{\est{O}_{\mathrm{mix}}^2} - \expect{\est{O}_{\mathrm{mix}}}^2\\
    &= \sum_{n} p_n \expect{\est{O}_n^2} - \expect{\est{O}_{\mathrm{mix}}}^2\\
    & = 1 - \order{e^{-2\beta \lambda}}
\end{align*}
where we have again used $\est{O}_n^2 =1$. This shows that when we are considering Pauli observable under Pauli circuit noise, the assumption $\var{\est{O}_{\mathrm{mix}}} \sim \var{\est{O}_{n}}$ mentioned in the last section is valid at large $\lambda$ even without needing to consider how the various $\est{O}_{n}$ are exactly constructed.

Since the error-mitigated estimator is $\est{O}_{\mathrm{em}} = A \est{O}_{\mathrm{mix}}$, we then have:
\begin{align*}
    \expect{\est{O}_{\mathrm{em}}}  &= A\expect{\est{O}_{\mathrm{mix}}} = \order{Ae^{-\beta \lambda}}
\end{align*}
In order to ensure the expectation value of the error-mitigated estimator does not decay with the noise level $\lambda$, we need to have $A = \order{e^{\beta \lambda}}$, which implies that the sampling cost using \cref{eqn:monte_cost} is
\begin{align*}
    C_{\mathrm{em}} = A^2 = \order{e^{2\beta \lambda}}
\end{align*}
i.e. it increases exponentially with $\lambda$.

\subsection{Pauli twirling}\label{sec:twirling}
Given a noise process $\mathcal{N}$, twirling it over a symmetry group $\mathbb{G}$ means conjugating $\mathcal{N}$ with random elements in $\mathbb{G}$:
\begin{align*}
    \mathbf{T}_{\mathbb{G}}(\mathcal{N}) = \frac{1}{\abs{\mathbb{G}}} \sum_{\mathcal{G} \in \mathbb{G}} \mathcal{G} \mathcal{N} \mathcal{G}. 
\end{align*}
Note that here we are using the super-operator formalism. Twirling over the Pauli group, which is called Pauli twirling, will remove all the off-diagonal elements of $\mathcal{N}$ in the Pauli basis, transforming the error channel into Pauli noise. 

Now suppose we have a noisy Clifford gate $\mathcal{C}_{\epsilon}$ that is simply the ideal Clifford Channel $\mathcal{C}$ followed by the noise channel $\mathcal{N}$: $\mathcal{C}_{\epsilon} = \mathcal{N} \mathcal{C}$. If we want to twirl the noise channel $\mathcal{N}$ of the noisy Clifford gate $\mathcal{C}_{\epsilon}$, we can apply $\mathcal{C}_{\epsilon}$ with a random Pauli $\mathcal{G}$ and its corresponding Pauli $\mathcal{C} \mathcal{G} \mathcal{C}^\dagger$ in each circuit run:
\begin{align*}
    \mathbf{T}_{\mathbb{G}}'(\mathcal{C}_{\epsilon}) &= \frac{1}{\abs{\mathbb{G}}} \sum_{\mathcal{G} \in \mathbb{G}} \left(\mathcal{C} \mathcal{G} \mathcal{C}^\dagger\right) \mathcal{C}_{\epsilon} \mathcal{G} = \frac{1}{\abs{\mathbb{G}}} \sum_{\mathcal{G}' \in \mathbb{G}} \mathcal{G}' \mathcal{C}_{\epsilon} \left(\mathcal{C}^\dagger \mathcal{G}' \mathcal{C}\right)\\
    &= \left(\frac{1}{\abs{\mathbb{G}}} \sum_{\mathcal{G} \in \mathbb{G}} \mathcal{G} \mathcal{N} \mathcal{G}\right) \mathcal{C}. 
\end{align*}
To twirl the error of a sequence of Clifford gates $\prod_{m=M}^{1} \mathcal{C}_m$, the random Pauli for twirling consecutive Clifford gates can be merged. 

To twirl a noisy $T$ gate: $\mathcal{T}_{\epsilon} = \mathcal{N} \mathcal{T}$, we then need to apply the following circuit
\begin{align*}
    \mathbf{T}_{\mathbb{G}}'(\mathcal{T}_{\epsilon}) &= \frac{1}{\abs{\mathbb{G}}} \sum_{\mathcal{G}' \in \mathbb{G}} \mathcal{G} \mathcal{T}_{\epsilon} \left(\mathcal{T}^\dagger \mathcal{G} \mathcal{T}\right) = \left(\frac{1}{\abs{\mathbb{G}}} \sum_{\mathcal{G} \in \mathbb{G}} \mathcal{G} \mathcal{N} \mathcal{G}\right) \mathcal{T},
\end{align*}
where $\mathcal{T}^\dagger \mathcal{G} \mathcal{T}$ is some Clifford gates.

\subsection{Measurement techniques}\label{sec:meas_tech}
To measure an arbitrary operator, we can always measure its Pauli basis and then combine the results. Hence, without loss of generality, we will be mostly focusing on Pauli measurements in this section.

\begin{figure}[htbp]
        \centering
        \includegraphics[width = 0.44\textwidth]{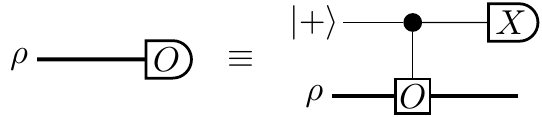}
        \caption{Hadamard test circuit for performing Pauli $O$ measurement. If $O$ is some general unitary, then the Hadamard test circuit above in which we measure $X\otimes I$ will output the expectation value $\Re(\Tr[O\rho])$, while measuring $Y\otimes I$ will output the expectation value $\Im(\Tr[O\rho])$.}
        \label{fig:Htest}
\end{figure}

In practical experiments, it is often the case that we can only perform single-qubit $Z$ measurements to high accuracy. Hence, one way to measure a Pauli operator is by transforming it into single-qubit $Z$ using Clifford circuits. Given linear qubit connectivity, the additional Clifford circuits needed will require long-range gates of depth $\order{\log(N)}$ or local gates of depth $\order{N}$, which is discussed in the context of symmetry verification by \textcite{bonet-monroigLowcostErrorMitigation2018}. Alternatively, for a given Pauli operator $O$, if we can implement controlled-$O$ gates, we can also \emph{indirectly} measure $O$ through the Hadamard test as shown in \cref{fig:Htest}. The controlled-$O$ gate can be implemented using a Clifford circuit that transforms $O$ to single-qubit $Z$ along with a controlled-$Z$ gate. The cost of this is similar to the direct measurement discussed above. 

Any given Pauli operator $O$ can be written as the tensor product of single-qubit Pauli operators $O = \bigotimes_{n=1}^N G_n$ where $G_n$ is the action of $O$ on the $n^{\text{th}}$ qubit. Hence, the controlled-$O$ in the Hadamard test can also be decomposed into low-weight controlled-$G_n$ gates instead. In fact, we can measure $O$ directly by performing single-qubit Pauli measurements of its components $\{G_n\}$ and multiplying the results. In this way, the additional circuitry needed for measuring $O$ is only one layer of single-qubit Clifford for changing the measurement basis.

Now suppose we want to measure \emph{two commuting} Pauli operators in the same circuit run, which can be useful in for example symmetry verification, we can use the single-qubit measurements plus post-processing scheme mentioned above, but we must make sure the two Pauli operators \emph{qubit-wise} commute~\cite{izmaylovRevisingMeasurementProcess2019}, i.e. for every qubit, the single-qubit Pauli components acting on it from different observables commute. For example, $XXI$ and $IXZ$ are qubit-wise commuting, but $XX$ and $YY$ are not. Note that for a set of operators to qubit-wise commute, their actions on a given qubit must be the same Pauli operator or the identity. More generally, the set of operators that are linear combinations of a set of qubit-wise commuting Pauli operators are also qubit-wise commuting and can be measured simultaneously using single-qubit Pauli measurements in the same circuit run. Whenever two Pauli operators commute, we can make them qubit-wise commuting using a suitable choice of Clifford circuit. Examples of this have been discussed in direct symmetry verification in \cref{sec:sym}. 

In some other cases, we are more interested in the expectation values of a set of commuting operators instead of their exact measurement results in any given circuit run. In these scenarios, instead of trying to measure multiple observables in every circuit run, it is possible to obtain these expectation values through \emph{shadow tomography}~\cite{huangPredictingManyProperties2020} using random single-qubit Pauli measurement and post-processing. 
 
\begin{figure}[htbp]
        \centering
        \includegraphics[width = 0.44\textwidth]{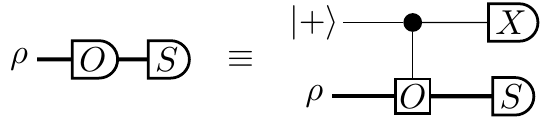}
        \caption{Circuits for measuring $OS_+ = \frac{OS + SO}{2}$ in which $O$ is Pauli and $S$ is Hermitian. On the left circuit, we first perform a non-destructive measurement of $O$ and then measure $S$ and take the product of the measurement results.  On the right circuit, the non-destructive measurement of $O$ is carried out using Hadamard tests and the expectation value of $OS_+$ is simply obtained by measuring $X \otimes S$ at the end.}
        \label{fig:HtestProd}
\end{figure}

Now let us move beyond measuring commuting Pauli operators and consider the case in which we want to measure the \emph{product} of a Pauli operator $O$ and some general Hermitian operator $S$ that does not necessarily commute with $O$. For the circuits shown in \cref{fig:HtestProd}, if we perform a projective measurement of the Pauli operator $O$ first, and then measure the operator $S$, then the latter measurement is equivalent to measuring the components of $S$ that commute with $O$, which is simply $S_+ = \frac{S + OSO}{2}$~\cite{mitaraiMethodologyReplacingIndirect2019,caiResourceefficientPurificationbasedQuantum2021,huoDualstatePurificationPractical2022}. Taking the product of the two measurements, we can obtain the expectation value of the symmetrised product $OS_+ = \frac{OS + SO}{2}$, which is useful in symmetry verification in \cref{sec:sym} (in which $S$ is the symmetry) and echo verification in \cref{sec:pur} (in which $S$ is the dual state). 

The component of $S$ that anti-commute with $O$, denoted as $S_- = \frac{S - OSO}{2}$, can be obtained by first applying the Pauli rotation $\exp(-i\frac{\pi}{4}O)$ and then measuring $S$~\cite{mitaraiMethodologyReplacingIndirect2019,caiResourceefficientPurificationbasedQuantum2021,huoDualstatePurificationPractical2022}. Combined with the measurement of $S_+$ above, we can obtain the expectation value of the product observable $OS$. Alternatively, if we can implement controlled-$O$ and controlled-$S$, then they can be composed to give controlled-$OS$, which can be used to measure $OS$ using the Hadamard test.


\begin{thebibliography}{247}%
\makeatletter
\providecommand \@ifxundefined [1]{%
 \@ifx{#1\undefined}
}%
\providecommand \@ifnum [1]{%
 \ifnum #1\expandafter \@firstoftwo
 \else \expandafter \@secondoftwo
 \fi
}%
\providecommand \@ifx [1]{%
 \ifx #1\expandafter \@firstoftwo
 \else \expandafter \@secondoftwo
 \fi
}%
\providecommand \natexlab [1]{#1}%
\providecommand \enquote  [1]{``#1''}%
\providecommand \bibnamefont  [1]{#1}%
\providecommand \bibfnamefont [1]{#1}%
\providecommand \citenamefont [1]{#1}%
\providecommand \href@noop [0]{\@secondoftwo}%
\providecommand \href [0]{\begingroup \@sanitize@url \@href}%
\providecommand \@href[1]{\@@startlink{#1}\@@href}%
\providecommand \@@href[1]{\endgroup#1\@@endlink}%
\providecommand \@sanitize@url [0]{\catcode `\\12\catcode `\$12\catcode
  `\&12\catcode `\#12\catcode `\^12\catcode `\_12\catcode `\%12\relax}%
\providecommand \@@startlink[1]{}%
\providecommand \@@endlink[0]{}%
\providecommand \url  [0]{\begingroup\@sanitize@url \@url }%
\providecommand \@url [1]{\endgroup\@href {#1}{\urlprefix }}%
\providecommand \urlprefix  [0]{URL }%
\providecommand \Eprint [0]{\href }%
\providecommand \doibase [0]{https://doi.org/}%
\providecommand \selectlanguage [0]{\@gobble}%
\providecommand \bibinfo  [0]{\@secondoftwo}%
\providecommand \bibfield  [0]{\@secondoftwo}%
\providecommand \translation [1]{[#1]}%
\providecommand \BibitemOpen [0]{}%
\providecommand \bibitemStop [0]{}%
\providecommand \bibitemNoStop [0]{.\EOS\space}%
\providecommand \EOS [0]{\spacefactor3000\relax}%
\providecommand \BibitemShut  [1]{\csname bibitem#1\endcsname}%
\let\auto@bib@innerbib\@empty
\bibitem [{\citenamefont {Abobeih}\ \emph {et~al.}(2022)\citenamefont
  {Abobeih}, \citenamefont {Wang}, \citenamefont {Randall}, \citenamefont
  {Loenen}, \citenamefont {Bradley}, \citenamefont {Markham}, \citenamefont
  {Twitchen}, \citenamefont {Terhal},\ and\ \citenamefont
  {Taminiau}}]{abobeihFaulttolerantOperationLogical2022}%
  \BibitemOpen
  \bibfield  {author} {\bibinfo {author} {\bibnamefont {Abobeih}, \bibfnamefont
  {M~H}}, \bibinfo {author} {\bibfnamefont {Y.}~\bibnamefont {Wang}}, \bibinfo
  {author} {\bibfnamefont {J.}~\bibnamefont {Randall}}, \bibinfo {author}
  {\bibfnamefont {S.~J.~H.}\ \bibnamefont {Loenen}}, \bibinfo {author}
  {\bibfnamefont {C.~E.}\ \bibnamefont {Bradley}}, \bibinfo {author}
  {\bibfnamefont {M.}~\bibnamefont {Markham}}, \bibinfo {author} {\bibfnamefont
  {D.~J.}\ \bibnamefont {Twitchen}}, \bibinfo {author} {\bibfnamefont {B.~M.}\
  \bibnamefont {Terhal}}, and\ \bibinfo {author} {\bibfnamefont {T.~H.}\
  \bibnamefont {Taminiau}}} (\bibinfo {year} {2022}),\ \bibfield  {title}
  {\enquote {\bibinfo {title} {Fault-tolerant operation of a logical qubit in a
  diamond quantum processor},}\ }\href
  {https://doi.org/10.1038/s41586-022-04819-6} {\bibfield  {journal} {\bibinfo
  {journal} {Nature}\ }\textbf {\bibinfo {volume} {606}}~(\bibinfo {number}
  {7916}),\ \bibinfo {pages} {884--889}}\BibitemShut {NoStop}%
\bibitem [{\citenamefont {Ac{\'i}n}\ \emph {et~al.}(2018)\citenamefont
  {Ac{\'i}n}, \citenamefont {Bloch}, \citenamefont {Buhrman}, \citenamefont
  {Calarco}, \citenamefont {Eichler}, \citenamefont {Eisert}, \citenamefont
  {Esteve}, \citenamefont {Gisin}, \citenamefont {Glaser}, \citenamefont
  {Jelezko}, \citenamefont {Kuhr}, \citenamefont {Lewenstein}, \citenamefont
  {Riedel}, \citenamefont {Schmidt}, \citenamefont {Thew}, \citenamefont
  {Wallraff}, \citenamefont {Walmsley},\ and\ \citenamefont
  {Wilhelm}}]{acinQuantumTechnologiesRoadmap2018}%
  \BibitemOpen
  \bibfield  {author} {\bibinfo {author} {\bibnamefont {Ac{\'i}n},
  \bibfnamefont {Antonio}}, \bibinfo {author} {\bibfnamefont {Immanuel}\
  \bibnamefont {Bloch}}, \bibinfo {author} {\bibfnamefont {Harry}\ \bibnamefont
  {Buhrman}}, \bibinfo {author} {\bibfnamefont {Tommaso}\ \bibnamefont
  {Calarco}}, \bibinfo {author} {\bibfnamefont {Christopher}\ \bibnamefont
  {Eichler}}, \bibinfo {author} {\bibfnamefont {Jens}\ \bibnamefont {Eisert}},
  \bibinfo {author} {\bibfnamefont {Daniel}\ \bibnamefont {Esteve}}, \bibinfo
  {author} {\bibfnamefont {Nicolas}\ \bibnamefont {Gisin}}, \bibinfo {author}
  {\bibfnamefont {Steffen~J.}\ \bibnamefont {Glaser}}, \bibinfo {author}
  {\bibfnamefont {Fedor}\ \bibnamefont {Jelezko}}, \bibinfo {author}
  {\bibfnamefont {Stefan}\ \bibnamefont {Kuhr}}, \bibinfo {author}
  {\bibfnamefont {Maciej}\ \bibnamefont {Lewenstein}}, \bibinfo {author}
  {\bibfnamefont {Max~F.}\ \bibnamefont {Riedel}}, \bibinfo {author}
  {\bibfnamefont {Piet~O.}\ \bibnamefont {Schmidt}}, \bibinfo {author}
  {\bibfnamefont {Rob}\ \bibnamefont {Thew}}, \bibinfo {author} {\bibfnamefont
  {Andreas}\ \bibnamefont {Wallraff}}, \bibinfo {author} {\bibfnamefont {Ian}\
  \bibnamefont {Walmsley}}, and\ \bibinfo {author} {\bibfnamefont {Frank~K.}\
  \bibnamefont {Wilhelm}}} (\bibinfo {year} {2018}),\ \bibfield  {title}
  {\enquote {\bibinfo {title} {The quantum technologies roadmap: A {{European}}
  community view},}\ }\href {https://doi.org/10.1088/1367-2630/aad1ea}
  {\bibfield  {journal} {\bibinfo  {journal} {New Journal of Physics}\ }\textbf
  {\bibinfo {volume} {20}}~(\bibinfo {number} {8}),\ \bibinfo {pages}
  {080201}}\BibitemShut {NoStop}%
\bibitem [{\citenamefont {Aharonov}\ and\ \citenamefont
  {{Ben-Or}}(1997)}]{aharonovFaulttolerantQuantumComputation1997}%
  \BibitemOpen
  \bibfield  {author} {\bibinfo {author} {\bibnamefont {Aharonov},
  \bibfnamefont {D}}, and\ \bibinfo {author} {\bibfnamefont {M.}~\bibnamefont
  {{Ben-Or}}}} (\bibinfo {year} {1997}),\ \bibfield  {title} {\enquote
  {\bibinfo {title} {Fault-tolerant {{Quantum Computation}} with {{Constant
  Error}}},}\ }in\ \href {https://doi.org/10.1145/258533.258579} {\emph
  {\bibinfo {booktitle} {Proceedings of the {{Twenty-ninth Annual ACM
  Symposium}} on {{Theory}} of {{Computing}}}}},\ \bibinfo {series and number}
  {{{STOC}} '97}\ (\bibinfo  {publisher} {{ACM}},\ \bibinfo {address} {{New
  York, NY, USA}})\ pp.\ \bibinfo {pages} {176--188}\BibitemShut {NoStop}%
\bibitem [{\citenamefont {Altman}\ \emph {et~al.}(2021)\citenamefont {Altman},
  \citenamefont {Brown}, \citenamefont {Carleo}, \citenamefont {Carr},
  \citenamefont {Demler}, \citenamefont {Chin}, \citenamefont {DeMarco},
  \citenamefont {Economou}, \citenamefont {Eriksson}, \citenamefont {Fu},
  \citenamefont {Greiner}, \citenamefont {Hazzard}, \citenamefont {Hulet},
  \citenamefont {Koll{\'a}r}, \citenamefont {Lev}, \citenamefont {Lukin},
  \citenamefont {Ma}, \citenamefont {Mi}, \citenamefont {Misra}, \citenamefont
  {Monroe}, \citenamefont {Murch}, \citenamefont {Nazario}, \citenamefont {Ni},
  \citenamefont {Potter}, \citenamefont {Roushan}, \citenamefont {Saffman},
  \citenamefont {{Schleier-Smith}}, \citenamefont {Siddiqi}, \citenamefont
  {Simmonds}, \citenamefont {Singh}, \citenamefont {Spielman}, \citenamefont
  {Temme}, \citenamefont {Weiss}, \citenamefont {Vu{\v c}kovi{\'c}},
  \citenamefont {Vuleti{\'c}}, \citenamefont {Ye},\ and\ \citenamefont
  {Zwierlein}}]{altmanQuantumSimulatorsArchitectures2021}%
  \BibitemOpen
  \bibfield  {author} {\bibinfo {author} {\bibnamefont {Altman}, \bibfnamefont
  {Ehud}}, \bibinfo {author} {\bibfnamefont {Kenneth~R.}\ \bibnamefont
  {Brown}}, \bibinfo {author} {\bibfnamefont {Giuseppe}\ \bibnamefont
  {Carleo}}, \bibinfo {author} {\bibfnamefont {Lincoln~D.}\ \bibnamefont
  {Carr}}, \bibinfo {author} {\bibfnamefont {Eugene}\ \bibnamefont {Demler}},
  \bibinfo {author} {\bibfnamefont {Cheng}\ \bibnamefont {Chin}}, \bibinfo
  {author} {\bibfnamefont {Brian}\ \bibnamefont {DeMarco}}, \bibinfo {author}
  {\bibfnamefont {Sophia~E.}\ \bibnamefont {Economou}}, \bibinfo {author}
  {\bibfnamefont {Mark~A.}\ \bibnamefont {Eriksson}}, \bibinfo {author}
  {\bibfnamefont {Kai-Mei~C.}\ \bibnamefont {Fu}}, \bibinfo {author}
  {\bibfnamefont {Markus}\ \bibnamefont {Greiner}}, \bibinfo {author}
  {\bibfnamefont {Kaden~R.A.}\ \bibnamefont {Hazzard}}, \bibinfo {author}
  {\bibfnamefont {Randall~G.}\ \bibnamefont {Hulet}}, \bibinfo {author}
  {\bibfnamefont {Alicia~J.}\ \bibnamefont {Koll{\'a}r}}, \bibinfo {author}
  {\bibfnamefont {Benjamin~L.}\ \bibnamefont {Lev}}, \bibinfo {author}
  {\bibfnamefont {Mikhail~D.}\ \bibnamefont {Lukin}}, \bibinfo {author}
  {\bibfnamefont {Ruichao}\ \bibnamefont {Ma}}, \bibinfo {author}
  {\bibfnamefont {Xiao}\ \bibnamefont {Mi}}, \bibinfo {author} {\bibfnamefont
  {Shashank}\ \bibnamefont {Misra}}, \bibinfo {author} {\bibfnamefont
  {Christopher}\ \bibnamefont {Monroe}}, \bibinfo {author} {\bibfnamefont
  {Kater}\ \bibnamefont {Murch}}, \bibinfo {author} {\bibfnamefont {Zaira}\
  \bibnamefont {Nazario}}, \bibinfo {author} {\bibfnamefont {Kang-Kuen}\
  \bibnamefont {Ni}}, \bibinfo {author} {\bibfnamefont {Andrew~C.}\
  \bibnamefont {Potter}}, \bibinfo {author} {\bibfnamefont {Pedram}\
  \bibnamefont {Roushan}}, \bibinfo {author} {\bibfnamefont {Mark}\
  \bibnamefont {Saffman}}, \bibinfo {author} {\bibfnamefont {Monika}\
  \bibnamefont {{Schleier-Smith}}}, \bibinfo {author} {\bibfnamefont {Irfan}\
  \bibnamefont {Siddiqi}}, \bibinfo {author} {\bibfnamefont {Raymond}\
  \bibnamefont {Simmonds}}, \bibinfo {author} {\bibfnamefont {Meenakshi}\
  \bibnamefont {Singh}}, \bibinfo {author} {\bibfnamefont {I.B.}\ \bibnamefont
  {Spielman}}, \bibinfo {author} {\bibfnamefont {Kristan}\ \bibnamefont
  {Temme}}, \bibinfo {author} {\bibfnamefont {David~S.}\ \bibnamefont {Weiss}},
  \bibinfo {author} {\bibfnamefont {Jelena}\ \bibnamefont {Vu{\v c}kovi{\'c}}},
  \bibinfo {author} {\bibfnamefont {Vladan}\ \bibnamefont {Vuleti{\'c}}},
  \bibinfo {author} {\bibfnamefont {Jun}\ \bibnamefont {Ye}}, and\ \bibinfo
  {author} {\bibfnamefont {Martin}\ \bibnamefont {Zwierlein}}} (\bibinfo {year}
  {2021}),\ \bibfield  {title} {\enquote {\bibinfo {title} {Quantum
  {{Simulators}}: {{Architectures}} and {{Opportunities}}},}\ }\href
  {https://doi.org/10.1103/PRXQuantum.2.017003} {\bibfield  {journal} {\bibinfo
   {journal} {PRX Quantum}\ }\textbf {\bibinfo {volume} {2}}~(\bibinfo {number}
  {1}),\ \bibinfo {pages} {017003}}\BibitemShut {NoStop}%
\bibitem [{\citenamefont {Arute}\ \emph {et~al.}(2019)\citenamefont {Arute},
  \citenamefont {Arya}, \citenamefont {Babbush}, \citenamefont {Bacon},
  \citenamefont {Bardin}, \citenamefont {Barends}, \citenamefont {Biswas},
  \citenamefont {Boixo}, \citenamefont {Brandao}, \citenamefont {Buell},
  \citenamefont {Burkett}, \citenamefont {Chen}, \citenamefont {Chen},
  \citenamefont {Chiaro}, \citenamefont {Collins}, \citenamefont {Courtney},
  \citenamefont {Dunsworth}, \citenamefont {Farhi}, \citenamefont {Foxen},
  \citenamefont {Fowler}, \citenamefont {Gidney}, \citenamefont {Giustina},
  \citenamefont {Graff}, \citenamefont {Guerin}, \citenamefont {Habegger},
  \citenamefont {Harrigan}, \citenamefont {Hartmann}, \citenamefont {Ho},
  \citenamefont {Hoffmann}, \citenamefont {Huang}, \citenamefont {Humble},
  \citenamefont {Isakov}, \citenamefont {Jeffrey}, \citenamefont {Jiang},
  \citenamefont {Kafri}, \citenamefont {Kechedzhi}, \citenamefont {Kelly},
  \citenamefont {Klimov}, \citenamefont {Knysh}, \citenamefont {Korotkov},
  \citenamefont {Kostritsa}, \citenamefont {Landhuis}, \citenamefont
  {Lindmark}, \citenamefont {Lucero}, \citenamefont {Lyakh}, \citenamefont
  {Mandr{\`a}}, \citenamefont {McClean}, \citenamefont {McEwen}, \citenamefont
  {Megrant}, \citenamefont {Mi}, \citenamefont {Michielsen}, \citenamefont
  {Mohseni}, \citenamefont {Mutus}, \citenamefont {Naaman}, \citenamefont
  {Neeley}, \citenamefont {Neill}, \citenamefont {Niu}, \citenamefont {Ostby},
  \citenamefont {Petukhov}, \citenamefont {Platt}, \citenamefont {Quintana},
  \citenamefont {Rieffel}, \citenamefont {Roushan}, \citenamefont {Rubin},
  \citenamefont {Sank}, \citenamefont {Satzinger}, \citenamefont {Smelyanskiy},
  \citenamefont {Sung}, \citenamefont {Trevithick}, \citenamefont
  {Vainsencher}, \citenamefont {Villalonga}, \citenamefont {White},
  \citenamefont {Yao}, \citenamefont {Yeh}, \citenamefont {Zalcman},
  \citenamefont {Neven},\ and\ \citenamefont
  {Martinis}}]{aruteQuantumSupremacyUsing2019}%
  \BibitemOpen
  \bibfield  {author} {\bibinfo {author} {\bibnamefont {Arute}, \bibfnamefont
  {Frank}}, \bibinfo {author} {\bibfnamefont {Kunal}\ \bibnamefont {Arya}},
  \bibinfo {author} {\bibfnamefont {Ryan}\ \bibnamefont {Babbush}}, \bibinfo
  {author} {\bibfnamefont {Dave}\ \bibnamefont {Bacon}}, \bibinfo {author}
  {\bibfnamefont {Joseph~C.}\ \bibnamefont {Bardin}}, \bibinfo {author}
  {\bibfnamefont {Rami}\ \bibnamefont {Barends}}, \bibinfo {author}
  {\bibfnamefont {Rupak}\ \bibnamefont {Biswas}}, \bibinfo {author}
  {\bibfnamefont {Sergio}\ \bibnamefont {Boixo}}, \bibinfo {author}
  {\bibfnamefont {Fernando G. S.~L.}\ \bibnamefont {Brandao}}, \bibinfo
  {author} {\bibfnamefont {David~A.}\ \bibnamefont {Buell}}, \bibinfo {author}
  {\bibfnamefont {Brian}\ \bibnamefont {Burkett}}, \bibinfo {author}
  {\bibfnamefont {Yu}~\bibnamefont {Chen}}, \bibinfo {author} {\bibfnamefont
  {Zijun}\ \bibnamefont {Chen}}, \bibinfo {author} {\bibfnamefont {Ben}\
  \bibnamefont {Chiaro}}, \bibinfo {author} {\bibfnamefont {Roberto}\
  \bibnamefont {Collins}}, \bibinfo {author} {\bibfnamefont {William}\
  \bibnamefont {Courtney}}, \bibinfo {author} {\bibfnamefont {Andrew}\
  \bibnamefont {Dunsworth}}, \bibinfo {author} {\bibfnamefont {Edward}\
  \bibnamefont {Farhi}}, \bibinfo {author} {\bibfnamefont {Brooks}\
  \bibnamefont {Foxen}}, \bibinfo {author} {\bibfnamefont {Austin}\
  \bibnamefont {Fowler}}, \bibinfo {author} {\bibfnamefont {Craig}\
  \bibnamefont {Gidney}}, \bibinfo {author} {\bibfnamefont {Marissa}\
  \bibnamefont {Giustina}}, \bibinfo {author} {\bibfnamefont {Rob}\
  \bibnamefont {Graff}}, \bibinfo {author} {\bibfnamefont {Keith}\ \bibnamefont
  {Guerin}}, \bibinfo {author} {\bibfnamefont {Steve}\ \bibnamefont
  {Habegger}}, \bibinfo {author} {\bibfnamefont {Matthew~P.}\ \bibnamefont
  {Harrigan}}, \bibinfo {author} {\bibfnamefont {Michael~J.}\ \bibnamefont
  {Hartmann}}, \bibinfo {author} {\bibfnamefont {Alan}\ \bibnamefont {Ho}},
  \bibinfo {author} {\bibfnamefont {Markus}\ \bibnamefont {Hoffmann}}, \bibinfo
  {author} {\bibfnamefont {Trent}\ \bibnamefont {Huang}}, \bibinfo {author}
  {\bibfnamefont {Travis~S.}\ \bibnamefont {Humble}}, \bibinfo {author}
  {\bibfnamefont {Sergei~V.}\ \bibnamefont {Isakov}}, \bibinfo {author}
  {\bibfnamefont {Evan}\ \bibnamefont {Jeffrey}}, \bibinfo {author}
  {\bibfnamefont {Zhang}\ \bibnamefont {Jiang}}, \bibinfo {author}
  {\bibfnamefont {Dvir}\ \bibnamefont {Kafri}}, \bibinfo {author}
  {\bibfnamefont {Kostyantyn}\ \bibnamefont {Kechedzhi}}, \bibinfo {author}
  {\bibfnamefont {Julian}\ \bibnamefont {Kelly}}, \bibinfo {author}
  {\bibfnamefont {Paul~V.}\ \bibnamefont {Klimov}}, \bibinfo {author}
  {\bibfnamefont {Sergey}\ \bibnamefont {Knysh}}, \bibinfo {author}
  {\bibfnamefont {Alexander}\ \bibnamefont {Korotkov}}, \bibinfo {author}
  {\bibfnamefont {Fedor}\ \bibnamefont {Kostritsa}}, \bibinfo {author}
  {\bibfnamefont {David}\ \bibnamefont {Landhuis}}, \bibinfo {author}
  {\bibfnamefont {Mike}\ \bibnamefont {Lindmark}}, \bibinfo {author}
  {\bibfnamefont {Erik}\ \bibnamefont {Lucero}}, \bibinfo {author}
  {\bibfnamefont {Dmitry}\ \bibnamefont {Lyakh}}, \bibinfo {author}
  {\bibfnamefont {Salvatore}\ \bibnamefont {Mandr{\`a}}}, \bibinfo {author}
  {\bibfnamefont {Jarrod~R.}\ \bibnamefont {McClean}}, \bibinfo {author}
  {\bibfnamefont {Matthew}\ \bibnamefont {McEwen}}, \bibinfo {author}
  {\bibfnamefont {Anthony}\ \bibnamefont {Megrant}}, \bibinfo {author}
  {\bibfnamefont {Xiao}\ \bibnamefont {Mi}}, \bibinfo {author} {\bibfnamefont
  {Kristel}\ \bibnamefont {Michielsen}}, \bibinfo {author} {\bibfnamefont
  {Masoud}\ \bibnamefont {Mohseni}}, \bibinfo {author} {\bibfnamefont {Josh}\
  \bibnamefont {Mutus}}, \bibinfo {author} {\bibfnamefont {Ofer}\ \bibnamefont
  {Naaman}}, \bibinfo {author} {\bibfnamefont {Matthew}\ \bibnamefont
  {Neeley}}, \bibinfo {author} {\bibfnamefont {Charles}\ \bibnamefont {Neill}},
  \bibinfo {author} {\bibfnamefont {Murphy~Yuezhen}\ \bibnamefont {Niu}},
  \bibinfo {author} {\bibfnamefont {Eric}\ \bibnamefont {Ostby}}, \bibinfo
  {author} {\bibfnamefont {Andre}\ \bibnamefont {Petukhov}}, \bibinfo {author}
  {\bibfnamefont {John~C.}\ \bibnamefont {Platt}}, \bibinfo {author}
  {\bibfnamefont {Chris}\ \bibnamefont {Quintana}}, \bibinfo {author}
  {\bibfnamefont {Eleanor~G.}\ \bibnamefont {Rieffel}}, \bibinfo {author}
  {\bibfnamefont {Pedram}\ \bibnamefont {Roushan}}, \bibinfo {author}
  {\bibfnamefont {Nicholas~C.}\ \bibnamefont {Rubin}}, \bibinfo {author}
  {\bibfnamefont {Daniel}\ \bibnamefont {Sank}}, \bibinfo {author}
  {\bibfnamefont {Kevin~J.}\ \bibnamefont {Satzinger}}, \bibinfo {author}
  {\bibfnamefont {Vadim}\ \bibnamefont {Smelyanskiy}}, \bibinfo {author}
  {\bibfnamefont {Kevin~J.}\ \bibnamefont {Sung}}, \bibinfo {author}
  {\bibfnamefont {Matthew~D.}\ \bibnamefont {Trevithick}}, \bibinfo {author}
  {\bibfnamefont {Amit}\ \bibnamefont {Vainsencher}}, \bibinfo {author}
  {\bibfnamefont {Benjamin}\ \bibnamefont {Villalonga}}, \bibinfo {author}
  {\bibfnamefont {Theodore}\ \bibnamefont {White}}, \bibinfo {author}
  {\bibfnamefont {Z.~Jamie}\ \bibnamefont {Yao}}, \bibinfo {author}
  {\bibfnamefont {Ping}\ \bibnamefont {Yeh}}, \bibinfo {author} {\bibfnamefont
  {Adam}\ \bibnamefont {Zalcman}}, \bibinfo {author} {\bibfnamefont {Hartmut}\
  \bibnamefont {Neven}}, and\ \bibinfo {author} {\bibfnamefont {John~M.}\
  \bibnamefont {Martinis}}} (\bibinfo {year} {2019}),\ \bibfield  {title}
  {\enquote {\bibinfo {title} {Quantum supremacy using a programmable
  superconducting processor},}\ }\href
  {https://doi.org/10.1038/s41586-019-1666-5} {\bibfield  {journal} {\bibinfo
  {journal} {Nature}\ }\textbf {\bibinfo {volume} {574}}~(\bibinfo {number}
  {7779}),\ \bibinfo {pages} {505--510}}\BibitemShut {NoStop}%
\bibitem [{\citenamefont {Asavanant}\ \emph {et~al.}(2019)\citenamefont
  {Asavanant}, \citenamefont {Shiozawa}, \citenamefont {Yokoyama},
  \citenamefont {Charoensombutamon}, \citenamefont {Emura}, \citenamefont
  {Alexander}, \citenamefont {Takeda}, \citenamefont {Yoshikawa}, \citenamefont
  {Menicucci}, \citenamefont {Yonezawa},\ and\ \citenamefont
  {Furusawa}}]{asavanantGenerationTimedomainmultiplexedTwodimensional2019}%
  \BibitemOpen
  \bibfield  {author} {\bibinfo {author} {\bibnamefont {Asavanant},
  \bibfnamefont {Warit}}, \bibinfo {author} {\bibfnamefont {Yu}~\bibnamefont
  {Shiozawa}}, \bibinfo {author} {\bibfnamefont {Shota}\ \bibnamefont
  {Yokoyama}}, \bibinfo {author} {\bibfnamefont {Baramee}\ \bibnamefont
  {Charoensombutamon}}, \bibinfo {author} {\bibfnamefont {Hiroki}\ \bibnamefont
  {Emura}}, \bibinfo {author} {\bibfnamefont {Rafael~N.}\ \bibnamefont
  {Alexander}}, \bibinfo {author} {\bibfnamefont {Shuntaro}\ \bibnamefont
  {Takeda}}, \bibinfo {author} {\bibfnamefont {Jun-ichi}\ \bibnamefont
  {Yoshikawa}}, \bibinfo {author} {\bibfnamefont {Nicolas~C.}\ \bibnamefont
  {Menicucci}}, \bibinfo {author} {\bibfnamefont {Hidehiro}\ \bibnamefont
  {Yonezawa}}, and\ \bibinfo {author} {\bibfnamefont {Akira}\ \bibnamefont
  {Furusawa}}} (\bibinfo {year} {2019}),\ \bibfield  {title} {\enquote
  {\bibinfo {title} {Generation of time-domain-multiplexed two-dimensional
  cluster state},}\ }\href {https://doi.org/10.1126/science.aay2645} {\bibfield
   {journal} {\bibinfo  {journal} {Science}\ }\textbf {\bibinfo {volume}
  {366}}~(\bibinfo {number} {6463}),\ \bibinfo {pages} {373--376}}\BibitemShut
  {NoStop}%
\bibitem [{\citenamefont {Baidu}(2023)}]{baiduQCompute2023}%
  \BibitemOpen
  \bibfield  {author} {\bibinfo {author} {\bibnamefont {Baidu},}} (\bibinfo
  {year} {2023}),\ \href {https://github.com/baidu/QCompute} {\enquote
  {\bibinfo {title} {{{QCompute}}},}\ }\bibinfo {howpublished}
  {Baidu}\BibitemShut {NoStop}%
\bibitem [{\citenamefont {Barron}\ and\ \citenamefont
  {Wood}(2020)}]{barronMeasurementErrorMitigation2020}%
  \BibitemOpen
  \bibfield  {author} {\bibinfo {author} {\bibnamefont {Barron}, \bibfnamefont
  {George~S}}, and\ \bibinfo {author} {\bibfnamefont {Christopher~J.}\
  \bibnamefont {Wood}}} (\bibinfo {year} {2020}),\ \href
  {http://arxiv.org/abs/2010.08520} {\enquote {\bibinfo {title} {Measurement
  {{Error Mitigation}} for {{Variational Quantum Algorithms}}},}\ }\bibinfo
  {howpublished} {arXiv:2010.08520 [quant-ph]}\BibitemShut {NoStop}%
\bibitem [{\citenamefont {Begu{\v s}i{\'c}}\ and\ \citenamefont
  {Chan}(2023)}]{begusicFastClassicalSimulation2023}%
  \BibitemOpen
  \bibfield  {author} {\bibinfo {author} {\bibnamefont {Begu{\v s}i{\'c}},
  \bibfnamefont {Tomislav}}, and\ \bibinfo {author} {\bibfnamefont {Garnet
  Kin-Lic}\ \bibnamefont {Chan}}} (\bibinfo {year} {2023}),\ \href
  {http://arxiv.org/abs/2306.16372} {\enquote {\bibinfo {title} {Fast classical
  simulation of evidence for the utility of quantum computing before fault
  tolerance},}\ }\bibinfo {howpublished} {arXiv:2306.16372
  [quant-ph]}\BibitemShut {NoStop}%
\bibitem [{\citenamefont {Bennett}\ \emph
  {et~al.}(1996{\natexlab{a}})\citenamefont {Bennett}, \citenamefont
  {Brassard}, \citenamefont {Popescu}, \citenamefont {Schumacher},
  \citenamefont {Smolin},\ and\ \citenamefont
  {Wootters}}]{bennettPurificationNoisyEntanglement1996}%
  \BibitemOpen
  \bibfield  {author} {\bibinfo {author} {\bibnamefont {Bennett}, \bibfnamefont
  {Charles~H}}, \bibinfo {author} {\bibfnamefont {Gilles}\ \bibnamefont
  {Brassard}}, \bibinfo {author} {\bibfnamefont {Sandu}\ \bibnamefont
  {Popescu}}, \bibinfo {author} {\bibfnamefont {Benjamin}\ \bibnamefont
  {Schumacher}}, \bibinfo {author} {\bibfnamefont {John~A.}\ \bibnamefont
  {Smolin}}, and\ \bibinfo {author} {\bibfnamefont {William~K.}\ \bibnamefont
  {Wootters}}} (\bibinfo {year} {1996}{\natexlab{a}}),\ \bibfield  {title}
  {\enquote {\bibinfo {title} {Purification of {{Noisy Entanglement}} and
  {{Faithful Teleportation}} via {{Noisy Channels}}},}\ }\href
  {https://doi.org/10.1103/PhysRevLett.76.722} {\bibfield  {journal} {\bibinfo
  {journal} {Physical Review Letters}\ }\textbf {\bibinfo {volume}
  {76}}~(\bibinfo {number} {5}),\ \bibinfo {pages} {722--725}}\BibitemShut
  {NoStop}%
\bibitem [{\citenamefont {Bennett}\ \emph
  {et~al.}(1996{\natexlab{b}})\citenamefont {Bennett}, \citenamefont
  {DiVincenzo}, \citenamefont {Smolin},\ and\ \citenamefont
  {Wootters}}]{bennettMixedstateEntanglementQuantum1996}%
  \BibitemOpen
  \bibfield  {author} {\bibinfo {author} {\bibnamefont {Bennett}, \bibfnamefont
  {Charles~H}}, \bibinfo {author} {\bibfnamefont {David~P.}\ \bibnamefont
  {DiVincenzo}}, \bibinfo {author} {\bibfnamefont {John~A.}\ \bibnamefont
  {Smolin}}, and\ \bibinfo {author} {\bibfnamefont {William~K.}\ \bibnamefont
  {Wootters}}} (\bibinfo {year} {1996}{\natexlab{b}}),\ \bibfield  {title}
  {\enquote {\bibinfo {title} {Mixed-state entanglement and quantum error
  correction},}\ }\href {https://doi.org/10.1103/PhysRevA.54.3824} {\bibfield
  {journal} {\bibinfo  {journal} {Physical Review A}\ }\textbf {\bibinfo
  {volume} {54}}~(\bibinfo {number} {5}),\ \bibinfo {pages}
  {3824--3851}}\BibitemShut {NoStop}%
\bibitem [{\citenamefont {Bergholm}\ \emph {et~al.}(2018)\citenamefont
  {Bergholm}, \citenamefont {Izaac}, \citenamefont {Schuld}, \citenamefont
  {Gogolin}, \citenamefont {Ahmed}, \citenamefont {Ajith}, \citenamefont
  {Alam}, \citenamefont {{Alonso-Linaje}}, \citenamefont {AkashNarayanan},
  \citenamefont {Asadi}, \citenamefont {Arrazola}, \citenamefont {Azad},
  \citenamefont {Banning}, \citenamefont {Blank}, \citenamefont {Bromley},
  \citenamefont {Cordier}, \citenamefont {Ceroni}, \citenamefont {Delgado},
  \citenamefont {Di~Matteo}, \citenamefont {Dusko}, \citenamefont {Garg},
  \citenamefont {Guala}, \citenamefont {Hayes}, \citenamefont {Hill},
  \citenamefont {Ijaz}, \citenamefont {Isacsson}, \citenamefont {Ittah},
  \citenamefont {Jahangiri}, \citenamefont {Jain}, \citenamefont {Jiang},
  \citenamefont {Khandelwal}, \citenamefont {Kottmann}, \citenamefont {Lang},
  \citenamefont {Lee}, \citenamefont {Loke}, \citenamefont {Lowe},
  \citenamefont {McKiernan}, \citenamefont {Meyer}, \citenamefont
  {{Monta{\~n}ez-Barrera}}, \citenamefont {Moyard}, \citenamefont {Niu},
  \citenamefont {O'Riordan}, \citenamefont {Oud}, \citenamefont {Panigrahi},
  \citenamefont {Park}, \citenamefont {Polatajko}, \citenamefont {Quesada},
  \citenamefont {Roberts}, \citenamefont {S{\'a}}, \citenamefont {Schoch},
  \citenamefont {Shi}, \citenamefont {Shu}, \citenamefont {Sim}, \citenamefont
  {Singh}, \citenamefont {Strandberg}, \citenamefont {Soni}, \citenamefont
  {Sz{\'a}va}, \citenamefont {Thabet}, \citenamefont {{Vargas-Hern{\'a}ndez}},
  \citenamefont {Vincent}, \citenamefont {Vitucci}, \citenamefont {Weber},
  \citenamefont {Wierichs}, \citenamefont {Wiersema}, \citenamefont {Willmann},
  \citenamefont {Wong}, \citenamefont {Zhang},\ and\ \citenamefont
  {Killoran}}]{bergholmPennyLaneAutomaticDifferentiation2018}%
  \BibitemOpen
  \bibfield  {author} {\bibinfo {author} {\bibnamefont {Bergholm},
  \bibfnamefont {Ville}}, \bibinfo {author} {\bibfnamefont {Josh}\ \bibnamefont
  {Izaac}}, \bibinfo {author} {\bibfnamefont {Maria}\ \bibnamefont {Schuld}},
  \bibinfo {author} {\bibfnamefont {Christian}\ \bibnamefont {Gogolin}},
  \bibinfo {author} {\bibfnamefont {Shahnawaz}\ \bibnamefont {Ahmed}}, \bibinfo
  {author} {\bibfnamefont {Vishnu}\ \bibnamefont {Ajith}}, \bibinfo {author}
  {\bibfnamefont {M.~Sohaib}\ \bibnamefont {Alam}}, \bibinfo {author}
  {\bibfnamefont {Guillermo}\ \bibnamefont {{Alonso-Linaje}}}, \bibinfo
  {author} {\bibfnamefont {B.}~\bibnamefont {AkashNarayanan}}, \bibinfo
  {author} {\bibfnamefont {Ali}\ \bibnamefont {Asadi}}, \bibinfo {author}
  {\bibfnamefont {Juan~Miguel}\ \bibnamefont {Arrazola}}, \bibinfo {author}
  {\bibfnamefont {Utkarsh}\ \bibnamefont {Azad}}, \bibinfo {author}
  {\bibfnamefont {Sam}\ \bibnamefont {Banning}}, \bibinfo {author}
  {\bibfnamefont {Carsten}\ \bibnamefont {Blank}}, \bibinfo {author}
  {\bibfnamefont {Thomas~R.}\ \bibnamefont {Bromley}}, \bibinfo {author}
  {\bibfnamefont {Benjamin~A.}\ \bibnamefont {Cordier}}, \bibinfo {author}
  {\bibfnamefont {Jack}\ \bibnamefont {Ceroni}}, \bibinfo {author}
  {\bibfnamefont {Alain}\ \bibnamefont {Delgado}}, \bibinfo {author}
  {\bibfnamefont {Olivia}\ \bibnamefont {Di~Matteo}}, \bibinfo {author}
  {\bibfnamefont {Amintor}\ \bibnamefont {Dusko}}, \bibinfo {author}
  {\bibfnamefont {Tanya}\ \bibnamefont {Garg}}, \bibinfo {author}
  {\bibfnamefont {Diego}\ \bibnamefont {Guala}}, \bibinfo {author}
  {\bibfnamefont {Anthony}\ \bibnamefont {Hayes}}, \bibinfo {author}
  {\bibfnamefont {Ryan}\ \bibnamefont {Hill}}, \bibinfo {author} {\bibfnamefont
  {Aroosa}\ \bibnamefont {Ijaz}}, \bibinfo {author} {\bibfnamefont {Theodor}\
  \bibnamefont {Isacsson}}, \bibinfo {author} {\bibfnamefont {David}\
  \bibnamefont {Ittah}}, \bibinfo {author} {\bibfnamefont {Soran}\ \bibnamefont
  {Jahangiri}}, \bibinfo {author} {\bibfnamefont {Prateek}\ \bibnamefont
  {Jain}}, \bibinfo {author} {\bibfnamefont {Edward}\ \bibnamefont {Jiang}},
  \bibinfo {author} {\bibfnamefont {Ankit}\ \bibnamefont {Khandelwal}},
  \bibinfo {author} {\bibfnamefont {Korbinian}\ \bibnamefont {Kottmann}},
  \bibinfo {author} {\bibfnamefont {Robert~A.}\ \bibnamefont {Lang}}, \bibinfo
  {author} {\bibfnamefont {Christina}\ \bibnamefont {Lee}}, \bibinfo {author}
  {\bibfnamefont {Thomas}\ \bibnamefont {Loke}}, \bibinfo {author}
  {\bibfnamefont {Angus}\ \bibnamefont {Lowe}}, \bibinfo {author}
  {\bibfnamefont {Keri}\ \bibnamefont {McKiernan}}, \bibinfo {author}
  {\bibfnamefont {Johannes~Jakob}\ \bibnamefont {Meyer}}, \bibinfo {author}
  {\bibfnamefont {J.~A.}\ \bibnamefont {{Monta{\~n}ez-Barrera}}}, \bibinfo
  {author} {\bibfnamefont {Romain}\ \bibnamefont {Moyard}}, \bibinfo {author}
  {\bibfnamefont {Zeyue}\ \bibnamefont {Niu}}, \bibinfo {author} {\bibfnamefont
  {Lee~James}\ \bibnamefont {O'Riordan}}, \bibinfo {author} {\bibfnamefont
  {Steven}\ \bibnamefont {Oud}}, \bibinfo {author} {\bibfnamefont {Ashish}\
  \bibnamefont {Panigrahi}}, \bibinfo {author} {\bibfnamefont {Chae-Yeun}\
  \bibnamefont {Park}}, \bibinfo {author} {\bibfnamefont {Daniel}\ \bibnamefont
  {Polatajko}}, \bibinfo {author} {\bibfnamefont {Nicol{\'a}s}\ \bibnamefont
  {Quesada}}, \bibinfo {author} {\bibfnamefont {Chase}\ \bibnamefont
  {Roberts}}, \bibinfo {author} {\bibfnamefont {Nahum}\ \bibnamefont {S{\'a}}},
  \bibinfo {author} {\bibfnamefont {Isidor}\ \bibnamefont {Schoch}}, \bibinfo
  {author} {\bibfnamefont {Borun}\ \bibnamefont {Shi}}, \bibinfo {author}
  {\bibfnamefont {Shuli}\ \bibnamefont {Shu}}, \bibinfo {author} {\bibfnamefont
  {Sukin}\ \bibnamefont {Sim}}, \bibinfo {author} {\bibfnamefont {Arshpreet}\
  \bibnamefont {Singh}}, \bibinfo {author} {\bibfnamefont {Ingrid}\
  \bibnamefont {Strandberg}}, \bibinfo {author} {\bibfnamefont {Jay}\
  \bibnamefont {Soni}}, \bibinfo {author} {\bibfnamefont {Antal}\ \bibnamefont
  {Sz{\'a}va}}, \bibinfo {author} {\bibfnamefont {Slimane}\ \bibnamefont
  {Thabet}}, \bibinfo {author} {\bibfnamefont {Rodrigo~A.}\ \bibnamefont
  {{Vargas-Hern{\'a}ndez}}}, \bibinfo {author} {\bibfnamefont {Trevor}\
  \bibnamefont {Vincent}}, \bibinfo {author} {\bibfnamefont {Nicola}\
  \bibnamefont {Vitucci}}, \bibinfo {author} {\bibfnamefont {Maurice}\
  \bibnamefont {Weber}}, \bibinfo {author} {\bibfnamefont {David}\ \bibnamefont
  {Wierichs}}, \bibinfo {author} {\bibfnamefont {Roeland}\ \bibnamefont
  {Wiersema}}, \bibinfo {author} {\bibfnamefont {Moritz}\ \bibnamefont
  {Willmann}}, \bibinfo {author} {\bibfnamefont {Vincent}\ \bibnamefont
  {Wong}}, \bibinfo {author} {\bibfnamefont {Shaoming}\ \bibnamefont {Zhang}},
  and\ \bibinfo {author} {\bibfnamefont {Nathan}\ \bibnamefont {Killoran}}}
  (\bibinfo {year} {2018}),\ \href {https://arxiv.org/abs/1811.04968v4}
  {\enquote {\bibinfo {title} {{{PennyLane}}: {{Automatic}} differentiation of
  hybrid quantum-classical computations},}\ }\BibitemShut {NoStop}%
\bibitem [{\citenamefont {Bergquist}\ \emph {et~al.}(1986)\citenamefont
  {Bergquist}, \citenamefont {Hulet}, \citenamefont {Itano},\ and\
  \citenamefont {Wineland}}]{bergquistObservationQuantumJumps1986}%
  \BibitemOpen
  \bibfield  {author} {\bibinfo {author} {\bibnamefont {Bergquist},
  \bibfnamefont {J~C}}, \bibinfo {author} {\bibfnamefont {Randall~G.}\
  \bibnamefont {Hulet}}, \bibinfo {author} {\bibfnamefont {Wayne~M.}\
  \bibnamefont {Itano}}, and\ \bibinfo {author} {\bibfnamefont {D.~J.}\
  \bibnamefont {Wineland}}} (\bibinfo {year} {1986}),\ \bibfield  {title}
  {\enquote {\bibinfo {title} {Observation of {{Quantum Jumps}} in a {{Single
  Atom}}},}\ }\href {https://doi.org/10.1103/PhysRevLett.57.1699} {\bibfield
  {journal} {\bibinfo  {journal} {Physical Review Letters}\ }\textbf {\bibinfo
  {volume} {57}}~(\bibinfo {number} {14}),\ \bibinfo {pages}
  {1699--1702}}\BibitemShut {NoStop}%
\bibitem [{\citenamefont {Biamonte}\ \emph {et~al.}(2017)\citenamefont
  {Biamonte}, \citenamefont {Wittek}, \citenamefont {Pancotti}, \citenamefont
  {Rebentrost}, \citenamefont {Wiebe},\ and\ \citenamefont
  {Lloyd}}]{biamonteQuantumMachineLearning2017}%
  \BibitemOpen
  \bibfield  {author} {\bibinfo {author} {\bibnamefont {Biamonte},
  \bibfnamefont {Jacob}}, \bibinfo {author} {\bibfnamefont {Peter}\
  \bibnamefont {Wittek}}, \bibinfo {author} {\bibfnamefont {Nicola}\
  \bibnamefont {Pancotti}}, \bibinfo {author} {\bibfnamefont {Patrick}\
  \bibnamefont {Rebentrost}}, \bibinfo {author} {\bibfnamefont {Nathan}\
  \bibnamefont {Wiebe}}, and\ \bibinfo {author} {\bibfnamefont {Seth}\
  \bibnamefont {Lloyd}}} (\bibinfo {year} {2017}),\ \bibfield  {title}
  {\enquote {\bibinfo {title} {Quantum machine learning},}\ }\href
  {https://doi.org/10.1038/nature23474} {\bibfield  {journal} {\bibinfo
  {journal} {Nature}\ }\textbf {\bibinfo {volume} {549}}~(\bibinfo {number}
  {7671}),\ \bibinfo {pages} {195--202}}\BibitemShut {NoStop}%
\bibitem [{\citenamefont {Blais}\ \emph {et~al.}(2004)\citenamefont {Blais},
  \citenamefont {Huang}, \citenamefont {Wallraff}, \citenamefont {Girvin},\
  and\ \citenamefont {Schoelkopf}}]{blaisCavityQuantumElectrodynamics2004}%
  \BibitemOpen
  \bibfield  {author} {\bibinfo {author} {\bibnamefont {Blais}, \bibfnamefont
  {Alexandre}}, \bibinfo {author} {\bibfnamefont {Ren-Shou}\ \bibnamefont
  {Huang}}, \bibinfo {author} {\bibfnamefont {Andreas}\ \bibnamefont
  {Wallraff}}, \bibinfo {author} {\bibfnamefont {S.~M.}\ \bibnamefont
  {Girvin}}, and\ \bibinfo {author} {\bibfnamefont {R.~J.}\ \bibnamefont
  {Schoelkopf}}} (\bibinfo {year} {2004}),\ \bibfield  {title} {\enquote
  {\bibinfo {title} {Cavity quantum electrodynamics for superconducting
  electrical circuits: {{An}} architecture for quantum computation},}\ }\href
  {https://doi.org/10.1103/PhysRevA.69.062320} {\bibfield  {journal} {\bibinfo
  {journal} {Physical Review A}\ }\textbf {\bibinfo {volume} {69}}~(\bibinfo
  {number} {6}),\ \bibinfo {pages} {062320}}\BibitemShut {NoStop}%
\bibitem [{\citenamefont {{Bonet-Monroig}}\ \emph {et~al.}(2018)\citenamefont
  {{Bonet-Monroig}}, \citenamefont {Sagastizabal}, \citenamefont {Singh},\ and\
  \citenamefont {O'Brien}}]{bonet-monroigLowcostErrorMitigation2018}%
  \BibitemOpen
  \bibfield  {author} {\bibinfo {author} {\bibnamefont {{Bonet-Monroig}},
  \bibfnamefont {X}}, \bibinfo {author} {\bibfnamefont {R.}~\bibnamefont
  {Sagastizabal}}, \bibinfo {author} {\bibfnamefont {M.}~\bibnamefont {Singh}},
  and\ \bibinfo {author} {\bibfnamefont {T.~E.}\ \bibnamefont {O'Brien}}}
  (\bibinfo {year} {2018}),\ \bibfield  {title} {\enquote {\bibinfo {title}
  {Low-cost error mitigation by symmetry verification},}\ }\href
  {https://doi.org/10.1103/PhysRevA.98.062339} {\bibfield  {journal} {\bibinfo
  {journal} {Physical Review A}\ }\textbf {\bibinfo {volume} {98}}~(\bibinfo
  {number} {6}),\ \bibinfo {pages} {062339}}\BibitemShut {NoStop}%
\bibitem [{\citenamefont {{Bonet-Monroig}}\ \emph {et~al.}(2020)\citenamefont
  {{Bonet-Monroig}}, \citenamefont {Babbush},\ and\ \citenamefont
  {O'Brien}}]{bonet-monroigNearlyOptimalMeasurement2020}%
  \BibitemOpen
  \bibfield  {author} {\bibinfo {author} {\bibnamefont {{Bonet-Monroig}},
  \bibfnamefont {Xavier}}, \bibinfo {author} {\bibfnamefont {Ryan}\
  \bibnamefont {Babbush}}, and\ \bibinfo {author} {\bibfnamefont {Thomas~E.}\
  \bibnamefont {O'Brien}}} (\bibinfo {year} {2020}),\ \bibfield  {title}
  {\enquote {\bibinfo {title} {Nearly {{Optimal Measurement Scheduling}} for
  {{Partial Tomography}} of {{Quantum States}}},}\ }\href
  {https://doi.org/10.1103/PhysRevX.10.031064} {\bibfield  {journal} {\bibinfo
  {journal} {Physical Review X}\ }\textbf {\bibinfo {volume} {10}}~(\bibinfo
  {number} {3}),\ \bibinfo {pages} {031064}}\BibitemShut {NoStop}%
\bibitem [{\citenamefont {Bravyi}\ and\ \citenamefont
  {Gosset}(2016)}]{bravyiImprovedClassicalSimulation2016}%
  \BibitemOpen
  \bibfield  {author} {\bibinfo {author} {\bibnamefont {Bravyi}, \bibfnamefont
  {Sergey}}, and\ \bibinfo {author} {\bibfnamefont {David}\ \bibnamefont
  {Gosset}}} (\bibinfo {year} {2016}),\ \bibfield  {title} {\enquote {\bibinfo
  {title} {Improved {{Classical Simulation}} of {{Quantum Circuits Dominated}}
  by {{Clifford Gates}}},}\ }\href
  {https://doi.org/10.1103/PhysRevLett.116.250501} {\bibfield  {journal}
  {\bibinfo  {journal} {Physical Review Letters}\ }\textbf {\bibinfo {volume}
  {116}}~(\bibinfo {number} {25}),\ \bibinfo {pages} {250501}}\BibitemShut
  {NoStop}%
\bibitem [{\citenamefont {Bravyi}\ and\ \citenamefont
  {Kitaev}(2002)}]{bravyiFermionicQuantumComputation2002}%
  \BibitemOpen
  \bibfield  {author} {\bibinfo {author} {\bibnamefont {Bravyi}, \bibfnamefont
  {Sergey}}, and\ \bibinfo {author} {\bibfnamefont {Alexei}\ \bibnamefont
  {Kitaev}}} (\bibinfo {year} {2002}),\ \bibfield  {title} {\enquote {\bibinfo
  {title} {Fermionic quantum computation},}\ }\href
  {https://doi.org/10.1006/aphy.2002.6254} {\bibfield  {journal} {\bibinfo
  {journal} {Annals of Physics}\ }\textbf {\bibinfo {volume} {298}}~(\bibinfo
  {number} {1}),\ \bibinfo {pages} {210--226}}\BibitemShut {NoStop}%
\bibitem [{\citenamefont {Bravyi}\ \emph {et~al.}(2021)\citenamefont {Bravyi},
  \citenamefont {Sheldon}, \citenamefont {Kandala}, \citenamefont {Mckay},\
  and\ \citenamefont {Gambetta}}]{bravyiMitigatingMeasurementErrors2021}%
  \BibitemOpen
  \bibfield  {author} {\bibinfo {author} {\bibnamefont {Bravyi}, \bibfnamefont
  {Sergey}}, \bibinfo {author} {\bibfnamefont {Sarah}\ \bibnamefont {Sheldon}},
  \bibinfo {author} {\bibfnamefont {Abhinav}\ \bibnamefont {Kandala}}, \bibinfo
  {author} {\bibfnamefont {David~C.}\ \bibnamefont {Mckay}}, and\ \bibinfo
  {author} {\bibfnamefont {Jay~M.}\ \bibnamefont {Gambetta}}} (\bibinfo {year}
  {2021}),\ \bibfield  {title} {\enquote {\bibinfo {title} {Mitigating
  measurement errors in multiqubit experiments},}\ }\href
  {https://doi.org/10.1103/PhysRevA.103.042605} {\bibfield  {journal} {\bibinfo
   {journal} {Physical Review A}\ }\textbf {\bibinfo {volume} {103}}~(\bibinfo
  {number} {4}),\ \bibinfo {pages} {042605}}\BibitemShut {NoStop}%
\bibitem [{\citenamefont {Breuckmann}\ and\ \citenamefont
  {Eberhardt}(2021)}]{breuckmannQuantumLowDensityParityCheck2021}%
  \BibitemOpen
  \bibfield  {author} {\bibinfo {author} {\bibnamefont {Breuckmann},
  \bibfnamefont {Nikolas~P}}, and\ \bibinfo {author} {\bibfnamefont
  {Jens~Niklas}\ \bibnamefont {Eberhardt}}} (\bibinfo {year} {2021}),\
  \bibfield  {title} {\enquote {\bibinfo {title} {Quantum {{Low-Density
  Parity-Check Codes}}},}\ }\href {https://doi.org/10.1103/PRXQuantum.2.040101}
  {\bibfield  {journal} {\bibinfo  {journal} {PRX Quantum}\ }\textbf {\bibinfo
  {volume} {2}}~(\bibinfo {number} {4}),\ \bibinfo {pages}
  {040101}}\BibitemShut {NoStop}%
\bibitem [{\citenamefont {Bridgeman}\ and\ \citenamefont
  {Chubb}(2017)}]{bridgemanHandwavingInterpretiveDance2017}%
  \BibitemOpen
  \bibfield  {author} {\bibinfo {author} {\bibnamefont {Bridgeman},
  \bibfnamefont {Jacob~C}}, and\ \bibinfo {author} {\bibfnamefont
  {Christopher~T}\ \bibnamefont {Chubb}}} (\bibinfo {year} {2017}),\ \bibfield
  {title} {\enquote {\bibinfo {title} {Hand-waving and interpretive dance: An
  introductory course on tensor networks},}\ }\href
  {https://doi.org/10.1088/1751-8121/aa6dc3} {\bibfield  {journal} {\bibinfo
  {journal} {Journal of Physics A: Mathematical and Theoretical}\ }\textbf
  {\bibinfo {volume} {50}}~(\bibinfo {number} {22}),\ \bibinfo {pages}
  {223001}}\BibitemShut {NoStop}%
\bibitem [{\citenamefont {Bultrini}\ \emph {et~al.}(2023)\citenamefont
  {Bultrini}, \citenamefont {Gordon}, \citenamefont {Czarnik}, \citenamefont
  {Arrasmith}, \citenamefont {Cerezo}, \citenamefont {Coles},\ and\
  \citenamefont {Cincio}}]{bultriniUnifyingBenchmarkingStateoftheart2023}%
  \BibitemOpen
  \bibfield  {author} {\bibinfo {author} {\bibnamefont {Bultrini},
  \bibfnamefont {Daniel}}, \bibinfo {author} {\bibfnamefont {Max~Hunter}\
  \bibnamefont {Gordon}}, \bibinfo {author} {\bibfnamefont {Piotr}\
  \bibnamefont {Czarnik}}, \bibinfo {author} {\bibfnamefont {Andrew}\
  \bibnamefont {Arrasmith}}, \bibinfo {author} {\bibfnamefont {M.}~\bibnamefont
  {Cerezo}}, \bibinfo {author} {\bibfnamefont {Patrick~J.}\ \bibnamefont
  {Coles}}, and\ \bibinfo {author} {\bibfnamefont {Lukasz}\ \bibnamefont
  {Cincio}}} (\bibinfo {year} {2023}),\ \bibfield  {title} {\enquote {\bibinfo
  {title} {Unifying and benchmarking state-of-the-art quantum error mitigation
  techniques},}\ }\href {https://doi.org/10.22331/q-2023-06-06-1034} {\bibfield
   {journal} {\bibinfo  {journal} {Quantum}\ }\textbf {\bibinfo {volume} {7}},\
  \bibinfo {pages} {1034}}\BibitemShut {NoStop}%
\bibitem [{\citenamefont {Burgarth}\ \emph {et~al.}(2019)\citenamefont
  {Burgarth}, \citenamefont {Facchi}, \citenamefont {Gramegna},\ and\
  \citenamefont {Pascazio}}]{burgarthGeneralizedProductFormulas2019}%
  \BibitemOpen
  \bibfield  {author} {\bibinfo {author} {\bibnamefont {Burgarth},
  \bibfnamefont {Daniel}}, \bibinfo {author} {\bibfnamefont {Paolo}\
  \bibnamefont {Facchi}}, \bibinfo {author} {\bibfnamefont {Giovanni}\
  \bibnamefont {Gramegna}}, and\ \bibinfo {author} {\bibfnamefont {Saverio}\
  \bibnamefont {Pascazio}}} (\bibinfo {year} {2019}),\ \bibfield  {title}
  {\enquote {\bibinfo {title} {Generalized product formulas and quantum
  control},}\ }\href {https://doi.org/10.1088/1751-8121/ab4403} {\bibfield
  {journal} {\bibinfo  {journal} {Journal of Physics A: Mathematical and
  Theoretical}\ }\textbf {\bibinfo {volume} {52}}~(\bibinfo {number} {43}),\
  \bibinfo {pages} {435301}}\BibitemShut {NoStop}%
\bibitem [{\citenamefont {Cai}(2020)}]{caiResourceEstimationQuantum2020}%
  \BibitemOpen
  \bibfield  {author} {\bibinfo {author} {\bibnamefont {Cai}, \bibfnamefont
  {Zhenyu}}} (\bibinfo {year} {2020}),\ \bibfield  {title} {\enquote {\bibinfo
  {title} {Resource {{Estimation}} for {{Quantum Variational Simulations}} of
  the {{Hubbard Model}}},}\ }\href
  {https://doi.org/10.1103/PhysRevApplied.14.014059} {\bibfield  {journal}
  {\bibinfo  {journal} {Physical Review Applied}\ }\textbf {\bibinfo {volume}
  {14}}~(\bibinfo {number} {1}),\ \bibinfo {pages} {014059}}\BibitemShut
  {NoStop}%
\bibitem [{\citenamefont
  {Cai}(2021{\natexlab{a}})}]{caiMultiexponentialErrorExtrapolation2021}%
  \BibitemOpen
  \bibfield  {author} {\bibinfo {author} {\bibnamefont {Cai}, \bibfnamefont
  {Zhenyu}}} (\bibinfo {year} {2021}{\natexlab{a}}),\ \bibfield  {title}
  {\enquote {\bibinfo {title} {Multi-exponential error extrapolation and
  combining error mitigation techniques for {{NISQ}} applications},}\ }\href
  {https://doi.org/10.1038/s41534-021-00404-3} {\bibfield  {journal} {\bibinfo
  {journal} {npj Quantum Information}\ }\textbf {\bibinfo {volume} {7}},\
  \bibinfo {pages} {80}}\BibitemShut {NoStop}%
\bibitem [{\citenamefont
  {Cai}(2021{\natexlab{b}})}]{caiPracticalFrameworkQuantum2021}%
  \BibitemOpen
  \bibfield  {author} {\bibinfo {author} {\bibnamefont {Cai}, \bibfnamefont
  {Zhenyu}}} (\bibinfo {year} {2021}{\natexlab{b}}),\ \href
  {http://arxiv.org/abs/2110.05389} {\enquote {\bibinfo {title} {A {{Practical
  Framework}} for {{Quantum Error Mitigation}}},}\ }\bibinfo {howpublished}
  {arXiv:2110.05389 [quant-ph]}\BibitemShut {NoStop}%
\bibitem [{\citenamefont
  {Cai}(2021{\natexlab{c}})}]{caiQuantumErrorMitigation2021}%
  \BibitemOpen
  \bibfield  {author} {\bibinfo {author} {\bibnamefont {Cai}, \bibfnamefont
  {Zhenyu}}} (\bibinfo {year} {2021}{\natexlab{c}}),\ \bibfield  {title}
  {\enquote {\bibinfo {title} {Quantum {{Error Mitigation}} using {{Symmetry
  Expansion}}},}\ }\href {https://doi.org/10.22331/q-2021-09-21-548} {\bibfield
   {journal} {\bibinfo  {journal} {Quantum}\ }\textbf {\bibinfo {volume} {5}},\
  \bibinfo {pages} {548}}\BibitemShut {NoStop}%
\bibitem [{\citenamefont
  {Cai}(2021{\natexlab{d}})}]{caiResourceefficientPurificationbasedQuantum2021}%
  \BibitemOpen
  \bibfield  {author} {\bibinfo {author} {\bibnamefont {Cai}, \bibfnamefont
  {Zhenyu}}} (\bibinfo {year} {2021}{\natexlab{d}}),\ \href
  {https://doi.org/10.48550/arXiv.2107.07279} {\enquote {\bibinfo {title}
  {Resource-efficient {{Purification-based Quantum Error Mitigation}}},}\
  }\bibinfo {howpublished} {arXiv:2107.07279 [quant-ph]}\BibitemShut {NoStop}%
\bibitem [{\citenamefont {Cai}\ \emph {et~al.}(2023)\citenamefont {Cai},
  \citenamefont {Siegel},\ and\ \citenamefont
  {Benjamin}}]{caiLoopedPipelinesEnabling2023}%
  \BibitemOpen
  \bibfield  {author} {\bibinfo {author} {\bibnamefont {Cai}, \bibfnamefont
  {Zhenyu}}, \bibinfo {author} {\bibfnamefont {Adam}\ \bibnamefont {Siegel}},
  and\ \bibinfo {author} {\bibfnamefont {Simon}\ \bibnamefont {Benjamin}}}
  (\bibinfo {year} {2023}),\ \bibfield  {title} {\enquote {\bibinfo {title}
  {Looped {{Pipelines Enabling Effective 3D Qubit Lattices}} in a {{Strictly 2D
  Device}}},}\ }\href {https://doi.org/10.1103/PRXQuantum.4.020345} {\bibfield
  {journal} {\bibinfo  {journal} {PRX Quantum}\ }\textbf {\bibinfo {volume}
  {4}}~(\bibinfo {number} {2}),\ \bibinfo {pages} {020345}}\BibitemShut
  {NoStop}%
\bibitem [{\citenamefont {Calderbank}\ and\ \citenamefont
  {Shor}(1996)}]{calderbankGoodQuantumErrorcorrecting1996}%
  \BibitemOpen
  \bibfield  {author} {\bibinfo {author} {\bibnamefont {Calderbank},
  \bibfnamefont {A~R}}, and\ \bibinfo {author} {\bibfnamefont {Peter~W.}\
  \bibnamefont {Shor}}} (\bibinfo {year} {1996}),\ \bibfield  {title} {\enquote
  {\bibinfo {title} {Good quantum error-correcting codes exist},}\ }\href
  {https://doi.org/10.1103/PhysRevA.54.1098} {\bibfield  {journal} {\bibinfo
  {journal} {Physical Review A}\ }\textbf {\bibinfo {volume} {54}}~(\bibinfo
  {number} {2}),\ \bibinfo {pages} {1098--1105}}\BibitemShut {NoStop}%
\bibitem [{\citenamefont {Campbell}(2019)}]{campbellRandomCompilerFast2019}%
  \BibitemOpen
  \bibfield  {author} {\bibinfo {author} {\bibnamefont {Campbell},
  \bibfnamefont {Earl}}} (\bibinfo {year} {2019}),\ \bibfield  {title}
  {\enquote {\bibinfo {title} {Random {{Compiler}} for {{Fast Hamiltonian
  Simulation}}},}\ }\href {https://doi.org/10.1103/PhysRevLett.123.070503}
  {\bibfield  {journal} {\bibinfo  {journal} {Physical Review Letters}\
  }\textbf {\bibinfo {volume} {123}}~(\bibinfo {number} {7}),\ \bibinfo {pages}
  {070503}}\BibitemShut {NoStop}%
\bibitem [{\citenamefont {Cao}\ \emph {et~al.}(2022{\natexlab{a}})\citenamefont
  {Cao}, \citenamefont {Yu}, \citenamefont {Wu}, \citenamefont {Shannon},
  \citenamefont {Zeng},\ and\ \citenamefont
  {Joynt}}]{caoMitigatingAlgorithmicErrors2022}%
  \BibitemOpen
  \bibfield  {author} {\bibinfo {author} {\bibnamefont {Cao}, \bibfnamefont
  {Chenfeng}}, \bibinfo {author} {\bibfnamefont {Yunlong}\ \bibnamefont {Yu}},
  \bibinfo {author} {\bibfnamefont {Zipeng}\ \bibnamefont {Wu}}, \bibinfo
  {author} {\bibfnamefont {Nic}\ \bibnamefont {Shannon}}, \bibinfo {author}
  {\bibfnamefont {Bei}\ \bibnamefont {Zeng}}, and\ \bibinfo {author}
  {\bibfnamefont {Robert}\ \bibnamefont {Joynt}}} (\bibinfo {year}
  {2022}{\natexlab{a}}),\ \bibfield  {title} {\enquote {\bibinfo {title}
  {Mitigating algorithmic errors in quantum optimization through energy
  extrapolation},}\ }\href {https://doi.org/10.1088/2058-9565/ac969c}
  {\bibfield  {journal} {\bibinfo  {journal} {Quantum Science and Technology}\
  }\textbf {\bibinfo {volume} {8}}~(\bibinfo {number} {1}),\ \bibinfo {pages}
  {015004}}\BibitemShut {NoStop}%
\bibitem [{\citenamefont {Cao}\ \emph {et~al.}(2022{\natexlab{b}})\citenamefont
  {Cao}, \citenamefont {Lin}, \citenamefont {Kribs}, \citenamefont {Poon},
  \citenamefont {Zeng},\ and\ \citenamefont
  {Laflamme}}]{caoNISQErrorCorrection2022}%
  \BibitemOpen
  \bibfield  {author} {\bibinfo {author} {\bibnamefont {Cao}, \bibfnamefont
  {Ningping}}, \bibinfo {author} {\bibfnamefont {Junan}\ \bibnamefont {Lin}},
  \bibinfo {author} {\bibfnamefont {David}\ \bibnamefont {Kribs}}, \bibinfo
  {author} {\bibfnamefont {Yiu-Tung}\ \bibnamefont {Poon}}, \bibinfo {author}
  {\bibfnamefont {Bei}\ \bibnamefont {Zeng}}, and\ \bibinfo {author}
  {\bibfnamefont {Raymond}\ \bibnamefont {Laflamme}}} (\bibinfo {year}
  {2022}{\natexlab{b}}),\ \href {https://doi.org/10.48550/arXiv.2111.02345}
  {\enquote {\bibinfo {title} {{{NISQ}}: {{Error Correction}}, {{Mitigation}},
  and {{Noise Simulation}}},}\ }\bibinfo {howpublished} {arXiv:2111.02345
  [quant-ph]}\BibitemShut {NoStop}%
\bibitem [{\citenamefont {Chen}\ \emph {et~al.}(2021)\citenamefont {Chen},
  \citenamefont {Yu}, \citenamefont {Zeng},\ and\ \citenamefont
  {Flammia}}]{chenRobustShadowEstimation2021}%
  \BibitemOpen
  \bibfield  {author} {\bibinfo {author} {\bibnamefont {Chen}, \bibfnamefont
  {Senrui}}, \bibinfo {author} {\bibfnamefont {Wenjun}\ \bibnamefont {Yu}},
  \bibinfo {author} {\bibfnamefont {Pei}\ \bibnamefont {Zeng}}, and\ \bibinfo
  {author} {\bibfnamefont {Steven~T.}\ \bibnamefont {Flammia}}} (\bibinfo
  {year} {2021}),\ \bibfield  {title} {\enquote {\bibinfo {title} {Robust
  {{Shadow Estimation}}},}\ }\href
  {https://doi.org/10.1103/PRXQuantum.2.030348} {\bibfield  {journal} {\bibinfo
   {journal} {PRX Quantum}\ }\textbf {\bibinfo {volume} {2}}~(\bibinfo {number}
  {3}),\ \bibinfo {pages} {030348}}\BibitemShut {NoStop}%
\bibitem [{\citenamefont {Chen}\ \emph {et~al.}(2023)\citenamefont {Chen},
  \citenamefont {Zhang}, \citenamefont {Zhang}, \citenamefont {Su},
  \citenamefont {Lu}, \citenamefont {Zhang}, \citenamefont {Qiao},
  \citenamefont {Li}, \citenamefont {Zhang},\ and\ \citenamefont
  {Kim}}]{chenErrorMitigatedQuantumSimulation2023}%
  \BibitemOpen
  \bibfield  {author} {\bibinfo {author} {\bibnamefont {Chen}, \bibfnamefont
  {Wentao}}, \bibinfo {author} {\bibfnamefont {Shuaining}\ \bibnamefont
  {Zhang}}, \bibinfo {author} {\bibfnamefont {Jialiang}\ \bibnamefont {Zhang}},
  \bibinfo {author} {\bibfnamefont {Xiaolu}\ \bibnamefont {Su}}, \bibinfo
  {author} {\bibfnamefont {Yao}\ \bibnamefont {Lu}}, \bibinfo {author}
  {\bibfnamefont {Kuan}\ \bibnamefont {Zhang}}, \bibinfo {author}
  {\bibfnamefont {Mu}~\bibnamefont {Qiao}}, \bibinfo {author} {\bibfnamefont
  {Ying}\ \bibnamefont {Li}}, \bibinfo {author} {\bibfnamefont {Jing-Ning}\
  \bibnamefont {Zhang}}, and\ \bibinfo {author} {\bibfnamefont {Kihwan}\
  \bibnamefont {Kim}}} (\bibinfo {year} {2023}),\ \href
  {https://doi.org/10.48550/arXiv.2302.10436} {\enquote {\bibinfo {title}
  {Error-{{Mitigated Quantum Simulation}} of {{Interacting Fermions}} with
  {{Trapped Ions}}},}\ }\bibinfo {howpublished} {arXiv:2302.10436 [cond-mat,
  physics:physics, physics:quant-ph]}\BibitemShut {NoStop}%
\bibitem [{\citenamefont {Chen}\ \emph {et~al.}(2019)\citenamefont {Chen},
  \citenamefont {Farahzad}, \citenamefont {Yoo},\ and\ \citenamefont
  {Wei}}]{chenDetectorTomographyIBM2019}%
  \BibitemOpen
  \bibfield  {author} {\bibinfo {author} {\bibnamefont {Chen}, \bibfnamefont
  {Yanzhu}}, \bibinfo {author} {\bibfnamefont {Maziar}\ \bibnamefont
  {Farahzad}}, \bibinfo {author} {\bibfnamefont {Shinjae}\ \bibnamefont {Yoo}},
  and\ \bibinfo {author} {\bibfnamefont {Tzu-Chieh}\ \bibnamefont {Wei}}}
  (\bibinfo {year} {2019}),\ \bibfield  {title} {\enquote {\bibinfo {title}
  {Detector tomography on {{IBM}} quantum computers and mitigation of an
  imperfect measurement},}\ }\href
  {https://doi.org/10.1103/PhysRevA.100.052315} {\bibfield  {journal} {\bibinfo
   {journal} {Physical Review A}\ }\textbf {\bibinfo {volume} {100}}~(\bibinfo
  {number} {5}),\ \bibinfo {pages} {052315}}\BibitemShut {NoStop}%
\bibitem [{\citenamefont {Childs}\ \emph {et~al.}(2019)\citenamefont {Childs},
  \citenamefont {Ostrander},\ and\ \citenamefont
  {Su}}]{childsFasterQuantumSimulation2019}%
  \BibitemOpen
  \bibfield  {author} {\bibinfo {author} {\bibnamefont {Childs}, \bibfnamefont
  {Andrew~M}}, \bibinfo {author} {\bibfnamefont {Aaron}\ \bibnamefont
  {Ostrander}}, and\ \bibinfo {author} {\bibfnamefont {Yuan}\ \bibnamefont
  {Su}}} (\bibinfo {year} {2019}),\ \bibfield  {title} {\enquote {\bibinfo
  {title} {Faster quantum simulation by randomization},}\ }\href
  {https://doi.org/10.22331/q-2019-09-02-182} {\bibfield  {journal} {\bibinfo
  {journal} {Quantum}\ }\textbf {\bibinfo {volume} {3}},\ \bibinfo {pages}
  {182}}\BibitemShut {NoStop}%
\bibitem [{\citenamefont {Childs}\ and\ \citenamefont
  {Wiebe}(2012)}]{childsHamiltonianSimulationUsing2012}%
  \BibitemOpen
  \bibfield  {author} {\bibinfo {author} {\bibnamefont {Childs}, \bibfnamefont
  {Andrew~M}}, and\ \bibinfo {author} {\bibfnamefont {Nathan}\ \bibnamefont
  {Wiebe}}} (\bibinfo {year} {2012}),\ \bibfield  {title} {\enquote {\bibinfo
  {title} {Hamiltonian simulation using linear combinations of unitary
  operations},}\ }\href {https://doi.org/10.26421/QIC12.11-12-1} {\bibfield
  {journal} {\bibinfo  {journal} {Quantum Information and Computation}\
  }\textbf {\bibinfo {volume} {12}}~(\bibinfo {number} {11\&12}),\ \bibinfo
  {pages} {901--924}}\BibitemShut {NoStop}%
\bibitem [{\citenamefont {Chow}\ \emph {et~al.}(2010)\citenamefont {Chow},
  \citenamefont {DiCarlo}, \citenamefont {Gambetta}, \citenamefont
  {Nunnenkamp}, \citenamefont {Bishop}, \citenamefont {Frunzio}, \citenamefont
  {Devoret}, \citenamefont {Girvin},\ and\ \citenamefont
  {Schoelkopf}}]{chowDetectingHighlyEntangled2010}%
  \BibitemOpen
  \bibfield  {author} {\bibinfo {author} {\bibnamefont {Chow}, \bibfnamefont
  {J~M}}, \bibinfo {author} {\bibfnamefont {L.}~\bibnamefont {DiCarlo}},
  \bibinfo {author} {\bibfnamefont {J.~M.}\ \bibnamefont {Gambetta}}, \bibinfo
  {author} {\bibfnamefont {A.}~\bibnamefont {Nunnenkamp}}, \bibinfo {author}
  {\bibfnamefont {Lev~S.}\ \bibnamefont {Bishop}}, \bibinfo {author}
  {\bibfnamefont {L.}~\bibnamefont {Frunzio}}, \bibinfo {author} {\bibfnamefont
  {M.~H.}\ \bibnamefont {Devoret}}, \bibinfo {author} {\bibfnamefont {S.~M.}\
  \bibnamefont {Girvin}}, and\ \bibinfo {author} {\bibfnamefont {R.~J.}\
  \bibnamefont {Schoelkopf}}} (\bibinfo {year} {2010}),\ \bibfield  {title}
  {\enquote {\bibinfo {title} {Detecting highly entangled states with a joint
  qubit readout},}\ }\href {https://doi.org/10.1103/PhysRevA.81.062325}
  {\bibfield  {journal} {\bibinfo  {journal} {Physical Review A}\ }\textbf
  {\bibinfo {volume} {81}}~(\bibinfo {number} {6}),\ \bibinfo {pages}
  {062325}}\BibitemShut {NoStop}%
\bibitem [{\citenamefont {Chow}\ \emph {et~al.}(2011)\citenamefont {Chow},
  \citenamefont {C{\'o}rcoles}, \citenamefont {Gambetta}, \citenamefont
  {Rigetti}, \citenamefont {Johnson}, \citenamefont {Smolin}, \citenamefont
  {Rozen}, \citenamefont {Keefe}, \citenamefont {Rothwell}, \citenamefont
  {Ketchen},\ and\ \citenamefont
  {Steffen}}]{chowSimpleAllMicrowaveEntangling2011}%
  \BibitemOpen
  \bibfield  {author} {\bibinfo {author} {\bibnamefont {Chow}, \bibfnamefont
  {Jerry~M}}, \bibinfo {author} {\bibfnamefont {A.~D.}\ \bibnamefont
  {C{\'o}rcoles}}, \bibinfo {author} {\bibfnamefont {Jay~M.}\ \bibnamefont
  {Gambetta}}, \bibinfo {author} {\bibfnamefont {Chad}\ \bibnamefont
  {Rigetti}}, \bibinfo {author} {\bibfnamefont {B.~R.}\ \bibnamefont
  {Johnson}}, \bibinfo {author} {\bibfnamefont {John~A.}\ \bibnamefont
  {Smolin}}, \bibinfo {author} {\bibfnamefont {J.~R.}\ \bibnamefont {Rozen}},
  \bibinfo {author} {\bibfnamefont {George~A.}\ \bibnamefont {Keefe}}, \bibinfo
  {author} {\bibfnamefont {Mary~B.}\ \bibnamefont {Rothwell}}, \bibinfo
  {author} {\bibfnamefont {Mark~B.}\ \bibnamefont {Ketchen}}, and\ \bibinfo
  {author} {\bibfnamefont {M.}~\bibnamefont {Steffen}}} (\bibinfo {year}
  {2011}),\ \bibfield  {title} {\enquote {\bibinfo {title} {Simple
  {{All-Microwave Entangling Gate}} for {{Fixed-Frequency Superconducting
  Qubits}}},}\ }\href {https://doi.org/10.1103/PhysRevLett.107.080502}
  {\bibfield  {journal} {\bibinfo  {journal} {Physical Review Letters}\
  }\textbf {\bibinfo {volume} {107}}~(\bibinfo {number} {8}),\ \bibinfo {pages}
  {080502}}\BibitemShut {NoStop}%
\bibitem [{\citenamefont {Cirstoiu}\ \emph {et~al.}(2023)\citenamefont
  {Cirstoiu}, \citenamefont {Dilkes}, \citenamefont {Mills}, \citenamefont
  {Sivarajah},\ and\ \citenamefont
  {Duncan}}]{cirstoiuVolumetricBenchmarkingError2023}%
  \BibitemOpen
  \bibfield  {author} {\bibinfo {author} {\bibnamefont {Cirstoiu},
  \bibfnamefont {Cristina}}, \bibinfo {author} {\bibfnamefont {Silas}\
  \bibnamefont {Dilkes}}, \bibinfo {author} {\bibfnamefont {Daniel}\
  \bibnamefont {Mills}}, \bibinfo {author} {\bibfnamefont {Seyon}\ \bibnamefont
  {Sivarajah}}, and\ \bibinfo {author} {\bibfnamefont {Ross}\ \bibnamefont
  {Duncan}}} (\bibinfo {year} {2023}),\ \bibfield  {title} {\enquote {\bibinfo
  {title} {Volumetric {{Benchmarking}} of {{Error Mitigation}} with
  {{Qermit}}},}\ }\href {https://doi.org/10.22331/q-2023-07-13-1059} {\bibfield
   {journal} {\bibinfo  {journal} {Quantum}\ }\textbf {\bibinfo {volume} {7}},\
  \bibinfo {pages} {1059}}\BibitemShut {NoStop}%
\bibitem [{\citenamefont {Colless}\ \emph {et~al.}(2018)\citenamefont
  {Colless}, \citenamefont {Ramasesh}, \citenamefont {Dahlen}, \citenamefont
  {Blok}, \citenamefont {{Kimchi-Schwartz}}, \citenamefont {McClean},
  \citenamefont {Carter}, \citenamefont {{de Jong}},\ and\ \citenamefont
  {Siddiqi}}]{collessComputationMolecularSpectra2018}%
  \BibitemOpen
  \bibfield  {author} {\bibinfo {author} {\bibnamefont {Colless}, \bibfnamefont
  {J~I}}, \bibinfo {author} {\bibfnamefont {V.~V.}\ \bibnamefont {Ramasesh}},
  \bibinfo {author} {\bibfnamefont {D.}~\bibnamefont {Dahlen}}, \bibinfo
  {author} {\bibfnamefont {M.~S.}\ \bibnamefont {Blok}}, \bibinfo {author}
  {\bibfnamefont {M.~E.}\ \bibnamefont {{Kimchi-Schwartz}}}, \bibinfo {author}
  {\bibfnamefont {J.~R.}\ \bibnamefont {McClean}}, \bibinfo {author}
  {\bibfnamefont {J.}~\bibnamefont {Carter}}, \bibinfo {author} {\bibfnamefont
  {W.~A.}\ \bibnamefont {{de Jong}}}, and\ \bibinfo {author} {\bibfnamefont
  {I.}~\bibnamefont {Siddiqi}}} (\bibinfo {year} {2018}),\ \bibfield  {title}
  {\enquote {\bibinfo {title} {Computation of {{Molecular Spectra}} on a
  {{Quantum Processor}} with an {{Error-Resilient Algorithm}}},}\ }\href
  {https://doi.org/10.1103/PhysRevX.8.011021} {\bibfield  {journal} {\bibinfo
  {journal} {Physical Review X}\ }\textbf {\bibinfo {volume} {8}}~(\bibinfo
  {number} {1}),\ \bibinfo {pages} {011021}}\BibitemShut {NoStop}%
\bibitem [{\citenamefont {Conlon}\ \emph {et~al.}(2023)\citenamefont {Conlon},
  \citenamefont {Vogl}, \citenamefont {Marciniak}, \citenamefont {Pogorelov},
  \citenamefont {Yung}, \citenamefont {Eilenberger}, \citenamefont {Berry},
  \citenamefont {Santana}, \citenamefont {Blatt}, \citenamefont {Monz},
  \citenamefont {Lam},\ and\ \citenamefont
  {Assad}}]{conlonApproachingOptimalEntangling2023}%
  \BibitemOpen
  \bibfield  {author} {\bibinfo {author} {\bibnamefont {Conlon}, \bibfnamefont
  {Lorc{\'a}n~O}}, \bibinfo {author} {\bibfnamefont {Tobias}\ \bibnamefont
  {Vogl}}, \bibinfo {author} {\bibfnamefont {Christian~D.}\ \bibnamefont
  {Marciniak}}, \bibinfo {author} {\bibfnamefont {Ivan}\ \bibnamefont
  {Pogorelov}}, \bibinfo {author} {\bibfnamefont {Simon~K.}\ \bibnamefont
  {Yung}}, \bibinfo {author} {\bibfnamefont {Falk}\ \bibnamefont
  {Eilenberger}}, \bibinfo {author} {\bibfnamefont {Dominic~W.}\ \bibnamefont
  {Berry}}, \bibinfo {author} {\bibfnamefont {Fabiana~S.}\ \bibnamefont
  {Santana}}, \bibinfo {author} {\bibfnamefont {Rainer}\ \bibnamefont {Blatt}},
  \bibinfo {author} {\bibfnamefont {Thomas}\ \bibnamefont {Monz}}, \bibinfo
  {author} {\bibfnamefont {Ping~Koy}\ \bibnamefont {Lam}}, and\ \bibinfo
  {author} {\bibfnamefont {Syed~M.}\ \bibnamefont {Assad}}} (\bibinfo {year}
  {2023}),\ \bibfield  {title} {\enquote {\bibinfo {title} {Approaching optimal
  entangling collective measurements on quantum computing platforms},}\ }\href
  {https://doi.org/10.1038/s41567-022-01875-7} {\bibfield  {journal} {\bibinfo
  {journal} {Nature Physics}\ }\textbf {\bibinfo {volume} {19}}~(\bibinfo
  {number} {3}),\ \bibinfo {pages} {351--357}}\BibitemShut {NoStop}%
\bibitem [{\citenamefont {C{\'o}rcoles}\ \emph {et~al.}(2015)\citenamefont
  {C{\'o}rcoles}, \citenamefont {Magesan}, \citenamefont {Srinivasan},
  \citenamefont {Cross}, \citenamefont {Steffen}, \citenamefont {Gambetta},\
  and\ \citenamefont {Chow}}]{corcolesDemonstrationQuantumError2015}%
  \BibitemOpen
  \bibfield  {author} {\bibinfo {author} {\bibnamefont {C{\'o}rcoles},
  \bibfnamefont {A~D}}, \bibinfo {author} {\bibfnamefont {Easwar}\ \bibnamefont
  {Magesan}}, \bibinfo {author} {\bibfnamefont {Srikanth~J.}\ \bibnamefont
  {Srinivasan}}, \bibinfo {author} {\bibfnamefont {Andrew~W.}\ \bibnamefont
  {Cross}}, \bibinfo {author} {\bibfnamefont {M.}~\bibnamefont {Steffen}},
  \bibinfo {author} {\bibfnamefont {Jay~M.}\ \bibnamefont {Gambetta}}, and\
  \bibinfo {author} {\bibfnamefont {Jerry~M.}\ \bibnamefont {Chow}}} (\bibinfo
  {year} {2015}),\ \bibfield  {title} {\enquote {\bibinfo {title}
  {Demonstration of a quantum error detection code using a square lattice of
  four superconducting qubits},}\ }\href {https://doi.org/10.1038/ncomms7979}
  {\bibfield  {journal} {\bibinfo  {journal} {Nature Communications}\ }\textbf
  {\bibinfo {volume} {6}},\ \bibinfo {pages} {6979}}\BibitemShut {NoStop}%
\bibitem [{\citenamefont {Czarnik}\ \emph {et~al.}(2021)\citenamefont
  {Czarnik}, \citenamefont {Arrasmith}, \citenamefont {Coles},\ and\
  \citenamefont {Cincio}}]{czarnikErrorMitigationClifford2021}%
  \BibitemOpen
  \bibfield  {author} {\bibinfo {author} {\bibnamefont {Czarnik}, \bibfnamefont
  {Piotr}}, \bibinfo {author} {\bibfnamefont {Andrew}\ \bibnamefont
  {Arrasmith}}, \bibinfo {author} {\bibfnamefont {Patrick~J.}\ \bibnamefont
  {Coles}}, and\ \bibinfo {author} {\bibfnamefont {Lukasz}\ \bibnamefont
  {Cincio}}} (\bibinfo {year} {2021}),\ \bibfield  {title} {\enquote {\bibinfo
  {title} {Error mitigation with {{Clifford}} quantum-circuit data},}\ }\href
  {https://doi.org/10.22331/q-2021-11-26-592} {\bibfield  {journal} {\bibinfo
  {journal} {Quantum}\ }\textbf {\bibinfo {volume} {5}},\ \bibinfo {pages}
  {592}}\BibitemShut {NoStop}%
\bibitem [{\citenamefont {Czarnik}\ \emph {et~al.}(2022)\citenamefont
  {Czarnik}, \citenamefont {McKerns}, \citenamefont {Sornborger},\ and\
  \citenamefont {Cincio}}]{czarnikImprovingEfficiencyLearningbased2022}%
  \BibitemOpen
  \bibfield  {author} {\bibinfo {author} {\bibnamefont {Czarnik}, \bibfnamefont
  {Piotr}}, \bibinfo {author} {\bibfnamefont {Michael}\ \bibnamefont
  {McKerns}}, \bibinfo {author} {\bibfnamefont {Andrew~T.}\ \bibnamefont
  {Sornborger}}, and\ \bibinfo {author} {\bibfnamefont {Lukasz}\ \bibnamefont
  {Cincio}}} (\bibinfo {year} {2022}),\ \href
  {https://doi.org/10.48550/arXiv.2204.07109} {\enquote {\bibinfo {title}
  {Improving the efficiency of learning-based error mitigation},}\ }\bibinfo
  {howpublished} {arXiv:2204.07109 [quant-ph]}\BibitemShut {NoStop}%
\bibitem [{\citenamefont {{Dallaire-Demers}}\ \emph {et~al.}(2019)\citenamefont
  {{Dallaire-Demers}}, \citenamefont {Romero}, \citenamefont {Veis},
  \citenamefont {Sim},\ and\ \citenamefont
  {{Aspuru-Guzik}}}]{dallaire-demersLowdepthCircuitAnsatz2019}%
  \BibitemOpen
  \bibfield  {author} {\bibinfo {author} {\bibnamefont {{Dallaire-Demers}},
  \bibfnamefont {Pierre-Luc}}, \bibinfo {author} {\bibfnamefont {Jonathan}\
  \bibnamefont {Romero}}, \bibinfo {author} {\bibfnamefont {Libor}\
  \bibnamefont {Veis}}, \bibinfo {author} {\bibfnamefont {Sukin}\ \bibnamefont
  {Sim}}, and\ \bibinfo {author} {\bibfnamefont {Al{\'a}n}\ \bibnamefont
  {{Aspuru-Guzik}}}} (\bibinfo {year} {2019}),\ \bibfield  {title} {\enquote
  {\bibinfo {title} {Low-depth circuit ansatz for preparing correlated
  fermionic states on a quantum computer},}\ }\href
  {https://doi.org/10.1088/2058-9565/ab3951} {\bibfield  {journal} {\bibinfo
  {journal} {Quantum Science and Technology}\ }\textbf {\bibinfo {volume}
  {4}}~(\bibinfo {number} {4}),\ \bibinfo {pages} {045005}}\BibitemShut
  {NoStop}%
\bibitem [{\citenamefont {Dawson}\ and\ \citenamefont
  {Nielsen}(2006)}]{dawsonSolovayKitaevAlgorithm2006}%
  \BibitemOpen
  \bibfield  {author} {\bibinfo {author} {\bibnamefont {Dawson}, \bibfnamefont
  {CM}}, and\ \bibinfo {author} {\bibfnamefont {M.A.}\ \bibnamefont {Nielsen}}}
  (\bibinfo {year} {2006}),\ \bibfield  {title} {\enquote {\bibinfo {title}
  {The {{Solovay-Kitaev}} algorithm},}\ }\href
  {https://doi.org/10.26421/QIC6.1-6} {\bibfield  {journal} {\bibinfo
  {journal} {Quantum Information and Computation}\ }\textbf {\bibinfo {volume}
  {6}}~(\bibinfo {number} {1}),\ \bibinfo {pages} {81--95}}\BibitemShut
  {NoStop}%
\bibitem [{\citenamefont {Dborin}\ \emph {et~al.}(2022)\citenamefont {Dborin},
  \citenamefont {Wimalaweera}, \citenamefont {Barratt}, \citenamefont {Ostby},
  \citenamefont {O'Brien},\ and\ \citenamefont
  {Green}}]{dborinSimulatingGroundstateDynamical2022}%
  \BibitemOpen
  \bibfield  {author} {\bibinfo {author} {\bibnamefont {Dborin}, \bibfnamefont
  {James}}, \bibinfo {author} {\bibfnamefont {Vinul}\ \bibnamefont
  {Wimalaweera}}, \bibinfo {author} {\bibfnamefont {F.}~\bibnamefont
  {Barratt}}, \bibinfo {author} {\bibfnamefont {Eric}\ \bibnamefont {Ostby}},
  \bibinfo {author} {\bibfnamefont {Thomas~E.}\ \bibnamefont {O'Brien}}, and\
  \bibinfo {author} {\bibfnamefont {A.~G.}\ \bibnamefont {Green}}} (\bibinfo
  {year} {2022}),\ \bibfield  {title} {\enquote {\bibinfo {title} {Simulating
  groundstate and dynamical quantum phase transitions on a superconducting
  quantum computer},}\ }\href {https://doi.org/10.1038/s41467-022-33737-4}
  {\bibfield  {journal} {\bibinfo  {journal} {Nature Communications}\ }\textbf
  {\bibinfo {volume} {13}}~(\bibinfo {number} {1}),\ \bibinfo {pages}
  {5977}}\BibitemShut {NoStop}%
\bibitem [{\citenamefont
  {DePrince}(2016)}]{deprinceVariationalOptimizationTwoelectron2016}%
  \BibitemOpen
  \bibfield  {author} {\bibinfo {author} {\bibnamefont {DePrince},
  \bibfnamefont {A~Eugene}}} (\bibinfo {year} {2016}),\ \bibfield  {title}
  {\enquote {\bibinfo {title} {Variational optimization of the two-electron
  reduced-density matrix under pure-state {{N-representability}} conditions},}\
  }\href {https://doi.org/10.1063/1.4965888} {\bibfield  {journal} {\bibinfo
  {journal} {The Journal of Chemical Physics}\ }\textbf {\bibinfo {volume}
  {145}}~(\bibinfo {number} {16}),\ \bibinfo {pages} {164109}}\BibitemShut
  {NoStop}%
\bibitem [{\citenamefont {Derby}\ and\ \citenamefont
  {Klassen}(2021)}]{derbyCompactFermionQubit2021a}%
  \BibitemOpen
  \bibfield  {author} {\bibinfo {author} {\bibnamefont {Derby}, \bibfnamefont
  {Charles}}, and\ \bibinfo {author} {\bibfnamefont {Joel}\ \bibnamefont
  {Klassen}}} (\bibinfo {year} {2021}),\ \href
  {http://arxiv.org/abs/2101.10735} {\enquote {\bibinfo {title} {A {{Compact
  Fermion}} to {{Qubit Mapping Part}} 2: {{Alternative Lattice Geometries}}},}\
  }\bibinfo {howpublished} {arXiv:2101.10735 [quant-ph]}\BibitemShut {NoStop}%
\bibitem [{\citenamefont {Dinur}\ \emph {et~al.}(2022)\citenamefont {Dinur},
  \citenamefont {Evra}, \citenamefont {Livne}, \citenamefont {Lubotzky},\ and\
  \citenamefont {Mozes}}]{dinurLocallyTestableCodes2022}%
  \BibitemOpen
  \bibfield  {author} {\bibinfo {author} {\bibnamefont {Dinur}, \bibfnamefont
  {Irit}}, \bibinfo {author} {\bibfnamefont {Shai}\ \bibnamefont {Evra}},
  \bibinfo {author} {\bibfnamefont {Ron}\ \bibnamefont {Livne}}, \bibinfo
  {author} {\bibfnamefont {Alexander}\ \bibnamefont {Lubotzky}}, and\ \bibinfo
  {author} {\bibfnamefont {Shahar}\ \bibnamefont {Mozes}}} (\bibinfo {year}
  {2022}),\ \bibfield  {title} {\enquote {\bibinfo {title} {Locally testable
  codes with constant rate, distance, and locality},}\ }in\ \href
  {https://doi.org/10.1145/3519935.3520024} {\emph {\bibinfo {booktitle}
  {Proceedings of the 54th {{Annual ACM SIGACT Symposium}} on {{Theory}} of
  {{Computing}}}}},\ \bibinfo {series and number} {{{STOC}} 2022}\ (\bibinfo
  {publisher} {{Association for Computing Machinery}},\ \bibinfo {address}
  {{New York, NY, USA}})\ pp.\ \bibinfo {pages} {357--374}\BibitemShut
  {NoStop}%
\bibitem [{\citenamefont {Dumitrescu}\ \emph {et~al.}(2018)\citenamefont
  {Dumitrescu}, \citenamefont {McCaskey}, \citenamefont {Hagen}, \citenamefont
  {Jansen}, \citenamefont {Morris}, \citenamefont {Papenbrock}, \citenamefont
  {Pooser}, \citenamefont {Dean},\ and\ \citenamefont
  {Lougovski}}]{dumitrescuCloudQuantumComputing2018}%
  \BibitemOpen
  \bibfield  {author} {\bibinfo {author} {\bibnamefont {Dumitrescu},
  \bibfnamefont {E~F}}, \bibinfo {author} {\bibfnamefont {A.~J.}\ \bibnamefont
  {McCaskey}}, \bibinfo {author} {\bibfnamefont {G.}~\bibnamefont {Hagen}},
  \bibinfo {author} {\bibfnamefont {G.~R.}\ \bibnamefont {Jansen}}, \bibinfo
  {author} {\bibfnamefont {T.~D.}\ \bibnamefont {Morris}}, \bibinfo {author}
  {\bibfnamefont {T.}~\bibnamefont {Papenbrock}}, \bibinfo {author}
  {\bibfnamefont {R.~C.}\ \bibnamefont {Pooser}}, \bibinfo {author}
  {\bibfnamefont {D.~J.}\ \bibnamefont {Dean}}, and\ \bibinfo {author}
  {\bibfnamefont {P.}~\bibnamefont {Lougovski}}} (\bibinfo {year} {2018}),\
  \bibfield  {title} {\enquote {\bibinfo {title} {Cloud {{Quantum Computing}}
  of an {{Atomic Nucleus}}},}\ }\href
  {https://doi.org/10.1103/PhysRevLett.120.210501} {\bibfield  {journal}
  {\bibinfo  {journal} {Physical Review Letters}\ }\textbf {\bibinfo {volume}
  {120}}~(\bibinfo {number} {21}),\ \bibinfo {pages} {210501}}\BibitemShut
  {NoStop}%
\bibitem [{\citenamefont {Eastin}\ and\ \citenamefont
  {Knill}(2009)}]{eastinRestrictionsTransversalEncoded2009}%
  \BibitemOpen
  \bibfield  {author} {\bibinfo {author} {\bibnamefont {Eastin}, \bibfnamefont
  {Bryan}}, and\ \bibinfo {author} {\bibfnamefont {Emanuel}\ \bibnamefont
  {Knill}}} (\bibinfo {year} {2009}),\ \bibfield  {title} {\enquote {\bibinfo
  {title} {Restrictions on {{Transversal Encoded Quantum Gate Sets}}},}\ }\href
  {https://doi.org/10.1103/PhysRevLett.102.110502} {\bibfield  {journal}
  {\bibinfo  {journal} {Physical Review Letters}\ }\textbf {\bibinfo {volume}
  {102}}~(\bibinfo {number} {11}),\ 10.1103/PhysRevLett.102.110502}\BibitemShut
  {NoStop}%
\bibitem [{\citenamefont {Ebadi}\ \emph {et~al.}(2021)\citenamefont {Ebadi},
  \citenamefont {Wang}, \citenamefont {Levine}, \citenamefont {Keesling},
  \citenamefont {Semeghini}, \citenamefont {Omran}, \citenamefont {Bluvstein},
  \citenamefont {Samajdar}, \citenamefont {Pichler}, \citenamefont {Ho},
  \citenamefont {Choi}, \citenamefont {Sachdev}, \citenamefont {Greiner},
  \citenamefont {Vuleti{\'c}},\ and\ \citenamefont
  {Lukin}}]{ebadiQuantumPhasesMatter2021}%
  \BibitemOpen
  \bibfield  {author} {\bibinfo {author} {\bibnamefont {Ebadi}, \bibfnamefont
  {Sepehr}}, \bibinfo {author} {\bibfnamefont {Tout~T.}\ \bibnamefont {Wang}},
  \bibinfo {author} {\bibfnamefont {Harry}\ \bibnamefont {Levine}}, \bibinfo
  {author} {\bibfnamefont {Alexander}\ \bibnamefont {Keesling}}, \bibinfo
  {author} {\bibfnamefont {Giulia}\ \bibnamefont {Semeghini}}, \bibinfo
  {author} {\bibfnamefont {Ahmed}\ \bibnamefont {Omran}}, \bibinfo {author}
  {\bibfnamefont {Dolev}\ \bibnamefont {Bluvstein}}, \bibinfo {author}
  {\bibfnamefont {Rhine}\ \bibnamefont {Samajdar}}, \bibinfo {author}
  {\bibfnamefont {Hannes}\ \bibnamefont {Pichler}}, \bibinfo {author}
  {\bibfnamefont {Wen~Wei}\ \bibnamefont {Ho}}, \bibinfo {author}
  {\bibfnamefont {Soonwon}\ \bibnamefont {Choi}}, \bibinfo {author}
  {\bibfnamefont {Subir}\ \bibnamefont {Sachdev}}, \bibinfo {author}
  {\bibfnamefont {Markus}\ \bibnamefont {Greiner}}, \bibinfo {author}
  {\bibfnamefont {Vladan}\ \bibnamefont {Vuleti{\'c}}}, and\ \bibinfo {author}
  {\bibfnamefont {Mikhail~D.}\ \bibnamefont {Lukin}}} (\bibinfo {year}
  {2021}),\ \bibfield  {title} {\enquote {\bibinfo {title} {Quantum phases of
  matter on a 256-atom programmable quantum simulator},}\ }\href
  {https://doi.org/10.1038/s41586-021-03582-4} {\bibfield  {journal} {\bibinfo
  {journal} {Nature}\ }\textbf {\bibinfo {volume} {595}}~(\bibinfo {number}
  {7866}),\ \bibinfo {pages} {227--232}}\BibitemShut {NoStop}%
\bibitem [{\citenamefont {Egan}\ \emph {et~al.}(2021)\citenamefont {Egan},
  \citenamefont {Debroy}, \citenamefont {Noel}, \citenamefont {Risinger},
  \citenamefont {Zhu}, \citenamefont {Biswas}, \citenamefont {Newman},
  \citenamefont {Li}, \citenamefont {Brown}, \citenamefont {Cetina},\ and\
  \citenamefont {Monroe}}]{eganFaulttolerantControlErrorcorrected2021}%
  \BibitemOpen
  \bibfield  {author} {\bibinfo {author} {\bibnamefont {Egan}, \bibfnamefont
  {Laird}}, \bibinfo {author} {\bibfnamefont {Dripto~M.}\ \bibnamefont
  {Debroy}}, \bibinfo {author} {\bibfnamefont {Crystal}\ \bibnamefont {Noel}},
  \bibinfo {author} {\bibfnamefont {Andrew}\ \bibnamefont {Risinger}}, \bibinfo
  {author} {\bibfnamefont {Daiwei}\ \bibnamefont {Zhu}}, \bibinfo {author}
  {\bibfnamefont {Debopriyo}\ \bibnamefont {Biswas}}, \bibinfo {author}
  {\bibfnamefont {Michael}\ \bibnamefont {Newman}}, \bibinfo {author}
  {\bibfnamefont {Muyuan}\ \bibnamefont {Li}}, \bibinfo {author} {\bibfnamefont
  {Kenneth~R.}\ \bibnamefont {Brown}}, \bibinfo {author} {\bibfnamefont
  {Marko}\ \bibnamefont {Cetina}}, and\ \bibinfo {author} {\bibfnamefont
  {Christopher}\ \bibnamefont {Monroe}}} (\bibinfo {year} {2021}),\ \bibfield
  {title} {\enquote {\bibinfo {title} {Fault-tolerant control of an
  error-corrected qubit},}\ }\href {https://doi.org/10.1038/s41586-021-03928-y}
  {\bibfield  {journal} {\bibinfo  {journal} {Nature}\ }\textbf {\bibinfo
  {volume} {598}}~(\bibinfo {number} {7880}),\ \bibinfo {pages}
  {281--286}}\BibitemShut {NoStop}%
\bibitem [{\citenamefont {Eisert}\ \emph {et~al.}(2020)\citenamefont {Eisert},
  \citenamefont {Hangleiter}, \citenamefont {Walk}, \citenamefont {Roth},
  \citenamefont {Markham}, \citenamefont {Parekh}, \citenamefont {Chabaud},\
  and\ \citenamefont {Kashefi}}]{eisertQuantumCertificationBenchmarking2020}%
  \BibitemOpen
  \bibfield  {author} {\bibinfo {author} {\bibnamefont {Eisert}, \bibfnamefont
  {Jens}}, \bibinfo {author} {\bibfnamefont {Dominik}\ \bibnamefont
  {Hangleiter}}, \bibinfo {author} {\bibfnamefont {Nathan}\ \bibnamefont
  {Walk}}, \bibinfo {author} {\bibfnamefont {Ingo}\ \bibnamefont {Roth}},
  \bibinfo {author} {\bibfnamefont {Damian}\ \bibnamefont {Markham}}, \bibinfo
  {author} {\bibfnamefont {Rhea}\ \bibnamefont {Parekh}}, \bibinfo {author}
  {\bibfnamefont {Ulysse}\ \bibnamefont {Chabaud}}, and\ \bibinfo {author}
  {\bibfnamefont {Elham}\ \bibnamefont {Kashefi}}} (\bibinfo {year} {2020}),\
  \bibfield  {title} {\enquote {\bibinfo {title} {Quantum certification and
  benchmarking},}\ }\href {https://doi.org/10.1038/s42254-020-0186-4}
  {\bibfield  {journal} {\bibinfo  {journal} {Nature Reviews Physics}\ }\textbf
  {\bibinfo {volume} {2}}~(\bibinfo {number} {7}),\ \bibinfo {pages}
  {382--390}}\BibitemShut {NoStop}%
\bibitem [{\citenamefont {Endo}\ \emph {et~al.}(2018)\citenamefont {Endo},
  \citenamefont {Benjamin},\ and\ \citenamefont
  {Li}}]{endoPracticalQuantumError2018}%
  \BibitemOpen
  \bibfield  {author} {\bibinfo {author} {\bibnamefont {Endo}, \bibfnamefont
  {Suguru}}, \bibinfo {author} {\bibfnamefont {Simon~C.}\ \bibnamefont
  {Benjamin}}, and\ \bibinfo {author} {\bibfnamefont {Ying}\ \bibnamefont
  {Li}}} (\bibinfo {year} {2018}),\ \bibfield  {title} {\enquote {\bibinfo
  {title} {Practical {{Quantum Error Mitigation}} for {{Near-Future
  Applications}}},}\ }\href {https://doi.org/10.1103/PhysRevX.8.031027}
  {\bibfield  {journal} {\bibinfo  {journal} {Physical Review X}\ }\textbf
  {\bibinfo {volume} {8}}~(\bibinfo {number} {3}),\ \bibinfo {pages}
  {031027}}\BibitemShut {NoStop}%
\bibitem [{\citenamefont {Endo}\ \emph {et~al.}(2022)\citenamefont {Endo},
  \citenamefont {Suzuki}, \citenamefont {Tsubouchi}, \citenamefont {Asaoka},
  \citenamefont {Yamamoto}, \citenamefont {Matsuzaki},\ and\ \citenamefont
  {Tokunaga}}]{endoQuantumErrorMitigation2022a}%
  \BibitemOpen
  \bibfield  {author} {\bibinfo {author} {\bibnamefont {Endo}, \bibfnamefont
  {Suguru}}, \bibinfo {author} {\bibfnamefont {Yasunari}\ \bibnamefont
  {Suzuki}}, \bibinfo {author} {\bibfnamefont {Kento}\ \bibnamefont
  {Tsubouchi}}, \bibinfo {author} {\bibfnamefont {Rui}\ \bibnamefont {Asaoka}},
  \bibinfo {author} {\bibfnamefont {Kaoru}\ \bibnamefont {Yamamoto}}, \bibinfo
  {author} {\bibfnamefont {Yuichiro}\ \bibnamefont {Matsuzaki}}, and\ \bibinfo
  {author} {\bibfnamefont {Yuuki}\ \bibnamefont {Tokunaga}}} (\bibinfo {year}
  {2022}),\ \href {https://doi.org/10.48550/arXiv.2211.06164} {\enquote
  {\bibinfo {title} {Quantum error mitigation for rotation symmetric bosonic
  codes with symmetry expansion},}\ }\bibinfo {howpublished} {arXiv:2211.06164
  [quant-ph]}\BibitemShut {NoStop}%
\bibitem [{\citenamefont {Endo}\ \emph {et~al.}(2019)\citenamefont {Endo},
  \citenamefont {Zhao}, \citenamefont {Li}, \citenamefont {Benjamin},\ and\
  \citenamefont {Yuan}}]{endoMitigatingAlgorithmicErrors2019}%
  \BibitemOpen
  \bibfield  {author} {\bibinfo {author} {\bibnamefont {Endo}, \bibfnamefont
  {Suguru}}, \bibinfo {author} {\bibfnamefont {Qi}~\bibnamefont {Zhao}},
  \bibinfo {author} {\bibfnamefont {Ying}\ \bibnamefont {Li}}, \bibinfo
  {author} {\bibfnamefont {Simon}\ \bibnamefont {Benjamin}}, and\ \bibinfo
  {author} {\bibfnamefont {Xiao}\ \bibnamefont {Yuan}}} (\bibinfo {year}
  {2019}),\ \bibfield  {title} {\enquote {\bibinfo {title} {Mitigating
  algorithmic errors in a {{Hamiltonian}} simulation},}\ }\href
  {https://doi.org/10.1103/PhysRevA.99.012334} {\bibfield  {journal} {\bibinfo
  {journal} {Physical Review A}\ }\textbf {\bibinfo {volume} {99}}~(\bibinfo
  {number} {1}),\ \bibinfo {pages} {012334}}\BibitemShut {NoStop}%
\bibitem [{\citenamefont {Facchi}\ \emph {et~al.}(2004)\citenamefont {Facchi},
  \citenamefont {Lidar},\ and\ \citenamefont
  {Pascazio}}]{facchiUnificationDynamicalDecoupling2004}%
  \BibitemOpen
  \bibfield  {author} {\bibinfo {author} {\bibnamefont {Facchi}, \bibfnamefont
  {P}}, \bibinfo {author} {\bibfnamefont {D.~A.}\ \bibnamefont {Lidar}}, and\
  \bibinfo {author} {\bibfnamefont {S.}~\bibnamefont {Pascazio}}} (\bibinfo
  {year} {2004}),\ \bibfield  {title} {\enquote {\bibinfo {title} {Unification
  of dynamical decoupling and the quantum {{Zeno}} effect},}\ }\href
  {https://doi.org/10.1103/PhysRevA.69.032314} {\bibfield  {journal} {\bibinfo
  {journal} {Physical Review A}\ }\textbf {\bibinfo {volume} {69}}~(\bibinfo
  {number} {3}),\ \bibinfo {pages} {032314}}\BibitemShut {NoStop}%
\bibitem [{\citenamefont {Farhi}\ \emph {et~al.}(2014)\citenamefont {Farhi},
  \citenamefont {Goldstone},\ and\ \citenamefont
  {Gutmann}}]{farhiQuantumApproximateOptimization2014}%
  \BibitemOpen
  \bibfield  {author} {\bibinfo {author} {\bibnamefont {Farhi}, \bibfnamefont
  {Edward}}, \bibinfo {author} {\bibfnamefont {Jeffrey}\ \bibnamefont
  {Goldstone}}, and\ \bibinfo {author} {\bibfnamefont {Sam}\ \bibnamefont
  {Gutmann}}} (\bibinfo {year} {2014}),\ \href {http://arxiv.org/abs/1411.4028}
  {\enquote {\bibinfo {title} {A {{Quantum Approximate Optimization
  Algorithm}}},}\ }\bibinfo {howpublished} {arXiv:1411.4028
  [quant-ph]}\BibitemShut {NoStop}%
\bibitem [{\citenamefont {Ferracin}\ \emph {et~al.}(2022)\citenamefont
  {Ferracin}, \citenamefont {Hashim}, \citenamefont {Ville}, \citenamefont
  {Naik}, \citenamefont {{Carignan-Dugas}}, \citenamefont {Qassim},
  \citenamefont {Morvan}, \citenamefont {Santiago}, \citenamefont {Siddiqi},\
  and\ \citenamefont {Wallman}}]{ferracinEfficientlyImprovingPerformance2022}%
  \BibitemOpen
  \bibfield  {author} {\bibinfo {author} {\bibnamefont {Ferracin},
  \bibfnamefont {Samuele}}, \bibinfo {author} {\bibfnamefont {Akel}\
  \bibnamefont {Hashim}}, \bibinfo {author} {\bibfnamefont {Jean-Loup}\
  \bibnamefont {Ville}}, \bibinfo {author} {\bibfnamefont {Ravi}\ \bibnamefont
  {Naik}}, \bibinfo {author} {\bibfnamefont {Arnaud}\ \bibnamefont
  {{Carignan-Dugas}}}, \bibinfo {author} {\bibfnamefont {Hammam}\ \bibnamefont
  {Qassim}}, \bibinfo {author} {\bibfnamefont {Alexis}\ \bibnamefont {Morvan}},
  \bibinfo {author} {\bibfnamefont {David~I.}\ \bibnamefont {Santiago}},
  \bibinfo {author} {\bibfnamefont {Irfan}\ \bibnamefont {Siddiqi}}, and\
  \bibinfo {author} {\bibfnamefont {Joel~J.}\ \bibnamefont {Wallman}}}
  (\bibinfo {year} {2022}),\ \href {https://doi.org/10.48550/arXiv.2201.10672}
  {\enquote {\bibinfo {title} {Efficiently improving the performance of noisy
  quantum computers},}\ }\bibinfo {howpublished} {arXiv:2201.10672
  [quant-ph]}\BibitemShut {NoStop}%
\bibitem [{\citenamefont
  {Feynman}(1982)}]{feynmanSimulatingPhysicsComputers1982}%
  \BibitemOpen
  \bibfield  {author} {\bibinfo {author} {\bibnamefont {Feynman}, \bibfnamefont
  {Richard~P}}} (\bibinfo {year} {1982}),\ \bibfield  {title} {\enquote
  {\bibinfo {title} {Simulating physics with computers},}\ }\href
  {https://doi.org/10.1007/BF02650179} {\bibfield  {journal} {\bibinfo
  {journal} {International Journal of Theoretical Physics}\ }\textbf {\bibinfo
  {volume} {21}}~(\bibinfo {number} {6-7}),\ \bibinfo {pages}
  {467--488}}\BibitemShut {NoStop}%
\bibitem [{\citenamefont {{Foss-Feig}}\ \emph {et~al.}(2022)\citenamefont
  {{Foss-Feig}}, \citenamefont {Ragole}, \citenamefont {Potter}, \citenamefont
  {Dreiling}, \citenamefont {Figgatt}, \citenamefont {Gaebler}, \citenamefont
  {Hall}, \citenamefont {Moses}, \citenamefont {Pino}, \citenamefont {Spaun},
  \citenamefont {Neyenhuis},\ and\ \citenamefont
  {Hayes}}]{foss-feigEntanglementTensorNetworks2022}%
  \BibitemOpen
  \bibfield  {author} {\bibinfo {author} {\bibnamefont {{Foss-Feig}},
  \bibfnamefont {Michael}}, \bibinfo {author} {\bibfnamefont {Stephen}\
  \bibnamefont {Ragole}}, \bibinfo {author} {\bibfnamefont {Andrew}\
  \bibnamefont {Potter}}, \bibinfo {author} {\bibfnamefont {Joan}\ \bibnamefont
  {Dreiling}}, \bibinfo {author} {\bibfnamefont {Caroline}\ \bibnamefont
  {Figgatt}}, \bibinfo {author} {\bibfnamefont {John}\ \bibnamefont {Gaebler}},
  \bibinfo {author} {\bibfnamefont {Alex}\ \bibnamefont {Hall}}, \bibinfo
  {author} {\bibfnamefont {Steven}\ \bibnamefont {Moses}}, \bibinfo {author}
  {\bibfnamefont {Juan}\ \bibnamefont {Pino}}, \bibinfo {author} {\bibfnamefont
  {Ben}\ \bibnamefont {Spaun}}, \bibinfo {author} {\bibfnamefont {Brian}\
  \bibnamefont {Neyenhuis}}, and\ \bibinfo {author} {\bibfnamefont {David}\
  \bibnamefont {Hayes}}} (\bibinfo {year} {2022}),\ \bibfield  {title}
  {\enquote {\bibinfo {title} {Entanglement from {{Tensor Networks}} on a
  {{Trapped-Ion Quantum Computer}}},}\ }\href
  {https://doi.org/10.1103/PhysRevLett.128.150504} {\bibfield  {journal}
  {\bibinfo  {journal} {Physical Review Letters}\ }\textbf {\bibinfo {volume}
  {128}}~(\bibinfo {number} {15}),\ \bibinfo {pages} {150504}}\BibitemShut
  {NoStop}%
\bibitem [{\citenamefont {Fowler}\ \emph {et~al.}(2012)\citenamefont {Fowler},
  \citenamefont {Mariantoni}, \citenamefont {Martinis},\ and\ \citenamefont
  {Cleland}}]{fowlerSurfaceCodesPractical2012}%
  \BibitemOpen
  \bibfield  {author} {\bibinfo {author} {\bibnamefont {Fowler}, \bibfnamefont
  {Austin~G}}, \bibinfo {author} {\bibfnamefont {Matteo}\ \bibnamefont
  {Mariantoni}}, \bibinfo {author} {\bibfnamefont {John~M.}\ \bibnamefont
  {Martinis}}, and\ \bibinfo {author} {\bibfnamefont {Andrew~N.}\ \bibnamefont
  {Cleland}}} (\bibinfo {year} {2012}),\ \bibfield  {title} {\enquote {\bibinfo
  {title} {Surface codes: {{Towards}} practical large-scale quantum
  computation},}\ }\href {https://doi.org/10.1103/PhysRevA.86.032324}
  {\bibfield  {journal} {\bibinfo  {journal} {Physical Review A}\ }\textbf
  {\bibinfo {volume} {86}}~(\bibinfo {number} {3}),\ \bibinfo {pages}
  {032324}}\BibitemShut {NoStop}%
\bibitem [{\citenamefont {Fuchs}\ and\ \citenamefont {{van de
  Graaf}}(1999)}]{fuchsCryptographicDistinguishabilityMeasures1999}%
  \BibitemOpen
  \bibfield  {author} {\bibinfo {author} {\bibnamefont {Fuchs}, \bibfnamefont
  {CA}}, and\ \bibinfo {author} {\bibfnamefont {J.}~\bibnamefont {{van de
  Graaf}}}} (\bibinfo {year} {1999}),\ \bibfield  {title} {\enquote {\bibinfo
  {title} {Cryptographic distinguishability measures for quantum-mechanical
  states},}\ }\href {https://doi.org/10.1109/18.761271} {\bibfield  {journal}
  {\bibinfo  {journal} {IEEE Transactions on Information Theory}\ }\textbf
  {\bibinfo {volume} {45}}~(\bibinfo {number} {4}),\ \bibinfo {pages}
  {1216--1227}}\BibitemShut {NoStop}%
\bibitem [{\citenamefont {Garmon}\ \emph {et~al.}(2020)\citenamefont {Garmon},
  \citenamefont {Pooser},\ and\ \citenamefont
  {Dumitrescu}}]{garmonBenchmarkingNoiseExtrapolation2020}%
  \BibitemOpen
  \bibfield  {author} {\bibinfo {author} {\bibnamefont {Garmon}, \bibfnamefont
  {J~W~O}}, \bibinfo {author} {\bibfnamefont {R.~C.}\ \bibnamefont {Pooser}},
  and\ \bibinfo {author} {\bibfnamefont {E.~F.}\ \bibnamefont {Dumitrescu}}}
  (\bibinfo {year} {2020}),\ \bibfield  {title} {\enquote {\bibinfo {title}
  {Benchmarking noise extrapolation with the {{OpenPulse}} control
  framework},}\ }\href {https://doi.org/10.1103/PhysRevA.101.042308} {\bibfield
   {journal} {\bibinfo  {journal} {Physical Review A}\ }\textbf {\bibinfo
  {volume} {101}}~(\bibinfo {number} {4}),\ \bibinfo {pages}
  {042308}}\BibitemShut {NoStop}%
\bibitem [{\citenamefont {Geller}(2020)}]{gellerRigorousMeasurementError2020}%
  \BibitemOpen
  \bibfield  {author} {\bibinfo {author} {\bibnamefont {Geller}, \bibfnamefont
  {Michael~R}}} (\bibinfo {year} {2020}),\ \bibfield  {title} {\enquote
  {\bibinfo {title} {Rigorous measurement error correction},}\ }\href
  {https://doi.org/10.1088/2058-9565/ab9591} {\bibfield  {journal} {\bibinfo
  {journal} {Quantum Science and Technology}\ }\textbf {\bibinfo {volume}
  {5}}~(\bibinfo {number} {3}),\ \bibinfo {pages} {03LT01}}\BibitemShut
  {NoStop}%
\bibitem [{\citenamefont {Gilchrist}\ \emph {et~al.}(2011)\citenamefont
  {Gilchrist}, \citenamefont {Terno},\ and\ \citenamefont
  {Wood}}]{gilchristVectorizationQuantumOperations2011}%
  \BibitemOpen
  \bibfield  {author} {\bibinfo {author} {\bibnamefont {Gilchrist},
  \bibfnamefont {Alexei}}, \bibinfo {author} {\bibfnamefont {Daniel~R.}\
  \bibnamefont {Terno}}, and\ \bibinfo {author} {\bibfnamefont
  {Christopher~J.}\ \bibnamefont {Wood}}} (\bibinfo {year} {2011}),\ \href
  {http://arxiv.org/abs/0911.2539} {\enquote {\bibinfo {title} {Vectorization
  of quantum operations and its use},}\ }\bibinfo {howpublished}
  {arXiv:0911.2539 [quant-ph]}\BibitemShut {NoStop}%
\bibitem [{\citenamefont {{Giurgica-Tiron}}\ \emph {et~al.}(2020)\citenamefont
  {{Giurgica-Tiron}}, \citenamefont {Hindy}, \citenamefont {LaRose},
  \citenamefont {Mari},\ and\ \citenamefont
  {Zeng}}]{giurgica-tironDigitalZeroNoise2020}%
  \BibitemOpen
  \bibfield  {author} {\bibinfo {author} {\bibnamefont {{Giurgica-Tiron}},
  \bibfnamefont {Tudor}}, \bibinfo {author} {\bibfnamefont {Yousef}\
  \bibnamefont {Hindy}}, \bibinfo {author} {\bibfnamefont {Ryan}\ \bibnamefont
  {LaRose}}, \bibinfo {author} {\bibfnamefont {Andrea}\ \bibnamefont {Mari}},
  and\ \bibinfo {author} {\bibfnamefont {William~J.}\ \bibnamefont {Zeng}}}
  (\bibinfo {year} {2020}),\ \bibfield  {title} {\enquote {\bibinfo {title}
  {Digital zero noise extrapolation for quantum error mitigation},}\ }in\ \href
  {https://doi.org/10.1109/QCE49297.2020.00045} {\emph {\bibinfo {booktitle}
  {2020 {{IEEE International Conference}} on {{Quantum Computing}} and
  {{Engineering}} ({{QCE}})}}},\ pp.\ \bibinfo {pages} {306--316}\BibitemShut
  {NoStop}%
\bibitem [{\citenamefont {{Google Quantum
  AI}}(2023)}]{googlequantumaiSuppressingQuantumErrors2023}%
  \BibitemOpen
  \bibfield  {author} {\bibinfo {author} {\bibnamefont {{Google Quantum AI}},}}
  (\bibinfo {year} {2023}),\ \bibfield  {title} {\enquote {\bibinfo {title}
  {Suppressing quantum errors by scaling a surface code logical qubit},}\
  }\href {https://doi.org/10.1038/s41586-022-05434-1} {\bibfield  {journal}
  {\bibinfo  {journal} {Nature}\ }\textbf {\bibinfo {volume} {614}}~(\bibinfo
  {number} {7949}),\ \bibinfo {pages} {676--681}}\BibitemShut {NoStop}%
\bibitem [{\citenamefont {{Google Quantum AI and
  Collaborators}}(2020{\natexlab{a}})}]{googlequantumaiandcollaboratorsHartreeFockSuperconductingQubit2020}%
  \BibitemOpen
  \bibfield  {author} {\bibinfo {author} {\bibnamefont {{Google Quantum AI and
  Collaborators}},}} (\bibinfo {year} {2020}{\natexlab{a}}),\ \bibfield
  {title} {\enquote {\bibinfo {title} {Hartree-{{Fock}} on a superconducting
  qubit quantum computer},}\ }\href {https://doi.org/10.1126/science.abb9811}
  {\bibfield  {journal} {\bibinfo  {journal} {Science}\ }\textbf {\bibinfo
  {volume} {369}}~(\bibinfo {number} {6507}),\ \bibinfo {pages}
  {1084--1089}}\BibitemShut {NoStop}%
\bibitem [{\citenamefont {{Google Quantum AI and
  Collaborators}}(2020{\natexlab{b}})}]{googlequantumaiandcollaboratorsObservationSeparatedDynamics2020}%
  \BibitemOpen
  \bibfield  {author} {\bibinfo {author} {\bibnamefont {{Google Quantum AI and
  Collaborators}},}} (\bibinfo {year} {2020}{\natexlab{b}}),\ \href
  {http://arxiv.org/abs/2010.07965} {\enquote {\bibinfo {title} {Observation of
  separated dynamics of charge and spin in the {{Fermi-Hubbard}} model},}\
  }\bibinfo {howpublished} {arXiv:2010.07965 [quant-ph]}\BibitemShut {NoStop}%
\bibitem [{\citenamefont {{Google Quantum AI and
  Collaborators}}(2022)}]{googlequantumaiandcollaboratorsFormationRobustBound2022}%
  \BibitemOpen
  \bibfield  {author} {\bibinfo {author} {\bibnamefont {{Google Quantum AI and
  Collaborators}},}} (\bibinfo {year} {2022}),\ \bibfield  {title} {\enquote
  {\bibinfo {title} {Formation of robust bound states of interacting microwave
  photons},}\ }\href {https://doi.org/10.1038/s41586-022-05348-y} {\bibfield
  {journal} {\bibinfo  {journal} {Nature}\ }\textbf {\bibinfo {volume}
  {612}}~(\bibinfo {number} {7939}),\ \bibinfo {pages} {240--245}}\BibitemShut
  {NoStop}%
\bibitem [{\citenamefont
  {Gottesman}(2009)}]{gottesmanIntroductionQuantumError2009}%
  \BibitemOpen
  \bibfield  {author} {\bibinfo {author} {\bibnamefont {Gottesman},
  \bibfnamefont {Daniel}}} (\bibinfo {year} {2009}),\ \href
  {http://arxiv.org/abs/0904.2557} {\enquote {\bibinfo {title} {An
  {{Introduction}} to {{Quantum Error Correction}} and {{Fault-Tolerant Quantum
  Computation}}},}\ }\bibinfo {howpublished} {arXiv:0904.2557
  [quant-ph]}\BibitemShut {NoStop}%
\bibitem [{\citenamefont
  {Gottesman}(2014)}]{gottesmanFaultTolerantQuantumComputation2014}%
  \BibitemOpen
  \bibfield  {author} {\bibinfo {author} {\bibnamefont {Gottesman},
  \bibfnamefont {Daniel}}} (\bibinfo {year} {2014}),\ \bibfield  {title}
  {\enquote {\bibinfo {title} {Fault-{{Tolerant}} quantum computation with
  constant overhead},}\ }\href {https://doi.org/10.26421/QIC14.15-16-5}
  {\bibfield  {journal} {\bibinfo  {journal} {Quantum Information and
  Computation}\ }\textbf {\bibinfo {volume} {14}}~(\bibinfo {number}
  {15\&16}),\ \bibinfo {pages} {1339--1371}}\BibitemShut {NoStop}%
\bibitem [{\citenamefont
  {Gottesman}(2016)}]{gottesmanQuantumFaultTolerance2016}%
  \BibitemOpen
  \bibfield  {author} {\bibinfo {author} {\bibnamefont {Gottesman},
  \bibfnamefont {Daniel}}} (\bibinfo {year} {2016}),\ \href
  {https://doi.org/10.48550/arXiv.1610.03507} {\enquote {\bibinfo {title}
  {Quantum fault tolerance in small experiments},}\ }\bibinfo {howpublished}
  {arXiv:1610.03507 [quant-ph]}\BibitemShut {NoStop}%
\bibitem [{\citenamefont
  {Gottesman}(1997)}]{gottesmanStabilizerCodesQuantum1997}%
  \BibitemOpen
  \bibfield  {author} {\bibinfo {author} {\bibnamefont {Gottesman},
  \bibfnamefont {Daniel~Eric}}} (\bibinfo {year} {1997}),\ \emph {\bibinfo
  {title} {Stabilizer {{Codes}} and {{Quantum Error Correction}}}},\ \href
  {https://doi.org/10.7907/rzr7-dt72} {Ph.D. thesis}\ (\bibinfo  {school}
  {California Institute of Technology})\BibitemShut {NoStop}%
\bibitem [{\citenamefont {Gu}\ \emph {et~al.}(2023)\citenamefont {Gu},
  \citenamefont {Ma}, \citenamefont {Forcellini},\ and\ \citenamefont
  {Liu}}]{guNoiseResilientPhaseEstimation2023}%
  \BibitemOpen
  \bibfield  {author} {\bibinfo {author} {\bibnamefont {Gu}, \bibfnamefont
  {Yanwu}}, \bibinfo {author} {\bibfnamefont {Yunheng}\ \bibnamefont {Ma}},
  \bibinfo {author} {\bibfnamefont {Nicol{\`o}}\ \bibnamefont {Forcellini}},
  and\ \bibinfo {author} {\bibfnamefont {Dong~E.}\ \bibnamefont {Liu}}}
  (\bibinfo {year} {2023}),\ \bibfield  {title} {\enquote {\bibinfo {title}
  {Noise-{{Resilient Phase Estimation}} with {{Randomized Compiling}}},}\
  }\href {https://doi.org/10.1103/PhysRevLett.130.250601} {\bibfield  {journal}
  {\bibinfo  {journal} {Physical Review Letters}\ }\textbf {\bibinfo {volume}
  {130}}~(\bibinfo {number} {25}),\ \bibinfo {pages} {250601}}\BibitemShut
  {NoStop}%
\bibitem [{\citenamefont {Guo}\ and\ \citenamefont
  {Yang}(2022)}]{guoQuantumErrorMitigation2022}%
  \BibitemOpen
  \bibfield  {author} {\bibinfo {author} {\bibnamefont {Guo}, \bibfnamefont
  {Yuchen}}, and\ \bibinfo {author} {\bibfnamefont {Shuo}\ \bibnamefont
  {Yang}}} (\bibinfo {year} {2022}),\ \bibfield  {title} {\enquote {\bibinfo
  {title} {Quantum {{Error Mitigation}} via {{Matrix Product Operators}}},}\
  }\href {https://doi.org/10.1103/PRXQuantum.3.040313} {\bibfield  {journal}
  {\bibinfo  {journal} {PRX Quantum}\ }\textbf {\bibinfo {volume}
  {3}}~(\bibinfo {number} {4}),\ \bibinfo {pages} {040313}}\BibitemShut
  {NoStop}%
\bibitem [{\citenamefont {Guo}\ and\ \citenamefont
  {Yang}(2023)}]{guoNoiseEffectsPurity2023}%
  \BibitemOpen
  \bibfield  {author} {\bibinfo {author} {\bibnamefont {Guo}, \bibfnamefont
  {Yuchen}}, and\ \bibinfo {author} {\bibfnamefont {Shuo}\ \bibnamefont
  {Yang}}} (\bibinfo {year} {2023}),\ \bibfield  {title} {\enquote {\bibinfo
  {title} {Noise effects on purity and quantum entanglement in terms of
  physical implementability},}\ }\href
  {https://doi.org/10.1038/s41534-023-00680-1} {\bibfield  {journal} {\bibinfo
  {journal} {npj Quantum Information}\ }\textbf {\bibinfo {volume}
  {9}}~(\bibinfo {number} {1}),\ \bibinfo {pages} {1--7}}\BibitemShut {NoStop}%
\bibitem [{\citenamefont {Guti{\'e}rrez}\ and\ \citenamefont
  {Brown}(2015)}]{gutierrezComparisonQuantumErrorcorrection2015}%
  \BibitemOpen
  \bibfield  {author} {\bibinfo {author} {\bibnamefont {Guti{\'e}rrez},
  \bibfnamefont {Mauricio}}, and\ \bibinfo {author} {\bibfnamefont
  {Kenneth~R.}\ \bibnamefont {Brown}}} (\bibinfo {year} {2015}),\ \bibfield
  {title} {\enquote {\bibinfo {title} {Comparison of a quantum error-correction
  threshold for exact and approximate errors},}\ }\href
  {https://doi.org/10.1103/PhysRevA.91.022335} {\bibfield  {journal} {\bibinfo
  {journal} {Physical Review A}\ }\textbf {\bibinfo {volume} {91}}~(\bibinfo
  {number} {2}),\ \bibinfo {pages} {022335}}\BibitemShut {NoStop}%
\bibitem [{\citenamefont {Hakoshima}\ \emph {et~al.}(2021)\citenamefont
  {Hakoshima}, \citenamefont {Matsuzaki},\ and\ \citenamefont
  {Endo}}]{hakoshimaRelationshipCostsQuantum2021}%
  \BibitemOpen
  \bibfield  {author} {\bibinfo {author} {\bibnamefont {Hakoshima},
  \bibfnamefont {Hideaki}}, \bibinfo {author} {\bibfnamefont {Yuichiro}\
  \bibnamefont {Matsuzaki}}, and\ \bibinfo {author} {\bibfnamefont {Suguru}\
  \bibnamefont {Endo}}} (\bibinfo {year} {2021}),\ \bibfield  {title} {\enquote
  {\bibinfo {title} {Relationship between costs for quantum error mitigation
  and non-{{Markovian}} measures},}\ }\href
  {https://doi.org/10.1103/PhysRevA.103.012611} {\bibfield  {journal} {\bibinfo
   {journal} {Physical Review A}\ }\textbf {\bibinfo {volume} {103}}~(\bibinfo
  {number} {1}),\ \bibinfo {pages} {012611}}\BibitemShut {NoStop}%
\bibitem [{\citenamefont {Hama}\ and\ \citenamefont
  {Nishi}(2022)}]{hamaQuantumErrorMitigation2022}%
  \BibitemOpen
  \bibfield  {author} {\bibinfo {author} {\bibnamefont {Hama}, \bibfnamefont
  {Yusuke}}, and\ \bibinfo {author} {\bibfnamefont {Hirofumi}\ \bibnamefont
  {Nishi}}} (\bibinfo {year} {2022}),\ \href
  {https://doi.org/10.48550/arXiv.2205.13907} {\enquote {\bibinfo {title}
  {Quantum {{Error Mitigation}} via {{Quantum-Noise-Effect Circuit Groups}}},}\
  }\bibinfo {howpublished} {arXiv:2205.13907 [quant-ph]}\BibitemShut {NoStop}%
\bibitem [{\citenamefont {Hamilton}\ \emph {et~al.}(2020)\citenamefont
  {Hamilton}, \citenamefont {Kharazi}, \citenamefont {Morris}, \citenamefont
  {McCaskey}, \citenamefont {Bennink},\ and\ \citenamefont
  {Pooser}}]{hamiltonScalableQuantumProcessor2020}%
  \BibitemOpen
  \bibfield  {author} {\bibinfo {author} {\bibnamefont {Hamilton},
  \bibfnamefont {Kathleen~E}}, \bibinfo {author} {\bibfnamefont {Tyler}\
  \bibnamefont {Kharazi}}, \bibinfo {author} {\bibfnamefont {Titus}\
  \bibnamefont {Morris}}, \bibinfo {author} {\bibfnamefont {Alexander~J.}\
  \bibnamefont {McCaskey}}, \bibinfo {author} {\bibfnamefont {Ryan~S.}\
  \bibnamefont {Bennink}}, and\ \bibinfo {author} {\bibfnamefont {Raphael~C.}\
  \bibnamefont {Pooser}}} (\bibinfo {year} {2020}),\ \bibfield  {title}
  {\enquote {\bibinfo {title} {Scalable quantum processor noise
  characterization},}\ }in\ \href {https://doi.org/10.1109/QCE49297.2020.00060}
  {\emph {\bibinfo {booktitle} {2020 {{IEEE International Conference}} on
  {{Quantum Computing}} and {{Engineering}} ({{QCE}})}}},\ pp.\ \bibinfo
  {pages} {430--440}\BibitemShut {NoStop}%
\bibitem [{\citenamefont {Haroche}\ and\ \citenamefont
  {Raimond}(2006)}]{harocheExploringQuantumAtoms2006}%
  \BibitemOpen
  \bibfield  {author} {\bibinfo {author} {\bibnamefont {Haroche}, \bibfnamefont
  {Serge}}, and\ \bibinfo {author} {\bibfnamefont {Jean-Michel}\ \bibnamefont
  {Raimond}}} (\bibinfo {year} {2006}),\ \href
  {https://doi.org/10.1093/acprof:oso/9780198509141.001.0001} {\emph {\bibinfo
  {title} {Exploring the {{Quantum}}: {{Atoms}}, {{Cavities}}, and
  {{Photons}}}}},\ Oxford {{Graduate Texts}}\ (\bibinfo  {publisher} {{Oxford
  University Press}},\ \bibinfo {address} {{Oxford}})\BibitemShut {NoStop}%
\bibitem [{\citenamefont {He}\ \emph {et~al.}(2020)\citenamefont {He},
  \citenamefont {Nachman}, \citenamefont {{de Jong}},\ and\ \citenamefont
  {Bauer}}]{heZeronoiseExtrapolationQuantumgate2020}%
  \BibitemOpen
  \bibfield  {author} {\bibinfo {author} {\bibnamefont {He}, \bibfnamefont
  {Andre}}, \bibinfo {author} {\bibfnamefont {Benjamin}\ \bibnamefont
  {Nachman}}, \bibinfo {author} {\bibfnamefont {Wibe~A.}\ \bibnamefont {{de
  Jong}}}, and\ \bibinfo {author} {\bibfnamefont {Christian~W.}\ \bibnamefont
  {Bauer}}} (\bibinfo {year} {2020}),\ \bibfield  {title} {\enquote {\bibinfo
  {title} {Zero-noise extrapolation for quantum-gate error mitigation with
  identity insertions},}\ }\href {https://doi.org/10.1103/PhysRevA.102.012426}
  {\bibfield  {journal} {\bibinfo  {journal} {Physical Review A}\ }\textbf
  {\bibinfo {volume} {102}}~(\bibinfo {number} {1}),\ \bibinfo {pages}
  {012426}}\BibitemShut {NoStop}%
\bibitem [{\citenamefont {Heinsoo}\ \emph {et~al.}(2018)\citenamefont
  {Heinsoo}, \citenamefont {Andersen}, \citenamefont {Remm}, \citenamefont
  {Krinner}, \citenamefont {Walter}, \citenamefont {Salath{\'e}}, \citenamefont
  {Gasparinetti}, \citenamefont {Besse}, \citenamefont {Poto{\v c}nik},
  \citenamefont {Wallraff},\ and\ \citenamefont
  {Eichler}}]{heinsooRapidHighfidelityMultiplexed2018}%
  \BibitemOpen
  \bibfield  {author} {\bibinfo {author} {\bibnamefont {Heinsoo}, \bibfnamefont
  {Johannes}}, \bibinfo {author} {\bibfnamefont {Christian~Kraglund}\
  \bibnamefont {Andersen}}, \bibinfo {author} {\bibfnamefont {Ants}\
  \bibnamefont {Remm}}, \bibinfo {author} {\bibfnamefont {Sebastian}\
  \bibnamefont {Krinner}}, \bibinfo {author} {\bibfnamefont {Theodore}\
  \bibnamefont {Walter}}, \bibinfo {author} {\bibfnamefont {Yves}\ \bibnamefont
  {Salath{\'e}}}, \bibinfo {author} {\bibfnamefont {Simone}\ \bibnamefont
  {Gasparinetti}}, \bibinfo {author} {\bibfnamefont {Jean-Claude}\ \bibnamefont
  {Besse}}, \bibinfo {author} {\bibfnamefont {Anton}\ \bibnamefont {Poto{\v
  c}nik}}, \bibinfo {author} {\bibfnamefont {Andreas}\ \bibnamefont
  {Wallraff}}, and\ \bibinfo {author} {\bibfnamefont {Christopher}\
  \bibnamefont {Eichler}}} (\bibinfo {year} {2018}),\ \bibfield  {title}
  {\enquote {\bibinfo {title} {Rapid {{High-fidelity Multiplexed Readout}} of
  {{Superconducting Qubits}}},}\ }\href
  {https://doi.org/10.1103/PhysRevApplied.10.034040} {\bibfield  {journal}
  {\bibinfo  {journal} {Physical Review Applied}\ }\textbf {\bibinfo {volume}
  {10}}~(\bibinfo {number} {3}),\ \bibinfo {pages} {034040}}\BibitemShut
  {NoStop}%
\bibitem [{\citenamefont {Helgaker}\ \emph {et~al.}(2000)\citenamefont
  {Helgaker}, \citenamefont {Jorgensen},\ and\ \citenamefont
  {Olsen}}]{helgakerMolecularElectronicStructureTheory2000}%
  \BibitemOpen
  \bibfield  {author} {\bibinfo {author} {\bibnamefont {Helgaker},
  \bibfnamefont {Trygve}}, \bibinfo {author} {\bibfnamefont {Poul}\
  \bibnamefont {Jorgensen}}, and\ \bibinfo {author} {\bibfnamefont {Jeppe}\
  \bibnamefont {Olsen}}} (\bibinfo {year} {2000}),\ \href
  {https://doi.org/10.1002/9781119019572} {\emph {\bibinfo {title} {Molecular
  {{Electronic-Structure Theory}}}}}\ (\bibinfo  {publisher} {{John Wiley \&
  Sons, Ltd}})\BibitemShut {NoStop}%
\bibitem [{\citenamefont {Henao}\ \emph {et~al.}(2023)\citenamefont {Henao},
  \citenamefont {Santos},\ and\ \citenamefont
  {Uzdin}}]{henaoAdaptiveQuantumError2023}%
  \BibitemOpen
  \bibfield  {author} {\bibinfo {author} {\bibnamefont {Henao}, \bibfnamefont
  {Ivan}}, \bibinfo {author} {\bibfnamefont {Jader~P.}\ \bibnamefont {Santos}},
  and\ \bibinfo {author} {\bibfnamefont {Raam}\ \bibnamefont {Uzdin}}}
  (\bibinfo {year} {2023}),\ \bibfield  {title} {\enquote {\bibinfo {title}
  {Adaptive quantum error mitigation using pulse-based inverse evolutions},}\
  }\href {https://doi.org/10.1038/s41534-023-00785-7} {\bibfield  {journal}
  {\bibinfo  {journal} {npj Quantum Information}\ }\textbf {\bibinfo {volume}
  {9}}~(\bibinfo {number} {1}),\ \bibinfo {pages} {1--10}}\BibitemShut
  {NoStop}%
\bibitem [{\citenamefont {Hiai}\ \emph {et~al.}(2008)\citenamefont {Hiai},
  \citenamefont {Ohya},\ and\ \citenamefont
  {Tsukada}}]{hiaiSufficiencyKmsCondition2008}%
  \BibitemOpen
  \bibfield  {author} {\bibinfo {author} {\bibnamefont {Hiai}, \bibfnamefont
  {Fumio}}, \bibinfo {author} {\bibfnamefont {Masanori}\ \bibnamefont {Ohya}},
  and\ \bibinfo {author} {\bibfnamefont {Makoto}\ \bibnamefont {Tsukada}}}
  (\bibinfo {year} {2008}),\ \bibfield  {title} {\enquote {\bibinfo {title}
  {Sufficiency, kms condition and relative entropy in von neumann algebras},}\
  }in\ \href {https://doi.org/10.1142/9789812794208_0030} {\emph {\bibinfo
  {booktitle} {Selected {{Papers}} of {{M Ohya}}}}}\ (\bibinfo  {publisher}
  {{WORLD SCIENTIFIC}})\ pp.\ \bibinfo {pages} {420--430}\BibitemShut {NoStop}%
\bibitem [{\citenamefont {Hicks}\ \emph {et~al.}(2022)\citenamefont {Hicks},
  \citenamefont {Kobrin}, \citenamefont {Bauer},\ and\ \citenamefont
  {Nachman}}]{hicksActiveReadouterrorMitigation2022}%
  \BibitemOpen
  \bibfield  {author} {\bibinfo {author} {\bibnamefont {Hicks}, \bibfnamefont
  {Rebecca}}, \bibinfo {author} {\bibfnamefont {Bryce}\ \bibnamefont {Kobrin}},
  \bibinfo {author} {\bibfnamefont {Christian~W.}\ \bibnamefont {Bauer}}, and\
  \bibinfo {author} {\bibfnamefont {Benjamin}\ \bibnamefont {Nachman}}}
  (\bibinfo {year} {2022}),\ \bibfield  {title} {\enquote {\bibinfo {title}
  {Active readout-error mitigation},}\ }\href
  {https://doi.org/10.1103/PhysRevA.105.012419} {\bibfield  {journal} {\bibinfo
   {journal} {Physical Review A}\ }\textbf {\bibinfo {volume} {105}}~(\bibinfo
  {number} {1}),\ \bibinfo {pages} {012419}}\BibitemShut {NoStop}%
\bibitem [{\citenamefont {Hu}\ \emph {et~al.}(2022)\citenamefont {Hu},
  \citenamefont {LaRose}, \citenamefont {You}, \citenamefont {Rieffel},\ and\
  \citenamefont {Wang}}]{huLogicalShadowTomography2022}%
  \BibitemOpen
  \bibfield  {author} {\bibinfo {author} {\bibnamefont {Hu}, \bibfnamefont
  {Hong-Ye}}, \bibinfo {author} {\bibfnamefont {Ryan}\ \bibnamefont {LaRose}},
  \bibinfo {author} {\bibfnamefont {Yi-Zhuang}\ \bibnamefont {You}}, \bibinfo
  {author} {\bibfnamefont {Eleanor}\ \bibnamefont {Rieffel}}, and\ \bibinfo
  {author} {\bibfnamefont {Zhihui}\ \bibnamefont {Wang}}} (\bibinfo {year}
  {2022}),\ \href {http://arxiv.org/abs/2203.07263} {\enquote {\bibinfo {title}
  {Logical shadow tomography: {{Efficient}} estimation of error-mitigated
  observables},}\ }\bibinfo {howpublished} {arXiv:2203.07263
  [quant-ph]}\BibitemShut {NoStop}%
\bibitem [{\citenamefont {Huang}\ \emph {et~al.}(2020)\citenamefont {Huang},
  \citenamefont {Kueng},\ and\ \citenamefont
  {Preskill}}]{huangPredictingManyProperties2020}%
  \BibitemOpen
  \bibfield  {author} {\bibinfo {author} {\bibnamefont {Huang}, \bibfnamefont
  {Hsin-Yuan}}, \bibinfo {author} {\bibfnamefont {Richard}\ \bibnamefont
  {Kueng}}, and\ \bibinfo {author} {\bibfnamefont {John}\ \bibnamefont
  {Preskill}}} (\bibinfo {year} {2020}),\ \bibfield  {title} {\enquote
  {\bibinfo {title} {Predicting many properties of a quantum system from very
  few measurements},}\ }\href {https://doi.org/10.1038/s41567-020-0932-7}
  {\bibfield  {journal} {\bibinfo  {journal} {Nature Physics}\ }\textbf
  {\bibinfo {volume} {16}}~(\bibinfo {number} {10}),\ \bibinfo {pages}
  {1050--1057}}\BibitemShut {NoStop}%
\bibitem [{\citenamefont {Huggins}\ \emph
  {et~al.}(2021{\natexlab{a}})\citenamefont {Huggins}, \citenamefont {McArdle},
  \citenamefont {O'Brien}, \citenamefont {Lee}, \citenamefont {Rubin},
  \citenamefont {Boixo}, \citenamefont {Whaley}, \citenamefont {Babbush},\ and\
  \citenamefont {McClean}}]{hugginsVirtualDistillationQuantum2021}%
  \BibitemOpen
  \bibfield  {author} {\bibinfo {author} {\bibnamefont {Huggins}, \bibfnamefont
  {William~J}}, \bibinfo {author} {\bibfnamefont {Sam}\ \bibnamefont
  {McArdle}}, \bibinfo {author} {\bibfnamefont {Thomas~E.}\ \bibnamefont
  {O'Brien}}, \bibinfo {author} {\bibfnamefont {Joonho}\ \bibnamefont {Lee}},
  \bibinfo {author} {\bibfnamefont {Nicholas~C.}\ \bibnamefont {Rubin}},
  \bibinfo {author} {\bibfnamefont {Sergio}\ \bibnamefont {Boixo}}, \bibinfo
  {author} {\bibfnamefont {K.~Birgitta}\ \bibnamefont {Whaley}}, \bibinfo
  {author} {\bibfnamefont {Ryan}\ \bibnamefont {Babbush}}, and\ \bibinfo
  {author} {\bibfnamefont {Jarrod~R.}\ \bibnamefont {McClean}}} (\bibinfo
  {year} {2021}{\natexlab{a}}),\ \bibfield  {title} {\enquote {\bibinfo {title}
  {Virtual {{Distillation}} for {{Quantum Error Mitigation}}},}\ }\href
  {https://doi.org/10.1103/PhysRevX.11.041036} {\bibfield  {journal} {\bibinfo
  {journal} {Physical Review X}\ }\textbf {\bibinfo {volume} {11}}~(\bibinfo
  {number} {4}),\ \bibinfo {pages} {041036}}\BibitemShut {NoStop}%
\bibitem [{\citenamefont {Huggins}\ \emph
  {et~al.}(2021{\natexlab{b}})\citenamefont {Huggins}, \citenamefont {McClean},
  \citenamefont {Rubin}, \citenamefont {Jiang}, \citenamefont {Wiebe},
  \citenamefont {Whaley},\ and\ \citenamefont
  {Babbush}}]{hugginsEfficientNoiseResilient2021}%
  \BibitemOpen
  \bibfield  {author} {\bibinfo {author} {\bibnamefont {Huggins}, \bibfnamefont
  {William~J}}, \bibinfo {author} {\bibfnamefont {Jarrod~R.}\ \bibnamefont
  {McClean}}, \bibinfo {author} {\bibfnamefont {Nicholas~C.}\ \bibnamefont
  {Rubin}}, \bibinfo {author} {\bibfnamefont {Zhang}\ \bibnamefont {Jiang}},
  \bibinfo {author} {\bibfnamefont {Nathan}\ \bibnamefont {Wiebe}}, \bibinfo
  {author} {\bibfnamefont {K.~Birgitta}\ \bibnamefont {Whaley}}, and\ \bibinfo
  {author} {\bibfnamefont {Ryan}\ \bibnamefont {Babbush}}} (\bibinfo {year}
  {2021}{\natexlab{b}}),\ \bibfield  {title} {\enquote {\bibinfo {title}
  {Efficient and noise resilient measurements for quantum chemistry on
  near-term quantum computers},}\ }\href
  {https://doi.org/10.1038/s41534-020-00341-7} {\bibfield  {journal} {\bibinfo
  {journal} {npj Quantum Information}\ }\textbf {\bibinfo {volume} {7}},\
  \bibinfo {pages} {23}}\BibitemShut {NoStop}%
\bibitem [{\citenamefont {Huo}\ and\ \citenamefont
  {Li}(2022)}]{huoDualstatePurificationPractical2022}%
  \BibitemOpen
  \bibfield  {author} {\bibinfo {author} {\bibnamefont {Huo}, \bibfnamefont
  {Mingxia}}, and\ \bibinfo {author} {\bibfnamefont {Ying}\ \bibnamefont {Li}}}
  (\bibinfo {year} {2022}),\ \bibfield  {title} {\enquote {\bibinfo {title}
  {Dual-state purification for practical quantum error mitigation},}\ }\href
  {https://doi.org/10.1103/PhysRevA.105.022427} {\bibfield  {journal} {\bibinfo
   {journal} {Physical Review A}\ }\textbf {\bibinfo {volume} {105}}~(\bibinfo
  {number} {2}),\ \bibinfo {pages} {022427}}\BibitemShut {NoStop}%
\bibitem [{\citenamefont {Izmaylov}\ \emph {et~al.}(2019)\citenamefont
  {Izmaylov}, \citenamefont {Yen},\ and\ \citenamefont
  {Ryabinkin}}]{izmaylovRevisingMeasurementProcess2019}%
  \BibitemOpen
  \bibfield  {author} {\bibinfo {author} {\bibnamefont {Izmaylov},
  \bibfnamefont {Artur~F}}, \bibinfo {author} {\bibfnamefont {Tzu-Ching}\
  \bibnamefont {Yen}}, and\ \bibinfo {author} {\bibfnamefont {Ilya~G.}\
  \bibnamefont {Ryabinkin}}} (\bibinfo {year} {2019}),\ \bibfield  {title}
  {\enquote {\bibinfo {title} {Revising the measurement process in the
  variational quantum eigensolver: Is it possible to reduce the number of
  separately measured operators?}}\ }\href {https://doi.org/10.1039/C8SC05592K}
  {\bibfield  {journal} {\bibinfo  {journal} {Chemical Science}\ }\textbf
  {\bibinfo {volume} {10}}~(\bibinfo {number} {13}),\ \bibinfo {pages}
  {3746--3755}}\BibitemShut {NoStop}%
\bibitem [{\citenamefont {Jiang}\ \emph {et~al.}(2021)\citenamefont {Jiang},
  \citenamefont {Wang},\ and\ \citenamefont
  {Wang}}]{jiangPhysicalImplementabilityLinear2021}%
  \BibitemOpen
  \bibfield  {author} {\bibinfo {author} {\bibnamefont {Jiang}, \bibfnamefont
  {Jiaqing}}, \bibinfo {author} {\bibfnamefont {Kun}\ \bibnamefont {Wang}},
  and\ \bibinfo {author} {\bibfnamefont {Xin}\ \bibnamefont {Wang}}} (\bibinfo
  {year} {2021}),\ \bibfield  {title} {\enquote {\bibinfo {title} {Physical
  {{Implementability}} of {{Linear Maps}} and {{Its Application}} in {{Error
  Mitigation}}},}\ }\href {https://doi.org/10.22331/q-2021-12-07-600}
  {\bibfield  {journal} {\bibinfo  {journal} {Quantum}\ }\textbf {\bibinfo
  {volume} {5}},\ \bibinfo {pages} {600}}\BibitemShut {NoStop}%
\bibitem [{\citenamefont {Jiang}\ \emph {et~al.}(2019)\citenamefont {Jiang},
  \citenamefont {McClean}, \citenamefont {Babbush},\ and\ \citenamefont
  {Neven}}]{jiangMajoranaLoopStabilizer2019}%
  \BibitemOpen
  \bibfield  {author} {\bibinfo {author} {\bibnamefont {Jiang}, \bibfnamefont
  {Zhang}}, \bibinfo {author} {\bibfnamefont {Jarrod}\ \bibnamefont {McClean}},
  \bibinfo {author} {\bibfnamefont {Ryan}\ \bibnamefont {Babbush}}, and\
  \bibinfo {author} {\bibfnamefont {Hartmut}\ \bibnamefont {Neven}}} (\bibinfo
  {year} {2019}),\ \bibfield  {title} {\enquote {\bibinfo {title} {Majorana
  {{Loop Stabilizer Codes}} for {{Error Mitigation}} in {{Fermionic Quantum
  Simulations}}},}\ }\href {https://doi.org/10.1103/PhysRevApplied.12.064041}
  {\bibfield  {journal} {\bibinfo  {journal} {Physical Review Applied}\
  }\textbf {\bibinfo {volume} {12}}~(\bibinfo {number} {6}),\ \bibinfo {pages}
  {064041}}\BibitemShut {NoStop}%
\bibitem [{\citenamefont {Jnane}\ \emph {et~al.}(2023)\citenamefont {Jnane},
  \citenamefont {Steinberg}, \citenamefont {Cai}, \citenamefont {Nguyen},\ and\
  \citenamefont {Koczor}}]{jnaneQuantumErrorMitigated2023}%
  \BibitemOpen
  \bibfield  {author} {\bibinfo {author} {\bibnamefont {Jnane}, \bibfnamefont
  {Hamza}}, \bibinfo {author} {\bibfnamefont {Jonathan}\ \bibnamefont
  {Steinberg}}, \bibinfo {author} {\bibfnamefont {Zhenyu}\ \bibnamefont {Cai}},
  \bibinfo {author} {\bibfnamefont {H.~Chau}\ \bibnamefont {Nguyen}}, and\
  \bibinfo {author} {\bibfnamefont {B{\'a}lint}\ \bibnamefont {Koczor}}}
  (\bibinfo {year} {2023}),\ \href {https://doi.org/10.48550/arXiv.2305.04956}
  {\enquote {\bibinfo {title} {Quantum {{Error Mitigated Classical
  Shadows}}},}\ }\bibinfo {howpublished} {arXiv:2305.04956
  [quant-ph]}\BibitemShut {NoStop}%
\bibitem [{\citenamefont {Jurcevic}\ \emph {et~al.}(2021)\citenamefont
  {Jurcevic}, \citenamefont {{Javadi-Abhari}}, \citenamefont {Bishop},
  \citenamefont {Lauer}, \citenamefont {Bogorin}, \citenamefont {Brink},
  \citenamefont {Capelluto}, \citenamefont {G{\"u}nl{\"u}k}, \citenamefont
  {Itoko}, \citenamefont {Kanazawa}, \citenamefont {Kandala}, \citenamefont
  {Keefe}, \citenamefont {Krsulich}, \citenamefont {Landers}, \citenamefont
  {Lewandowski}, \citenamefont {McClure}, \citenamefont {Nannicini},
  \citenamefont {Narasgond}, \citenamefont {Nayfeh}, \citenamefont {Pritchett},
  \citenamefont {Rothwell}, \citenamefont {Srinivasan}, \citenamefont
  {Sundaresan}, \citenamefont {Wang}, \citenamefont {Wei}, \citenamefont
  {Wood}, \citenamefont {Yau}, \citenamefont {Zhang}, \citenamefont {Dial},
  \citenamefont {Chow},\ and\ \citenamefont
  {Gambetta}}]{jurcevicDemonstrationQuantumVolume2021}%
  \BibitemOpen
  \bibfield  {author} {\bibinfo {author} {\bibnamefont {Jurcevic},
  \bibfnamefont {Petar}}, \bibinfo {author} {\bibfnamefont {Ali}\ \bibnamefont
  {{Javadi-Abhari}}}, \bibinfo {author} {\bibfnamefont {Lev~S.}\ \bibnamefont
  {Bishop}}, \bibinfo {author} {\bibfnamefont {Isaac}\ \bibnamefont {Lauer}},
  \bibinfo {author} {\bibfnamefont {Daniela~F.}\ \bibnamefont {Bogorin}},
  \bibinfo {author} {\bibfnamefont {Markus}\ \bibnamefont {Brink}}, \bibinfo
  {author} {\bibfnamefont {Lauren}\ \bibnamefont {Capelluto}}, \bibinfo
  {author} {\bibfnamefont {Oktay}\ \bibnamefont {G{\"u}nl{\"u}k}}, \bibinfo
  {author} {\bibfnamefont {Toshinari}\ \bibnamefont {Itoko}}, \bibinfo {author}
  {\bibfnamefont {Naoki}\ \bibnamefont {Kanazawa}}, \bibinfo {author}
  {\bibfnamefont {Abhinav}\ \bibnamefont {Kandala}}, \bibinfo {author}
  {\bibfnamefont {George~A.}\ \bibnamefont {Keefe}}, \bibinfo {author}
  {\bibfnamefont {Kevin}\ \bibnamefont {Krsulich}}, \bibinfo {author}
  {\bibfnamefont {William}\ \bibnamefont {Landers}}, \bibinfo {author}
  {\bibfnamefont {Eric~P.}\ \bibnamefont {Lewandowski}}, \bibinfo {author}
  {\bibfnamefont {Douglas~T.}\ \bibnamefont {McClure}}, \bibinfo {author}
  {\bibfnamefont {Giacomo}\ \bibnamefont {Nannicini}}, \bibinfo {author}
  {\bibfnamefont {Adinath}\ \bibnamefont {Narasgond}}, \bibinfo {author}
  {\bibfnamefont {Hasan~M.}\ \bibnamefont {Nayfeh}}, \bibinfo {author}
  {\bibfnamefont {Emily}\ \bibnamefont {Pritchett}}, \bibinfo {author}
  {\bibfnamefont {Mary~Beth}\ \bibnamefont {Rothwell}}, \bibinfo {author}
  {\bibfnamefont {Srikanth}\ \bibnamefont {Srinivasan}}, \bibinfo {author}
  {\bibfnamefont {Neereja}\ \bibnamefont {Sundaresan}}, \bibinfo {author}
  {\bibfnamefont {Cindy}\ \bibnamefont {Wang}}, \bibinfo {author}
  {\bibfnamefont {Ken~X.}\ \bibnamefont {Wei}}, \bibinfo {author}
  {\bibfnamefont {Christopher~J.}\ \bibnamefont {Wood}}, \bibinfo {author}
  {\bibfnamefont {Jeng-Bang}\ \bibnamefont {Yau}}, \bibinfo {author}
  {\bibfnamefont {Eric~J.}\ \bibnamefont {Zhang}}, \bibinfo {author}
  {\bibfnamefont {Oliver~E.}\ \bibnamefont {Dial}}, \bibinfo {author}
  {\bibfnamefont {Jerry~M.}\ \bibnamefont {Chow}}, and\ \bibinfo {author}
  {\bibfnamefont {Jay~M.}\ \bibnamefont {Gambetta}}} (\bibinfo {year} {2021}),\
  \bibfield  {title} {\enquote {\bibinfo {title} {Demonstration of quantum
  volume 64 on a superconducting quantum computing system},}\ }\href
  {https://doi.org/10.1088/2058-9565/abe519} {\bibfield  {journal} {\bibinfo
  {journal} {Quantum Science and Technology}\ }\textbf {\bibinfo {volume}
  {6}}~(\bibinfo {number} {2}),\ \bibinfo {pages} {025020}}\BibitemShut
  {NoStop}%
\bibitem [{\citenamefont {Kandala}\ \emph {et~al.}(2017)\citenamefont
  {Kandala}, \citenamefont {Mezzacapo}, \citenamefont {Temme}, \citenamefont
  {Takita}, \citenamefont {Brink}, \citenamefont {Chow},\ and\ \citenamefont
  {Gambetta}}]{kandalaHardwareefficientVariationalQuantum2017}%
  \BibitemOpen
  \bibfield  {author} {\bibinfo {author} {\bibnamefont {Kandala}, \bibfnamefont
  {Abhinav}}, \bibinfo {author} {\bibfnamefont {Antonio}\ \bibnamefont
  {Mezzacapo}}, \bibinfo {author} {\bibfnamefont {Kristan}\ \bibnamefont
  {Temme}}, \bibinfo {author} {\bibfnamefont {Maika}\ \bibnamefont {Takita}},
  \bibinfo {author} {\bibfnamefont {Markus}\ \bibnamefont {Brink}}, \bibinfo
  {author} {\bibfnamefont {Jerry~M.}\ \bibnamefont {Chow}}, and\ \bibinfo
  {author} {\bibfnamefont {Jay~M.}\ \bibnamefont {Gambetta}}} (\bibinfo {year}
  {2017}),\ \bibfield  {title} {\enquote {\bibinfo {title} {Hardware-efficient
  variational quantum eigensolver for small molecules and quantum magnets},}\
  }\href {https://doi.org/10.1038/nature23879} {\bibfield  {journal} {\bibinfo
  {journal} {Nature}\ }\textbf {\bibinfo {volume} {549}}~(\bibinfo {number}
  {7671}),\ \bibinfo {pages} {242--246}}\BibitemShut {NoStop}%
\bibitem [{\citenamefont {Kandala}\ \emph {et~al.}(2019)\citenamefont
  {Kandala}, \citenamefont {Temme}, \citenamefont {C{\'o}rcoles}, \citenamefont
  {Mezzacapo}, \citenamefont {Chow},\ and\ \citenamefont
  {Gambetta}}]{kandalaErrorMitigationExtends2019}%
  \BibitemOpen
  \bibfield  {author} {\bibinfo {author} {\bibnamefont {Kandala}, \bibfnamefont
  {Abhinav}}, \bibinfo {author} {\bibfnamefont {Kristan}\ \bibnamefont
  {Temme}}, \bibinfo {author} {\bibfnamefont {Antonio~D.}\ \bibnamefont
  {C{\'o}rcoles}}, \bibinfo {author} {\bibfnamefont {Antonio}\ \bibnamefont
  {Mezzacapo}}, \bibinfo {author} {\bibfnamefont {Jerry~M.}\ \bibnamefont
  {Chow}}, and\ \bibinfo {author} {\bibfnamefont {Jay~M.}\ \bibnamefont
  {Gambetta}}} (\bibinfo {year} {2019}),\ \bibfield  {title} {\enquote
  {\bibinfo {title} {Error mitigation extends the computational reach of a
  noisy quantum processor},}\ }\href
  {https://doi.org/10.1038/s41586-019-1040-7} {\bibfield  {journal} {\bibinfo
  {journal} {Nature}\ }\textbf {\bibinfo {volume} {567}}~(\bibinfo {number}
  {7749}),\ \bibinfo {pages} {491--495}}\BibitemShut {NoStop}%
\bibitem [{\citenamefont {Kastoryano}\ and\ \citenamefont
  {Temme}(2013)}]{kastoryanoQuantumLogarithmicSobolev2013}%
  \BibitemOpen
  \bibfield  {author} {\bibinfo {author} {\bibnamefont {Kastoryano},
  \bibfnamefont {Michael~J}}, and\ \bibinfo {author} {\bibfnamefont {Kristan}\
  \bibnamefont {Temme}}} (\bibinfo {year} {2013}),\ \bibfield  {title}
  {\enquote {\bibinfo {title} {Quantum logarithmic {{Sobolev}} inequalities and
  rapid mixing},}\ }\href {https://doi.org/10.1063/1.4804995} {\bibfield
  {journal} {\bibinfo  {journal} {Journal of Mathematical Physics}\ }\textbf
  {\bibinfo {volume} {54}}~(\bibinfo {number} {5}),\ \bibinfo {pages}
  {052202}}\BibitemShut {NoStop}%
\bibitem [{\citenamefont {Kechedzhi}\ \emph {et~al.}(2023)\citenamefont
  {Kechedzhi}, \citenamefont {Isakov}, \citenamefont {Mandr{\`a}},
  \citenamefont {Villalonga}, \citenamefont {Mi}, \citenamefont {Boixo},\ and\
  \citenamefont {Smelyanskiy}}]{kechedzhiEffectiveQuantumVolume2023}%
  \BibitemOpen
  \bibfield  {author} {\bibinfo {author} {\bibnamefont {Kechedzhi},
  \bibfnamefont {K}}, \bibinfo {author} {\bibfnamefont {S.~V.}\ \bibnamefont
  {Isakov}}, \bibinfo {author} {\bibfnamefont {S.}~\bibnamefont {Mandr{\`a}}},
  \bibinfo {author} {\bibfnamefont {B.}~\bibnamefont {Villalonga}}, \bibinfo
  {author} {\bibfnamefont {X.}~\bibnamefont {Mi}}, \bibinfo {author}
  {\bibfnamefont {S.}~\bibnamefont {Boixo}}, and\ \bibinfo {author}
  {\bibfnamefont {V.}~\bibnamefont {Smelyanskiy}}} (\bibinfo {year} {2023}),\
  \href {https://doi.org/10.48550/arXiv.2306.15970} {\enquote {\bibinfo {title}
  {Effective quantum volume, fidelity and computational cost of noisy quantum
  processing experiments},}\ }\bibinfo {howpublished} {arXiv:2306.15970
  [quant-ph]}\BibitemShut {NoStop}%
\bibitem [{\citenamefont {Keen}\ \emph {et~al.}(2020)\citenamefont {Keen},
  \citenamefont {Maier}, \citenamefont {Johnston},\ and\ \citenamefont
  {Lougovski}}]{keenQuantumclassicalSimulationTwosite2020}%
  \BibitemOpen
  \bibfield  {author} {\bibinfo {author} {\bibnamefont {Keen}, \bibfnamefont
  {Trevor}}, \bibinfo {author} {\bibfnamefont {Thomas}\ \bibnamefont {Maier}},
  \bibinfo {author} {\bibfnamefont {Steven}\ \bibnamefont {Johnston}}, and\
  \bibinfo {author} {\bibfnamefont {Pavel}\ \bibnamefont {Lougovski}}}
  (\bibinfo {year} {2020}),\ \bibfield  {title} {\enquote {\bibinfo {title}
  {Quantum-classical simulation of two-site dynamical mean-field theory on
  noisy quantum hardware},}\ }\href {https://doi.org/10.1088/2058-9565/ab7d4c}
  {\bibfield  {journal} {\bibinfo  {journal} {Quantum Science and Technology}\
  }\textbf {\bibinfo {volume} {5}}~(\bibinfo {number} {3}),\ \bibinfo {pages}
  {035001}}\BibitemShut {NoStop}%
\bibitem [{\citenamefont {Kelly}\ \emph {et~al.}(2015)\citenamefont {Kelly},
  \citenamefont {Barends}, \citenamefont {Fowler}, \citenamefont {Megrant},
  \citenamefont {Jeffrey}, \citenamefont {White}, \citenamefont {Sank},
  \citenamefont {Mutus}, \citenamefont {Campbell}, \citenamefont {Chen},
  \citenamefont {Chen}, \citenamefont {Chiaro}, \citenamefont {Dunsworth},
  \citenamefont {Hoi}, \citenamefont {Neill}, \citenamefont {O'Malley},
  \citenamefont {Quintana}, \citenamefont {Roushan}, \citenamefont
  {Vainsencher}, \citenamefont {Wenner}, \citenamefont {Cleland},\ and\
  \citenamefont {Martinis}}]{kellyStatePreservationRepetitive2015}%
  \BibitemOpen
  \bibfield  {author} {\bibinfo {author} {\bibnamefont {Kelly}, \bibfnamefont
  {J}}, \bibinfo {author} {\bibfnamefont {R.}~\bibnamefont {Barends}}, \bibinfo
  {author} {\bibfnamefont {A.~G.}\ \bibnamefont {Fowler}}, \bibinfo {author}
  {\bibfnamefont {A.}~\bibnamefont {Megrant}}, \bibinfo {author} {\bibfnamefont
  {E.}~\bibnamefont {Jeffrey}}, \bibinfo {author} {\bibfnamefont {T.~C.}\
  \bibnamefont {White}}, \bibinfo {author} {\bibfnamefont {D.}~\bibnamefont
  {Sank}}, \bibinfo {author} {\bibfnamefont {J.~Y.}\ \bibnamefont {Mutus}},
  \bibinfo {author} {\bibfnamefont {B.}~\bibnamefont {Campbell}}, \bibinfo
  {author} {\bibfnamefont {Yu}~\bibnamefont {Chen}}, \bibinfo {author}
  {\bibfnamefont {Z.}~\bibnamefont {Chen}}, \bibinfo {author} {\bibfnamefont
  {B.}~\bibnamefont {Chiaro}}, \bibinfo {author} {\bibfnamefont
  {A.}~\bibnamefont {Dunsworth}}, \bibinfo {author} {\bibfnamefont {I.-C.}\
  \bibnamefont {Hoi}}, \bibinfo {author} {\bibfnamefont {C.}~\bibnamefont
  {Neill}}, \bibinfo {author} {\bibfnamefont {P.~J.~J.}\ \bibnamefont
  {O'Malley}}, \bibinfo {author} {\bibfnamefont {C.}~\bibnamefont {Quintana}},
  \bibinfo {author} {\bibfnamefont {P.}~\bibnamefont {Roushan}}, \bibinfo
  {author} {\bibfnamefont {A.}~\bibnamefont {Vainsencher}}, \bibinfo {author}
  {\bibfnamefont {J.}~\bibnamefont {Wenner}}, \bibinfo {author} {\bibfnamefont
  {A.~N.}\ \bibnamefont {Cleland}}, and\ \bibinfo {author} {\bibfnamefont
  {John~M.}\ \bibnamefont {Martinis}}} (\bibinfo {year} {2015}),\ \bibfield
  {title} {\enquote {\bibinfo {title} {State preservation by repetitive error
  detection in a superconducting quantum circuit},}\ }\href
  {https://doi.org/10.1038/nature14270} {\bibfield  {journal} {\bibinfo
  {journal} {Nature}\ }\textbf {\bibinfo {volume} {519}}~(\bibinfo {number}
  {7541}),\ \bibinfo {pages} {66--69}}\BibitemShut {NoStop}%
\bibitem [{\citenamefont {Khamoshi}\ \emph {et~al.}(2020)\citenamefont
  {Khamoshi}, \citenamefont {Evangelista},\ and\ \citenamefont
  {Scuseria}}]{khamoshiCorrelatingAGPQuantum2020}%
  \BibitemOpen
  \bibfield  {author} {\bibinfo {author} {\bibnamefont {Khamoshi},
  \bibfnamefont {Armin}}, \bibinfo {author} {\bibfnamefont {Francesco~A.}\
  \bibnamefont {Evangelista}}, and\ \bibinfo {author} {\bibfnamefont
  {Gustavo~E.}\ \bibnamefont {Scuseria}}} (\bibinfo {year} {2020}),\ \bibfield
  {title} {\enquote {\bibinfo {title} {Correlating {{AGP}} on a quantum
  computer},}\ }\href {https://doi.org/10.1088/2058-9565/abc1bb} {\bibfield
  {journal} {\bibinfo  {journal} {Quantum Science and Technology}\ }\textbf
  {\bibinfo {volume} {6}}~(\bibinfo {number} {1}),\ \bibinfo {pages}
  {014004}}\BibitemShut {NoStop}%
\bibitem [{\citenamefont {Kim}\ \emph {et~al.}(2023{\natexlab{a}})\citenamefont
  {Kim}, \citenamefont {Eddins}, \citenamefont {Anand}, \citenamefont {Wei},
  \citenamefont {{van den Berg}}, \citenamefont {Rosenblatt}, \citenamefont
  {Nayfeh}, \citenamefont {Wu}, \citenamefont {Zaletel}, \citenamefont
  {Temme},\ and\ \citenamefont {Kandala}}]{kimEvidenceUtilityQuantum2023}%
  \BibitemOpen
  \bibfield  {author} {\bibinfo {author} {\bibnamefont {Kim}, \bibfnamefont
  {Youngseok}}, \bibinfo {author} {\bibfnamefont {Andrew}\ \bibnamefont
  {Eddins}}, \bibinfo {author} {\bibfnamefont {Sajant}\ \bibnamefont {Anand}},
  \bibinfo {author} {\bibfnamefont {Ken~Xuan}\ \bibnamefont {Wei}}, \bibinfo
  {author} {\bibfnamefont {Ewout}\ \bibnamefont {{van den Berg}}}, \bibinfo
  {author} {\bibfnamefont {Sami}\ \bibnamefont {Rosenblatt}}, \bibinfo {author}
  {\bibfnamefont {Hasan}\ \bibnamefont {Nayfeh}}, \bibinfo {author}
  {\bibfnamefont {Yantao}\ \bibnamefont {Wu}}, \bibinfo {author} {\bibfnamefont
  {Michael}\ \bibnamefont {Zaletel}}, \bibinfo {author} {\bibfnamefont
  {Kristan}\ \bibnamefont {Temme}}, and\ \bibinfo {author} {\bibfnamefont
  {Abhinav}\ \bibnamefont {Kandala}}} (\bibinfo {year} {2023}{\natexlab{a}}),\
  \bibfield  {title} {\enquote {\bibinfo {title} {Evidence for the utility of
  quantum computing before fault tolerance},}\ }\href
  {https://doi.org/10.1038/s41586-023-06096-3} {\bibfield  {journal} {\bibinfo
  {journal} {Nature}\ }\textbf {\bibinfo {volume} {618}}~(\bibinfo {number}
  {7965}),\ \bibinfo {pages} {500--505}}\BibitemShut {NoStop}%
\bibitem [{\citenamefont {Kim}\ \emph {et~al.}(2023{\natexlab{b}})\citenamefont
  {Kim}, \citenamefont {Wood}, \citenamefont {Yoder}, \citenamefont {Merkel},
  \citenamefont {Gambetta}, \citenamefont {Temme},\ and\ \citenamefont
  {Kandala}}]{kimScalableErrorMitigation2023}%
  \BibitemOpen
  \bibfield  {author} {\bibinfo {author} {\bibnamefont {Kim}, \bibfnamefont
  {Youngseok}}, \bibinfo {author} {\bibfnamefont {Christopher~J.}\ \bibnamefont
  {Wood}}, \bibinfo {author} {\bibfnamefont {Theodore~J.}\ \bibnamefont
  {Yoder}}, \bibinfo {author} {\bibfnamefont {Seth~T.}\ \bibnamefont {Merkel}},
  \bibinfo {author} {\bibfnamefont {Jay~M.}\ \bibnamefont {Gambetta}}, \bibinfo
  {author} {\bibfnamefont {Kristan}\ \bibnamefont {Temme}}, and\ \bibinfo
  {author} {\bibfnamefont {Abhinav}\ \bibnamefont {Kandala}}} (\bibinfo {year}
  {2023}{\natexlab{b}}),\ \bibfield  {title} {\enquote {\bibinfo {title}
  {Scalable error mitigation for noisy quantum circuits produces competitive
  expectation values},}\ }\href {https://doi.org/10.1038/s41567-022-01914-3}
  {\bibinfo  {journal} {Nature Physics}\ ,\ \bibinfo {pages}
  {1--8}}\BibitemShut {NoStop}%
\bibitem [{\citenamefont
  {Kitaev}(1997)}]{kitaevQuantumComputationsAlgorithms1997}%
  \BibitemOpen
\bibfield  {journal} {  }\bibfield  {author} {\bibinfo {author} {\bibnamefont
  {Kitaev}, \bibfnamefont {A~Yu}}} (\bibinfo {year} {1997}),\ \bibfield
  {title} {\enquote {\bibinfo {title} {Quantum computations: Algorithms and
  error correction},}\ }\href {https://doi.org/10.1070/RM1997v052n06ABEH002155}
  {\bibfield  {journal} {\bibinfo  {journal} {Russian Mathematical Surveys}\
  }\textbf {\bibinfo {volume} {52}}~(\bibinfo {number} {6}),\ \bibinfo {pages}
  {1191}}\BibitemShut {NoStop}%
\bibitem [{\citenamefont {Kivlichan}\ \emph {et~al.}(2020)\citenamefont
  {Kivlichan}, \citenamefont {Gidney}, \citenamefont {Berry}, \citenamefont
  {Wiebe}, \citenamefont {McClean}, \citenamefont {Sun}, \citenamefont {Jiang},
  \citenamefont {Rubin}, \citenamefont {Fowler}, \citenamefont
  {{Aspuru-Guzik}}, \citenamefont {Neven},\ and\ \citenamefont
  {Babbush}}]{kivlichanImprovedFaultTolerantQuantum2020}%
  \BibitemOpen
  \bibfield  {author} {\bibinfo {author} {\bibnamefont {Kivlichan},
  \bibfnamefont {Ian~D}}, \bibinfo {author} {\bibfnamefont {Craig}\
  \bibnamefont {Gidney}}, \bibinfo {author} {\bibfnamefont {Dominic~W.}\
  \bibnamefont {Berry}}, \bibinfo {author} {\bibfnamefont {Nathan}\
  \bibnamefont {Wiebe}}, \bibinfo {author} {\bibfnamefont {Jarrod}\
  \bibnamefont {McClean}}, \bibinfo {author} {\bibfnamefont {Wei}\ \bibnamefont
  {Sun}}, \bibinfo {author} {\bibfnamefont {Zhang}\ \bibnamefont {Jiang}},
  \bibinfo {author} {\bibfnamefont {Nicholas}\ \bibnamefont {Rubin}}, \bibinfo
  {author} {\bibfnamefont {Austin}\ \bibnamefont {Fowler}}, \bibinfo {author}
  {\bibfnamefont {Al{\'a}n}\ \bibnamefont {{Aspuru-Guzik}}}, \bibinfo {author}
  {\bibfnamefont {Hartmut}\ \bibnamefont {Neven}}, and\ \bibinfo {author}
  {\bibfnamefont {Ryan}\ \bibnamefont {Babbush}}} (\bibinfo {year} {2020}),\
  \bibfield  {title} {\enquote {\bibinfo {title} {Improved {{Fault-Tolerant
  Quantum Simulation}} of {{Condensed-Phase Correlated Electrons}} via
  {{Trotterization}}},}\ }\href {https://doi.org/10.22331/q-2020-07-16-296}
  {\bibfield  {journal} {\bibinfo  {journal} {Quantum}\ }\textbf {\bibinfo
  {volume} {4}},\ \bibinfo {pages} {296}}\BibitemShut {NoStop}%
\bibitem [{\citenamefont {Klco}\ \emph {et~al.}(2018)\citenamefont {Klco},
  \citenamefont {Dumitrescu}, \citenamefont {McCaskey}, \citenamefont {Morris},
  \citenamefont {Pooser}, \citenamefont {Sanz}, \citenamefont {Solano},
  \citenamefont {Lougovski},\ and\ \citenamefont
  {Savage}}]{klcoQuantumclassicalComputationSchwinger2018}%
  \BibitemOpen
  \bibfield  {author} {\bibinfo {author} {\bibnamefont {Klco}, \bibfnamefont
  {N}}, \bibinfo {author} {\bibfnamefont {E.~F.}\ \bibnamefont {Dumitrescu}},
  \bibinfo {author} {\bibfnamefont {A.~J.}\ \bibnamefont {McCaskey}}, \bibinfo
  {author} {\bibfnamefont {T.~D.}\ \bibnamefont {Morris}}, \bibinfo {author}
  {\bibfnamefont {R.~C.}\ \bibnamefont {Pooser}}, \bibinfo {author}
  {\bibfnamefont {M.}~\bibnamefont {Sanz}}, \bibinfo {author} {\bibfnamefont
  {E.}~\bibnamefont {Solano}}, \bibinfo {author} {\bibfnamefont
  {P.}~\bibnamefont {Lougovski}}, and\ \bibinfo {author} {\bibfnamefont
  {M.~J.}\ \bibnamefont {Savage}}} (\bibinfo {year} {2018}),\ \bibfield
  {title} {\enquote {\bibinfo {title} {Quantum-classical computation of
  {{Schwinger}} model dynamics using quantum computers},}\ }\href
  {https://doi.org/10.1103/PhysRevA.98.032331} {\bibfield  {journal} {\bibinfo
  {journal} {Physical Review A}\ }\textbf {\bibinfo {volume} {98}}~(\bibinfo
  {number} {3}),\ \bibinfo {pages} {032331}}\BibitemShut {NoStop}%
\bibitem [{\citenamefont {Kliesch}\ and\ \citenamefont
  {Roth}(2021)}]{klieschTheoryQuantumSystem2021}%
  \BibitemOpen
  \bibfield  {author} {\bibinfo {author} {\bibnamefont {Kliesch}, \bibfnamefont
  {Martin}}, and\ \bibinfo {author} {\bibfnamefont {Ingo}\ \bibnamefont
  {Roth}}} (\bibinfo {year} {2021}),\ \bibfield  {title} {\enquote {\bibinfo
  {title} {Theory of {{Quantum System Certification}}},}\ }\href
  {https://doi.org/10.1103/PRXQuantum.2.010201} {\bibfield  {journal} {\bibinfo
   {journal} {PRX Quantum}\ }\textbf {\bibinfo {volume} {2}}~(\bibinfo {number}
  {1}),\ \bibinfo {pages} {010201}}\BibitemShut {NoStop}%
\bibitem [{\citenamefont {Kliuchnikov}\ \emph {et~al.}(2013)\citenamefont
  {Kliuchnikov}, \citenamefont {Maslov},\ and\ \citenamefont
  {Mosca}}]{kliuchnikovAsymptoticallyOptimalApproximation2013}%
  \BibitemOpen
  \bibfield  {author} {\bibinfo {author} {\bibnamefont {Kliuchnikov},
  \bibfnamefont {Vadym}}, \bibinfo {author} {\bibfnamefont {Dmitri}\
  \bibnamefont {Maslov}}, and\ \bibinfo {author} {\bibfnamefont {Michele}\
  \bibnamefont {Mosca}}} (\bibinfo {year} {2013}),\ \bibfield  {title}
  {\enquote {\bibinfo {title} {Asymptotically {{Optimal Approximation}} of
  {{Single Qubit Unitaries}} by {{Clifford}} and \${{T}}\$ {{Circuits Using}} a
  {{Constant Number}} of {{Ancillary Qubits}}},}\ }\href
  {https://doi.org/10.1103/PhysRevLett.110.190502} {\bibfield  {journal}
  {\bibinfo  {journal} {Physical Review Letters}\ }\textbf {\bibinfo {volume}
  {110}}~(\bibinfo {number} {19}),\ \bibinfo {pages} {190502}}\BibitemShut
  {NoStop}%
\bibitem [{\citenamefont
  {Klyachko}(2006)}]{klyachkoQuantumMarginalProblem2006}%
  \BibitemOpen
  \bibfield  {author} {\bibinfo {author} {\bibnamefont {Klyachko},
  \bibfnamefont {Alexander~A}}} (\bibinfo {year} {2006}),\ \bibfield  {title}
  {\enquote {\bibinfo {title} {Quantum marginal problem and
  {{N-representability}}},}\ }\href
  {https://doi.org/10.1088/1742-6596/36/1/014} {\bibfield  {journal} {\bibinfo
  {journal} {Journal of Physics: Conference Series}\ }\textbf {\bibinfo
  {volume} {36}},\ \bibinfo {pages} {72--86}}\BibitemShut {NoStop}%
\bibitem [{\citenamefont
  {Knill}(2004)}]{knillFaultTolerantPostselectedQuantum2004}%
  \BibitemOpen
  \bibfield  {author} {\bibinfo {author} {\bibnamefont {Knill}, \bibfnamefont
  {E}}} (\bibinfo {year} {2004}),\ \href
  {http://arxiv.org/abs/quant-ph/0404104} {\enquote {\bibinfo {title}
  {Fault-{{Tolerant Postselected Quantum Computation}}: {{Threshold
  Analysis}}},}\ }\bibinfo {howpublished} {arXiv:quant-ph/0404104}\BibitemShut
  {NoStop}%
\bibitem [{\citenamefont
  {Koczor}(2021{\natexlab{a}})}]{koczorDominantEigenvectorNoisy2021}%
  \BibitemOpen
  \bibfield  {author} {\bibinfo {author} {\bibnamefont {Koczor}, \bibfnamefont
  {B{\'a}lint}}} (\bibinfo {year} {2021}{\natexlab{a}}),\ \bibfield  {title}
  {\enquote {\bibinfo {title} {The dominant eigenvector of a noisy quantum
  state},}\ }\href {https://doi.org/10.1088/1367-2630/ac37ae} {\bibfield
  {journal} {\bibinfo  {journal} {New Journal of Physics}\ }\textbf {\bibinfo
  {volume} {23}}~(\bibinfo {number} {12}),\ \bibinfo {pages}
  {123047}}\BibitemShut {NoStop}%
\bibitem [{\citenamefont
  {Koczor}(2021{\natexlab{b}})}]{koczorExponentialErrorSuppression2021}%
  \BibitemOpen
  \bibfield  {author} {\bibinfo {author} {\bibnamefont {Koczor}, \bibfnamefont
  {B{\'a}lint}}} (\bibinfo {year} {2021}{\natexlab{b}}),\ \bibfield  {title}
  {\enquote {\bibinfo {title} {Exponential {{Error Suppression}} for
  {{Near-Term Quantum Devices}}},}\ }\href
  {https://doi.org/10.1103/PhysRevX.11.031057} {\bibfield  {journal} {\bibinfo
  {journal} {Physical Review X}\ }\textbf {\bibinfo {volume} {11}}~(\bibinfo
  {number} {3}),\ \bibinfo {pages} {031057}}\BibitemShut {NoStop}%
\bibitem [{\citenamefont {Krebsbach}\ \emph {et~al.}(2022)\citenamefont
  {Krebsbach}, \citenamefont {Trauzettel},\ and\ \citenamefont
  {Calzona}}]{krebsbachOptimizationRichardsonExtrapolation2022}%
  \BibitemOpen
  \bibfield  {author} {\bibinfo {author} {\bibnamefont {Krebsbach},
  \bibfnamefont {Michael}}, \bibinfo {author} {\bibfnamefont {Bj{\"o}rn}\
  \bibnamefont {Trauzettel}}, and\ \bibinfo {author} {\bibfnamefont {Alessio}\
  \bibnamefont {Calzona}}} (\bibinfo {year} {2022}),\ \bibfield  {title}
  {\enquote {\bibinfo {title} {Optimization of {{Richardson}} extrapolation for
  quantum error mitigation},}\ }\href
  {https://doi.org/10.1103/PhysRevA.106.062436} {\bibfield  {journal} {\bibinfo
   {journal} {Physical Review A}\ }\textbf {\bibinfo {volume} {106}}~(\bibinfo
  {number} {6}),\ \bibinfo {pages} {062436}}\BibitemShut {NoStop}%
\bibitem [{\citenamefont {Krinner}\ \emph {et~al.}(2022)\citenamefont
  {Krinner}, \citenamefont {Lacroix}, \citenamefont {Remm}, \citenamefont
  {Di~Paolo}, \citenamefont {Genois}, \citenamefont {Leroux}, \citenamefont
  {Hellings}, \citenamefont {Lazar}, \citenamefont {Swiadek}, \citenamefont
  {Herrmann}, \citenamefont {Norris}, \citenamefont {Andersen}, \citenamefont
  {M{\"u}ller}, \citenamefont {Blais}, \citenamefont {Eichler},\ and\
  \citenamefont {Wallraff}}]{krinnerRealizingRepeatedQuantum2022}%
  \BibitemOpen
  \bibfield  {author} {\bibinfo {author} {\bibnamefont {Krinner}, \bibfnamefont
  {Sebastian}}, \bibinfo {author} {\bibfnamefont {Nathan}\ \bibnamefont
  {Lacroix}}, \bibinfo {author} {\bibfnamefont {Ants}\ \bibnamefont {Remm}},
  \bibinfo {author} {\bibfnamefont {Agustin}\ \bibnamefont {Di~Paolo}},
  \bibinfo {author} {\bibfnamefont {Elie}\ \bibnamefont {Genois}}, \bibinfo
  {author} {\bibfnamefont {Catherine}\ \bibnamefont {Leroux}}, \bibinfo
  {author} {\bibfnamefont {Christoph}\ \bibnamefont {Hellings}}, \bibinfo
  {author} {\bibfnamefont {Stefania}\ \bibnamefont {Lazar}}, \bibinfo {author}
  {\bibfnamefont {Francois}\ \bibnamefont {Swiadek}}, \bibinfo {author}
  {\bibfnamefont {Johannes}\ \bibnamefont {Herrmann}}, \bibinfo {author}
  {\bibfnamefont {Graham~J.}\ \bibnamefont {Norris}}, \bibinfo {author}
  {\bibfnamefont {Christian~Kraglund}\ \bibnamefont {Andersen}}, \bibinfo
  {author} {\bibfnamefont {Markus}\ \bibnamefont {M{\"u}ller}}, \bibinfo
  {author} {\bibfnamefont {Alexandre}\ \bibnamefont {Blais}}, \bibinfo {author}
  {\bibfnamefont {Christopher}\ \bibnamefont {Eichler}}, and\ \bibinfo {author}
  {\bibfnamefont {Andreas}\ \bibnamefont {Wallraff}}} (\bibinfo {year}
  {2022}),\ \bibfield  {title} {\enquote {\bibinfo {title} {Realizing repeated
  quantum error correction in a distance-three surface code},}\ }\href
  {https://doi.org/10.1038/s41586-022-04566-8} {\bibfield  {journal} {\bibinfo
  {journal} {Nature}\ }\textbf {\bibinfo {volume} {605}}~(\bibinfo {number}
  {7911}),\ \bibinfo {pages} {669--674}}\BibitemShut {NoStop}%
\bibitem [{\citenamefont {Kueng}\ \emph {et~al.}(2016)\citenamefont {Kueng},
  \citenamefont {Long}, \citenamefont {Doherty},\ and\ \citenamefont
  {Flammia}}]{kuengComparingExperimentsFaultTolerance2016}%
  \BibitemOpen
  \bibfield  {author} {\bibinfo {author} {\bibnamefont {Kueng}, \bibfnamefont
  {Richard}}, \bibinfo {author} {\bibfnamefont {David~M.}\ \bibnamefont
  {Long}}, \bibinfo {author} {\bibfnamefont {Andrew~C.}\ \bibnamefont
  {Doherty}}, and\ \bibinfo {author} {\bibfnamefont {Steven~T.}\ \bibnamefont
  {Flammia}}} (\bibinfo {year} {2016}),\ \bibfield  {title} {\enquote {\bibinfo
  {title} {Comparing {{Experiments}} to the {{Fault-Tolerance Threshold}}},}\
  }\href {https://doi.org/10.1103/PhysRevLett.117.170502} {\bibfield  {journal}
  {\bibinfo  {journal} {Physical Review Letters}\ }\textbf {\bibinfo {volume}
  {117}}~(\bibinfo {number} {17}),\ \bibinfo {pages} {170502}}\BibitemShut
  {NoStop}%
\bibitem [{\citenamefont {Kwon}\ and\ \citenamefont
  {Bae}(2021)}]{kwonHybridQuantumClassicalApproach2021}%
  \BibitemOpen
  \bibfield  {author} {\bibinfo {author} {\bibnamefont {Kwon}, \bibfnamefont
  {Hyeokjea}}, and\ \bibinfo {author} {\bibfnamefont {Joonwoo}\ \bibnamefont
  {Bae}}} (\bibinfo {year} {2021}),\ \bibfield  {title} {\enquote {\bibinfo
  {title} {A {{Hybrid Quantum-Classical Approach}} to {{Mitigating Measurement
  Errors}} in {{Quantum Algorithms}}},}\ }\href
  {https://doi.org/10.1109/TC.2020.3009664} {\bibfield  {journal} {\bibinfo
  {journal} {IEEE Transactions on Computers}\ }\textbf {\bibinfo {volume}
  {70}}~(\bibinfo {number} {09}),\ \bibinfo {pages} {1401--1411}}\BibitemShut
  {NoStop}%
\bibitem [{\citenamefont {LaRose}\ \emph {et~al.}(2021)\citenamefont {LaRose},
  \citenamefont {Mari}, \citenamefont {Kaiser}, \citenamefont {Karalekas},
  \citenamefont {Alves}, \citenamefont {Czarnik}, \citenamefont {El~Mandouh},
  \citenamefont {Gordon}, \citenamefont {Hindy}, \citenamefont {Robertson},
  \citenamefont {Thakre}, \citenamefont {Shammah},\ and\ \citenamefont
  {Zeng}}]{laroseMitiqSoftwarePackage2021}%
  \BibitemOpen
  \bibfield  {author} {\bibinfo {author} {\bibnamefont {LaRose}, \bibfnamefont
  {Ryan}}, \bibinfo {author} {\bibfnamefont {Andrea}\ \bibnamefont {Mari}},
  \bibinfo {author} {\bibfnamefont {Sarah}\ \bibnamefont {Kaiser}}, \bibinfo
  {author} {\bibfnamefont {Peter}\ \bibnamefont {Karalekas}}, \bibinfo {author}
  {\bibfnamefont {Andre}\ \bibnamefont {Alves}}, \bibinfo {author}
  {\bibfnamefont {Piotr}\ \bibnamefont {Czarnik}}, \bibinfo {author}
  {\bibfnamefont {Mohamed}\ \bibnamefont {El~Mandouh}}, \bibinfo {author}
  {\bibfnamefont {Max}\ \bibnamefont {Gordon}}, \bibinfo {author}
  {\bibfnamefont {Yousef}\ \bibnamefont {Hindy}}, \bibinfo {author}
  {\bibfnamefont {Aaron}\ \bibnamefont {Robertson}}, \bibinfo {author}
  {\bibfnamefont {Purva}\ \bibnamefont {Thakre}}, \bibinfo {author}
  {\bibfnamefont {Nathan}\ \bibnamefont {Shammah}}, and\ \bibinfo {author}
  {\bibfnamefont {William}\ \bibnamefont {Zeng}}} (\bibinfo {year} {2021}),\
  \href {https://github.com/unitaryfund/mitiq} {\enquote {\bibinfo {title}
  {Mitiq: {{A}} software package for error mitigation on noisy quantum
  computers},}\ }\BibitemShut {NoStop}%
\bibitem [{\citenamefont
  {Le~Cam}(1960)}]{lecamApproximationTheoremPoisson1960}%
  \BibitemOpen
  \bibfield  {author} {\bibinfo {author} {\bibnamefont {Le~Cam}, \bibfnamefont
  {Lucien}}} (\bibinfo {year} {1960}),\ \bibfield  {title} {\enquote {\bibinfo
  {title} {An approximation theorem for the {{Poisson}} binomial
  distribution},}\ }\href {https://doi.org/10.2140/pjm.1960.10.1181} {\bibfield
   {journal} {\bibinfo  {journal} {Pacific Journal of Mathematics}\ }\textbf
  {\bibinfo {volume} {10}}~(\bibinfo {number} {4}),\ \bibinfo {pages}
  {1181--1197}}\BibitemShut {NoStop}%
\bibitem [{\citenamefont {Lee}\ \emph {et~al.}(2021)\citenamefont {Lee},
  \citenamefont {Berry}, \citenamefont {Gidney}, \citenamefont {Huggins},
  \citenamefont {McClean}, \citenamefont {Wiebe},\ and\ \citenamefont
  {Babbush}}]{leeEvenMoreEfficient2021}%
  \BibitemOpen
  \bibfield  {author} {\bibinfo {author} {\bibnamefont {Lee}, \bibfnamefont
  {Joonho}}, \bibinfo {author} {\bibfnamefont {Dominic~W.}\ \bibnamefont
  {Berry}}, \bibinfo {author} {\bibfnamefont {Craig}\ \bibnamefont {Gidney}},
  \bibinfo {author} {\bibfnamefont {William~J.}\ \bibnamefont {Huggins}},
  \bibinfo {author} {\bibfnamefont {Jarrod~R.}\ \bibnamefont {McClean}},
  \bibinfo {author} {\bibfnamefont {Nathan}\ \bibnamefont {Wiebe}}, and\
  \bibinfo {author} {\bibfnamefont {Ryan}\ \bibnamefont {Babbush}}} (\bibinfo
  {year} {2021}),\ \bibfield  {title} {\enquote {\bibinfo {title} {Even {{More
  Efficient Quantum Computations}} of {{Chemistry Through Tensor
  Hypercontraction}}},}\ }\href {https://doi.org/10.1103/PRXQuantum.2.030305}
  {\bibfield  {journal} {\bibinfo  {journal} {PRX Quantum}\ }\textbf {\bibinfo
  {volume} {2}}~(\bibinfo {number} {3}),\ \bibinfo {pages}
  {030305}}\BibitemShut {NoStop}%
\bibitem [{\citenamefont {Li}\ and\ \citenamefont
  {Benjamin}(2017)}]{liEfficientVariationalQuantum2017}%
  \BibitemOpen
  \bibfield  {author} {\bibinfo {author} {\bibnamefont {Li}, \bibfnamefont
  {Ying}}, and\ \bibinfo {author} {\bibfnamefont {Simon~C.}\ \bibnamefont
  {Benjamin}}} (\bibinfo {year} {2017}),\ \bibfield  {title} {\enquote
  {\bibinfo {title} {Efficient {{Variational Quantum Simulator Incorporating
  Active Error Minimization}}},}\ }\href
  {https://doi.org/10.1103/PhysRevX.7.021050} {\bibfield  {journal} {\bibinfo
  {journal} {Physical Review X}\ }\textbf {\bibinfo {volume} {7}}~(\bibinfo
  {number} {2}),\ \bibinfo {pages} {021050}}\BibitemShut {NoStop}%
\bibitem [{\citenamefont
  {Lidar}(2014)}]{lidarReviewDecoherenceFreeSubspaces2014}%
  \BibitemOpen
  \bibfield  {author} {\bibinfo {author} {\bibnamefont {Lidar}, \bibfnamefont
  {Daniel~A}}} (\bibinfo {year} {2014}),\ \bibfield  {title} {\enquote
  {\bibinfo {title} {Review of {{Decoherence-Free Subspaces}}, {{Noiseless
  Subsystems}}, and {{Dynamical Decoupling}}},}\ }in\ \href
  {https://doi.org/10.1002/9781118742631.ch11} {\emph {\bibinfo {booktitle}
  {Quantum {{Information}} and {{Computation}} for {{Chemistry}}}}}\ (\bibinfo
  {publisher} {{John Wiley \& Sons, Ltd}})\ pp.\ \bibinfo {pages}
  {295--354}\BibitemShut {NoStop}%
\bibitem [{\citenamefont {Lidar}\ and\ \citenamefont
  {Brun}(2013)}]{lidarIntroductionDecoherenceNoise2013}%
  \BibitemOpen
  \bibfield  {author} {\bibinfo {author} {\bibnamefont {Lidar}, \bibfnamefont
  {Daniel~A}}, and\ \bibinfo {author} {\bibfnamefont {Todd~A.}\ \bibnamefont
  {Brun}}} (\bibinfo {year} {2013}),\ \bibfield  {title} {\enquote {\bibinfo
  {title} {Introduction to decoherence and noise in open quantum systems},}\
  }in\ \href {https://doi.org/10.1017/CBO9781139034807.003} {\emph {\bibinfo
  {booktitle} {Quantum {{Error Correction}}}}},\ \bibinfo {editor} {edited by\
  \bibinfo {editor} {\bibfnamefont {Daniel~A.}\ \bibnamefont {Lidar}}\ and\
  \bibinfo {editor} {\bibfnamefont {Todd~A.}\ \bibnamefont {Brun}}}\ (\bibinfo
  {publisher} {{Cambridge University Press}},\ \bibinfo {address}
  {{Cambridge}})\ pp.\ \bibinfo {pages} {3--45}\BibitemShut {NoStop}%
\bibitem [{\citenamefont {Lin}\ \emph {et~al.}(2021)\citenamefont {Lin},
  \citenamefont {Wallman}, \citenamefont {Hincks},\ and\ \citenamefont
  {Laflamme}}]{linIndependentStateMeasurement2021}%
  \BibitemOpen
  \bibfield  {author} {\bibinfo {author} {\bibnamefont {Lin}, \bibfnamefont
  {Junan}}, \bibinfo {author} {\bibfnamefont {Joel~J.}\ \bibnamefont
  {Wallman}}, \bibinfo {author} {\bibfnamefont {Ian}\ \bibnamefont {Hincks}},
  and\ \bibinfo {author} {\bibfnamefont {Raymond}\ \bibnamefont {Laflamme}}}
  (\bibinfo {year} {2021}),\ \bibfield  {title} {\enquote {\bibinfo {title}
  {Independent state and measurement characterization for quantum computers},}\
  }\href {https://doi.org/10.1103/PhysRevResearch.3.033285} {\bibfield
  {journal} {\bibinfo  {journal} {Physical Review Research}\ }\textbf {\bibinfo
  {volume} {3}}~(\bibinfo {number} {3}),\ \bibinfo {pages}
  {033285}}\BibitemShut {NoStop}%
\bibitem [{\citenamefont {Linke}\ \emph {et~al.}(2017)\citenamefont {Linke},
  \citenamefont {Gutierrez}, \citenamefont {Landsman}, \citenamefont {Figgatt},
  \citenamefont {Debnath}, \citenamefont {Brown},\ and\ \citenamefont
  {Monroe}}]{linkeFaulttolerantQuantumError2017}%
  \BibitemOpen
  \bibfield  {author} {\bibinfo {author} {\bibnamefont {Linke}, \bibfnamefont
  {Norbert~M}}, \bibinfo {author} {\bibfnamefont {Mauricio}\ \bibnamefont
  {Gutierrez}}, \bibinfo {author} {\bibfnamefont {Kevin~A.}\ \bibnamefont
  {Landsman}}, \bibinfo {author} {\bibfnamefont {Caroline}\ \bibnamefont
  {Figgatt}}, \bibinfo {author} {\bibfnamefont {Shantanu}\ \bibnamefont
  {Debnath}}, \bibinfo {author} {\bibfnamefont {Kenneth~R.}\ \bibnamefont
  {Brown}}, and\ \bibinfo {author} {\bibfnamefont {Christopher}\ \bibnamefont
  {Monroe}}} (\bibinfo {year} {2017}),\ \bibfield  {title} {\enquote {\bibinfo
  {title} {Fault-tolerant quantum error detection},}\ }\href
  {https://doi.org/10.1126/sciadv.1701074} {\bibfield  {journal} {\bibinfo
  {journal} {Science Advances}\ }\textbf {\bibinfo {volume} {3}}~(\bibinfo
  {number} {10}),\ \bibinfo {pages} {e1701074}}\BibitemShut {NoStop}%
\bibitem [{\citenamefont
  {Litinski}(2019)}]{litinskiMagicStateDistillation2019}%
  \BibitemOpen
  \bibfield  {author} {\bibinfo {author} {\bibnamefont {Litinski},
  \bibfnamefont {Daniel}}} (\bibinfo {year} {2019}),\ \bibfield  {title}
  {\enquote {\bibinfo {title} {Magic {{State Distillation}}: {{Not}} as
  {{Costly}} as {{You Think}}},}\ }\href
  {https://doi.org/10.22331/q-2019-12-02-205} {\bibfield  {journal} {\bibinfo
  {journal} {Quantum}\ }\textbf {\bibinfo {volume} {3}},\ \bibinfo {pages}
  {205}}\BibitemShut {NoStop}%
\bibitem [{\citenamefont {Liu}\ \emph {et~al.}(2007)\citenamefont {Liu},
  \citenamefont {Christandl},\ and\ \citenamefont
  {Verstraete}}]{liuQuantumComputationalComplexity2007}%
  \BibitemOpen
  \bibfield  {author} {\bibinfo {author} {\bibnamefont {Liu}, \bibfnamefont
  {Yi-Kai}}, \bibinfo {author} {\bibfnamefont {Matthias}\ \bibnamefont
  {Christandl}}, and\ \bibinfo {author} {\bibfnamefont {F.}~\bibnamefont
  {Verstraete}}} (\bibinfo {year} {2007}),\ \bibfield  {title} {\enquote
  {\bibinfo {title} {Quantum {{Computational Complexity}} of the
  \${{N}}\$-{{Representability Problem}}: {{QMA Complete}}},}\ }\href
  {https://doi.org/10.1103/PhysRevLett.98.110503} {\bibfield  {journal}
  {\bibinfo  {journal} {Physical Review Letters}\ }\textbf {\bibinfo {volume}
  {98}}~(\bibinfo {number} {11}),\ \bibinfo {pages} {110503}}\BibitemShut
  {NoStop}%
\bibitem [{\citenamefont {Lloyd}\ \emph {et~al.}(2014)\citenamefont {Lloyd},
  \citenamefont {Mohseni},\ and\ \citenamefont
  {Rebentrost}}]{lloydQuantumPrincipalComponent2014}%
  \BibitemOpen
  \bibfield  {author} {\bibinfo {author} {\bibnamefont {Lloyd}, \bibfnamefont
  {Seth}}, \bibinfo {author} {\bibfnamefont {Masoud}\ \bibnamefont {Mohseni}},
  and\ \bibinfo {author} {\bibfnamefont {Patrick}\ \bibnamefont {Rebentrost}}}
  (\bibinfo {year} {2014}),\ \bibfield  {title} {\enquote {\bibinfo {title}
  {Quantum principal component analysis},}\ }\href
  {https://doi.org/10.1038/nphys3029} {\bibfield  {journal} {\bibinfo
  {journal} {Nature Physics}\ }\textbf {\bibinfo {volume} {10}}~(\bibinfo
  {number} {9}),\ \bibinfo {pages} {631--633}}\BibitemShut {NoStop}%
\bibitem [{\citenamefont {Lostaglio}\ and\ \citenamefont
  {Ciani}(2021)}]{lostaglioErrorMitigationQuantumAssisted2021}%
  \BibitemOpen
  \bibfield  {author} {\bibinfo {author} {\bibnamefont {Lostaglio},
  \bibfnamefont {M}}, and\ \bibinfo {author} {\bibfnamefont {A.}~\bibnamefont
  {Ciani}}} (\bibinfo {year} {2021}),\ \bibfield  {title} {\enquote {\bibinfo
  {title} {Error {{Mitigation}} and {{Quantum-Assisted Simulation}} in the
  {{Error Corrected Regime}}},}\ }\href
  {https://doi.org/10.1103/PhysRevLett.127.200506} {\bibfield  {journal}
  {\bibinfo  {journal} {Physical Review Letters}\ }\textbf {\bibinfo {volume}
  {127}}~(\bibinfo {number} {20}),\ \bibinfo {pages} {200506}}\BibitemShut
  {NoStop}%
\bibitem [{\citenamefont {Lowe}\ \emph {et~al.}(2021)\citenamefont {Lowe},
  \citenamefont {Gordon}, \citenamefont {Czarnik}, \citenamefont {Arrasmith},
  \citenamefont {Coles},\ and\ \citenamefont
  {Cincio}}]{loweUnifiedApproachDatadriven2021}%
  \BibitemOpen
  \bibfield  {author} {\bibinfo {author} {\bibnamefont {Lowe}, \bibfnamefont
  {Angus}}, \bibinfo {author} {\bibfnamefont {Max~Hunter}\ \bibnamefont
  {Gordon}}, \bibinfo {author} {\bibfnamefont {Piotr}\ \bibnamefont {Czarnik}},
  \bibinfo {author} {\bibfnamefont {Andrew}\ \bibnamefont {Arrasmith}},
  \bibinfo {author} {\bibfnamefont {Patrick~J.}\ \bibnamefont {Coles}}, and\
  \bibinfo {author} {\bibfnamefont {Lukasz}\ \bibnamefont {Cincio}}} (\bibinfo
  {year} {2021}),\ \bibfield  {title} {\enquote {\bibinfo {title} {Unified
  approach to data-driven quantum error mitigation},}\ }\href
  {https://doi.org/10.1103/PhysRevResearch.3.033098} {\bibfield  {journal}
  {\bibinfo  {journal} {Physical Review Research}\ }\textbf {\bibinfo {volume}
  {3}}~(\bibinfo {number} {3}),\ \bibinfo {pages} {033098}}\BibitemShut
  {NoStop}%
\bibitem [{\citenamefont {Lu}\ \emph {et~al.}(2021)\citenamefont {Lu},
  \citenamefont {Ba{\~n}uls},\ and\ \citenamefont
  {Cirac}}]{luAlgorithmsQuantumSimulation2021}%
  \BibitemOpen
  \bibfield  {author} {\bibinfo {author} {\bibnamefont {Lu}, \bibfnamefont
  {Sirui}}, \bibinfo {author} {\bibfnamefont {Mari~Carmen}\ \bibnamefont
  {Ba{\~n}uls}}, and\ \bibinfo {author} {\bibfnamefont {J.~Ignacio}\
  \bibnamefont {Cirac}}} (\bibinfo {year} {2021}),\ \bibfield  {title}
  {\enquote {\bibinfo {title} {Algorithms for {{Quantum Simulation}} at
  {{Finite Energies}}},}\ }\href {https://doi.org/10.1103/PRXQuantum.2.020321}
  {\bibfield  {journal} {\bibinfo  {journal} {PRX Quantum}\ }\textbf {\bibinfo
  {volume} {2}}~(\bibinfo {number} {2}),\ \bibinfo {pages}
  {020321}}\BibitemShut {NoStop}%
\bibitem [{\citenamefont {Maciejewski}\ \emph
  {et~al.}(2020{\natexlab{a}})\citenamefont {Maciejewski}, \citenamefont
  {Rybotycki}, \citenamefont {Slowik},\ and\ \citenamefont
  {Tuziemski}}]{maciejewskiQuantumReadoutErrors2020}%
  \BibitemOpen
  \bibfield  {author} {\bibinfo {author} {\bibnamefont {Maciejewski},
  \bibfnamefont {Filip~B}}, \bibinfo {author} {\bibfnamefont {Tomasz}\
  \bibnamefont {Rybotycki}}, \bibinfo {author} {\bibfnamefont {Oskar}\
  \bibnamefont {Slowik}}, and\ \bibinfo {author} {\bibfnamefont {Jan}\
  \bibnamefont {Tuziemski}}} (\bibinfo {year} {2020}{\natexlab{a}}),\ \href
  {https://github.com/fbm2718/QREM} {\enquote {\bibinfo {title} {Quantum
  {{Readout Errors Mitigation}} ({{QREM}}) -- open source {{GitHub}}
  repository},}\ }\BibitemShut {NoStop}%
\bibitem [{\citenamefont {Maciejewski}\ \emph
  {et~al.}(2020{\natexlab{b}})\citenamefont {Maciejewski}, \citenamefont
  {Zimbor{\'a}s},\ and\ \citenamefont
  {Oszmaniec}}]{maciejewskiMitigationReadoutNoise2020}%
  \BibitemOpen
  \bibfield  {author} {\bibinfo {author} {\bibnamefont {Maciejewski},
  \bibfnamefont {Filip~B}}, \bibinfo {author} {\bibfnamefont {Zolt{\'a}n}\
  \bibnamefont {Zimbor{\'a}s}}, and\ \bibinfo {author} {\bibfnamefont
  {Micha{\l}}\ \bibnamefont {Oszmaniec}}} (\bibinfo {year}
  {2020}{\natexlab{b}}),\ \bibfield  {title} {\enquote {\bibinfo {title}
  {Mitigation of readout noise in near-term quantum devices by classical
  post-processing based on detector tomography},}\ }\href
  {https://doi.org/10.22331/q-2020-04-24-257} {\bibfield  {journal} {\bibinfo
  {journal} {Quantum}\ }\textbf {\bibinfo {volume} {4}},\ \bibinfo {pages}
  {257}}\BibitemShut {NoStop}%
\bibitem [{\citenamefont {Madjarov}\ \emph {et~al.}(2020)\citenamefont
  {Madjarov}, \citenamefont {Covey}, \citenamefont {Shaw}, \citenamefont
  {Choi}, \citenamefont {Kale}, \citenamefont {Cooper}, \citenamefont
  {Pichler}, \citenamefont {Schkolnik}, \citenamefont {Williams},\ and\
  \citenamefont {Endres}}]{madjarovHighfidelityEntanglementDetection2020}%
  \BibitemOpen
  \bibfield  {author} {\bibinfo {author} {\bibnamefont {Madjarov},
  \bibfnamefont {Ivaylo~S}}, \bibinfo {author} {\bibfnamefont {Jacob~P.}\
  \bibnamefont {Covey}}, \bibinfo {author} {\bibfnamefont {Adam~L.}\
  \bibnamefont {Shaw}}, \bibinfo {author} {\bibfnamefont {Joonhee}\
  \bibnamefont {Choi}}, \bibinfo {author} {\bibfnamefont {Anant}\ \bibnamefont
  {Kale}}, \bibinfo {author} {\bibfnamefont {Alexandre}\ \bibnamefont
  {Cooper}}, \bibinfo {author} {\bibfnamefont {Hannes}\ \bibnamefont
  {Pichler}}, \bibinfo {author} {\bibfnamefont {Vladimir}\ \bibnamefont
  {Schkolnik}}, \bibinfo {author} {\bibfnamefont {Jason~R.}\ \bibnamefont
  {Williams}}, and\ \bibinfo {author} {\bibfnamefont {Manuel}\ \bibnamefont
  {Endres}}} (\bibinfo {year} {2020}),\ \bibfield  {title} {\enquote {\bibinfo
  {title} {High-fidelity entanglement and detection of alkaline-earth
  {{Rydberg}} atoms},}\ }\href {https://doi.org/10.1038/s41567-020-0903-z}
  {\bibfield  {journal} {\bibinfo  {journal} {Nature Physics}\ }\textbf
  {\bibinfo {volume} {16}}~(\bibinfo {number} {8}),\ \bibinfo {pages}
  {857--861}}\BibitemShut {NoStop}%
\bibitem [{\citenamefont {Mari}\ \emph {et~al.}(2021)\citenamefont {Mari},
  \citenamefont {Shammah},\ and\ \citenamefont
  {Zeng}}]{mariExtendingQuantumProbabilistic2021}%
  \BibitemOpen
  \bibfield  {author} {\bibinfo {author} {\bibnamefont {Mari}, \bibfnamefont
  {Andrea}}, \bibinfo {author} {\bibfnamefont {Nathan}\ \bibnamefont
  {Shammah}}, and\ \bibinfo {author} {\bibfnamefont {William~J.}\ \bibnamefont
  {Zeng}}} (\bibinfo {year} {2021}),\ \bibfield  {title} {\enquote {\bibinfo
  {title} {Extending quantum probabilistic error cancellation by noise
  scaling},}\ }\href {https://doi.org/10.1103/PhysRevA.104.052607} {\bibfield
  {journal} {\bibinfo  {journal} {Physical Review A}\ }\textbf {\bibinfo
  {volume} {104}}~(\bibinfo {number} {5}),\ \bibinfo {pages}
  {052607}}\BibitemShut {NoStop}%
\bibitem [{\citenamefont
  {Mazziotti}(2016)}]{mazziottiPureRepresentabilityConditions2016}%
  \BibitemOpen
  \bibfield  {author} {\bibinfo {author} {\bibnamefont {Mazziotti},
  \bibfnamefont {David~A}}} (\bibinfo {year} {2016}),\ \bibfield  {title}
  {\enquote {\bibinfo {title} {Pure-\${{N}}\$-representability conditions of
  two-fermion reduced density matrices},}\ }\href
  {https://doi.org/10.1103/PhysRevA.94.032516} {\bibfield  {journal} {\bibinfo
  {journal} {Physical Review A}\ }\textbf {\bibinfo {volume} {94}}~(\bibinfo
  {number} {3}),\ \bibinfo {pages} {032516}}\BibitemShut {NoStop}%
\bibitem [{\citenamefont {McArdle}\ \emph {et~al.}(2019)\citenamefont
  {McArdle}, \citenamefont {Yuan},\ and\ \citenamefont
  {Benjamin}}]{mcardleErrorMitigatedDigitalQuantum2019}%
  \BibitemOpen
  \bibfield  {author} {\bibinfo {author} {\bibnamefont {McArdle}, \bibfnamefont
  {Sam}}, \bibinfo {author} {\bibfnamefont {Xiao}\ \bibnamefont {Yuan}}, and\
  \bibinfo {author} {\bibfnamefont {Simon}\ \bibnamefont {Benjamin}}} (\bibinfo
  {year} {2019}),\ \bibfield  {title} {\enquote {\bibinfo {title}
  {Error-{{Mitigated Digital Quantum Simulation}}},}\ }\href
  {https://doi.org/10.1103/PhysRevLett.122.180501} {\bibfield  {journal}
  {\bibinfo  {journal} {Physical Review Letters}\ }\textbf {\bibinfo {volume}
  {122}}~(\bibinfo {number} {18}),\ \bibinfo {pages} {180501}}\BibitemShut
  {NoStop}%
\bibitem [{\citenamefont {McCaskey}\ \emph {et~al.}(2019)\citenamefont
  {McCaskey}, \citenamefont {Parks}, \citenamefont {Jakowski}, \citenamefont
  {Moore}, \citenamefont {Morris}, \citenamefont {Humble},\ and\ \citenamefont
  {Pooser}}]{mccaskeyQuantumChemistryBenchmark2019}%
  \BibitemOpen
  \bibfield  {author} {\bibinfo {author} {\bibnamefont {McCaskey},
  \bibfnamefont {Alexander~J}}, \bibinfo {author} {\bibfnamefont {Zachary~P.}\
  \bibnamefont {Parks}}, \bibinfo {author} {\bibfnamefont {Jacek}\ \bibnamefont
  {Jakowski}}, \bibinfo {author} {\bibfnamefont {Shirley~V.}\ \bibnamefont
  {Moore}}, \bibinfo {author} {\bibfnamefont {Titus~D.}\ \bibnamefont
  {Morris}}, \bibinfo {author} {\bibfnamefont {Travis~S.}\ \bibnamefont
  {Humble}}, and\ \bibinfo {author} {\bibfnamefont {Raphael~C.}\ \bibnamefont
  {Pooser}}} (\bibinfo {year} {2019}),\ \bibfield  {title} {\enquote {\bibinfo
  {title} {Quantum chemistry as a benchmark for near-term quantum computers},}\
  }\href {https://doi.org/10.1038/s41534-019-0209-0} {\bibfield  {journal}
  {\bibinfo  {journal} {npj Quantum Information}\ }\textbf {\bibinfo {volume}
  {5}},\ \bibinfo {pages} {99}}\BibitemShut {NoStop}%
\bibitem [{\citenamefont {McClean}\ \emph {et~al.}(2020)\citenamefont
  {McClean}, \citenamefont {Jiang}, \citenamefont {Rubin}, \citenamefont
  {Babbush},\ and\ \citenamefont {Neven}}]{mccleanDecodingQuantumErrors2020}%
  \BibitemOpen
  \bibfield  {author} {\bibinfo {author} {\bibnamefont {McClean}, \bibfnamefont
  {Jarrod~R}}, \bibinfo {author} {\bibfnamefont {Zhang}\ \bibnamefont {Jiang}},
  \bibinfo {author} {\bibfnamefont {Nicholas~C.}\ \bibnamefont {Rubin}},
  \bibinfo {author} {\bibfnamefont {Ryan}\ \bibnamefont {Babbush}}, and\
  \bibinfo {author} {\bibfnamefont {Hartmut}\ \bibnamefont {Neven}}} (\bibinfo
  {year} {2020}),\ \bibfield  {title} {\enquote {\bibinfo {title} {Decoding
  quantum errors with subspace expansions},}\ }\href
  {https://doi.org/10.1038/s41467-020-14341-w} {\bibfield  {journal} {\bibinfo
  {journal} {Nature Communications}\ }\textbf {\bibinfo {volume} {11}},\
  \bibinfo {pages} {636}}\BibitemShut {NoStop}%
\bibitem [{\citenamefont {McClean}\ \emph {et~al.}(2017)\citenamefont
  {McClean}, \citenamefont {{Kimchi-Schwartz}}, \citenamefont {Carter},\ and\
  \citenamefont {{de Jong}}}]{mccleanHybridQuantumclassicalHierarchy2017}%
  \BibitemOpen
  \bibfield  {author} {\bibinfo {author} {\bibnamefont {McClean}, \bibfnamefont
  {Jarrod~R}}, \bibinfo {author} {\bibfnamefont {Mollie~E.}\ \bibnamefont
  {{Kimchi-Schwartz}}}, \bibinfo {author} {\bibfnamefont {Jonathan}\
  \bibnamefont {Carter}}, and\ \bibinfo {author} {\bibfnamefont {Wibe~A.}\
  \bibnamefont {{de Jong}}}} (\bibinfo {year} {2017}),\ \bibfield  {title}
  {\enquote {\bibinfo {title} {Hybrid quantum-classical hierarchy for
  mitigation of decoherence and determination of excited states},}\ }\href
  {https://doi.org/10.1103/PhysRevA.95.042308} {\bibfield  {journal} {\bibinfo
  {journal} {Physical Review A}\ }\textbf {\bibinfo {volume} {95}}~(\bibinfo
  {number} {4}),\ \bibinfo {pages} {042308}}\BibitemShut {NoStop}%
\bibitem [{\citenamefont {McClean}\ \emph {et~al.}(2016)\citenamefont
  {McClean}, \citenamefont {Romero}, \citenamefont {Babbush},\ and\
  \citenamefont {{Aspuru-Guzik}}}]{mccleanTheoryVariationalHybrid2016}%
  \BibitemOpen
  \bibfield  {author} {\bibinfo {author} {\bibnamefont {McClean}, \bibfnamefont
  {Jarrod~R}}, \bibinfo {author} {\bibfnamefont {Jonathan}\ \bibnamefont
  {Romero}}, \bibinfo {author} {\bibfnamefont {Ryan}\ \bibnamefont {Babbush}},
  and\ \bibinfo {author} {\bibfnamefont {Al{\'a}n}\ \bibnamefont
  {{Aspuru-Guzik}}}} (\bibinfo {year} {2016}),\ \bibfield  {title} {\enquote
  {\bibinfo {title} {The theory of variational hybrid quantum-classical
  algorithms},}\ }\href {https://doi.org/10.1088/1367-2630/18/2/023023}
  {\bibfield  {journal} {\bibinfo  {journal} {New Journal of Physics}\ }\textbf
  {\bibinfo {volume} {18}}~(\bibinfo {number} {2}),\ \bibinfo {pages}
  {023023}}\BibitemShut {NoStop}%
\bibitem [{\citenamefont {McWeeny}(1960)}]{mcweenyRecentAdvancesDensity1960}%
  \BibitemOpen
  \bibfield  {author} {\bibinfo {author} {\bibnamefont {McWeeny}, \bibfnamefont
  {R}}} (\bibinfo {year} {1960}),\ \bibfield  {title} {\enquote {\bibinfo
  {title} {Some {{Recent Advances}} in {{Density Matrix Theory}}},}\ }\href
  {https://doi.org/10.1103/RevModPhys.32.335} {\bibfield  {journal} {\bibinfo
  {journal} {Reviews of Modern Physics}\ }\textbf {\bibinfo {volume}
  {32}}~(\bibinfo {number} {2}),\ \bibinfo {pages} {335--369}}\BibitemShut
  {NoStop}%
\bibitem [{\citenamefont {Merkel}\ \emph {et~al.}(2013)\citenamefont {Merkel},
  \citenamefont {Gambetta}, \citenamefont {Smolin}, \citenamefont {Poletto},
  \citenamefont {C{\'o}rcoles}, \citenamefont {Johnson}, \citenamefont {Ryan},\
  and\ \citenamefont {Steffen}}]{merkelSelfconsistentQuantumProcess2013}%
  \BibitemOpen
  \bibfield  {author} {\bibinfo {author} {\bibnamefont {Merkel}, \bibfnamefont
  {Seth~T}}, \bibinfo {author} {\bibfnamefont {Jay~M.}\ \bibnamefont
  {Gambetta}}, \bibinfo {author} {\bibfnamefont {John~A.}\ \bibnamefont
  {Smolin}}, \bibinfo {author} {\bibfnamefont {Stefano}\ \bibnamefont
  {Poletto}}, \bibinfo {author} {\bibfnamefont {Antonio~D.}\ \bibnamefont
  {C{\'o}rcoles}}, \bibinfo {author} {\bibfnamefont {Blake~R.}\ \bibnamefont
  {Johnson}}, \bibinfo {author} {\bibfnamefont {Colm~A.}\ \bibnamefont {Ryan}},
  and\ \bibinfo {author} {\bibfnamefont {Matthias}\ \bibnamefont {Steffen}}}
  (\bibinfo {year} {2013}),\ \bibfield  {title} {\enquote {\bibinfo {title}
  {Self-consistent quantum process tomography},}\ }\href
  {https://doi.org/10.1103/PhysRevA.87.062119} {\bibfield  {journal} {\bibinfo
  {journal} {Physical Review A}\ }\textbf {\bibinfo {volume} {87}}~(\bibinfo
  {number} {6}),\ \bibinfo {pages} {062119}}\BibitemShut {NoStop}%
\bibitem [{\citenamefont {Mi}\ \emph {et~al.}(2021)\citenamefont {Mi},
  \citenamefont {Roushan}, \citenamefont {Quintana}, \citenamefont
  {Mandr{\`a}}, \citenamefont {Marshall}, \citenamefont {Neill}, \citenamefont
  {Arute}, \citenamefont {Arya}, \citenamefont {Atalaya}, \citenamefont
  {Babbush}, \citenamefont {Bardin}, \citenamefont {Barends}, \citenamefont
  {Basso}, \citenamefont {Bengtsson}, \citenamefont {Boixo}, \citenamefont
  {Bourassa}, \citenamefont {Broughton}, \citenamefont {Buckley}, \citenamefont
  {Buell}, \citenamefont {Burkett}, \citenamefont {Bushnell}, \citenamefont
  {Chen}, \citenamefont {Chiaro}, \citenamefont {Collins}, \citenamefont
  {Courtney}, \citenamefont {Demura}, \citenamefont {Derk}, \citenamefont
  {Dunsworth}, \citenamefont {Eppens}, \citenamefont {Erickson}, \citenamefont
  {Farhi}, \citenamefont {Fowler}, \citenamefont {Foxen}, \citenamefont
  {Gidney}, \citenamefont {Giustina}, \citenamefont {Gross}, \citenamefont
  {Harrigan}, \citenamefont {Harrington}, \citenamefont {Hilton}, \citenamefont
  {Ho}, \citenamefont {Hong}, \citenamefont {Huang}, \citenamefont {Huggins},
  \citenamefont {Ioffe}, \citenamefont {Isakov}, \citenamefont {Jeffrey},
  \citenamefont {Jiang}, \citenamefont {Jones}, \citenamefont {Kafri},
  \citenamefont {Kelly}, \citenamefont {Kim}, \citenamefont {Kitaev},
  \citenamefont {Klimov}, \citenamefont {Korotkov}, \citenamefont {Kostritsa},
  \citenamefont {Landhuis}, \citenamefont {Laptev}, \citenamefont {Lucero},
  \citenamefont {Martin}, \citenamefont {McClean}, \citenamefont {McCourt},
  \citenamefont {McEwen}, \citenamefont {Megrant}, \citenamefont {Miao},
  \citenamefont {Mohseni}, \citenamefont {Montazeri}, \citenamefont
  {Mruczkiewicz}, \citenamefont {Mutus}, \citenamefont {Naaman}, \citenamefont
  {Neeley}, \citenamefont {Newman}, \citenamefont {Niu}, \citenamefont
  {O'Brien}, \citenamefont {Opremcak}, \citenamefont {Ostby}, \citenamefont
  {Pato}, \citenamefont {Petukhov}, \citenamefont {Redd}, \citenamefont
  {Rubin}, \citenamefont {Sank}, \citenamefont {Satzinger}, \citenamefont
  {Shvarts}, \citenamefont {Strain}, \citenamefont {Szalay}, \citenamefont
  {Trevithick}, \citenamefont {Villalonga}, \citenamefont {White},
  \citenamefont {Yao}, \citenamefont {Yeh}, \citenamefont {Zalcman},
  \citenamefont {Neven}, \citenamefont {Aleiner}, \citenamefont {Kechedzhi},
  \citenamefont {Smelyanskiy},\ and\ \citenamefont
  {Chen}}]{miInformationScramblingQuantum2021}%
  \BibitemOpen
  \bibfield  {author} {\bibinfo {author} {\bibnamefont {Mi}, \bibfnamefont
  {Xiao}}, \bibinfo {author} {\bibfnamefont {Pedram}\ \bibnamefont {Roushan}},
  \bibinfo {author} {\bibfnamefont {Chris}\ \bibnamefont {Quintana}}, \bibinfo
  {author} {\bibfnamefont {Salvatore}\ \bibnamefont {Mandr{\`a}}}, \bibinfo
  {author} {\bibfnamefont {Jeffrey}\ \bibnamefont {Marshall}}, \bibinfo
  {author} {\bibfnamefont {Charles}\ \bibnamefont {Neill}}, \bibinfo {author}
  {\bibfnamefont {Frank}\ \bibnamefont {Arute}}, \bibinfo {author}
  {\bibfnamefont {Kunal}\ \bibnamefont {Arya}}, \bibinfo {author}
  {\bibfnamefont {Juan}\ \bibnamefont {Atalaya}}, \bibinfo {author}
  {\bibfnamefont {Ryan}\ \bibnamefont {Babbush}}, \bibinfo {author}
  {\bibfnamefont {Joseph~C.}\ \bibnamefont {Bardin}}, \bibinfo {author}
  {\bibfnamefont {Rami}\ \bibnamefont {Barends}}, \bibinfo {author}
  {\bibfnamefont {Joao}\ \bibnamefont {Basso}}, \bibinfo {author}
  {\bibfnamefont {Andreas}\ \bibnamefont {Bengtsson}}, \bibinfo {author}
  {\bibfnamefont {Sergio}\ \bibnamefont {Boixo}}, \bibinfo {author}
  {\bibfnamefont {Alexandre}\ \bibnamefont {Bourassa}}, \bibinfo {author}
  {\bibfnamefont {Michael}\ \bibnamefont {Broughton}}, \bibinfo {author}
  {\bibfnamefont {Bob~B.}\ \bibnamefont {Buckley}}, \bibinfo {author}
  {\bibfnamefont {David~A.}\ \bibnamefont {Buell}}, \bibinfo {author}
  {\bibfnamefont {Brian}\ \bibnamefont {Burkett}}, \bibinfo {author}
  {\bibfnamefont {Nicholas}\ \bibnamefont {Bushnell}}, \bibinfo {author}
  {\bibfnamefont {Zijun}\ \bibnamefont {Chen}}, \bibinfo {author}
  {\bibfnamefont {Benjamin}\ \bibnamefont {Chiaro}}, \bibinfo {author}
  {\bibfnamefont {Roberto}\ \bibnamefont {Collins}}, \bibinfo {author}
  {\bibfnamefont {William}\ \bibnamefont {Courtney}}, \bibinfo {author}
  {\bibfnamefont {Sean}\ \bibnamefont {Demura}}, \bibinfo {author}
  {\bibfnamefont {Alan~R.}\ \bibnamefont {Derk}}, \bibinfo {author}
  {\bibfnamefont {Andrew}\ \bibnamefont {Dunsworth}}, \bibinfo {author}
  {\bibfnamefont {Daniel}\ \bibnamefont {Eppens}}, \bibinfo {author}
  {\bibfnamefont {Catherine}\ \bibnamefont {Erickson}}, \bibinfo {author}
  {\bibfnamefont {Edward}\ \bibnamefont {Farhi}}, \bibinfo {author}
  {\bibfnamefont {Austin~G.}\ \bibnamefont {Fowler}}, \bibinfo {author}
  {\bibfnamefont {Brooks}\ \bibnamefont {Foxen}}, \bibinfo {author}
  {\bibfnamefont {Craig}\ \bibnamefont {Gidney}}, \bibinfo {author}
  {\bibfnamefont {Marissa}\ \bibnamefont {Giustina}}, \bibinfo {author}
  {\bibfnamefont {Jonathan~A.}\ \bibnamefont {Gross}}, \bibinfo {author}
  {\bibfnamefont {Matthew~P.}\ \bibnamefont {Harrigan}}, \bibinfo {author}
  {\bibfnamefont {Sean~D.}\ \bibnamefont {Harrington}}, \bibinfo {author}
  {\bibfnamefont {Jeremy}\ \bibnamefont {Hilton}}, \bibinfo {author}
  {\bibfnamefont {Alan}\ \bibnamefont {Ho}}, \bibinfo {author} {\bibfnamefont
  {Sabrina}\ \bibnamefont {Hong}}, \bibinfo {author} {\bibfnamefont {Trent}\
  \bibnamefont {Huang}}, \bibinfo {author} {\bibfnamefont {William~J.}\
  \bibnamefont {Huggins}}, \bibinfo {author} {\bibfnamefont {L.~B.}\
  \bibnamefont {Ioffe}}, \bibinfo {author} {\bibfnamefont {Sergei~V.}\
  \bibnamefont {Isakov}}, \bibinfo {author} {\bibfnamefont {Evan}\ \bibnamefont
  {Jeffrey}}, \bibinfo {author} {\bibfnamefont {Zhang}\ \bibnamefont {Jiang}},
  \bibinfo {author} {\bibfnamefont {Cody}\ \bibnamefont {Jones}}, \bibinfo
  {author} {\bibfnamefont {Dvir}\ \bibnamefont {Kafri}}, \bibinfo {author}
  {\bibfnamefont {Julian}\ \bibnamefont {Kelly}}, \bibinfo {author}
  {\bibfnamefont {Seon}\ \bibnamefont {Kim}}, \bibinfo {author} {\bibfnamefont
  {Alexei}\ \bibnamefont {Kitaev}}, \bibinfo {author} {\bibfnamefont {Paul~V.}\
  \bibnamefont {Klimov}}, \bibinfo {author} {\bibfnamefont {Alexander~N.}\
  \bibnamefont {Korotkov}}, \bibinfo {author} {\bibfnamefont {Fedor}\
  \bibnamefont {Kostritsa}}, \bibinfo {author} {\bibfnamefont {David}\
  \bibnamefont {Landhuis}}, \bibinfo {author} {\bibfnamefont {Pavel}\
  \bibnamefont {Laptev}}, \bibinfo {author} {\bibfnamefont {Erik}\ \bibnamefont
  {Lucero}}, \bibinfo {author} {\bibfnamefont {Orion}\ \bibnamefont {Martin}},
  \bibinfo {author} {\bibfnamefont {Jarrod~R.}\ \bibnamefont {McClean}},
  \bibinfo {author} {\bibfnamefont {Trevor}\ \bibnamefont {McCourt}}, \bibinfo
  {author} {\bibfnamefont {Matt}\ \bibnamefont {McEwen}}, \bibinfo {author}
  {\bibfnamefont {Anthony}\ \bibnamefont {Megrant}}, \bibinfo {author}
  {\bibfnamefont {Kevin~C.}\ \bibnamefont {Miao}}, \bibinfo {author}
  {\bibfnamefont {Masoud}\ \bibnamefont {Mohseni}}, \bibinfo {author}
  {\bibfnamefont {Shirin}\ \bibnamefont {Montazeri}}, \bibinfo {author}
  {\bibfnamefont {Wojciech}\ \bibnamefont {Mruczkiewicz}}, \bibinfo {author}
  {\bibfnamefont {Josh}\ \bibnamefont {Mutus}}, \bibinfo {author}
  {\bibfnamefont {Ofer}\ \bibnamefont {Naaman}}, \bibinfo {author}
  {\bibfnamefont {Matthew}\ \bibnamefont {Neeley}}, \bibinfo {author}
  {\bibfnamefont {Michael}\ \bibnamefont {Newman}}, \bibinfo {author}
  {\bibfnamefont {Murphy~Yuezhen}\ \bibnamefont {Niu}}, \bibinfo {author}
  {\bibfnamefont {Thomas~E.}\ \bibnamefont {O'Brien}}, \bibinfo {author}
  {\bibfnamefont {Alex}\ \bibnamefont {Opremcak}}, \bibinfo {author}
  {\bibfnamefont {Eric}\ \bibnamefont {Ostby}}, \bibinfo {author}
  {\bibfnamefont {Balint}\ \bibnamefont {Pato}}, \bibinfo {author}
  {\bibfnamefont {Andre}\ \bibnamefont {Petukhov}}, \bibinfo {author}
  {\bibfnamefont {Nicholas}\ \bibnamefont {Redd}}, \bibinfo {author}
  {\bibfnamefont {Nicholas~C.}\ \bibnamefont {Rubin}}, \bibinfo {author}
  {\bibfnamefont {Daniel}\ \bibnamefont {Sank}}, \bibinfo {author}
  {\bibfnamefont {Kevin~J.}\ \bibnamefont {Satzinger}}, \bibinfo {author}
  {\bibfnamefont {Vladimir}\ \bibnamefont {Shvarts}}, \bibinfo {author}
  {\bibfnamefont {Doug}\ \bibnamefont {Strain}}, \bibinfo {author}
  {\bibfnamefont {Marco}\ \bibnamefont {Szalay}}, \bibinfo {author}
  {\bibfnamefont {Matthew~D.}\ \bibnamefont {Trevithick}}, \bibinfo {author}
  {\bibfnamefont {Benjamin}\ \bibnamefont {Villalonga}}, \bibinfo {author}
  {\bibfnamefont {Theodore}\ \bibnamefont {White}}, \bibinfo {author}
  {\bibfnamefont {Z.~Jamie}\ \bibnamefont {Yao}}, \bibinfo {author}
  {\bibfnamefont {Ping}\ \bibnamefont {Yeh}}, \bibinfo {author} {\bibfnamefont
  {Adam}\ \bibnamefont {Zalcman}}, \bibinfo {author} {\bibfnamefont {Hartmut}\
  \bibnamefont {Neven}}, \bibinfo {author} {\bibfnamefont {Igor}\ \bibnamefont
  {Aleiner}}, \bibinfo {author} {\bibfnamefont {Kostyantyn}\ \bibnamefont
  {Kechedzhi}}, \bibinfo {author} {\bibfnamefont {Vadim}\ \bibnamefont
  {Smelyanskiy}}, and\ \bibinfo {author} {\bibfnamefont {Yu}~\bibnamefont
  {Chen}}} (\bibinfo {year} {2021}),\ \bibfield  {title} {\enquote {\bibinfo
  {title} {Information scrambling in quantum circuits},}\ }\href
  {https://doi.org/10.1126/science.abg5029} {\bibfield  {journal} {\bibinfo
  {journal} {Science}\ }\textbf {\bibinfo {volume} {374}}~(\bibinfo {number}
  {6574}),\ \bibinfo {pages} {1479--1483}}\BibitemShut {NoStop}%
\bibitem [{\citenamefont {Mitarai}\ and\ \citenamefont
  {Fujii}(2019)}]{mitaraiMethodologyReplacingIndirect2019}%
  \BibitemOpen
  \bibfield  {author} {\bibinfo {author} {\bibnamefont {Mitarai}, \bibfnamefont
  {Kosuke}}, and\ \bibinfo {author} {\bibfnamefont {Keisuke}\ \bibnamefont
  {Fujii}}} (\bibinfo {year} {2019}),\ \bibfield  {title} {\enquote {\bibinfo
  {title} {Methodology for replacing indirect measurements with direct
  measurements},}\ }\href {https://doi.org/10.1103/PhysRevResearch.1.013006}
  {\bibfield  {journal} {\bibinfo  {journal} {Physical Review Research}\
  }\textbf {\bibinfo {volume} {1}}~(\bibinfo {number} {1}),\ \bibinfo {pages}
  {013006}}\BibitemShut {NoStop}%
\bibitem [{\citenamefont {Montanaro}\ and\ \citenamefont
  {Stanisic}(2021)}]{montanaroErrorMitigationTraining2021}%
  \BibitemOpen
  \bibfield  {author} {\bibinfo {author} {\bibnamefont {Montanaro},
  \bibfnamefont {Ashley}}, and\ \bibinfo {author} {\bibfnamefont {Stasja}\
  \bibnamefont {Stanisic}}} (\bibinfo {year} {2021}),\ \href
  {http://arxiv.org/abs/2102.02120} {\enquote {\bibinfo {title} {Error
  mitigation by training with fermionic linear optics},}\ }\bibinfo
  {howpublished} {arXiv:2102.02120 [quant-ph]}\BibitemShut {NoStop}%
\bibitem [{\citenamefont {Motta}\ \emph {et~al.}(2020)\citenamefont {Motta},
  \citenamefont {Sun}, \citenamefont {Tan}, \citenamefont {O'Rourke},
  \citenamefont {Ye}, \citenamefont {Minnich}, \citenamefont {Brand{\~a}o},\
  and\ \citenamefont {Chan}}]{mottaDeterminingEigenstatesThermal2020}%
  \BibitemOpen
  \bibfield  {author} {\bibinfo {author} {\bibnamefont {Motta}, \bibfnamefont
  {Mario}}, \bibinfo {author} {\bibfnamefont {Chong}\ \bibnamefont {Sun}},
  \bibinfo {author} {\bibfnamefont {Adrian T.~K.}\ \bibnamefont {Tan}},
  \bibinfo {author} {\bibfnamefont {Matthew~J.}\ \bibnamefont {O'Rourke}},
  \bibinfo {author} {\bibfnamefont {Erika}\ \bibnamefont {Ye}}, \bibinfo
  {author} {\bibfnamefont {Austin~J.}\ \bibnamefont {Minnich}}, \bibinfo
  {author} {\bibfnamefont {Fernando G. S.~L.}\ \bibnamefont {Brand{\~a}o}},
  and\ \bibinfo {author} {\bibfnamefont {Garnet Kin-Lic}\ \bibnamefont {Chan}}}
  (\bibinfo {year} {2020}),\ \bibfield  {title} {\enquote {\bibinfo {title}
  {Determining eigenstates and thermal states on a quantum computer using
  quantum imaginary time evolution},}\ }\href
  {https://doi.org/10.1038/s41567-019-0704-4} {\bibfield  {journal} {\bibinfo
  {journal} {Nature Physics}\ }\textbf {\bibinfo {volume} {16}}~(\bibinfo
  {number} {2}),\ \bibinfo {pages} {205--210}}\BibitemShut {NoStop}%
\bibitem [{\citenamefont {{M{\"u}ller-Hermes}}\ \emph
  {et~al.}(2016)\citenamefont {{M{\"u}ller-Hermes}}, \citenamefont
  {Stilck~Fran{\c c}a},\ and\ \citenamefont
  {Wolf}}]{muller-hermesEntropyProductionDoubly2016}%
  \BibitemOpen
  \bibfield  {author} {\bibinfo {author} {\bibnamefont {{M{\"u}ller-Hermes}},
  \bibfnamefont {Alexander}}, \bibinfo {author} {\bibfnamefont {Daniel}\
  \bibnamefont {Stilck~Fran{\c c}a}}, and\ \bibinfo {author} {\bibfnamefont
  {Michael~M.}\ \bibnamefont {Wolf}}} (\bibinfo {year} {2016}),\ \bibfield
  {title} {\enquote {\bibinfo {title} {Entropy production of doubly stochastic
  quantum channels},}\ }\href {https://doi.org/10.1063/1.4941136} {\bibfield
  {journal} {\bibinfo  {journal} {Journal of Mathematical Physics}\ }\textbf
  {\bibinfo {volume} {57}}~(\bibinfo {number} {2}),\ \bibinfo {pages}
  {022203}}\BibitemShut {NoStop}%
\bibitem [{\citenamefont {Nachman}\ \emph {et~al.}(2020)\citenamefont
  {Nachman}, \citenamefont {Urbanek}, \citenamefont {{de Jong}},\ and\
  \citenamefont {Bauer}}]{nachmanUnfoldingQuantumComputer2020}%
  \BibitemOpen
  \bibfield  {author} {\bibinfo {author} {\bibnamefont {Nachman}, \bibfnamefont
  {Benjamin}}, \bibinfo {author} {\bibfnamefont {Miroslav}\ \bibnamefont
  {Urbanek}}, \bibinfo {author} {\bibfnamefont {Wibe~A.}\ \bibnamefont {{de
  Jong}}}, and\ \bibinfo {author} {\bibfnamefont {Christian~W.}\ \bibnamefont
  {Bauer}}} (\bibinfo {year} {2020}),\ \bibfield  {title} {\enquote {\bibinfo
  {title} {Unfolding quantum computer readout noise},}\ }\href
  {https://doi.org/10.1038/s41534-020-00309-7} {\bibfield  {journal} {\bibinfo
  {journal} {npj Quantum Information}\ }\textbf {\bibinfo {volume} {6}},\
  \bibinfo {pages} {84}}\BibitemShut {NoStop}%
\bibitem [{\citenamefont {Nagourney}\ \emph {et~al.}(1986)\citenamefont
  {Nagourney}, \citenamefont {Sandberg},\ and\ \citenamefont
  {Dehmelt}}]{nagourneyShelvedOpticalElectron1986}%
  \BibitemOpen
  \bibfield  {author} {\bibinfo {author} {\bibnamefont {Nagourney},
  \bibfnamefont {Warren}}, \bibinfo {author} {\bibfnamefont {Jon}\ \bibnamefont
  {Sandberg}}, and\ \bibinfo {author} {\bibfnamefont {Hans}\ \bibnamefont
  {Dehmelt}}} (\bibinfo {year} {1986}),\ \bibfield  {title} {\enquote {\bibinfo
  {title} {Shelved optical electron amplifier: {{Observation}} of quantum
  jumps},}\ }\href {https://doi.org/10.1103/PhysRevLett.56.2797} {\bibfield
  {journal} {\bibinfo  {journal} {Physical Review Letters}\ }\textbf {\bibinfo
  {volume} {56}}~(\bibinfo {number} {26}),\ \bibinfo {pages}
  {2797--2799}}\BibitemShut {NoStop}%
\bibitem [{\citenamefont {Nation}\ \emph {et~al.}(2021)\citenamefont {Nation},
  \citenamefont {Kang}, \citenamefont {Sundaresan},\ and\ \citenamefont
  {Gambetta}}]{nationScalableMitigationMeasurement2021}%
  \BibitemOpen
  \bibfield  {author} {\bibinfo {author} {\bibnamefont {Nation}, \bibfnamefont
  {Paul~D}}, \bibinfo {author} {\bibfnamefont {Hwajung}\ \bibnamefont {Kang}},
  \bibinfo {author} {\bibfnamefont {Neereja}\ \bibnamefont {Sundaresan}}, and\
  \bibinfo {author} {\bibfnamefont {Jay~M.}\ \bibnamefont {Gambetta}}}
  (\bibinfo {year} {2021}),\ \bibfield  {title} {\enquote {\bibinfo {title}
  {Scalable {{Mitigation}} of {{Measurement Errors}} on {{Quantum
  Computers}}},}\ }\href {https://doi.org/10.1103/PRXQuantum.2.040326}
  {\bibfield  {journal} {\bibinfo  {journal} {PRX Quantum}\ }\textbf {\bibinfo
  {volume} {2}}~(\bibinfo {number} {4}),\ \bibinfo {pages}
  {040326}}\BibitemShut {NoStop}%
\bibitem [{\citenamefont {Neill}\ \emph {et~al.}(2021)\citenamefont {Neill},
  \citenamefont {McCourt}, \citenamefont {Mi}, \citenamefont {Jiang},
  \citenamefont {Niu}, \citenamefont {Mruczkiewicz}, \citenamefont {Aleiner},
  \citenamefont {Arute}, \citenamefont {Arya}, \citenamefont {Atalaya},
  \citenamefont {Babbush}, \citenamefont {Bardin}, \citenamefont {Barends},
  \citenamefont {Bengtsson}, \citenamefont {Bourassa}, \citenamefont
  {Broughton}, \citenamefont {Buckley}, \citenamefont {Buell}, \citenamefont
  {Burkett}, \citenamefont {Bushnell}, \citenamefont {Campero}, \citenamefont
  {Chen}, \citenamefont {Chiaro}, \citenamefont {Collins}, \citenamefont
  {Courtney}, \citenamefont {Demura}, \citenamefont {Derk}, \citenamefont
  {Dunsworth}, \citenamefont {Eppens}, \citenamefont {Erickson}, \citenamefont
  {Farhi}, \citenamefont {Fowler}, \citenamefont {Foxen}, \citenamefont
  {Gidney}, \citenamefont {Giustina}, \citenamefont {Gross}, \citenamefont
  {Harrigan}, \citenamefont {Harrington}, \citenamefont {Hilton}, \citenamefont
  {Ho}, \citenamefont {Hong}, \citenamefont {Huang}, \citenamefont {Huggins},
  \citenamefont {Isakov}, \citenamefont {{Jacob-Mitos}}, \citenamefont
  {Jeffrey}, \citenamefont {Jones}, \citenamefont {Kafri}, \citenamefont
  {Kechedzhi}, \citenamefont {Kelly}, \citenamefont {Kim}, \citenamefont
  {Klimov}, \citenamefont {Korotkov}, \citenamefont {Kostritsa}, \citenamefont
  {Landhuis}, \citenamefont {Laptev}, \citenamefont {Lucero}, \citenamefont
  {Martin}, \citenamefont {McClean}, \citenamefont {McEwen}, \citenamefont
  {Megrant}, \citenamefont {Miao}, \citenamefont {Mohseni}, \citenamefont
  {Mutus}, \citenamefont {Naaman}, \citenamefont {Neeley}, \citenamefont
  {Newman}, \citenamefont {O'Brien}, \citenamefont {Opremcak}, \citenamefont
  {Ostby}, \citenamefont {Pat{\'o}}, \citenamefont {Petukhov}, \citenamefont
  {Quintana}, \citenamefont {Redd}, \citenamefont {Rubin}, \citenamefont
  {Sank}, \citenamefont {Satzinger}, \citenamefont {Shvarts}, \citenamefont
  {Strain}, \citenamefont {Szalay}, \citenamefont {Trevithick}, \citenamefont
  {Villalonga}, \citenamefont {White}, \citenamefont {Yao}, \citenamefont
  {Yeh}, \citenamefont {Zalcman}, \citenamefont {Neven}, \citenamefont {Boixo},
  \citenamefont {Ioffe}, \citenamefont {Roushan}, \citenamefont {Chen},\ and\
  \citenamefont {Smelyanskiy}}]{neillAccuratelyComputingElectronic2021}%
  \BibitemOpen
  \bibfield  {author} {\bibinfo {author} {\bibnamefont {Neill}, \bibfnamefont
  {C}}, \bibinfo {author} {\bibfnamefont {T.}~\bibnamefont {McCourt}}, \bibinfo
  {author} {\bibfnamefont {X.}~\bibnamefont {Mi}}, \bibinfo {author}
  {\bibfnamefont {Z.}~\bibnamefont {Jiang}}, \bibinfo {author} {\bibfnamefont
  {M.~Y.}\ \bibnamefont {Niu}}, \bibinfo {author} {\bibfnamefont
  {W.}~\bibnamefont {Mruczkiewicz}}, \bibinfo {author} {\bibfnamefont
  {I.}~\bibnamefont {Aleiner}}, \bibinfo {author} {\bibfnamefont
  {F.}~\bibnamefont {Arute}}, \bibinfo {author} {\bibfnamefont
  {K.}~\bibnamefont {Arya}}, \bibinfo {author} {\bibfnamefont {J.}~\bibnamefont
  {Atalaya}}, \bibinfo {author} {\bibfnamefont {R.}~\bibnamefont {Babbush}},
  \bibinfo {author} {\bibfnamefont {J.~C.}\ \bibnamefont {Bardin}}, \bibinfo
  {author} {\bibfnamefont {R.}~\bibnamefont {Barends}}, \bibinfo {author}
  {\bibfnamefont {A.}~\bibnamefont {Bengtsson}}, \bibinfo {author}
  {\bibfnamefont {A.}~\bibnamefont {Bourassa}}, \bibinfo {author}
  {\bibfnamefont {M.}~\bibnamefont {Broughton}}, \bibinfo {author}
  {\bibfnamefont {B.~B.}\ \bibnamefont {Buckley}}, \bibinfo {author}
  {\bibfnamefont {D.~A.}\ \bibnamefont {Buell}}, \bibinfo {author}
  {\bibfnamefont {B.}~\bibnamefont {Burkett}}, \bibinfo {author} {\bibfnamefont
  {N.}~\bibnamefont {Bushnell}}, \bibinfo {author} {\bibfnamefont
  {J.}~\bibnamefont {Campero}}, \bibinfo {author} {\bibfnamefont
  {Z.}~\bibnamefont {Chen}}, \bibinfo {author} {\bibfnamefont {B.}~\bibnamefont
  {Chiaro}}, \bibinfo {author} {\bibfnamefont {R.}~\bibnamefont {Collins}},
  \bibinfo {author} {\bibfnamefont {W.}~\bibnamefont {Courtney}}, \bibinfo
  {author} {\bibfnamefont {S.}~\bibnamefont {Demura}}, \bibinfo {author}
  {\bibfnamefont {A.~R.}\ \bibnamefont {Derk}}, \bibinfo {author}
  {\bibfnamefont {A.}~\bibnamefont {Dunsworth}}, \bibinfo {author}
  {\bibfnamefont {D.}~\bibnamefont {Eppens}}, \bibinfo {author} {\bibfnamefont
  {C.}~\bibnamefont {Erickson}}, \bibinfo {author} {\bibfnamefont
  {E.}~\bibnamefont {Farhi}}, \bibinfo {author} {\bibfnamefont {A.~G.}\
  \bibnamefont {Fowler}}, \bibinfo {author} {\bibfnamefont {B.}~\bibnamefont
  {Foxen}}, \bibinfo {author} {\bibfnamefont {C.}~\bibnamefont {Gidney}},
  \bibinfo {author} {\bibfnamefont {M.}~\bibnamefont {Giustina}}, \bibinfo
  {author} {\bibfnamefont {J.~A.}\ \bibnamefont {Gross}}, \bibinfo {author}
  {\bibfnamefont {M.~P.}\ \bibnamefont {Harrigan}}, \bibinfo {author}
  {\bibfnamefont {S.~D.}\ \bibnamefont {Harrington}}, \bibinfo {author}
  {\bibfnamefont {J.}~\bibnamefont {Hilton}}, \bibinfo {author} {\bibfnamefont
  {A.}~\bibnamefont {Ho}}, \bibinfo {author} {\bibfnamefont {S.}~\bibnamefont
  {Hong}}, \bibinfo {author} {\bibfnamefont {T.}~\bibnamefont {Huang}},
  \bibinfo {author} {\bibfnamefont {W.~J.}\ \bibnamefont {Huggins}}, \bibinfo
  {author} {\bibfnamefont {S.~V.}\ \bibnamefont {Isakov}}, \bibinfo {author}
  {\bibfnamefont {M.}~\bibnamefont {{Jacob-Mitos}}}, \bibinfo {author}
  {\bibfnamefont {E.}~\bibnamefont {Jeffrey}}, \bibinfo {author} {\bibfnamefont
  {C.}~\bibnamefont {Jones}}, \bibinfo {author} {\bibfnamefont
  {D.}~\bibnamefont {Kafri}}, \bibinfo {author} {\bibfnamefont
  {K.}~\bibnamefont {Kechedzhi}}, \bibinfo {author} {\bibfnamefont
  {J.}~\bibnamefont {Kelly}}, \bibinfo {author} {\bibfnamefont
  {S.}~\bibnamefont {Kim}}, \bibinfo {author} {\bibfnamefont {P.~V.}\
  \bibnamefont {Klimov}}, \bibinfo {author} {\bibfnamefont {A.~N.}\
  \bibnamefont {Korotkov}}, \bibinfo {author} {\bibfnamefont {F.}~\bibnamefont
  {Kostritsa}}, \bibinfo {author} {\bibfnamefont {D.}~\bibnamefont {Landhuis}},
  \bibinfo {author} {\bibfnamefont {P.}~\bibnamefont {Laptev}}, \bibinfo
  {author} {\bibfnamefont {E.}~\bibnamefont {Lucero}}, \bibinfo {author}
  {\bibfnamefont {O.}~\bibnamefont {Martin}}, \bibinfo {author} {\bibfnamefont
  {J.~R.}\ \bibnamefont {McClean}}, \bibinfo {author} {\bibfnamefont
  {M.}~\bibnamefont {McEwen}}, \bibinfo {author} {\bibfnamefont
  {A.}~\bibnamefont {Megrant}}, \bibinfo {author} {\bibfnamefont {K.~C.}\
  \bibnamefont {Miao}}, \bibinfo {author} {\bibfnamefont {M.}~\bibnamefont
  {Mohseni}}, \bibinfo {author} {\bibfnamefont {J.}~\bibnamefont {Mutus}},
  \bibinfo {author} {\bibfnamefont {O.}~\bibnamefont {Naaman}}, \bibinfo
  {author} {\bibfnamefont {M.}~\bibnamefont {Neeley}}, \bibinfo {author}
  {\bibfnamefont {M.}~\bibnamefont {Newman}}, \bibinfo {author} {\bibfnamefont
  {T.~E.}\ \bibnamefont {O'Brien}}, \bibinfo {author} {\bibfnamefont
  {A.}~\bibnamefont {Opremcak}}, \bibinfo {author} {\bibfnamefont
  {E.}~\bibnamefont {Ostby}}, \bibinfo {author} {\bibfnamefont
  {B.}~\bibnamefont {Pat{\'o}}}, \bibinfo {author} {\bibfnamefont
  {A.}~\bibnamefont {Petukhov}}, \bibinfo {author} {\bibfnamefont
  {C.}~\bibnamefont {Quintana}}, \bibinfo {author} {\bibfnamefont
  {N.}~\bibnamefont {Redd}}, \bibinfo {author} {\bibfnamefont {N.~C.}\
  \bibnamefont {Rubin}}, \bibinfo {author} {\bibfnamefont {D.}~\bibnamefont
  {Sank}}, \bibinfo {author} {\bibfnamefont {K.~J.}\ \bibnamefont {Satzinger}},
  \bibinfo {author} {\bibfnamefont {V.}~\bibnamefont {Shvarts}}, \bibinfo
  {author} {\bibfnamefont {D.}~\bibnamefont {Strain}}, \bibinfo {author}
  {\bibfnamefont {M.}~\bibnamefont {Szalay}}, \bibinfo {author} {\bibfnamefont
  {M.~D.}\ \bibnamefont {Trevithick}}, \bibinfo {author} {\bibfnamefont
  {B.}~\bibnamefont {Villalonga}}, \bibinfo {author} {\bibfnamefont {T.~C.}\
  \bibnamefont {White}}, \bibinfo {author} {\bibfnamefont {Z.}~\bibnamefont
  {Yao}}, \bibinfo {author} {\bibfnamefont {P.}~\bibnamefont {Yeh}}, \bibinfo
  {author} {\bibfnamefont {A.}~\bibnamefont {Zalcman}}, \bibinfo {author}
  {\bibfnamefont {H.}~\bibnamefont {Neven}}, \bibinfo {author} {\bibfnamefont
  {S.}~\bibnamefont {Boixo}}, \bibinfo {author} {\bibfnamefont {L.~B.}\
  \bibnamefont {Ioffe}}, \bibinfo {author} {\bibfnamefont {P.}~\bibnamefont
  {Roushan}}, \bibinfo {author} {\bibfnamefont {Y.}~\bibnamefont {Chen}}, and\
  \bibinfo {author} {\bibfnamefont {V.}~\bibnamefont {Smelyanskiy}}} (\bibinfo
  {year} {2021}),\ \bibfield  {title} {\enquote {\bibinfo {title} {Accurately
  computing the electronic properties of a quantum ring},}\ }\href
  {https://doi.org/10.1038/s41586-021-03576-2} {\bibfield  {journal} {\bibinfo
  {journal} {Nature}\ }\textbf {\bibinfo {volume} {594}}~(\bibinfo {number}
  {7864}),\ \bibinfo {pages} {508--512}}\BibitemShut {NoStop}%
\bibitem [{\citenamefont {Nielsen}\ and\ \citenamefont
  {Chuang}(2010)}]{nielsenQuantumComputationQuantum2010}%
  \BibitemOpen
  \bibfield  {author} {\bibinfo {author} {\bibnamefont {Nielsen}, \bibfnamefont
  {Michael~A}}, and\ \bibinfo {author} {\bibfnamefont {Isaac~L.}\ \bibnamefont
  {Chuang}}} (\bibinfo {year} {2010}),\ \href
  {https://doi.org/10.1017/CBO9780511976667} {\emph {\bibinfo {title} {Quantum
  {{Computation}} and {{Quantum Information}}: 10th {{Anniversary Edition}}}}}\
  (\bibinfo  {publisher} {{Cambridge University Press}},\ \bibinfo {address}
  {{Cambridge}})\BibitemShut {NoStop}%
\bibitem [{\citenamefont {Nigg}\ \emph {et~al.}(2014)\citenamefont {Nigg},
  \citenamefont {M{\"u}ller}, \citenamefont {Martinez}, \citenamefont
  {Schindler}, \citenamefont {Hennrich}, \citenamefont {Monz}, \citenamefont
  {{Martin-Delgado}},\ and\ \citenamefont
  {Blatt}}]{niggQuantumComputationsTopologically2014}%
  \BibitemOpen
  \bibfield  {author} {\bibinfo {author} {\bibnamefont {Nigg}, \bibfnamefont
  {D}}, \bibinfo {author} {\bibfnamefont {M.}~\bibnamefont {M{\"u}ller}},
  \bibinfo {author} {\bibfnamefont {E.~A.}\ \bibnamefont {Martinez}}, \bibinfo
  {author} {\bibfnamefont {P.}~\bibnamefont {Schindler}}, \bibinfo {author}
  {\bibfnamefont {M.}~\bibnamefont {Hennrich}}, \bibinfo {author}
  {\bibfnamefont {T.}~\bibnamefont {Monz}}, \bibinfo {author} {\bibfnamefont
  {M.~A.}\ \bibnamefont {{Martin-Delgado}}}, and\ \bibinfo {author}
  {\bibfnamefont {R.}~\bibnamefont {Blatt}}} (\bibinfo {year} {2014}),\
  \bibfield  {title} {\enquote {\bibinfo {title} {Quantum computations on a
  topologically encoded qubit},}\ }\href
  {https://doi.org/10.1126/science.1253742} {\bibfield  {journal} {\bibinfo
  {journal} {Science}\ }\textbf {\bibinfo {volume} {345}}~(\bibinfo {number}
  {6194}),\ \bibinfo {pages} {302--305}}\BibitemShut {NoStop}%
\bibitem [{\citenamefont {O'Brien}\ \emph {et~al.}(2023)\citenamefont
  {O'Brien}, \citenamefont {Anselmetti}, \citenamefont {Gkritsis},
  \citenamefont {Elfving}, \citenamefont {Polla}, \citenamefont {Huggins},
  \citenamefont {Oumarou}, \citenamefont {Kechedzhi}, \citenamefont {Abanin},
  \citenamefont {Acharya}, \citenamefont {Aleiner}, \citenamefont {Allen},
  \citenamefont {Andersen}, \citenamefont {Anderson}, \citenamefont {Ansmann},
  \citenamefont {Arute}, \citenamefont {Arya}, \citenamefont {Asfaw},
  \citenamefont {Atalaya}, \citenamefont {Bardin}, \citenamefont {Bengtsson},
  \citenamefont {Bortoli}, \citenamefont {Bourassa}, \citenamefont {Bovaird},
  \citenamefont {Brill}, \citenamefont {Broughton}, \citenamefont {Buckley},
  \citenamefont {Buell}, \citenamefont {Burger}, \citenamefont {Burkett},
  \citenamefont {Bushnell}, \citenamefont {Campero}, \citenamefont {Chen},
  \citenamefont {Chiaro}, \citenamefont {Chik}, \citenamefont {Cogan},
  \citenamefont {Collins}, \citenamefont {Conner}, \citenamefont {Courtney},
  \citenamefont {Crook}, \citenamefont {Curtin}, \citenamefont {Debroy},
  \citenamefont {Demura}, \citenamefont {Drozdov}, \citenamefont {Dunsworth},
  \citenamefont {Erickson}, \citenamefont {Faoro}, \citenamefont {Farhi},
  \citenamefont {Fatemi}, \citenamefont {Ferreira}, \citenamefont
  {Flores~Burgos}, \citenamefont {Forati}, \citenamefont {Fowler},
  \citenamefont {Foxen}, \citenamefont {Giang}, \citenamefont {Gidney},
  \citenamefont {Gilboa}, \citenamefont {Giustina}, \citenamefont {Gosula},
  \citenamefont {Grajales~Dau}, \citenamefont {Gross}, \citenamefont
  {Habegger}, \citenamefont {Hamilton}, \citenamefont {Hansen}, \citenamefont
  {Harrigan}, \citenamefont {Harrington}, \citenamefont {Heu}, \citenamefont
  {Hoffmann}, \citenamefont {Hong}, \citenamefont {Huang}, \citenamefont
  {Huff}, \citenamefont {Ioffe}, \citenamefont {Isakov}, \citenamefont
  {Iveland}, \citenamefont {Jeffrey}, \citenamefont {Jiang}, \citenamefont
  {Jones}, \citenamefont {Juhas}, \citenamefont {Kafri}, \citenamefont
  {Khattar}, \citenamefont {Khezri}, \citenamefont {Kieferov{\'a}},
  \citenamefont {Kim}, \citenamefont {Klimov}, \citenamefont {Klots},
  \citenamefont {Korotkov}, \citenamefont {Kostritsa}, \citenamefont
  {Kreikebaum}, \citenamefont {Landhuis}, \citenamefont {Laptev}, \citenamefont
  {Lau}, \citenamefont {Laws}, \citenamefont {Lee}, \citenamefont {Lee},
  \citenamefont {Lester}, \citenamefont {Lill}, \citenamefont {Liu},
  \citenamefont {Livingston}, \citenamefont {Locharla}, \citenamefont {Malone},
  \citenamefont {Mandr{\`a}}, \citenamefont {Martin}, \citenamefont {Martin},
  \citenamefont {McClean}, \citenamefont {McCourt}, \citenamefont {McEwen},
  \citenamefont {Mi}, \citenamefont {Mieszala}, \citenamefont {Miao},
  \citenamefont {Mohseni}, \citenamefont {Montazeri}, \citenamefont {Morvan},
  \citenamefont {Movassagh}, \citenamefont {Mruczkiewicz}, \citenamefont
  {Naaman}, \citenamefont {Neeley}, \citenamefont {Neill}, \citenamefont
  {Nersisyan}, \citenamefont {Newman}, \citenamefont {Ng}, \citenamefont
  {Nguyen}, \citenamefont {Nguyen}, \citenamefont {Niu}, \citenamefont
  {Omonije}, \citenamefont {Opremcak}, \citenamefont {Petukhov}, \citenamefont
  {Potter}, \citenamefont {Pryadko}, \citenamefont {Quintana}, \citenamefont
  {Rocque}, \citenamefont {Roushan}, \citenamefont {Saei}, \citenamefont
  {Sank}, \citenamefont {Sankaragomathi}, \citenamefont {Satzinger},
  \citenamefont {Schurkus}, \citenamefont {Schuster}, \citenamefont {Shearn},
  \citenamefont {Shorter}, \citenamefont {Shutty}, \citenamefont {Shvarts},
  \citenamefont {Skruzny}, \citenamefont {Smith}, \citenamefont {Somma},
  \citenamefont {Sterling}, \citenamefont {Strain}, \citenamefont {Szalay},
  \citenamefont {Thor}, \citenamefont {Torres}, \citenamefont {Vidal},
  \citenamefont {Villalonga}, \citenamefont {Vollgraff~Heidweiller},
  \citenamefont {White}, \citenamefont {Woo}, \citenamefont {Xing},
  \citenamefont {Yao}, \citenamefont {Yeh}, \citenamefont {Yoo}, \citenamefont
  {Young}, \citenamefont {Zalcman}, \citenamefont {Zhang}, \citenamefont {Zhu},
  \citenamefont {Zobrist}, \citenamefont {Bacon}, \citenamefont {Boixo},
  \citenamefont {Chen}, \citenamefont {Hilton}, \citenamefont {Kelly},
  \citenamefont {Lucero}, \citenamefont {Megrant}, \citenamefont {Neven},
  \citenamefont {Smelyanskiy}, \citenamefont {Gogolin}, \citenamefont
  {Babbush},\ and\ \citenamefont
  {Rubin}}]{obrienPurificationbasedQuantumError2023}%
  \BibitemOpen
  \bibfield  {author} {\bibinfo {author} {\bibnamefont {O'Brien}, \bibfnamefont
  {T~E}}, \bibinfo {author} {\bibfnamefont {G.}~\bibnamefont {Anselmetti}},
  \bibinfo {author} {\bibfnamefont {F.}~\bibnamefont {Gkritsis}}, \bibinfo
  {author} {\bibfnamefont {V.~E.}\ \bibnamefont {Elfving}}, \bibinfo {author}
  {\bibfnamefont {S.}~\bibnamefont {Polla}}, \bibinfo {author} {\bibfnamefont
  {W.~J.}\ \bibnamefont {Huggins}}, \bibinfo {author} {\bibfnamefont
  {O.}~\bibnamefont {Oumarou}}, \bibinfo {author} {\bibfnamefont
  {K.}~\bibnamefont {Kechedzhi}}, \bibinfo {author} {\bibfnamefont
  {D.}~\bibnamefont {Abanin}}, \bibinfo {author} {\bibfnamefont
  {R.}~\bibnamefont {Acharya}}, \bibinfo {author} {\bibfnamefont
  {I.}~\bibnamefont {Aleiner}}, \bibinfo {author} {\bibfnamefont
  {R.}~\bibnamefont {Allen}}, \bibinfo {author} {\bibfnamefont {T.~I.}\
  \bibnamefont {Andersen}}, \bibinfo {author} {\bibfnamefont {K.}~\bibnamefont
  {Anderson}}, \bibinfo {author} {\bibfnamefont {M.}~\bibnamefont {Ansmann}},
  \bibinfo {author} {\bibfnamefont {F.}~\bibnamefont {Arute}}, \bibinfo
  {author} {\bibfnamefont {K.}~\bibnamefont {Arya}}, \bibinfo {author}
  {\bibfnamefont {A.}~\bibnamefont {Asfaw}}, \bibinfo {author} {\bibfnamefont
  {J.}~\bibnamefont {Atalaya}}, \bibinfo {author} {\bibfnamefont {J.~C.}\
  \bibnamefont {Bardin}}, \bibinfo {author} {\bibfnamefont {A.}~\bibnamefont
  {Bengtsson}}, \bibinfo {author} {\bibfnamefont {G.}~\bibnamefont {Bortoli}},
  \bibinfo {author} {\bibfnamefont {A.}~\bibnamefont {Bourassa}}, \bibinfo
  {author} {\bibfnamefont {J.}~\bibnamefont {Bovaird}}, \bibinfo {author}
  {\bibfnamefont {L.}~\bibnamefont {Brill}}, \bibinfo {author} {\bibfnamefont
  {M.}~\bibnamefont {Broughton}}, \bibinfo {author} {\bibfnamefont
  {B.}~\bibnamefont {Buckley}}, \bibinfo {author} {\bibfnamefont {D.~A.}\
  \bibnamefont {Buell}}, \bibinfo {author} {\bibfnamefont {T.}~\bibnamefont
  {Burger}}, \bibinfo {author} {\bibfnamefont {B.}~\bibnamefont {Burkett}},
  \bibinfo {author} {\bibfnamefont {N.}~\bibnamefont {Bushnell}}, \bibinfo
  {author} {\bibfnamefont {J.}~\bibnamefont {Campero}}, \bibinfo {author}
  {\bibfnamefont {Z.}~\bibnamefont {Chen}}, \bibinfo {author} {\bibfnamefont
  {B.}~\bibnamefont {Chiaro}}, \bibinfo {author} {\bibfnamefont
  {D.}~\bibnamefont {Chik}}, \bibinfo {author} {\bibfnamefont {J.}~\bibnamefont
  {Cogan}}, \bibinfo {author} {\bibfnamefont {R.}~\bibnamefont {Collins}},
  \bibinfo {author} {\bibfnamefont {P.}~\bibnamefont {Conner}}, \bibinfo
  {author} {\bibfnamefont {W.}~\bibnamefont {Courtney}}, \bibinfo {author}
  {\bibfnamefont {A.~L.}\ \bibnamefont {Crook}}, \bibinfo {author}
  {\bibfnamefont {B.}~\bibnamefont {Curtin}}, \bibinfo {author} {\bibfnamefont
  {D.~M.}\ \bibnamefont {Debroy}}, \bibinfo {author} {\bibfnamefont
  {S.}~\bibnamefont {Demura}}, \bibinfo {author} {\bibfnamefont
  {I.}~\bibnamefont {Drozdov}}, \bibinfo {author} {\bibfnamefont
  {A.}~\bibnamefont {Dunsworth}}, \bibinfo {author} {\bibfnamefont
  {C.}~\bibnamefont {Erickson}}, \bibinfo {author} {\bibfnamefont
  {L.}~\bibnamefont {Faoro}}, \bibinfo {author} {\bibfnamefont
  {E.}~\bibnamefont {Farhi}}, \bibinfo {author} {\bibfnamefont
  {R.}~\bibnamefont {Fatemi}}, \bibinfo {author} {\bibfnamefont {V.~S.}\
  \bibnamefont {Ferreira}}, \bibinfo {author} {\bibfnamefont {L.}~\bibnamefont
  {Flores~Burgos}}, \bibinfo {author} {\bibfnamefont {E.}~\bibnamefont
  {Forati}}, \bibinfo {author} {\bibfnamefont {A.~G.}\ \bibnamefont {Fowler}},
  \bibinfo {author} {\bibfnamefont {B.}~\bibnamefont {Foxen}}, \bibinfo
  {author} {\bibfnamefont {W.}~\bibnamefont {Giang}}, \bibinfo {author}
  {\bibfnamefont {C.}~\bibnamefont {Gidney}}, \bibinfo {author} {\bibfnamefont
  {D.}~\bibnamefont {Gilboa}}, \bibinfo {author} {\bibfnamefont
  {M.}~\bibnamefont {Giustina}}, \bibinfo {author} {\bibfnamefont
  {R.}~\bibnamefont {Gosula}}, \bibinfo {author} {\bibfnamefont
  {A.}~\bibnamefont {Grajales~Dau}}, \bibinfo {author} {\bibfnamefont {J.~A.}\
  \bibnamefont {Gross}}, \bibinfo {author} {\bibfnamefont {S.}~\bibnamefont
  {Habegger}}, \bibinfo {author} {\bibfnamefont {M.~C.}\ \bibnamefont
  {Hamilton}}, \bibinfo {author} {\bibfnamefont {M.}~\bibnamefont {Hansen}},
  \bibinfo {author} {\bibfnamefont {M.~P.}\ \bibnamefont {Harrigan}}, \bibinfo
  {author} {\bibfnamefont {S.~D.}\ \bibnamefont {Harrington}}, \bibinfo
  {author} {\bibfnamefont {P.}~\bibnamefont {Heu}}, \bibinfo {author}
  {\bibfnamefont {M.~R.}\ \bibnamefont {Hoffmann}}, \bibinfo {author}
  {\bibfnamefont {S.}~\bibnamefont {Hong}}, \bibinfo {author} {\bibfnamefont
  {T.}~\bibnamefont {Huang}}, \bibinfo {author} {\bibfnamefont
  {A.}~\bibnamefont {Huff}}, \bibinfo {author} {\bibfnamefont {L.~B.}\
  \bibnamefont {Ioffe}}, \bibinfo {author} {\bibfnamefont {S.~V.}\ \bibnamefont
  {Isakov}}, \bibinfo {author} {\bibfnamefont {J.}~\bibnamefont {Iveland}},
  \bibinfo {author} {\bibfnamefont {E.}~\bibnamefont {Jeffrey}}, \bibinfo
  {author} {\bibfnamefont {Z.}~\bibnamefont {Jiang}}, \bibinfo {author}
  {\bibfnamefont {C.}~\bibnamefont {Jones}}, \bibinfo {author} {\bibfnamefont
  {P.}~\bibnamefont {Juhas}}, \bibinfo {author} {\bibfnamefont
  {D.}~\bibnamefont {Kafri}}, \bibinfo {author} {\bibfnamefont
  {T.}~\bibnamefont {Khattar}}, \bibinfo {author} {\bibfnamefont
  {M.}~\bibnamefont {Khezri}}, \bibinfo {author} {\bibfnamefont
  {M.}~\bibnamefont {Kieferov{\'a}}}, \bibinfo {author} {\bibfnamefont
  {S.}~\bibnamefont {Kim}}, \bibinfo {author} {\bibfnamefont {P.~V.}\
  \bibnamefont {Klimov}}, \bibinfo {author} {\bibfnamefont {A.~R.}\
  \bibnamefont {Klots}}, \bibinfo {author} {\bibfnamefont {A.~N.}\ \bibnamefont
  {Korotkov}}, \bibinfo {author} {\bibfnamefont {F.}~\bibnamefont {Kostritsa}},
  \bibinfo {author} {\bibfnamefont {J.~M.}\ \bibnamefont {Kreikebaum}},
  \bibinfo {author} {\bibfnamefont {D.}~\bibnamefont {Landhuis}}, \bibinfo
  {author} {\bibfnamefont {P.}~\bibnamefont {Laptev}}, \bibinfo {author}
  {\bibfnamefont {K.-M.}\ \bibnamefont {Lau}}, \bibinfo {author} {\bibfnamefont
  {L.}~\bibnamefont {Laws}}, \bibinfo {author} {\bibfnamefont {J.}~\bibnamefont
  {Lee}}, \bibinfo {author} {\bibfnamefont {K.}~\bibnamefont {Lee}}, \bibinfo
  {author} {\bibfnamefont {B.~J.}\ \bibnamefont {Lester}}, \bibinfo {author}
  {\bibfnamefont {A.~T.}\ \bibnamefont {Lill}}, \bibinfo {author}
  {\bibfnamefont {W.}~\bibnamefont {Liu}}, \bibinfo {author} {\bibfnamefont
  {W.~P.}\ \bibnamefont {Livingston}}, \bibinfo {author} {\bibfnamefont
  {A.}~\bibnamefont {Locharla}}, \bibinfo {author} {\bibfnamefont {F.~D.}\
  \bibnamefont {Malone}}, \bibinfo {author} {\bibfnamefont {S.}~\bibnamefont
  {Mandr{\`a}}}, \bibinfo {author} {\bibfnamefont {O.}~\bibnamefont {Martin}},
  \bibinfo {author} {\bibfnamefont {S.}~\bibnamefont {Martin}}, \bibinfo
  {author} {\bibfnamefont {J.~R.}\ \bibnamefont {McClean}}, \bibinfo {author}
  {\bibfnamefont {T.}~\bibnamefont {McCourt}}, \bibinfo {author} {\bibfnamefont
  {M.}~\bibnamefont {McEwen}}, \bibinfo {author} {\bibfnamefont
  {X.}~\bibnamefont {Mi}}, \bibinfo {author} {\bibfnamefont {A.}~\bibnamefont
  {Mieszala}}, \bibinfo {author} {\bibfnamefont {K.~C.}\ \bibnamefont {Miao}},
  \bibinfo {author} {\bibfnamefont {M.}~\bibnamefont {Mohseni}}, \bibinfo
  {author} {\bibfnamefont {S.}~\bibnamefont {Montazeri}}, \bibinfo {author}
  {\bibfnamefont {A.}~\bibnamefont {Morvan}}, \bibinfo {author} {\bibfnamefont
  {R.}~\bibnamefont {Movassagh}}, \bibinfo {author} {\bibfnamefont
  {W.}~\bibnamefont {Mruczkiewicz}}, \bibinfo {author} {\bibfnamefont
  {O.}~\bibnamefont {Naaman}}, \bibinfo {author} {\bibfnamefont
  {M.}~\bibnamefont {Neeley}}, \bibinfo {author} {\bibfnamefont
  {C.}~\bibnamefont {Neill}}, \bibinfo {author} {\bibfnamefont
  {A.}~\bibnamefont {Nersisyan}}, \bibinfo {author} {\bibfnamefont
  {M.}~\bibnamefont {Newman}}, \bibinfo {author} {\bibfnamefont {J.~H.}\
  \bibnamefont {Ng}}, \bibinfo {author} {\bibfnamefont {A.}~\bibnamefont
  {Nguyen}}, \bibinfo {author} {\bibfnamefont {M.}~\bibnamefont {Nguyen}},
  \bibinfo {author} {\bibfnamefont {M.~Y.}\ \bibnamefont {Niu}}, \bibinfo
  {author} {\bibfnamefont {S.}~\bibnamefont {Omonije}}, \bibinfo {author}
  {\bibfnamefont {A.}~\bibnamefont {Opremcak}}, \bibinfo {author}
  {\bibfnamefont {A.}~\bibnamefont {Petukhov}}, \bibinfo {author}
  {\bibfnamefont {R.}~\bibnamefont {Potter}}, \bibinfo {author} {\bibfnamefont
  {L.~P.}\ \bibnamefont {Pryadko}}, \bibinfo {author} {\bibfnamefont
  {C.}~\bibnamefont {Quintana}}, \bibinfo {author} {\bibfnamefont
  {C.}~\bibnamefont {Rocque}}, \bibinfo {author} {\bibfnamefont
  {P.}~\bibnamefont {Roushan}}, \bibinfo {author} {\bibfnamefont
  {N.}~\bibnamefont {Saei}}, \bibinfo {author} {\bibfnamefont {D.}~\bibnamefont
  {Sank}}, \bibinfo {author} {\bibfnamefont {K.}~\bibnamefont
  {Sankaragomathi}}, \bibinfo {author} {\bibfnamefont {K.~J.}\ \bibnamefont
  {Satzinger}}, \bibinfo {author} {\bibfnamefont {H.~F.}\ \bibnamefont
  {Schurkus}}, \bibinfo {author} {\bibfnamefont {C.}~\bibnamefont {Schuster}},
  \bibinfo {author} {\bibfnamefont {M.~J.}\ \bibnamefont {Shearn}}, \bibinfo
  {author} {\bibfnamefont {A.}~\bibnamefont {Shorter}}, \bibinfo {author}
  {\bibfnamefont {N.}~\bibnamefont {Shutty}}, \bibinfo {author} {\bibfnamefont
  {V.}~\bibnamefont {Shvarts}}, \bibinfo {author} {\bibfnamefont
  {J.}~\bibnamefont {Skruzny}}, \bibinfo {author} {\bibfnamefont {W.~C.}\
  \bibnamefont {Smith}}, \bibinfo {author} {\bibfnamefont {R.~D.}\ \bibnamefont
  {Somma}}, \bibinfo {author} {\bibfnamefont {G.}~\bibnamefont {Sterling}},
  \bibinfo {author} {\bibfnamefont {D.}~\bibnamefont {Strain}}, \bibinfo
  {author} {\bibfnamefont {M.}~\bibnamefont {Szalay}}, \bibinfo {author}
  {\bibfnamefont {D.}~\bibnamefont {Thor}}, \bibinfo {author} {\bibfnamefont
  {A.}~\bibnamefont {Torres}}, \bibinfo {author} {\bibfnamefont
  {G.}~\bibnamefont {Vidal}}, \bibinfo {author} {\bibfnamefont
  {B.}~\bibnamefont {Villalonga}}, \bibinfo {author} {\bibfnamefont
  {C.}~\bibnamefont {Vollgraff~Heidweiller}}, \bibinfo {author} {\bibfnamefont
  {T.}~\bibnamefont {White}}, \bibinfo {author} {\bibfnamefont {B.~W.~K.}\
  \bibnamefont {Woo}}, \bibinfo {author} {\bibfnamefont {C.}~\bibnamefont
  {Xing}}, \bibinfo {author} {\bibfnamefont {Z.~J.}\ \bibnamefont {Yao}},
  \bibinfo {author} {\bibfnamefont {P.}~\bibnamefont {Yeh}}, \bibinfo {author}
  {\bibfnamefont {J.}~\bibnamefont {Yoo}}, \bibinfo {author} {\bibfnamefont
  {G.}~\bibnamefont {Young}}, \bibinfo {author} {\bibfnamefont
  {A.}~\bibnamefont {Zalcman}}, \bibinfo {author} {\bibfnamefont
  {Y.}~\bibnamefont {Zhang}}, \bibinfo {author} {\bibfnamefont
  {N.}~\bibnamefont {Zhu}}, \bibinfo {author} {\bibfnamefont {N.}~\bibnamefont
  {Zobrist}}, \bibinfo {author} {\bibfnamefont {D.}~\bibnamefont {Bacon}},
  \bibinfo {author} {\bibfnamefont {S.}~\bibnamefont {Boixo}}, \bibinfo
  {author} {\bibfnamefont {Y.}~\bibnamefont {Chen}}, \bibinfo {author}
  {\bibfnamefont {J.}~\bibnamefont {Hilton}}, \bibinfo {author} {\bibfnamefont
  {J.}~\bibnamefont {Kelly}}, \bibinfo {author} {\bibfnamefont
  {E.}~\bibnamefont {Lucero}}, \bibinfo {author} {\bibfnamefont
  {A.}~\bibnamefont {Megrant}}, \bibinfo {author} {\bibfnamefont
  {H.}~\bibnamefont {Neven}}, \bibinfo {author} {\bibfnamefont
  {V.}~\bibnamefont {Smelyanskiy}}, \bibinfo {author} {\bibfnamefont
  {C.}~\bibnamefont {Gogolin}}, \bibinfo {author} {\bibfnamefont
  {R.}~\bibnamefont {Babbush}}, and\ \bibinfo {author} {\bibfnamefont {N.~C.}\
  \bibnamefont {Rubin}}} (\bibinfo {year} {2023}),\ \bibfield  {title}
  {\enquote {\bibinfo {title} {Purification-based quantum error mitigation of
  pair-correlated electron simulations},}\ }\href
  {https://doi.org/10.1038/s41567-023-02240-y} {\bibinfo  {journal} {Nature
  Physics}\ ,\ \bibinfo {pages} {1--6}}\BibitemShut {NoStop}%
\bibitem [{\citenamefont {O'Brien}\ \emph {et~al.}(2021)\citenamefont
  {O'Brien}, \citenamefont {Polla}, \citenamefont {Rubin}, \citenamefont
  {Huggins}, \citenamefont {McArdle}, \citenamefont {Boixo}, \citenamefont
  {McClean},\ and\ \citenamefont
  {Babbush}}]{obrienErrorMitigationVerified2021}%
  \BibitemOpen
\bibfield  {journal} {  }\bibfield  {author} {\bibinfo {author} {\bibnamefont
  {O'Brien}, \bibfnamefont {Thomas~E}}, \bibinfo {author} {\bibfnamefont
  {Stefano}\ \bibnamefont {Polla}}, \bibinfo {author} {\bibfnamefont
  {Nicholas~C.}\ \bibnamefont {Rubin}}, \bibinfo {author} {\bibfnamefont
  {William~J.}\ \bibnamefont {Huggins}}, \bibinfo {author} {\bibfnamefont
  {Sam}\ \bibnamefont {McArdle}}, \bibinfo {author} {\bibfnamefont {Sergio}\
  \bibnamefont {Boixo}}, \bibinfo {author} {\bibfnamefont {Jarrod~R.}\
  \bibnamefont {McClean}}, and\ \bibinfo {author} {\bibfnamefont {Ryan}\
  \bibnamefont {Babbush}}} (\bibinfo {year} {2021}),\ \bibfield  {title}
  {\enquote {\bibinfo {title} {Error {{Mitigation}} via {{Verified Phase
  Estimation}}},}\ }\href {https://doi.org/10.1103/PRXQuantum.2.020317}
  {\bibfield  {journal} {\bibinfo  {journal} {PRX Quantum}\ }\textbf {\bibinfo
  {volume} {2}}~(\bibinfo {number} {2}),\ \bibinfo {pages}
  {020317}}\BibitemShut {NoStop}%
\bibitem [{\citenamefont {O'Gorman}\ and\ \citenamefont
  {Campbell}(2017)}]{ogormanQuantumComputationRealistic2017}%
  \BibitemOpen
  \bibfield  {author} {\bibinfo {author} {\bibnamefont {O'Gorman},
  \bibfnamefont {Joe}}, and\ \bibinfo {author} {\bibfnamefont {Earl~T.}\
  \bibnamefont {Campbell}}} (\bibinfo {year} {2017}),\ \bibfield  {title}
  {\enquote {\bibinfo {title} {Quantum computation with realistic magic-state
  factories},}\ }\href {https://doi.org/10.1103/PhysRevA.95.032338} {\bibfield
  {journal} {\bibinfo  {journal} {Physical Review A}\ }\textbf {\bibinfo
  {volume} {95}}~(\bibinfo {number} {3}),\ \bibinfo {pages}
  {032338}}\BibitemShut {NoStop}%
\bibitem [{\citenamefont {Otten}\ and\ \citenamefont
  {Gray}(2019)}]{ottenAccountingErrorsQuantum2019}%
  \BibitemOpen
  \bibfield  {author} {\bibinfo {author} {\bibnamefont {Otten}, \bibfnamefont
  {Matthew}}, and\ \bibinfo {author} {\bibfnamefont {Stephen~K.}\ \bibnamefont
  {Gray}}} (\bibinfo {year} {2019}),\ \bibfield  {title} {\enquote {\bibinfo
  {title} {Accounting for errors in quantum algorithms via individual error
  reduction},}\ }\href {https://doi.org/10.1038/s41534-019-0125-3} {\bibfield
  {journal} {\bibinfo  {journal} {npj Quantum Information}\ }\textbf {\bibinfo
  {volume} {5}},\ \bibinfo {pages} {11}}\BibitemShut {NoStop}%
\bibitem [{\citenamefont {Ouyang}\ \emph {et~al.}(2020)\citenamefont {Ouyang},
  \citenamefont {White},\ and\ \citenamefont
  {Campbell}}]{ouyangCompilationStochasticHamiltonian2020}%
  \BibitemOpen
  \bibfield  {author} {\bibinfo {author} {\bibnamefont {Ouyang}, \bibfnamefont
  {Yingkai}}, \bibinfo {author} {\bibfnamefont {David~R.}\ \bibnamefont
  {White}}, and\ \bibinfo {author} {\bibfnamefont {Earl~T.}\ \bibnamefont
  {Campbell}}} (\bibinfo {year} {2020}),\ \bibfield  {title} {\enquote
  {\bibinfo {title} {Compilation by stochastic {{Hamiltonian}}
  sparsification},}\ }\href {https://doi.org/10.22331/q-2020-02-27-235}
  {\bibfield  {journal} {\bibinfo  {journal} {Quantum}\ }\textbf {\bibinfo
  {volume} {4}},\ \bibinfo {pages} {235}}\BibitemShut {NoStop}%
\bibitem [{\citenamefont {Palmieri}\ \emph {et~al.}(2020)\citenamefont
  {Palmieri}, \citenamefont {Kovlakov}, \citenamefont {Bianchi}, \citenamefont
  {Yudin}, \citenamefont {Straupe}, \citenamefont {Biamonte},\ and\
  \citenamefont {Kulik}}]{palmieriExperimentalNeuralNetwork2020}%
  \BibitemOpen
  \bibfield  {author} {\bibinfo {author} {\bibnamefont {Palmieri},
  \bibfnamefont {Adriano~Macarone}}, \bibinfo {author} {\bibfnamefont {Egor}\
  \bibnamefont {Kovlakov}}, \bibinfo {author} {\bibfnamefont {Federico}\
  \bibnamefont {Bianchi}}, \bibinfo {author} {\bibfnamefont {Dmitry}\
  \bibnamefont {Yudin}}, \bibinfo {author} {\bibfnamefont {Stanislav}\
  \bibnamefont {Straupe}}, \bibinfo {author} {\bibfnamefont {Jacob~D.}\
  \bibnamefont {Biamonte}}, and\ \bibinfo {author} {\bibfnamefont {Sergei}\
  \bibnamefont {Kulik}}} (\bibinfo {year} {2020}),\ \bibfield  {title}
  {\enquote {\bibinfo {title} {Experimental neural network enhanced quantum
  tomography},}\ }\href {https://doi.org/10.1038/s41534-020-0248-6} {\bibfield
  {journal} {\bibinfo  {journal} {npj Quantum Information}\ }\textbf {\bibinfo
  {volume} {6}},\ \bibinfo {pages} {20}}\BibitemShut {NoStop}%
\bibitem [{\citenamefont {Panteleev}\ and\ \citenamefont
  {Kalachev}(2022)}]{panteleevAsymptoticallyGoodQuantum2022}%
  \BibitemOpen
  \bibfield  {author} {\bibinfo {author} {\bibnamefont {Panteleev},
  \bibfnamefont {Pavel}}, and\ \bibinfo {author} {\bibfnamefont {Gleb}\
  \bibnamefont {Kalachev}}} (\bibinfo {year} {2022}),\ \bibfield  {title}
  {\enquote {\bibinfo {title} {Asymptotically good {{Quantum}} and locally
  testable classical {{LDPC}} codes},}\ }in\ \href
  {https://doi.org/10.1145/3519935.3520017} {\emph {\bibinfo {booktitle}
  {Proceedings of the 54th {{Annual ACM SIGACT Symposium}} on {{Theory}} of
  {{Computing}}}}},\ \bibinfo {series and number} {{{STOC}} 2022}\ (\bibinfo
  {publisher} {{Association for Computing Machinery}},\ \bibinfo {address}
  {{New York, NY, USA}})\ pp.\ \bibinfo {pages} {375--388}\BibitemShut
  {NoStop}%
\bibitem [{\citenamefont {Peruzzo}\ \emph {et~al.}(2014)\citenamefont
  {Peruzzo}, \citenamefont {McClean}, \citenamefont {Shadbolt}, \citenamefont
  {Yung}, \citenamefont {Zhou}, \citenamefont {Love}, \citenamefont
  {{Aspuru-Guzik}},\ and\ \citenamefont
  {O'Brien}}]{peruzzoVariationalEigenvalueSolver2014}%
  \BibitemOpen
  \bibfield  {author} {\bibinfo {author} {\bibnamefont {Peruzzo}, \bibfnamefont
  {Alberto}}, \bibinfo {author} {\bibfnamefont {Jarrod}\ \bibnamefont
  {McClean}}, \bibinfo {author} {\bibfnamefont {Peter}\ \bibnamefont
  {Shadbolt}}, \bibinfo {author} {\bibfnamefont {Man-Hong}\ \bibnamefont
  {Yung}}, \bibinfo {author} {\bibfnamefont {Xiao-Qi}\ \bibnamefont {Zhou}},
  \bibinfo {author} {\bibfnamefont {Peter~J.}\ \bibnamefont {Love}}, \bibinfo
  {author} {\bibfnamefont {Al{\'a}n}\ \bibnamefont {{Aspuru-Guzik}}}, and\
  \bibinfo {author} {\bibfnamefont {Jeremy~L.}\ \bibnamefont {O'Brien}}}
  (\bibinfo {year} {2014}),\ \bibfield  {title} {\enquote {\bibinfo {title} {A
  variational eigenvalue solver on a photonic quantum processor},}\ }\href
  {https://doi.org/10.1038/ncomms5213} {\bibfield  {journal} {\bibinfo
  {journal} {Nature Communications}\ }\textbf {\bibinfo {volume} {5}},\
  \bibinfo {pages} {4213}}\BibitemShut {NoStop}%
\bibitem [{\citenamefont {Piveteau}\ \emph {et~al.}(2021)\citenamefont
  {Piveteau}, \citenamefont {Sutter}, \citenamefont {Bravyi}, \citenamefont
  {Gambetta},\ and\ \citenamefont
  {Temme}}]{piveteauErrorMitigationUniversal2021}%
  \BibitemOpen
  \bibfield  {author} {\bibinfo {author} {\bibnamefont {Piveteau},
  \bibfnamefont {Christophe}}, \bibinfo {author} {\bibfnamefont {David}\
  \bibnamefont {Sutter}}, \bibinfo {author} {\bibfnamefont {Sergey}\
  \bibnamefont {Bravyi}}, \bibinfo {author} {\bibfnamefont {Jay~M.}\
  \bibnamefont {Gambetta}}, and\ \bibinfo {author} {\bibfnamefont {Kristan}\
  \bibnamefont {Temme}}} (\bibinfo {year} {2021}),\ \bibfield  {title}
  {\enquote {\bibinfo {title} {Error {{Mitigation}} for {{Universal Gates}} on
  {{Encoded Qubits}}},}\ }\href
  {https://doi.org/10.1103/PhysRevLett.127.200505} {\bibfield  {journal}
  {\bibinfo  {journal} {Physical Review Letters}\ }\textbf {\bibinfo {volume}
  {127}}~(\bibinfo {number} {20}),\ \bibinfo {pages} {200505}}\BibitemShut
  {NoStop}%
\bibitem [{\citenamefont {Piveteau}\ \emph {et~al.}(2022)\citenamefont
  {Piveteau}, \citenamefont {Sutter},\ and\ \citenamefont
  {Woerner}}]{piveteauQuasiprobabilityDecompositionsReduced2022}%
  \BibitemOpen
  \bibfield  {author} {\bibinfo {author} {\bibnamefont {Piveteau},
  \bibfnamefont {Christophe}}, \bibinfo {author} {\bibfnamefont {David}\
  \bibnamefont {Sutter}}, and\ \bibinfo {author} {\bibfnamefont {Stefan}\
  \bibnamefont {Woerner}}} (\bibinfo {year} {2022}),\ \bibfield  {title}
  {\enquote {\bibinfo {title} {Quasiprobability decompositions with reduced
  sampling overhead},}\ }\href {https://doi.org/10.1038/s41534-022-00517-3}
  {\bibfield  {journal} {\bibinfo  {journal} {npj Quantum Information}\
  }\textbf {\bibinfo {volume} {8}},\ \bibinfo {pages} {12}}\BibitemShut
  {NoStop}%
\bibitem [{\citenamefont {Polla}\ \emph {et~al.}(2023)\citenamefont {Polla},
  \citenamefont {Anselmetti},\ and\ \citenamefont
  {O'Brien}}]{pollaOptimizingInformationExtracted2023}%
  \BibitemOpen
  \bibfield  {author} {\bibinfo {author} {\bibnamefont {Polla}, \bibfnamefont
  {Stefano}}, \bibinfo {author} {\bibfnamefont {Gian-Luca~R.}\ \bibnamefont
  {Anselmetti}}, and\ \bibinfo {author} {\bibfnamefont {Thomas~E.}\
  \bibnamefont {O'Brien}}} (\bibinfo {year} {2023}),\ \bibfield  {title}
  {\enquote {\bibinfo {title} {Optimizing the information extracted by a single
  qubit measurement},}\ }\href {https://doi.org/10.1103/PhysRevA.108.012403}
  {\bibfield  {journal} {\bibinfo  {journal} {Physical Review A}\ }\textbf
  {\bibinfo {volume} {108}}~(\bibinfo {number} {1}),\ \bibinfo {pages}
  {012403}}\BibitemShut {NoStop}%
\bibitem [{\citenamefont {Postler}\ \emph {et~al.}(2022)\citenamefont
  {Postler}, \citenamefont {Heu{$\beta$}en}, \citenamefont {Pogorelov},
  \citenamefont {Rispler}, \citenamefont {Feldker}, \citenamefont {Meth},
  \citenamefont {Marciniak}, \citenamefont {Stricker}, \citenamefont
  {Ringbauer}, \citenamefont {Blatt}, \citenamefont {Schindler}, \citenamefont
  {M{\"u}ller},\ and\ \citenamefont
  {Monz}}]{postlerDemonstrationFaulttolerantUniversal2022}%
  \BibitemOpen
  \bibfield  {author} {\bibinfo {author} {\bibnamefont {Postler}, \bibfnamefont
  {Lukas}}, \bibinfo {author} {\bibfnamefont {Sascha}\ \bibnamefont
  {Heu{$\beta$}en}}, \bibinfo {author} {\bibfnamefont {Ivan}\ \bibnamefont
  {Pogorelov}}, \bibinfo {author} {\bibfnamefont {Manuel}\ \bibnamefont
  {Rispler}}, \bibinfo {author} {\bibfnamefont {Thomas}\ \bibnamefont
  {Feldker}}, \bibinfo {author} {\bibfnamefont {Michael}\ \bibnamefont {Meth}},
  \bibinfo {author} {\bibfnamefont {Christian~D.}\ \bibnamefont {Marciniak}},
  \bibinfo {author} {\bibfnamefont {Roman}\ \bibnamefont {Stricker}}, \bibinfo
  {author} {\bibfnamefont {Martin}\ \bibnamefont {Ringbauer}}, \bibinfo
  {author} {\bibfnamefont {Rainer}\ \bibnamefont {Blatt}}, \bibinfo {author}
  {\bibfnamefont {Philipp}\ \bibnamefont {Schindler}}, \bibinfo {author}
  {\bibfnamefont {Markus}\ \bibnamefont {M{\"u}ller}}, and\ \bibinfo {author}
  {\bibfnamefont {Thomas}\ \bibnamefont {Monz}}} (\bibinfo {year} {2022}),\
  \bibfield  {title} {\enquote {\bibinfo {title} {Demonstration of
  fault-tolerant universal quantum gate operations},}\ }\href
  {https://doi.org/10.1038/s41586-022-04721-1} {\bibfield  {journal} {\bibinfo
  {journal} {Nature}\ }\textbf {\bibinfo {volume} {605}}~(\bibinfo {number}
  {7911}),\ \bibinfo {pages} {675--680}}\BibitemShut {NoStop}%
\bibitem [{\citenamefont {Qin}\ \emph {et~al.}(2023)\citenamefont {Qin},
  \citenamefont {Chen},\ and\ \citenamefont
  {Li}}]{qinErrorStatisticsScalability2023}%
  \BibitemOpen
  \bibfield  {author} {\bibinfo {author} {\bibnamefont {Qin}, \bibfnamefont
  {Dayue}}, \bibinfo {author} {\bibfnamefont {Yanzhu}\ \bibnamefont {Chen}},
  and\ \bibinfo {author} {\bibfnamefont {Ying}\ \bibnamefont {Li}}} (\bibinfo
  {year} {2023}),\ \bibfield  {title} {\enquote {\bibinfo {title} {Error
  statistics and scalability of quantum error mitigation formulas},}\ }\href
  {https://doi.org/10.1038/s41534-023-00707-7} {\bibfield  {journal} {\bibinfo
  {journal} {npj Quantum Information}\ }\textbf {\bibinfo {volume}
  {9}}~(\bibinfo {number} {1}),\ \bibinfo {pages} {1--14}}\BibitemShut
  {NoStop}%
\bibitem [{\citenamefont {{Qiskit
  contributors}}(2023)}]{qiskitcontributorsQiskitOpensourceFramework2023}%
  \BibitemOpen
  \bibfield  {author} {\bibinfo {author} {\bibnamefont {{Qiskit
  contributors}},}} (\bibinfo {year} {2023}),\ \href
  {https://doi.org/10.5281/zenodo.2573505} {\enquote {\bibinfo {title} {Qiskit:
  {{An Open-source Framework}} for {{Quantum Computing}}},}\ }\bibinfo
  {howpublished} {Zenodo}\BibitemShut {NoStop}%
\bibitem [{\citenamefont {Quek}\ \emph {et~al.}(2022)\citenamefont {Quek},
  \citenamefont {Fran{\c c}a}, \citenamefont {Khatri}, \citenamefont {Meyer},\
  and\ \citenamefont {Eisert}}]{quekExponentiallyTighterBounds2022}%
  \BibitemOpen
  \bibfield  {author} {\bibinfo {author} {\bibnamefont {Quek}, \bibfnamefont
  {Yihui}}, \bibinfo {author} {\bibfnamefont {Daniel~Stilck}\ \bibnamefont
  {Fran{\c c}a}}, \bibinfo {author} {\bibfnamefont {Sumeet}\ \bibnamefont
  {Khatri}}, \bibinfo {author} {\bibfnamefont {Johannes~Jakob}\ \bibnamefont
  {Meyer}}, and\ \bibinfo {author} {\bibfnamefont {Jens}\ \bibnamefont
  {Eisert}}} (\bibinfo {year} {2022}),\ \href
  {https://doi.org/10.48550/arXiv.2210.11505} {\enquote {\bibinfo {title}
  {Exponentially tighter bounds on limitations of quantum error mitigation},}\
  }\bibinfo {howpublished} {arXiv:2210.11505 [math-ph,
  physics:quant-ph]}\BibitemShut {NoStop}%
\bibitem [{\citenamefont {Regula}\ \emph {et~al.}(2021)\citenamefont {Regula},
  \citenamefont {Takagi},\ and\ \citenamefont
  {Gu}}]{regulaOperationalApplicationsDiamond2021}%
  \BibitemOpen
  \bibfield  {author} {\bibinfo {author} {\bibnamefont {Regula}, \bibfnamefont
  {Bartosz}}, \bibinfo {author} {\bibfnamefont {Ryuji}\ \bibnamefont {Takagi}},
  and\ \bibinfo {author} {\bibfnamefont {Mile}\ \bibnamefont {Gu}}} (\bibinfo
  {year} {2021}),\ \bibfield  {title} {\enquote {\bibinfo {title} {Operational
  applications of the diamond norm and related measures in quantifying the
  non-physicality of quantum maps},}\ }\href
  {https://doi.org/10.22331/q-2021-08-09-522} {\bibfield  {journal} {\bibinfo
  {journal} {Quantum}\ }\textbf {\bibinfo {volume} {5}},\ \bibinfo {pages}
  {522}}\BibitemShut {NoStop}%
\bibitem [{\citenamefont {Rosenberg}\ \emph {et~al.}(2022)\citenamefont
  {Rosenberg}, \citenamefont {Ginsparg},\ and\ \citenamefont
  {McMahon}}]{rosenbergExperimentalErrorMitigation2022}%
  \BibitemOpen
  \bibfield  {author} {\bibinfo {author} {\bibnamefont {Rosenberg},
  \bibfnamefont {Eliott}}, \bibinfo {author} {\bibfnamefont {Paul}\
  \bibnamefont {Ginsparg}}, and\ \bibinfo {author} {\bibfnamefont {Peter~L.}\
  \bibnamefont {McMahon}}} (\bibinfo {year} {2022}),\ \bibfield  {title}
  {\enquote {\bibinfo {title} {Experimental error mitigation using linear
  rescaling for variational quantum eigensolving with up to 20 qubits},}\
  }\href {https://doi.org/10.1088/2058-9565/ac3b37} {\bibfield  {journal}
  {\bibinfo  {journal} {Quantum Science and Technology}\ }\textbf {\bibinfo
  {volume} {7}}~(\bibinfo {number} {1}),\ \bibinfo {pages}
  {015024}}\BibitemShut {NoStop}%
\bibitem [{\citenamefont {Ross}\ and\ \citenamefont
  {Selinger}(2016)}]{rossOptimalAncillafreeClifford2016}%
  \BibitemOpen
  \bibfield  {author} {\bibinfo {author} {\bibnamefont {Ross}, \bibfnamefont
  {Neil~J}}, and\ \bibinfo {author} {\bibfnamefont {Peter}\ \bibnamefont
  {Selinger}}} (\bibinfo {year} {2016}),\ \bibfield  {title} {\enquote
  {\bibinfo {title} {Optimal ancilla-free {{Clifford}}+{{T}} approximation of
  z-rotations},}\ }\href@noop {} {\bibfield  {journal} {\bibinfo  {journal}
  {Quantum Information \& Computation}\ }\textbf {\bibinfo {volume}
  {16}}~(\bibinfo {number} {11-12}),\ \bibinfo {pages} {901--953}}\BibitemShut
  {NoStop}%
\bibitem [{\citenamefont {Rubin}\ \emph {et~al.}(2018)\citenamefont {Rubin},
  \citenamefont {Babbush},\ and\ \citenamefont
  {McClean}}]{rubinApplicationFermionicMarginal2018}%
  \BibitemOpen
  \bibfield  {author} {\bibinfo {author} {\bibnamefont {Rubin}, \bibfnamefont
  {Nicholas~C}}, \bibinfo {author} {\bibfnamefont {Ryan}\ \bibnamefont
  {Babbush}}, and\ \bibinfo {author} {\bibfnamefont {Jarrod}\ \bibnamefont
  {McClean}}} (\bibinfo {year} {2018}),\ \bibfield  {title} {\enquote {\bibinfo
  {title} {Application of fermionic marginal constraints to hybrid quantum
  algorithms},}\ }\href {https://doi.org/10.1088/1367-2630/aab919} {\bibfield
  {journal} {\bibinfo  {journal} {New Journal of Physics}\ }\textbf {\bibinfo
  {volume} {20}}~(\bibinfo {number} {5}),\ \bibinfo {pages}
  {053020}}\BibitemShut {NoStop}%
\bibitem [{\citenamefont {Russo}\ \emph {et~al.}(2021)\citenamefont {Russo},
  \citenamefont {Rudinger}, \citenamefont {Morrison},\ and\ \citenamefont
  {Baczewski}}]{russoEvaluatingEnergyDifferences2021}%
  \BibitemOpen
  \bibfield  {author} {\bibinfo {author} {\bibnamefont {Russo}, \bibfnamefont
  {A~E}}, \bibinfo {author} {\bibfnamefont {K.~M.}\ \bibnamefont {Rudinger}},
  \bibinfo {author} {\bibfnamefont {B.~C.~A.}\ \bibnamefont {Morrison}}, and\
  \bibinfo {author} {\bibfnamefont {A.~D.}\ \bibnamefont {Baczewski}}}
  (\bibinfo {year} {2021}),\ \bibfield  {title} {\enquote {\bibinfo {title}
  {Evaluating {{Energy Differences}} on a {{Quantum Computer}} with {{Robust
  Phase Estimation}}},}\ }\href
  {https://doi.org/10.1103/PhysRevLett.126.210501} {\bibfield  {journal}
  {\bibinfo  {journal} {Physical Review Letters}\ }\textbf {\bibinfo {volume}
  {126}}~(\bibinfo {number} {21}),\ \bibinfo {pages} {210501}}\BibitemShut
  {NoStop}%
\bibitem [{\citenamefont {{Ryan-Anderson}}\ \emph {et~al.}(2022)\citenamefont
  {{Ryan-Anderson}}, \citenamefont {Brown}, \citenamefont {Allman},
  \citenamefont {Arkin}, \citenamefont {{Asa-Attuah}}, \citenamefont {Baldwin},
  \citenamefont {Berg}, \citenamefont {Bohnet}, \citenamefont {Braxton},
  \citenamefont {Burdick}, \citenamefont {Campora}, \citenamefont
  {Chernoguzov}, \citenamefont {Esposito}, \citenamefont {Evans}, \citenamefont
  {Francois}, \citenamefont {Gaebler}, \citenamefont {Gatterman}, \citenamefont
  {Gerber}, \citenamefont {Gilmore}, \citenamefont {Gresh}, \citenamefont
  {Hall}, \citenamefont {Hankin}, \citenamefont {Hostetter}, \citenamefont
  {Lucchetti}, \citenamefont {Mayer}, \citenamefont {Myers}, \citenamefont
  {Neyenhuis}, \citenamefont {Santiago}, \citenamefont {Sedlacek},
  \citenamefont {Skripka}, \citenamefont {Slattery}, \citenamefont {Stutz},
  \citenamefont {Tait}, \citenamefont {Tobey}, \citenamefont {Vittorini},
  \citenamefont {Walker},\ and\ \citenamefont
  {Hayes}}]{ryan-andersonImplementingFaulttolerantEntangling2022}%
  \BibitemOpen
  \bibfield  {author} {\bibinfo {author} {\bibnamefont {{Ryan-Anderson}},
  \bibfnamefont {C}}, \bibinfo {author} {\bibfnamefont {N.~C.}\ \bibnamefont
  {Brown}}, \bibinfo {author} {\bibfnamefont {M.~S.}\ \bibnamefont {Allman}},
  \bibinfo {author} {\bibfnamefont {B.}~\bibnamefont {Arkin}}, \bibinfo
  {author} {\bibfnamefont {G.}~\bibnamefont {{Asa-Attuah}}}, \bibinfo {author}
  {\bibfnamefont {C.}~\bibnamefont {Baldwin}}, \bibinfo {author} {\bibfnamefont
  {J.}~\bibnamefont {Berg}}, \bibinfo {author} {\bibfnamefont {J.~G.}\
  \bibnamefont {Bohnet}}, \bibinfo {author} {\bibfnamefont {S.}~\bibnamefont
  {Braxton}}, \bibinfo {author} {\bibfnamefont {N.}~\bibnamefont {Burdick}},
  \bibinfo {author} {\bibfnamefont {J.~P.}\ \bibnamefont {Campora}}, \bibinfo
  {author} {\bibfnamefont {A.}~\bibnamefont {Chernoguzov}}, \bibinfo {author}
  {\bibfnamefont {J.}~\bibnamefont {Esposito}}, \bibinfo {author}
  {\bibfnamefont {B.}~\bibnamefont {Evans}}, \bibinfo {author} {\bibfnamefont
  {D.}~\bibnamefont {Francois}}, \bibinfo {author} {\bibfnamefont {J.~P.}\
  \bibnamefont {Gaebler}}, \bibinfo {author} {\bibfnamefont {T.~M.}\
  \bibnamefont {Gatterman}}, \bibinfo {author} {\bibfnamefont {J.}~\bibnamefont
  {Gerber}}, \bibinfo {author} {\bibfnamefont {K.}~\bibnamefont {Gilmore}},
  \bibinfo {author} {\bibfnamefont {D.}~\bibnamefont {Gresh}}, \bibinfo
  {author} {\bibfnamefont {A.}~\bibnamefont {Hall}}, \bibinfo {author}
  {\bibfnamefont {A.}~\bibnamefont {Hankin}}, \bibinfo {author} {\bibfnamefont
  {J.}~\bibnamefont {Hostetter}}, \bibinfo {author} {\bibfnamefont
  {D.}~\bibnamefont {Lucchetti}}, \bibinfo {author} {\bibfnamefont
  {K.}~\bibnamefont {Mayer}}, \bibinfo {author} {\bibfnamefont
  {J.}~\bibnamefont {Myers}}, \bibinfo {author} {\bibfnamefont
  {B.}~\bibnamefont {Neyenhuis}}, \bibinfo {author} {\bibfnamefont
  {J.}~\bibnamefont {Santiago}}, \bibinfo {author} {\bibfnamefont
  {J.}~\bibnamefont {Sedlacek}}, \bibinfo {author} {\bibfnamefont
  {T.}~\bibnamefont {Skripka}}, \bibinfo {author} {\bibfnamefont
  {A.}~\bibnamefont {Slattery}}, \bibinfo {author} {\bibfnamefont {R.~P.}\
  \bibnamefont {Stutz}}, \bibinfo {author} {\bibfnamefont {J.}~\bibnamefont
  {Tait}}, \bibinfo {author} {\bibfnamefont {R.}~\bibnamefont {Tobey}},
  \bibinfo {author} {\bibfnamefont {G.}~\bibnamefont {Vittorini}}, \bibinfo
  {author} {\bibfnamefont {J.}~\bibnamefont {Walker}}, and\ \bibinfo {author}
  {\bibfnamefont {D.}~\bibnamefont {Hayes}}} (\bibinfo {year} {2022}),\ \href
  {http://arxiv.org/abs/2208.01863} {\enquote {\bibinfo {title} {Implementing
  {{Fault-tolerant Entangling Gates}} on the {{Five-qubit Code}} and the
  {{Color Code}}},}\ }\bibinfo {howpublished} {arXiv:2208.01863
  [quant-ph]}\BibitemShut {NoStop}%
\bibitem [{\citenamefont {Sagastizabal}\ \emph {et~al.}(2019)\citenamefont
  {Sagastizabal}, \citenamefont {{Bonet-Monroig}}, \citenamefont {Singh},
  \citenamefont {Rol}, \citenamefont {Bultink}, \citenamefont {Fu},
  \citenamefont {Price}, \citenamefont {Ostroukh}, \citenamefont
  {Muthusubramanian}, \citenamefont {Bruno}, \citenamefont {Beekman},
  \citenamefont {Haider}, \citenamefont {O'Brien},\ and\ \citenamefont
  {DiCarlo}}]{sagastizabalExperimentalErrorMitigation2019}%
  \BibitemOpen
  \bibfield  {author} {\bibinfo {author} {\bibnamefont {Sagastizabal},
  \bibfnamefont {R}}, \bibinfo {author} {\bibfnamefont {X.}~\bibnamefont
  {{Bonet-Monroig}}}, \bibinfo {author} {\bibfnamefont {M.}~\bibnamefont
  {Singh}}, \bibinfo {author} {\bibfnamefont {M.~A.}\ \bibnamefont {Rol}},
  \bibinfo {author} {\bibfnamefont {C.~C.}\ \bibnamefont {Bultink}}, \bibinfo
  {author} {\bibfnamefont {X.}~\bibnamefont {Fu}}, \bibinfo {author}
  {\bibfnamefont {C.~H.}\ \bibnamefont {Price}}, \bibinfo {author}
  {\bibfnamefont {V.~P.}\ \bibnamefont {Ostroukh}}, \bibinfo {author}
  {\bibfnamefont {N.}~\bibnamefont {Muthusubramanian}}, \bibinfo {author}
  {\bibfnamefont {A.}~\bibnamefont {Bruno}}, \bibinfo {author} {\bibfnamefont
  {M.}~\bibnamefont {Beekman}}, \bibinfo {author} {\bibfnamefont
  {N.}~\bibnamefont {Haider}}, \bibinfo {author} {\bibfnamefont {T.~E.}\
  \bibnamefont {O'Brien}}, and\ \bibinfo {author} {\bibfnamefont
  {L.}~\bibnamefont {DiCarlo}}} (\bibinfo {year} {2019}),\ \bibfield  {title}
  {\enquote {\bibinfo {title} {Experimental error mitigation via symmetry
  verification in a variational quantum eigensolver},}\ }\href
  {https://doi.org/10.1103/PhysRevA.100.010302} {\bibfield  {journal} {\bibinfo
   {journal} {Physical Review A}\ }\textbf {\bibinfo {volume} {100}}~(\bibinfo
  {number} {1}),\ \bibinfo {pages} {010302}}\BibitemShut {NoStop}%
\bibitem [{\citenamefont {Sanders}\ \emph {et~al.}(2015)\citenamefont
  {Sanders}, \citenamefont {Wallman},\ and\ \citenamefont
  {Sanders}}]{sandersBoundingQuantumGate2015}%
  \BibitemOpen
  \bibfield  {author} {\bibinfo {author} {\bibnamefont {Sanders}, \bibfnamefont
  {Yuval~R}}, \bibinfo {author} {\bibfnamefont {Joel~J.}\ \bibnamefont
  {Wallman}}, and\ \bibinfo {author} {\bibfnamefont {Barry~C.}\ \bibnamefont
  {Sanders}}} (\bibinfo {year} {2015}),\ \bibfield  {title} {\enquote {\bibinfo
  {title} {Bounding quantum gate error rate based on reported average
  fidelity},}\ }\href {https://doi.org/10.1088/1367-2630/18/1/012002}
  {\bibfield  {journal} {\bibinfo  {journal} {New Journal of Physics}\ }\textbf
  {\bibinfo {volume} {18}}~(\bibinfo {number} {1}),\ \bibinfo {pages}
  {012002}}\BibitemShut {NoStop}%
\bibitem [{\citenamefont {Sauter}\ \emph {et~al.}(1986)\citenamefont {Sauter},
  \citenamefont {Neuhauser}, \citenamefont {Blatt},\ and\ \citenamefont
  {Toschek}}]{sauterObservationQuantumJumps1986}%
  \BibitemOpen
  \bibfield  {author} {\bibinfo {author} {\bibnamefont {Sauter}, \bibfnamefont
  {{\relax Th}}}, \bibinfo {author} {\bibfnamefont {W.}~\bibnamefont
  {Neuhauser}}, \bibinfo {author} {\bibfnamefont {R.}~\bibnamefont {Blatt}},
  and\ \bibinfo {author} {\bibfnamefont {P.~E.}\ \bibnamefont {Toschek}}}
  (\bibinfo {year} {1986}),\ \bibfield  {title} {\enquote {\bibinfo {title}
  {Observation of {{Quantum Jumps}}},}\ }\href
  {https://doi.org/10.1103/PhysRevLett.57.1696} {\bibfield  {journal} {\bibinfo
   {journal} {Physical Review Letters}\ }\textbf {\bibinfo {volume}
  {57}}~(\bibinfo {number} {14}),\ \bibinfo {pages} {1696--1698}}\BibitemShut
  {NoStop}%
\bibitem [{\citenamefont {Seif}\ \emph {et~al.}(2023)\citenamefont {Seif},
  \citenamefont {Cian}, \citenamefont {Zhou}, \citenamefont {Chen},\ and\
  \citenamefont {Jiang}}]{seifShadowDistillationQuantum2023}%
  \BibitemOpen
  \bibfield  {author} {\bibinfo {author} {\bibnamefont {Seif}, \bibfnamefont
  {Alireza}}, \bibinfo {author} {\bibfnamefont {Ze-Pei}\ \bibnamefont {Cian}},
  \bibinfo {author} {\bibfnamefont {Sisi}\ \bibnamefont {Zhou}}, \bibinfo
  {author} {\bibfnamefont {Senrui}\ \bibnamefont {Chen}}, and\ \bibinfo
  {author} {\bibfnamefont {Liang}\ \bibnamefont {Jiang}}} (\bibinfo {year}
  {2023}),\ \bibfield  {title} {\enquote {\bibinfo {title} {Shadow
  {{Distillation}}: {{Quantum Error Mitigation}} with {{Classical Shadows}} for
  {{Near-Term Quantum Processors}}},}\ }\href
  {https://doi.org/10.1103/PRXQuantum.4.010303} {\bibfield  {journal} {\bibinfo
   {journal} {PRX Quantum}\ }\textbf {\bibinfo {volume} {4}}~(\bibinfo {number}
  {1}),\ \bibinfo {pages} {010303}}\BibitemShut {NoStop}%
\bibitem [{\citenamefont {Setia}\ \emph {et~al.}(2019)\citenamefont {Setia},
  \citenamefont {Bravyi}, \citenamefont {Mezzacapo},\ and\ \citenamefont
  {Whitfield}}]{setiaSuperfastEncodingsFermionic2019}%
  \BibitemOpen
  \bibfield  {author} {\bibinfo {author} {\bibnamefont {Setia}, \bibfnamefont
  {Kanav}}, \bibinfo {author} {\bibfnamefont {Sergey}\ \bibnamefont {Bravyi}},
  \bibinfo {author} {\bibfnamefont {Antonio}\ \bibnamefont {Mezzacapo}}, and\
  \bibinfo {author} {\bibfnamefont {James~D.}\ \bibnamefont {Whitfield}}}
  (\bibinfo {year} {2019}),\ \bibfield  {title} {\enquote {\bibinfo {title}
  {Superfast encodings for fermionic quantum simulation},}\ }\href
  {https://doi.org/10.1103/PhysRevResearch.1.033033} {\bibfield  {journal}
  {\bibinfo  {journal} {Physical Review Research}\ }\textbf {\bibinfo {volume}
  {1}}~(\bibinfo {number} {3}),\ \bibinfo {pages} {033033}}\BibitemShut
  {NoStop}%
\bibitem [{\citenamefont {Setia}\ and\ \citenamefont
  {Whitfield}(2018)}]{setiaBravyiKitaevSuperfastSimulation2018}%
  \BibitemOpen
  \bibfield  {author} {\bibinfo {author} {\bibnamefont {Setia}, \bibfnamefont
  {Kanav}}, and\ \bibinfo {author} {\bibfnamefont {James~D.}\ \bibnamefont
  {Whitfield}}} (\bibinfo {year} {2018}),\ \bibfield  {title} {\enquote
  {\bibinfo {title} {Bravyi-{{Kitaev Superfast}} simulation of fermions on a
  quantum computer},}\ }\href {https://doi.org/10.1063/1.5019371} {\bibfield
  {journal} {\bibinfo  {journal} {The Journal of Chemical Physics}\ }\textbf
  {\bibinfo {volume} {148}}~(\bibinfo {number} {16}),\ \bibinfo {pages}
  {164104}}\BibitemShut {NoStop}%
\bibitem [{\citenamefont {Shor}(1999)}]{shorPolynomialTimeAlgorithmsPrime1999}%
  \BibitemOpen
  \bibfield  {author} {\bibinfo {author} {\bibnamefont {Shor}, \bibfnamefont
  {P}}} (\bibinfo {year} {1999}),\ \bibfield  {title} {\enquote {\bibinfo
  {title} {Polynomial-{{Time Algorithms}} for {{Prime Factorization}} and
  {{Discrete Logarithms}} on a {{Quantum Computer}}},}\ }\href
  {https://doi.org/10.1137/S0036144598347011} {\bibfield  {journal} {\bibinfo
  {journal} {SIAM Review}\ }\textbf {\bibinfo {volume} {41}}~(\bibinfo {number}
  {2}),\ \bibinfo {pages} {303--332}}\BibitemShut {NoStop}%
\bibitem [{\citenamefont {Shor}(1995)}]{shorSchemeReducingDecoherence1995}%
  \BibitemOpen
  \bibfield  {author} {\bibinfo {author} {\bibnamefont {Shor}, \bibfnamefont
  {Peter~W}}} (\bibinfo {year} {1995}),\ \bibfield  {title} {\enquote {\bibinfo
  {title} {Scheme for reducing decoherence in quantum computer memory},}\
  }\href {https://doi.org/10.1103/PhysRevA.52.R2493} {\bibfield  {journal}
  {\bibinfo  {journal} {Physical Review A}\ }\textbf {\bibinfo {volume}
  {52}}~(\bibinfo {number} {4}),\ \bibinfo {pages} {R2493--R2496}}\BibitemShut
  {NoStop}%
\bibitem [{\citenamefont
  {Shor}(1996)}]{shorFaulttolerantQuantumComputation1996}%
  \BibitemOpen
  \bibfield  {author} {\bibinfo {author} {\bibnamefont {Shor}, \bibfnamefont
  {PW}}} (\bibinfo {year} {1996}),\ \bibfield  {title} {\enquote {\bibinfo
  {title} {Fault-tolerant quantum computation},}\ }in\ \href
  {https://doi.org/10.1109/SFCS.1996.548464} {\emph {\bibinfo {booktitle}
  {Proceedings of 37th {{Conference}} on {{Foundations}} of {{Computer
  Science}}}}},\ pp.\ \bibinfo {pages} {56--65}\BibitemShut {NoStop}%
\bibitem [{\citenamefont {Sidi}(2003)}]{sidiPracticalExtrapolationMethods2003}%
  \BibitemOpen
  \bibfield  {author} {\bibinfo {author} {\bibnamefont {Sidi}, \bibfnamefont
  {Avram}}} (\bibinfo {year} {2003}),\ \href
  {https://doi.org/10.1017/CBO9780511546815} {\emph {\bibinfo {title}
  {Practical {{Extrapolation Methods}}: {{Theory}} and {{Applications}}}}},\
  Cambridge {{Monographs}} on {{Applied}} and {{Computational Mathematics}}\
  (\bibinfo  {publisher} {{Cambridge University Press}},\ \bibinfo {address}
  {{Cambridge}})\BibitemShut {NoStop}%
\bibitem [{\citenamefont {Smart}\ and\ \citenamefont
  {Mazziotti}(2019)}]{smartQuantumclassicalHybridAlgorithm2019}%
  \BibitemOpen
  \bibfield  {author} {\bibinfo {author} {\bibnamefont {Smart}, \bibfnamefont
  {Scott~E}}, and\ \bibinfo {author} {\bibfnamefont {David~A.}\ \bibnamefont
  {Mazziotti}}} (\bibinfo {year} {2019}),\ \bibfield  {title} {\enquote
  {\bibinfo {title} {Quantum-classical hybrid algorithm using an
  error-mitigating \${{N}}\$-representability condition to compute the {{Mott}}
  metal-insulator transition},}\ }\href
  {https://doi.org/10.1103/PhysRevA.100.022517} {\bibfield  {journal} {\bibinfo
   {journal} {Physical Review A}\ }\textbf {\bibinfo {volume} {100}}~(\bibinfo
  {number} {2}),\ \bibinfo {pages} {022517}}\BibitemShut {NoStop}%
\bibitem [{\citenamefont {Smart}\ and\ \citenamefont
  {Mazziotti}(2020)}]{smartEfficientTwoelectronAnsatz2020}%
  \BibitemOpen
  \bibfield  {author} {\bibinfo {author} {\bibnamefont {Smart}, \bibfnamefont
  {Scott~E}}, and\ \bibinfo {author} {\bibfnamefont {David~A.}\ \bibnamefont
  {Mazziotti}}} (\bibinfo {year} {2020}),\ \bibfield  {title} {\enquote
  {\bibinfo {title} {Efficient two-electron ansatz for benchmarking quantum
  chemistry on a quantum computer},}\ }\href
  {https://doi.org/10.1103/PhysRevResearch.2.023048} {\bibfield  {journal}
  {\bibinfo  {journal} {Physical Review Research}\ }\textbf {\bibinfo {volume}
  {2}}~(\bibinfo {number} {2}),\ \bibinfo {pages} {023048}}\BibitemShut
  {NoStop}%
\bibitem [{\citenamefont {Song}\ \emph {et~al.}(2019)\citenamefont {Song},
  \citenamefont {Cui}, \citenamefont {Wang}, \citenamefont {Hao}, \citenamefont
  {Feng},\ and\ \citenamefont {Li}}]{songQuantumComputationUniversal2019}%
  \BibitemOpen
  \bibfield  {author} {\bibinfo {author} {\bibnamefont {Song}, \bibfnamefont
  {Chao}}, \bibinfo {author} {\bibfnamefont {Jing}\ \bibnamefont {Cui}},
  \bibinfo {author} {\bibfnamefont {H.}~\bibnamefont {Wang}}, \bibinfo {author}
  {\bibfnamefont {J.}~\bibnamefont {Hao}}, \bibinfo {author} {\bibfnamefont
  {H.}~\bibnamefont {Feng}}, and\ \bibinfo {author} {\bibfnamefont {Ying}\
  \bibnamefont {Li}}} (\bibinfo {year} {2019}),\ \bibfield  {title} {\enquote
  {\bibinfo {title} {Quantum computation with universal error mitigation on a
  superconducting quantum processor},}\ }\href
  {https://doi.org/10.1126/sciadv.aaw5686} {\bibfield  {journal} {\bibinfo
  {journal} {Science Advances}\ }\textbf {\bibinfo {volume} {5}}~(\bibinfo
  {number} {9}),\ \bibinfo {pages} {eaaw5686}}\BibitemShut {NoStop}%
\bibitem [{\citenamefont {Stanisic}\ \emph {et~al.}(2022)\citenamefont
  {Stanisic}, \citenamefont {Bosse}, \citenamefont {Gambetta}, \citenamefont
  {Santos}, \citenamefont {Mruczkiewicz}, \citenamefont {O'Brien},
  \citenamefont {Ostby},\ and\ \citenamefont
  {Montanaro}}]{stanisicObservingGroundstateProperties2022}%
  \BibitemOpen
  \bibfield  {author} {\bibinfo {author} {\bibnamefont {Stanisic},
  \bibfnamefont {Stasja}}, \bibinfo {author} {\bibfnamefont {Jan~Lukas}\
  \bibnamefont {Bosse}}, \bibinfo {author} {\bibfnamefont {Filippo~Maria}\
  \bibnamefont {Gambetta}}, \bibinfo {author} {\bibfnamefont {Raul~A.}\
  \bibnamefont {Santos}}, \bibinfo {author} {\bibfnamefont {Wojciech}\
  \bibnamefont {Mruczkiewicz}}, \bibinfo {author} {\bibfnamefont {Thomas~E.}\
  \bibnamefont {O'Brien}}, \bibinfo {author} {\bibfnamefont {Eric}\
  \bibnamefont {Ostby}}, and\ \bibinfo {author} {\bibfnamefont {Ashley}\
  \bibnamefont {Montanaro}}} (\bibinfo {year} {2022}),\ \bibfield  {title}
  {\enquote {\bibinfo {title} {Observing ground-state properties of the
  {{Fermi-Hubbard}} model using a scalable algorithm on a quantum computer},}\
  }\href {https://doi.org/10.1038/s41467-022-33335-4} {\bibfield  {journal}
  {\bibinfo  {journal} {Nature Communications}\ }\textbf {\bibinfo {volume}
  {13}}~(\bibinfo {number} {1}),\ \bibinfo {pages} {5743}}\BibitemShut
  {NoStop}%
\bibitem [{\citenamefont {Steane}(1996)}]{steaneErrorCorrectingCodes1996}%
  \BibitemOpen
  \bibfield  {author} {\bibinfo {author} {\bibnamefont {Steane}, \bibfnamefont
  {A~M}}} (\bibinfo {year} {1996}),\ \bibfield  {title} {\enquote {\bibinfo
  {title} {Error {{Correcting Codes}} in {{Quantum Theory}}},}\ }\href
  {https://doi.org/10.1103/PhysRevLett.77.793} {\bibfield  {journal} {\bibinfo
  {journal} {Physical Review Letters}\ }\textbf {\bibinfo {volume}
  {77}}~(\bibinfo {number} {5}),\ \bibinfo {pages} {793--797}}\BibitemShut
  {NoStop}%
\bibitem [{\citenamefont {Steudtner}\ and\ \citenamefont
  {Wehner}(2018)}]{steudtnerFermiontoqubitMappingsVarying2018}%
  \BibitemOpen
  \bibfield  {author} {\bibinfo {author} {\bibnamefont {Steudtner},
  \bibfnamefont {Mark}}, and\ \bibinfo {author} {\bibfnamefont {Stephanie}\
  \bibnamefont {Wehner}}} (\bibinfo {year} {2018}),\ \bibfield  {title}
  {\enquote {\bibinfo {title} {Fermion-to-qubit mappings with varying resource
  requirements for quantum simulation},}\ }\href
  {https://doi.org/10.1088/1367-2630/aac54f} {\bibfield  {journal} {\bibinfo
  {journal} {New Journal of Physics}\ }\textbf {\bibinfo {volume}
  {20}}~(\bibinfo {number} {6}),\ \bibinfo {pages} {063010}}\BibitemShut
  {NoStop}%
\bibitem [{\citenamefont {Steudtner}\ and\ \citenamefont
  {Wehner}(2019)}]{steudtnerQuantumCodesQuantum2019}%
  \BibitemOpen
  \bibfield  {author} {\bibinfo {author} {\bibnamefont {Steudtner},
  \bibfnamefont {Mark}}, and\ \bibinfo {author} {\bibfnamefont {Stephanie}\
  \bibnamefont {Wehner}}} (\bibinfo {year} {2019}),\ \bibfield  {title}
  {\enquote {\bibinfo {title} {Quantum codes for quantum simulation of fermions
  on a square lattice of qubits},}\ }\href
  {https://doi.org/10.1103/PhysRevA.99.022308} {\bibfield  {journal} {\bibinfo
  {journal} {Physical Review A}\ }\textbf {\bibinfo {volume} {99}}~(\bibinfo
  {number} {2}),\ \bibinfo {pages} {022308}}\BibitemShut {NoStop}%
\bibitem [{\citenamefont {Strikis}\ \emph {et~al.}(2021)\citenamefont
  {Strikis}, \citenamefont {Qin}, \citenamefont {Chen}, \citenamefont
  {Benjamin},\ and\ \citenamefont {Li}}]{strikisLearningBasedQuantumError2021}%
  \BibitemOpen
  \bibfield  {author} {\bibinfo {author} {\bibnamefont {Strikis}, \bibfnamefont
  {Armands}}, \bibinfo {author} {\bibfnamefont {Dayue}\ \bibnamefont {Qin}},
  \bibinfo {author} {\bibfnamefont {Yanzhu}\ \bibnamefont {Chen}}, \bibinfo
  {author} {\bibfnamefont {Simon~C.}\ \bibnamefont {Benjamin}}, and\ \bibinfo
  {author} {\bibfnamefont {Ying}\ \bibnamefont {Li}}} (\bibinfo {year}
  {2021}),\ \bibfield  {title} {\enquote {\bibinfo {title} {Learning-{{Based
  Quantum Error Mitigation}}},}\ }\href
  {https://doi.org/10.1103/PRXQuantum.2.040330} {\bibfield  {journal} {\bibinfo
   {journal} {PRX Quantum}\ }\textbf {\bibinfo {volume} {2}}~(\bibinfo {number}
  {4}),\ \bibinfo {pages} {040330}}\BibitemShut {NoStop}%
\bibitem [{\citenamefont {Suchsland}\ \emph {et~al.}(2021)\citenamefont
  {Suchsland}, \citenamefont {Tacchino}, \citenamefont {Fischer}, \citenamefont
  {Neupert}, \citenamefont {Barkoutsos},\ and\ \citenamefont
  {Tavernelli}}]{suchslandAlgorithmicErrorMitigation2021}%
  \BibitemOpen
  \bibfield  {author} {\bibinfo {author} {\bibnamefont {Suchsland},
  \bibfnamefont {Philippe}}, \bibinfo {author} {\bibfnamefont {Francesco}\
  \bibnamefont {Tacchino}}, \bibinfo {author} {\bibfnamefont {Mark~H.}\
  \bibnamefont {Fischer}}, \bibinfo {author} {\bibfnamefont {Titus}\
  \bibnamefont {Neupert}}, \bibinfo {author} {\bibfnamefont {Panagiotis~Kl}\
  \bibnamefont {Barkoutsos}}, and\ \bibinfo {author} {\bibfnamefont {Ivano}\
  \bibnamefont {Tavernelli}}} (\bibinfo {year} {2021}),\ \bibfield  {title}
  {\enquote {\bibinfo {title} {Algorithmic {{Error Mitigation Scheme}} for
  {{Current Quantum Processors}}},}\ }\href
  {https://doi.org/10.22331/q-2021-07-01-492} {\bibfield  {journal} {\bibinfo
  {journal} {Quantum}\ }\textbf {\bibinfo {volume} {5}},\ \bibinfo {pages}
  {492}}\BibitemShut {NoStop}%
\bibitem [{\citenamefont {Sun}\ \emph {et~al.}(2021)\citenamefont {Sun},
  \citenamefont {Yuan}, \citenamefont {Tsunoda}, \citenamefont {Vedral},
  \citenamefont {Benjamin},\ and\ \citenamefont
  {Endo}}]{sunMitigatingRealisticNoise2021}%
  \BibitemOpen
  \bibfield  {author} {\bibinfo {author} {\bibnamefont {Sun}, \bibfnamefont
  {Jinzhao}}, \bibinfo {author} {\bibfnamefont {Xiao}\ \bibnamefont {Yuan}},
  \bibinfo {author} {\bibfnamefont {Takahiro}\ \bibnamefont {Tsunoda}},
  \bibinfo {author} {\bibfnamefont {Vlatko}\ \bibnamefont {Vedral}}, \bibinfo
  {author} {\bibfnamefont {Simon~C.}\ \bibnamefont {Benjamin}}, and\ \bibinfo
  {author} {\bibfnamefont {Suguru}\ \bibnamefont {Endo}}} (\bibinfo {year}
  {2021}),\ \bibfield  {title} {\enquote {\bibinfo {title} {Mitigating
  {{Realistic Noise}} in {{Practical Noisy Intermediate-Scale Quantum
  Devices}}},}\ }\href {https://doi.org/10.1103/PhysRevApplied.15.034026}
  {\bibfield  {journal} {\bibinfo  {journal} {Physical Review Applied}\
  }\textbf {\bibinfo {volume} {15}}~(\bibinfo {number} {3}),\ \bibinfo {pages}
  {034026}}\BibitemShut {NoStop}%
\bibitem [{\citenamefont {Suter}\ and\ \citenamefont
  {{\'A}lvarez}(2016)}]{suterColloquiumProtectingQuantum2016}%
  \BibitemOpen
  \bibfield  {author} {\bibinfo {author} {\bibnamefont {Suter}, \bibfnamefont
  {Dieter}}, and\ \bibinfo {author} {\bibfnamefont {Gonzalo~A.}\ \bibnamefont
  {{\'A}lvarez}}} (\bibinfo {year} {2016}),\ \bibfield  {title} {\enquote
  {\bibinfo {title} {Colloquium: {{Protecting}} quantum information against
  environmental noise},}\ }\href {https://doi.org/10.1103/RevModPhys.88.041001}
  {\bibfield  {journal} {\bibinfo  {journal} {Reviews of Modern Physics}\
  }\textbf {\bibinfo {volume} {88}}~(\bibinfo {number} {4}),\ \bibinfo {pages}
  {041001}}\BibitemShut {NoStop}%
\bibitem [{\citenamefont
  {Suzuki}(1990)}]{suzukiFractalDecompositionExponential1990}%
  \BibitemOpen
  \bibfield  {author} {\bibinfo {author} {\bibnamefont {Suzuki}, \bibfnamefont
  {Masuo}}} (\bibinfo {year} {1990}),\ \bibfield  {title} {\enquote {\bibinfo
  {title} {Fractal decomposition of exponential operators with applications to
  many-body theories and {{Monte Carlo}} simulations},}\ }\href
  {https://doi.org/10.1016/0375-9601(90)90962-N} {\bibfield  {journal}
  {\bibinfo  {journal} {Physics Letters A}\ }\textbf {\bibinfo {volume}
  {146}}~(\bibinfo {number} {6}),\ \bibinfo {pages} {319--323}}\BibitemShut
  {NoStop}%
\bibitem [{\citenamefont {Suzuki}(1991)}]{suzukiGeneralTheoryFractal1991}%
  \BibitemOpen
  \bibfield  {author} {\bibinfo {author} {\bibnamefont {Suzuki}, \bibfnamefont
  {Masuo}}} (\bibinfo {year} {1991}),\ \bibfield  {title} {\enquote {\bibinfo
  {title} {General theory of fractal path integrals with applications to
  many-body theories and statistical physics},}\ }\href
  {https://doi.org/10.1063/1.529425} {\bibfield  {journal} {\bibinfo  {journal}
  {Journal of Mathematical Physics}\ }\textbf {\bibinfo {volume}
  {32}}~(\bibinfo {number} {2}),\ \bibinfo {pages} {400--407}}\BibitemShut
  {NoStop}%
\bibitem [{\citenamefont {Suzuki}\ \emph {et~al.}(2022)\citenamefont {Suzuki},
  \citenamefont {Endo}, \citenamefont {Fujii},\ and\ \citenamefont
  {Tokunaga}}]{suzukiQuantumErrorMitigation2022}%
  \BibitemOpen
  \bibfield  {author} {\bibinfo {author} {\bibnamefont {Suzuki}, \bibfnamefont
  {Yasunari}}, \bibinfo {author} {\bibfnamefont {Suguru}\ \bibnamefont {Endo}},
  \bibinfo {author} {\bibfnamefont {Keisuke}\ \bibnamefont {Fujii}}, and\
  \bibinfo {author} {\bibfnamefont {Yuuki}\ \bibnamefont {Tokunaga}}} (\bibinfo
  {year} {2022}),\ \bibfield  {title} {\enquote {\bibinfo {title} {Quantum
  {{Error Mitigation}} as a {{Universal Error Reduction Technique}}:
  {{Applications}} from the {{NISQ}} to the {{Fault-Tolerant Quantum Computing
  Eras}}},}\ }\href {https://doi.org/10.1103/PRXQuantum.3.010345} {\bibfield
  {journal} {\bibinfo  {journal} {PRX Quantum}\ }\textbf {\bibinfo {volume}
  {3}}~(\bibinfo {number} {1}),\ \bibinfo {pages} {010345}}\BibitemShut
  {NoStop}%
\bibitem [{\citenamefont {Tacchino}\ \emph {et~al.}(2020)\citenamefont
  {Tacchino}, \citenamefont {Barkoutsos}, \citenamefont {Macchiavello},
  \citenamefont {Tavernelli}, \citenamefont {Gerace},\ and\ \citenamefont
  {Bajoni}}]{tacchinoQuantumImplementationArtificial2020}%
  \BibitemOpen
  \bibfield  {author} {\bibinfo {author} {\bibnamefont {Tacchino},
  \bibfnamefont {Francesco}}, \bibinfo {author} {\bibfnamefont {Panagiotis}\
  \bibnamefont {Barkoutsos}}, \bibinfo {author} {\bibfnamefont {Chiara}\
  \bibnamefont {Macchiavello}}, \bibinfo {author} {\bibfnamefont {Ivano}\
  \bibnamefont {Tavernelli}}, \bibinfo {author} {\bibfnamefont {Dario}\
  \bibnamefont {Gerace}}, and\ \bibinfo {author} {\bibfnamefont {Daniele}\
  \bibnamefont {Bajoni}}} (\bibinfo {year} {2020}),\ \bibfield  {title}
  {\enquote {\bibinfo {title} {Quantum implementation of an artificial
  feed-forward neural network},}\ }\href
  {https://doi.org/10.1088/2058-9565/abb8e4} {\bibfield  {journal} {\bibinfo
  {journal} {Quantum Science and Technology}\ }\textbf {\bibinfo {volume}
  {5}}~(\bibinfo {number} {4}),\ \bibinfo {pages} {044010}}\BibitemShut
  {NoStop}%
\bibitem [{\citenamefont {Takagi}(2021)}]{takagiOptimalResourceCost2021}%
  \BibitemOpen
  \bibfield  {author} {\bibinfo {author} {\bibnamefont {Takagi}, \bibfnamefont
  {Ryuji}}} (\bibinfo {year} {2021}),\ \bibfield  {title} {\enquote {\bibinfo
  {title} {Optimal resource cost for error mitigation},}\ }\href
  {https://doi.org/10.1103/PhysRevResearch.3.033178} {\bibfield  {journal}
  {\bibinfo  {journal} {Physical Review Research}\ }\textbf {\bibinfo {volume}
  {3}}~(\bibinfo {number} {3}),\ \bibinfo {pages} {033178}}\BibitemShut
  {NoStop}%
\bibitem [{\citenamefont {Takagi}\ \emph {et~al.}(2022)\citenamefont {Takagi},
  \citenamefont {Endo}, \citenamefont {Minagawa},\ and\ \citenamefont
  {Gu}}]{takagiFundamentalLimitsQuantum2022}%
  \BibitemOpen
  \bibfield  {author} {\bibinfo {author} {\bibnamefont {Takagi}, \bibfnamefont
  {Ryuji}}, \bibinfo {author} {\bibfnamefont {Suguru}\ \bibnamefont {Endo}},
  \bibinfo {author} {\bibfnamefont {Shintaro}\ \bibnamefont {Minagawa}}, and\
  \bibinfo {author} {\bibfnamefont {Mile}\ \bibnamefont {Gu}}} (\bibinfo {year}
  {2022}),\ \bibfield  {title} {\enquote {\bibinfo {title} {Fundamental limits
  of quantum error mitigation},}\ }\href
  {https://doi.org/10.1038/s41534-022-00618-z} {\bibfield  {journal} {\bibinfo
  {journal} {npj Quantum Information}\ }\textbf {\bibinfo {volume} {8}},\
  \bibinfo {pages} {114}}\BibitemShut {NoStop}%
\bibitem [{\citenamefont {Takagi}\ \emph {et~al.}(2023)\citenamefont {Takagi},
  \citenamefont {Tajima},\ and\ \citenamefont
  {Gu}}]{takagiUniversalSamplingLower2023}%
  \BibitemOpen
  \bibfield  {author} {\bibinfo {author} {\bibnamefont {Takagi}, \bibfnamefont
  {Ryuji}}, \bibinfo {author} {\bibfnamefont {Hiroyasu}\ \bibnamefont
  {Tajima}}, and\ \bibinfo {author} {\bibfnamefont {Mile}\ \bibnamefont {Gu}}}
  (\bibinfo {year} {2023}),\ \bibfield  {title} {\enquote {\bibinfo {title}
  {Universal {{Sampling Lower Bounds}} for {{Quantum Error Mitigation}}},}\
  }\href {https://doi.org/10.1103/PhysRevLett.131.210602} {\bibfield  {journal}
  {\bibinfo  {journal} {Physical Review Letters}\ }\textbf {\bibinfo {volume}
  {131}}~(\bibinfo {number} {21}),\ \bibinfo {pages} {210602}}\BibitemShut
  {NoStop}%
\bibitem [{\citenamefont {Takeda}\ \emph {et~al.}(2022)\citenamefont {Takeda},
  \citenamefont {Noiri}, \citenamefont {Nakajima}, \citenamefont {Kobayashi},\
  and\ \citenamefont {Tarucha}}]{takedaQuantumErrorCorrection2022}%
  \BibitemOpen
  \bibfield  {author} {\bibinfo {author} {\bibnamefont {Takeda}, \bibfnamefont
  {Kenta}}, \bibinfo {author} {\bibfnamefont {Akito}\ \bibnamefont {Noiri}},
  \bibinfo {author} {\bibfnamefont {Takashi}\ \bibnamefont {Nakajima}},
  \bibinfo {author} {\bibfnamefont {Takashi}\ \bibnamefont {Kobayashi}}, and\
  \bibinfo {author} {\bibfnamefont {Seigo}\ \bibnamefont {Tarucha}}} (\bibinfo
  {year} {2022}),\ \bibfield  {title} {\enquote {\bibinfo {title} {Quantum
  error correction with silicon spin qubits},}\ }\href
  {https://doi.org/10.1038/s41586-022-04986-6} {\bibfield  {journal} {\bibinfo
  {journal} {Nature}\ }\textbf {\bibinfo {volume} {608}}~(\bibinfo {number}
  {7924}),\ \bibinfo {pages} {682--686}}\BibitemShut {NoStop}%
\bibitem [{\citenamefont {Takeshita}\ \emph {et~al.}(2020)\citenamefont
  {Takeshita}, \citenamefont {Rubin}, \citenamefont {Jiang}, \citenamefont
  {Lee}, \citenamefont {Babbush},\ and\ \citenamefont
  {McClean}}]{takeshitaIncreasingRepresentationAccuracy2020}%
  \BibitemOpen
  \bibfield  {author} {\bibinfo {author} {\bibnamefont {Takeshita},
  \bibfnamefont {Tyler}}, \bibinfo {author} {\bibfnamefont {Nicholas~C.}\
  \bibnamefont {Rubin}}, \bibinfo {author} {\bibfnamefont {Zhang}\ \bibnamefont
  {Jiang}}, \bibinfo {author} {\bibfnamefont {Eunseok}\ \bibnamefont {Lee}},
  \bibinfo {author} {\bibfnamefont {Ryan}\ \bibnamefont {Babbush}}, and\
  \bibinfo {author} {\bibfnamefont {Jarrod~R.}\ \bibnamefont {McClean}}}
  (\bibinfo {year} {2020}),\ \bibfield  {title} {\enquote {\bibinfo {title}
  {Increasing the {{Representation Accuracy}} of {{Quantum Simulations}} of
  {{Chemistry}} without {{Extra Quantum Resources}}},}\ }\href
  {https://doi.org/10.1103/PhysRevX.10.011004} {\bibfield  {journal} {\bibinfo
  {journal} {Physical Review X}\ }\textbf {\bibinfo {volume} {10}}~(\bibinfo
  {number} {1}),\ \bibinfo {pages} {011004}}\BibitemShut {NoStop}%
\bibitem [{\citenamefont {Tannu}\ and\ \citenamefont
  {Qureshi}(2019)}]{tannuMitigatingMeasurementErrors2019}%
  \BibitemOpen
  \bibfield  {author} {\bibinfo {author} {\bibnamefont {Tannu}, \bibfnamefont
  {Swamit~S}}, and\ \bibinfo {author} {\bibfnamefont {Moinuddin~K.}\
  \bibnamefont {Qureshi}}} (\bibinfo {year} {2019}),\ \bibfield  {title}
  {\enquote {\bibinfo {title} {Mitigating {{Measurement Errors}} in {{Quantum
  Computers}} by {{Exploiting State-Dependent Bias}}},}\ }in\ \href
  {https://doi.org/10.1145/3352460.3358265} {\emph {\bibinfo {booktitle}
  {Proceedings of the 52nd {{Annual IEEE}}/{{ACM International Symposium}} on
  {{Microarchitecture}}}}},\ \bibinfo {series and number} {{{MICRO}} '52}\
  (\bibinfo  {publisher} {{Association for Computing Machinery}},\ \bibinfo
  {address} {{New York, NY, USA}})\ pp.\ \bibinfo {pages}
  {279--290}\BibitemShut {NoStop}%
\bibitem [{\citenamefont {Temme}\ \emph {et~al.}(2017)\citenamefont {Temme},
  \citenamefont {Bravyi},\ and\ \citenamefont
  {Gambetta}}]{temmeErrorMitigationShortDepth2017}%
  \BibitemOpen
  \bibfield  {author} {\bibinfo {author} {\bibnamefont {Temme}, \bibfnamefont
  {Kristan}}, \bibinfo {author} {\bibfnamefont {Sergey}\ \bibnamefont
  {Bravyi}}, and\ \bibinfo {author} {\bibfnamefont {Jay~M.}\ \bibnamefont
  {Gambetta}}} (\bibinfo {year} {2017}),\ \bibfield  {title} {\enquote
  {\bibinfo {title} {Error {{Mitigation}} for {{Short-Depth Quantum
  Circuits}}},}\ }\href {https://doi.org/10.1103/PhysRevLett.119.180509}
  {\bibfield  {journal} {\bibinfo  {journal} {Physical Review Letters}\
  }\textbf {\bibinfo {volume} {119}}~(\bibinfo {number} {18}),\ \bibinfo
  {pages} {180509}}\BibitemShut {NoStop}%
\bibitem [{\citenamefont {Terhal}(2015)}]{terhalQuantumErrorCorrection2015}%
  \BibitemOpen
  \bibfield  {author} {\bibinfo {author} {\bibnamefont {Terhal}, \bibfnamefont
  {Barbara~M}}} (\bibinfo {year} {2015}),\ \bibfield  {title} {\enquote
  {\bibinfo {title} {Quantum error correction for quantum memories},}\ }\href
  {https://doi.org/10.1103/RevModPhys.87.307} {\bibfield  {journal} {\bibinfo
  {journal} {Reviews of Modern Physics}\ }\textbf {\bibinfo {volume}
  {87}}~(\bibinfo {number} {2}),\ \bibinfo {pages} {307--346}}\BibitemShut
  {NoStop}%
\bibitem [{\citenamefont {Tindall}\ \emph {et~al.}(2023)\citenamefont
  {Tindall}, \citenamefont {Fishman}, \citenamefont {Stoudenmire},\ and\
  \citenamefont {Sels}}]{tindallEfficientTensorNetwork2023}%
  \BibitemOpen
  \bibfield  {author} {\bibinfo {author} {\bibnamefont {Tindall}, \bibfnamefont
  {Joseph}}, \bibinfo {author} {\bibfnamefont {Matt}\ \bibnamefont {Fishman}},
  \bibinfo {author} {\bibfnamefont {Miles}\ \bibnamefont {Stoudenmire}}, and\
  \bibinfo {author} {\bibfnamefont {Dries}\ \bibnamefont {Sels}}} (\bibinfo
  {year} {2023}),\ \href {http://arxiv.org/abs/2306.14887} {\enquote {\bibinfo
  {title} {Efficient tensor network simulation of {{IBM}}'s kicked {{Ising}}
  experiment},}\ }\bibinfo {howpublished} {arXiv:2306.14887
  [quant-ph]}\BibitemShut {NoStop}%
\bibitem [{\citenamefont {Tran}\ \emph {et~al.}(2021)\citenamefont {Tran},
  \citenamefont {Su}, \citenamefont {Carney},\ and\ \citenamefont
  {Taylor}}]{tranFasterDigitalQuantum2021}%
  \BibitemOpen
  \bibfield  {author} {\bibinfo {author} {\bibnamefont {Tran}, \bibfnamefont
  {Minh~C}}, \bibinfo {author} {\bibfnamefont {Yuan}\ \bibnamefont {Su}},
  \bibinfo {author} {\bibfnamefont {Daniel}\ \bibnamefont {Carney}}, and\
  \bibinfo {author} {\bibfnamefont {Jacob~M.}\ \bibnamefont {Taylor}}}
  (\bibinfo {year} {2021}),\ \bibfield  {title} {\enquote {\bibinfo {title}
  {Faster {{Digital Quantum Simulation}} by {{Symmetry Protection}}},}\ }\href
  {https://doi.org/10.1103/PRXQuantum.2.010323} {\bibfield  {journal} {\bibinfo
   {journal} {PRX Quantum}\ }\textbf {\bibinfo {volume} {2}}~(\bibinfo {number}
  {1}),\ \bibinfo {pages} {010323}}\BibitemShut {NoStop}%
\bibitem [{\citenamefont {Tsubouchi}\ \emph
  {et~al.}(2023{\natexlab{a}})\citenamefont {Tsubouchi}, \citenamefont
  {Sagawa},\ and\ \citenamefont {Yoshioka}}]{tsubouchiUniversalCostBound2023}%
  \BibitemOpen
  \bibfield  {author} {\bibinfo {author} {\bibnamefont {Tsubouchi},
  \bibfnamefont {Kento}}, \bibinfo {author} {\bibfnamefont {Takahiro}\
  \bibnamefont {Sagawa}}, and\ \bibinfo {author} {\bibfnamefont {Nobuyuki}\
  \bibnamefont {Yoshioka}}} (\bibinfo {year} {2023}{\natexlab{a}}),\ \bibfield
  {title} {\enquote {\bibinfo {title} {Universal {{Cost Bound}} of {{Quantum
  Error Mitigation Based}} on {{Quantum Estimation Theory}}},}\ }\href
  {https://doi.org/10.1103/PhysRevLett.131.210601} {\bibfield  {journal}
  {\bibinfo  {journal} {Physical Review Letters}\ }\textbf {\bibinfo {volume}
  {131}}~(\bibinfo {number} {21}),\ \bibinfo {pages} {210601}}\BibitemShut
  {NoStop}%
\bibitem [{\citenamefont {Tsubouchi}\ \emph
  {et~al.}(2023{\natexlab{b}})\citenamefont {Tsubouchi}, \citenamefont
  {Suzuki}, \citenamefont {Tokunaga}, \citenamefont {Yoshioka},\ and\
  \citenamefont {Endo}}]{tsubouchiVirtualQuantumError2023}%
  \BibitemOpen
  \bibfield  {author} {\bibinfo {author} {\bibnamefont {Tsubouchi},
  \bibfnamefont {Kento}}, \bibinfo {author} {\bibfnamefont {Yasunari}\
  \bibnamefont {Suzuki}}, \bibinfo {author} {\bibfnamefont {Yuuki}\
  \bibnamefont {Tokunaga}}, \bibinfo {author} {\bibfnamefont {Nobuyuki}\
  \bibnamefont {Yoshioka}}, and\ \bibinfo {author} {\bibfnamefont {Suguru}\
  \bibnamefont {Endo}}} (\bibinfo {year} {2023}{\natexlab{b}}),\ \bibfield
  {title} {\enquote {\bibinfo {title} {Virtual quantum error detection},}\
  }\href {https://doi.org/10.1103/PhysRevA.108.042426} {\bibfield  {journal}
  {\bibinfo  {journal} {Physical Review A}\ }\textbf {\bibinfo {volume}
  {108}}~(\bibinfo {number} {4}),\ \bibinfo {pages} {042426}}\BibitemShut
  {NoStop}%
\bibitem [{\citenamefont {Tsuchimochi}\ \emph {et~al.}(2020)\citenamefont
  {Tsuchimochi}, \citenamefont {Mori},\ and\ \citenamefont
  {{Ten-no}}}]{tsuchimochiSpinprojectionQuantumComputation2020}%
  \BibitemOpen
  \bibfield  {author} {\bibinfo {author} {\bibnamefont {Tsuchimochi},
  \bibfnamefont {Takashi}}, \bibinfo {author} {\bibfnamefont {Yuto}\
  \bibnamefont {Mori}}, and\ \bibinfo {author} {\bibfnamefont {Seiichiro~L.}\
  \bibnamefont {{Ten-no}}}} (\bibinfo {year} {2020}),\ \bibfield  {title}
  {\enquote {\bibinfo {title} {Spin-projection for quantum computation: {{A}}
  low-depth approach to strong correlation},}\ }\href
  {https://doi.org/10.1103/PhysRevResearch.2.043142} {\bibfield  {journal}
  {\bibinfo  {journal} {Physical Review Research}\ }\textbf {\bibinfo {volume}
  {2}}~(\bibinfo {number} {4}),\ \bibinfo {pages} {043142}}\BibitemShut
  {NoStop}%
\bibitem [{\citenamefont {Unruh}(1995)}]{unruhMaintainingCoherenceQuantum1995}%
  \BibitemOpen
  \bibfield  {author} {\bibinfo {author} {\bibnamefont {Unruh}, \bibfnamefont
  {W~G}}} (\bibinfo {year} {1995}),\ \bibfield  {title} {\enquote {\bibinfo
  {title} {Maintaining coherence in quantum computers},}\ }\href
  {https://doi.org/10.1103/PhysRevA.51.992} {\bibfield  {journal} {\bibinfo
  {journal} {Physical Review A}\ }\textbf {\bibinfo {volume} {51}}~(\bibinfo
  {number} {2}),\ \bibinfo {pages} {992--997}}\BibitemShut {NoStop}%
\bibitem [{\citenamefont {Urbanek}\ \emph {et~al.}(2020)\citenamefont
  {Urbanek}, \citenamefont {Camps}, \citenamefont {Van~Beeumen},\ and\
  \citenamefont {{de Jong}}}]{urbanekChemistryQuantumComputers2020}%
  \BibitemOpen
  \bibfield  {author} {\bibinfo {author} {\bibnamefont {Urbanek}, \bibfnamefont
  {Miroslav}}, \bibinfo {author} {\bibfnamefont {Daan}\ \bibnamefont {Camps}},
  \bibinfo {author} {\bibfnamefont {Roel}\ \bibnamefont {Van~Beeumen}}, and\
  \bibinfo {author} {\bibfnamefont {Wibe~A.}\ \bibnamefont {{de Jong}}}}
  (\bibinfo {year} {2020}),\ \bibfield  {title} {\enquote {\bibinfo {title}
  {Chemistry on {{Quantum Computers}} with {{Virtual Quantum Subspace
  Expansion}}},}\ }\href {https://doi.org/10.1021/acs.jctc.0c00447} {\bibfield
  {journal} {\bibinfo  {journal} {Journal of Chemical Theory and Computation}\
  }\textbf {\bibinfo {volume} {16}}~(\bibinfo {number} {9}),\ \bibinfo {pages}
  {5425--5431}}\BibitemShut {NoStop}%
\bibitem [{\citenamefont {Urbanek}\ \emph {et~al.}(2021)\citenamefont
  {Urbanek}, \citenamefont {Nachman}, \citenamefont {Pascuzzi}, \citenamefont
  {He}, \citenamefont {Bauer},\ and\ \citenamefont {{de
  Jong}}}]{urbanekMitigatingDepolarizingNoise2021}%
  \BibitemOpen
  \bibfield  {author} {\bibinfo {author} {\bibnamefont {Urbanek}, \bibfnamefont
  {Miroslav}}, \bibinfo {author} {\bibfnamefont {Benjamin}\ \bibnamefont
  {Nachman}}, \bibinfo {author} {\bibfnamefont {Vincent~R.}\ \bibnamefont
  {Pascuzzi}}, \bibinfo {author} {\bibfnamefont {Andre}\ \bibnamefont {He}},
  \bibinfo {author} {\bibfnamefont {Christian~W.}\ \bibnamefont {Bauer}}, and\
  \bibinfo {author} {\bibfnamefont {Wibe~A.}\ \bibnamefont {{de Jong}}}}
  (\bibinfo {year} {2021}),\ \bibfield  {title} {\enquote {\bibinfo {title}
  {Mitigating {{Depolarizing Noise}} on {{Quantum Computers}} with
  {{Noise-Estimation Circuits}}},}\ }\href
  {https://doi.org/10.1103/PhysRevLett.127.270502} {\bibfield  {journal}
  {\bibinfo  {journal} {Physical Review Letters}\ }\textbf {\bibinfo {volume}
  {127}}~(\bibinfo {number} {27}),\ \bibinfo {pages} {270502}}\BibitemShut
  {NoStop}%
\bibitem [{\citenamefont {{van den Berg}}\ \emph {et~al.}(2023)\citenamefont
  {{van den Berg}}, \citenamefont {Minev}, \citenamefont {Kandala},\ and\
  \citenamefont {Temme}}]{vandenbergProbabilisticErrorCancellation2023}%
  \BibitemOpen
  \bibfield  {author} {\bibinfo {author} {\bibnamefont {{van den Berg}},
  \bibfnamefont {Ewout}}, \bibinfo {author} {\bibfnamefont {Zlatko~K.}\
  \bibnamefont {Minev}}, \bibinfo {author} {\bibfnamefont {Abhinav}\
  \bibnamefont {Kandala}}, and\ \bibinfo {author} {\bibfnamefont {Kristan}\
  \bibnamefont {Temme}}} (\bibinfo {year} {2023}),\ \bibfield  {title}
  {\enquote {\bibinfo {title} {Probabilistic error cancellation with sparse
  {{Pauli}}{\textendash}{{Lindblad}} models on noisy quantum processors},}\
  }\href {https://doi.org/10.1038/s41567-023-02042-2} {\bibinfo  {journal}
  {Nature Physics}\ ,\ \bibinfo {pages} {1--6}}\BibitemShut {NoStop}%
\bibitem [{\citenamefont {{van den Berg}}\ \emph {et~al.}(2022)\citenamefont
  {{van den Berg}}, \citenamefont {Minev},\ and\ \citenamefont
  {Temme}}]{vandenbergModelfreeReadouterrorMitigation2022}%
  \BibitemOpen
\bibfield  {journal} {  }\bibfield  {author} {\bibinfo {author} {\bibnamefont
  {{van den Berg}}, \bibfnamefont {Ewout}}, \bibinfo {author} {\bibfnamefont
  {Zlatko~K.}\ \bibnamefont {Minev}}, and\ \bibinfo {author} {\bibfnamefont
  {Kristan}\ \bibnamefont {Temme}}} (\bibinfo {year} {2022}),\ \bibfield
  {title} {\enquote {\bibinfo {title} {Model-free readout-error mitigation for
  quantum expectation values},}\ }\href
  {https://doi.org/10.1103/PhysRevA.105.032620} {\bibfield  {journal} {\bibinfo
   {journal} {Physical Review A}\ }\textbf {\bibinfo {volume} {105}}~(\bibinfo
  {number} {3}),\ \bibinfo {pages} {032620}}\BibitemShut {NoStop}%
\bibitem [{\citenamefont {Vazquez}\ \emph {et~al.}(2022)\citenamefont
  {Vazquez}, \citenamefont {Hiptmair},\ and\ \citenamefont
  {Woerner}}]{vazquezEnhancingQuantumLinear2022}%
  \BibitemOpen
  \bibfield  {author} {\bibinfo {author} {\bibnamefont {Vazquez}, \bibfnamefont
  {Almudena~Carrera}}, \bibinfo {author} {\bibfnamefont {Ralf}\ \bibnamefont
  {Hiptmair}}, and\ \bibinfo {author} {\bibfnamefont {Stefan}\ \bibnamefont
  {Woerner}}} (\bibinfo {year} {2022}),\ \bibfield  {title} {\enquote {\bibinfo
  {title} {Enhancing the {{Quantum Linear Systems Algorithm Using Richardson
  Extrapolation}}},}\ }\href {https://doi.org/10.1145/3490631} {\bibfield
  {journal} {\bibinfo  {journal} {ACM Transactions on Quantum Computing}\
  }\textbf {\bibinfo {volume} {3}}~(\bibinfo {number} {1}),\ \bibinfo {pages}
  {2:1--2:37}}\BibitemShut {NoStop}%
\bibitem [{\citenamefont {Vovrosh}\ \emph {et~al.}(2021)\citenamefont
  {Vovrosh}, \citenamefont {Khosla}, \citenamefont {Greenaway}, \citenamefont
  {Self}, \citenamefont {Kim},\ and\ \citenamefont
  {Knolle}}]{vovroshSimpleMitigationGlobal2021}%
  \BibitemOpen
  \bibfield  {author} {\bibinfo {author} {\bibnamefont {Vovrosh}, \bibfnamefont
  {Joseph}}, \bibinfo {author} {\bibfnamefont {Kiran~E.}\ \bibnamefont
  {Khosla}}, \bibinfo {author} {\bibfnamefont {Sean}\ \bibnamefont
  {Greenaway}}, \bibinfo {author} {\bibfnamefont {Christopher}\ \bibnamefont
  {Self}}, \bibinfo {author} {\bibfnamefont {M.~S.}\ \bibnamefont {Kim}}, and\
  \bibinfo {author} {\bibfnamefont {Johannes}\ \bibnamefont {Knolle}}}
  (\bibinfo {year} {2021}),\ \bibfield  {title} {\enquote {\bibinfo {title}
  {Simple mitigation of global depolarizing errors in quantum simulations},}\
  }\href {https://doi.org/10.1103/PhysRevE.104.035309} {\bibfield  {journal}
  {\bibinfo  {journal} {Physical Review E}\ }\textbf {\bibinfo {volume}
  {104}}~(\bibinfo {number} {3}),\ \bibinfo {pages} {035309}}\BibitemShut
  {NoStop}%
\bibitem [{\citenamefont {Vuillot}(2018)}]{vuillotErrorDetectionHelpful2018}%
  \BibitemOpen
  \bibfield  {author} {\bibinfo {author} {\bibnamefont {Vuillot}, \bibfnamefont
  {Christophe}}} (\bibinfo {year} {2018}),\ \bibfield  {title} {\enquote
  {\bibinfo {title} {Is error detection helpful on {{IBM 5Q}} chips?}}\ }\href
  {https://doi.org/10.26421/QIC18.11-12-4} {\bibfield  {journal} {\bibinfo
  {journal} {Quantum Information and Computation}\ }\textbf {\bibinfo {volume}
  {18}}~(\bibinfo {number} {11\&12}),\ \bibinfo {pages} {949--964}}\BibitemShut
  {NoStop}%
\bibitem [{\citenamefont {Wallman}\ \emph {et~al.}(2015)\citenamefont
  {Wallman}, \citenamefont {Granade}, \citenamefont {Harper},\ and\
  \citenamefont {Flammia}}]{wallmanEstimatingCoherenceNoise2015}%
  \BibitemOpen
  \bibfield  {author} {\bibinfo {author} {\bibnamefont {Wallman}, \bibfnamefont
  {Joel}}, \bibinfo {author} {\bibfnamefont {Chris}\ \bibnamefont {Granade}},
  \bibinfo {author} {\bibfnamefont {Robin}\ \bibnamefont {Harper}}, and\
  \bibinfo {author} {\bibfnamefont {Steven~T.}\ \bibnamefont {Flammia}}}
  (\bibinfo {year} {2015}),\ \bibfield  {title} {\enquote {\bibinfo {title}
  {Estimating the coherence of noise},}\ }\href
  {https://doi.org/10.1088/1367-2630/17/11/113020} {\bibfield  {journal}
  {\bibinfo  {journal} {New Journal of Physics}\ }\textbf {\bibinfo {volume}
  {17}}~(\bibinfo {number} {11}),\ \bibinfo {pages} {113020}}\BibitemShut
  {NoStop}%
\bibitem [{\citenamefont {Wallman}\ and\ \citenamefont
  {Emerson}(2016)}]{wallmanNoiseTailoringScalable2016}%
  \BibitemOpen
  \bibfield  {author} {\bibinfo {author} {\bibnamefont {Wallman}, \bibfnamefont
  {Joel~J}}, and\ \bibinfo {author} {\bibfnamefont {Joseph}\ \bibnamefont
  {Emerson}}} (\bibinfo {year} {2016}),\ \bibfield  {title} {\enquote {\bibinfo
  {title} {Noise tailoring for scalable quantum computation via randomized
  compiling},}\ }\href {https://doi.org/10.1103/PhysRevA.94.052325} {\bibfield
  {journal} {\bibinfo  {journal} {Physical Review A}\ }\textbf {\bibinfo
  {volume} {94}}~(\bibinfo {number} {5}),\ \bibinfo {pages}
  {052325}}\BibitemShut {NoStop}%
\bibitem [{\citenamefont {Wallraff}\ \emph {et~al.}(2004)\citenamefont
  {Wallraff}, \citenamefont {Schuster}, \citenamefont {Blais}, \citenamefont
  {Frunzio}, \citenamefont {Huang}, \citenamefont {Majer}, \citenamefont
  {Kumar}, \citenamefont {Girvin},\ and\ \citenamefont
  {Schoelkopf}}]{wallraffStrongCouplingSingle2004}%
  \BibitemOpen
  \bibfield  {author} {\bibinfo {author} {\bibnamefont {Wallraff},
  \bibfnamefont {A}}, \bibinfo {author} {\bibfnamefont {D.~I.}\ \bibnamefont
  {Schuster}}, \bibinfo {author} {\bibfnamefont {A.}~\bibnamefont {Blais}},
  \bibinfo {author} {\bibfnamefont {L.}~\bibnamefont {Frunzio}}, \bibinfo
  {author} {\bibfnamefont {R.-S.}\ \bibnamefont {Huang}}, \bibinfo {author}
  {\bibfnamefont {J.}~\bibnamefont {Majer}}, \bibinfo {author} {\bibfnamefont
  {S.}~\bibnamefont {Kumar}}, \bibinfo {author} {\bibfnamefont {S.~M.}\
  \bibnamefont {Girvin}}, and\ \bibinfo {author} {\bibfnamefont {R.~J.}\
  \bibnamefont {Schoelkopf}}} (\bibinfo {year} {2004}),\ \bibfield  {title}
  {\enquote {\bibinfo {title} {Strong coupling of a single photon to a
  superconducting qubit using circuit quantum electrodynamics},}\ }\href
  {https://doi.org/10.1038/nature02851} {\bibfield  {journal} {\bibinfo
  {journal} {Nature}\ }\textbf {\bibinfo {volume} {431}}~(\bibinfo {number}
  {7005}),\ \bibinfo {pages} {162--167}}\BibitemShut {NoStop}%
\bibitem [{\citenamefont {Wang}\ \emph
  {et~al.}(2021{\natexlab{a}})\citenamefont {Wang}, \citenamefont {Czarnik},
  \citenamefont {Arrasmith}, \citenamefont {Cerezo}, \citenamefont {Cincio},\
  and\ \citenamefont {Coles}}]{wangCanErrorMitigation2021}%
  \BibitemOpen
  \bibfield  {author} {\bibinfo {author} {\bibnamefont {Wang}, \bibfnamefont
  {Samson}}, \bibinfo {author} {\bibfnamefont {Piotr}\ \bibnamefont {Czarnik}},
  \bibinfo {author} {\bibfnamefont {Andrew}\ \bibnamefont {Arrasmith}},
  \bibinfo {author} {\bibfnamefont {M.}~\bibnamefont {Cerezo}}, \bibinfo
  {author} {\bibfnamefont {Lukasz}\ \bibnamefont {Cincio}}, and\ \bibinfo
  {author} {\bibfnamefont {Patrick~J.}\ \bibnamefont {Coles}}} (\bibinfo {year}
  {2021}{\natexlab{a}}),\ \href {http://arxiv.org/abs/2109.01051} {\enquote
  {\bibinfo {title} {Can {{Error Mitigation Improve Trainability}} of {{Noisy
  Variational Quantum Algorithms}}?}}\ }\bibinfo {howpublished}
  {arXiv:2109.01051 [quant-ph]}\BibitemShut {NoStop}%
\bibitem [{\citenamefont {Wang}\ \emph
  {et~al.}(2021{\natexlab{b}})\citenamefont {Wang}, \citenamefont {Fontana},
  \citenamefont {Cerezo}, \citenamefont {Sharma}, \citenamefont {Sone},
  \citenamefont {Cincio},\ and\ \citenamefont
  {Coles}}]{wangNoiseinducedBarrenPlateaus2021}%
  \BibitemOpen
  \bibfield  {author} {\bibinfo {author} {\bibnamefont {Wang}, \bibfnamefont
  {Samson}}, \bibinfo {author} {\bibfnamefont {Enrico}\ \bibnamefont
  {Fontana}}, \bibinfo {author} {\bibfnamefont {M.}~\bibnamefont {Cerezo}},
  \bibinfo {author} {\bibfnamefont {Kunal}\ \bibnamefont {Sharma}}, \bibinfo
  {author} {\bibfnamefont {Akira}\ \bibnamefont {Sone}}, \bibinfo {author}
  {\bibfnamefont {Lukasz}\ \bibnamefont {Cincio}}, and\ \bibinfo {author}
  {\bibfnamefont {Patrick~J.}\ \bibnamefont {Coles}}} (\bibinfo {year}
  {2021}{\natexlab{b}}),\ \bibfield  {title} {\enquote {\bibinfo {title}
  {Noise-induced barren plateaus in variational quantum algorithms},}\ }\href
  {https://doi.org/10.1038/s41467-021-27045-6} {\bibfield  {journal} {\bibinfo
  {journal} {Nature Communications}\ }\textbf {\bibinfo {volume} {12}},\
  \bibinfo {pages} {6961}}\BibitemShut {NoStop}%
\bibitem [{\citenamefont {Wang}\ \emph
  {et~al.}(2021{\natexlab{c}})\citenamefont {Wang}, \citenamefont {Chen},
  \citenamefont {Song}, \citenamefont {Qin}, \citenamefont {Li}, \citenamefont
  {Guo}, \citenamefont {Wang}, \citenamefont {Song},\ and\ \citenamefont
  {Li}}]{wangScalableEvaluationQuantumCircuit2021}%
  \BibitemOpen
  \bibfield  {author} {\bibinfo {author} {\bibnamefont {Wang}, \bibfnamefont
  {Zhen}}, \bibinfo {author} {\bibfnamefont {Yanzhu}\ \bibnamefont {Chen}},
  \bibinfo {author} {\bibfnamefont {Zixuan}\ \bibnamefont {Song}}, \bibinfo
  {author} {\bibfnamefont {Dayue}\ \bibnamefont {Qin}}, \bibinfo {author}
  {\bibfnamefont {Hekang}\ \bibnamefont {Li}}, \bibinfo {author} {\bibfnamefont
  {Qiujiang}\ \bibnamefont {Guo}}, \bibinfo {author} {\bibfnamefont
  {H.}~\bibnamefont {Wang}}, \bibinfo {author} {\bibfnamefont {Chao}\
  \bibnamefont {Song}}, and\ \bibinfo {author} {\bibfnamefont {Ying}\
  \bibnamefont {Li}}} (\bibinfo {year} {2021}{\natexlab{c}}),\ \bibfield
  {title} {\enquote {\bibinfo {title} {Scalable {{Evaluation}} of
  {{Quantum-Circuit Error Loss Using Clifford Sampling}}},}\ }\href
  {https://doi.org/10.1103/PhysRevLett.126.080501} {\bibfield  {journal}
  {\bibinfo  {journal} {Physical Review Letters}\ }\textbf {\bibinfo {volume}
  {126}}~(\bibinfo {number} {8}),\ \bibinfo {pages} {080501}}\BibitemShut
  {NoStop}%
\bibitem [{\citenamefont {Watanabe}\ \emph {et~al.}(2010)\citenamefont
  {Watanabe}, \citenamefont {Sagawa},\ and\ \citenamefont
  {Ueda}}]{watanabeOptimalMeasurementNoisy2010}%
  \BibitemOpen
  \bibfield  {author} {\bibinfo {author} {\bibnamefont {Watanabe},
  \bibfnamefont {Yu}}, \bibinfo {author} {\bibfnamefont {Takahiro}\
  \bibnamefont {Sagawa}}, and\ \bibinfo {author} {\bibfnamefont {Masahito}\
  \bibnamefont {Ueda}}} (\bibinfo {year} {2010}),\ \bibfield  {title} {\enquote
  {\bibinfo {title} {Optimal {{Measurement}} on {{Noisy Quantum Systems}}},}\
  }\href {https://doi.org/10.1103/PhysRevLett.104.020401} {\bibfield  {journal}
  {\bibinfo  {journal} {Physical Review Letters}\ }\textbf {\bibinfo {volume}
  {104}}~(\bibinfo {number} {2}),\ \bibinfo {pages} {020401}}\BibitemShut
  {NoStop}%
\bibitem [{\citenamefont {Wecker}\ \emph {et~al.}(2015)\citenamefont {Wecker},
  \citenamefont {Hastings},\ and\ \citenamefont
  {Troyer}}]{weckerProgressPracticalQuantum2015}%
  \BibitemOpen
  \bibfield  {author} {\bibinfo {author} {\bibnamefont {Wecker}, \bibfnamefont
  {Dave}}, \bibinfo {author} {\bibfnamefont {Matthew~B.}\ \bibnamefont
  {Hastings}}, and\ \bibinfo {author} {\bibfnamefont {Matthias}\ \bibnamefont
  {Troyer}}} (\bibinfo {year} {2015}),\ \bibfield  {title} {\enquote {\bibinfo
  {title} {Progress towards practical quantum variational algorithms},}\ }\href
  {https://doi.org/10.1103/PhysRevA.92.042303} {\bibfield  {journal} {\bibinfo
  {journal} {Physical Review A}\ }\textbf {\bibinfo {volume} {92}}~(\bibinfo
  {number} {4}),\ \bibinfo {pages} {042303}}\BibitemShut {NoStop}%
\bibitem [{\citenamefont {Wu}\ \emph {et~al.}(2021)\citenamefont {Wu},
  \citenamefont {Bao}, \citenamefont {Cao}, \citenamefont {Chen}, \citenamefont
  {Chen}, \citenamefont {Chen}, \citenamefont {Chung}, \citenamefont {Deng},
  \citenamefont {Du}, \citenamefont {Fan}, \citenamefont {Gong}, \citenamefont
  {Guo}, \citenamefont {Guo}, \citenamefont {Guo}, \citenamefont {Han},
  \citenamefont {Hong}, \citenamefont {Huang}, \citenamefont {Huo},
  \citenamefont {Li}, \citenamefont {Li}, \citenamefont {Li}, \citenamefont
  {Li}, \citenamefont {Liang}, \citenamefont {Lin}, \citenamefont {Lin},
  \citenamefont {Qian}, \citenamefont {Qiao}, \citenamefont {Rong},
  \citenamefont {Su}, \citenamefont {Sun}, \citenamefont {Wang}, \citenamefont
  {Wang}, \citenamefont {Wu}, \citenamefont {Xu}, \citenamefont {Yan},
  \citenamefont {Yang}, \citenamefont {Yang}, \citenamefont {Ye}, \citenamefont
  {Yin}, \citenamefont {Ying}, \citenamefont {Yu}, \citenamefont {Zha},
  \citenamefont {Zhang}, \citenamefont {Zhang}, \citenamefont {Zhang},
  \citenamefont {Zhang}, \citenamefont {Zhao}, \citenamefont {Zhao},
  \citenamefont {Zhou}, \citenamefont {Zhu}, \citenamefont {Lu}, \citenamefont
  {Peng}, \citenamefont {Zhu},\ and\ \citenamefont
  {Pan}}]{wuStrongQuantumComputational2021}%
  \BibitemOpen
  \bibfield  {author} {\bibinfo {author} {\bibnamefont {Wu}, \bibfnamefont
  {Yulin}}, \bibinfo {author} {\bibfnamefont {Wan-Su}\ \bibnamefont {Bao}},
  \bibinfo {author} {\bibfnamefont {Sirui}\ \bibnamefont {Cao}}, \bibinfo
  {author} {\bibfnamefont {Fusheng}\ \bibnamefont {Chen}}, \bibinfo {author}
  {\bibfnamefont {Ming-Cheng}\ \bibnamefont {Chen}}, \bibinfo {author}
  {\bibfnamefont {Xiawei}\ \bibnamefont {Chen}}, \bibinfo {author}
  {\bibfnamefont {Tung-Hsun}\ \bibnamefont {Chung}}, \bibinfo {author}
  {\bibfnamefont {Hui}\ \bibnamefont {Deng}}, \bibinfo {author} {\bibfnamefont
  {Yajie}\ \bibnamefont {Du}}, \bibinfo {author} {\bibfnamefont {Daojin}\
  \bibnamefont {Fan}}, \bibinfo {author} {\bibfnamefont {Ming}\ \bibnamefont
  {Gong}}, \bibinfo {author} {\bibfnamefont {Cheng}\ \bibnamefont {Guo}},
  \bibinfo {author} {\bibfnamefont {Chu}\ \bibnamefont {Guo}}, \bibinfo
  {author} {\bibfnamefont {Shaojun}\ \bibnamefont {Guo}}, \bibinfo {author}
  {\bibfnamefont {Lianchen}\ \bibnamefont {Han}}, \bibinfo {author}
  {\bibfnamefont {Linyin}\ \bibnamefont {Hong}}, \bibinfo {author}
  {\bibfnamefont {He-Liang}\ \bibnamefont {Huang}}, \bibinfo {author}
  {\bibfnamefont {Yong-Heng}\ \bibnamefont {Huo}}, \bibinfo {author}
  {\bibfnamefont {Liping}\ \bibnamefont {Li}}, \bibinfo {author} {\bibfnamefont
  {Na}~\bibnamefont {Li}}, \bibinfo {author} {\bibfnamefont {Shaowei}\
  \bibnamefont {Li}}, \bibinfo {author} {\bibfnamefont {Yuan}\ \bibnamefont
  {Li}}, \bibinfo {author} {\bibfnamefont {Futian}\ \bibnamefont {Liang}},
  \bibinfo {author} {\bibfnamefont {Chun}\ \bibnamefont {Lin}}, \bibinfo
  {author} {\bibfnamefont {Jin}\ \bibnamefont {Lin}}, \bibinfo {author}
  {\bibfnamefont {Haoran}\ \bibnamefont {Qian}}, \bibinfo {author}
  {\bibfnamefont {Dan}\ \bibnamefont {Qiao}}, \bibinfo {author} {\bibfnamefont
  {Hao}\ \bibnamefont {Rong}}, \bibinfo {author} {\bibfnamefont {Hong}\
  \bibnamefont {Su}}, \bibinfo {author} {\bibfnamefont {Lihua}\ \bibnamefont
  {Sun}}, \bibinfo {author} {\bibfnamefont {Liangyuan}\ \bibnamefont {Wang}},
  \bibinfo {author} {\bibfnamefont {Shiyu}\ \bibnamefont {Wang}}, \bibinfo
  {author} {\bibfnamefont {Dachao}\ \bibnamefont {Wu}}, \bibinfo {author}
  {\bibfnamefont {Yu}~\bibnamefont {Xu}}, \bibinfo {author} {\bibfnamefont
  {Kai}\ \bibnamefont {Yan}}, \bibinfo {author} {\bibfnamefont {Weifeng}\
  \bibnamefont {Yang}}, \bibinfo {author} {\bibfnamefont {Yang}\ \bibnamefont
  {Yang}}, \bibinfo {author} {\bibfnamefont {Yangsen}\ \bibnamefont {Ye}},
  \bibinfo {author} {\bibfnamefont {Jianghan}\ \bibnamefont {Yin}}, \bibinfo
  {author} {\bibfnamefont {Chong}\ \bibnamefont {Ying}}, \bibinfo {author}
  {\bibfnamefont {Jiale}\ \bibnamefont {Yu}}, \bibinfo {author} {\bibfnamefont
  {Chen}\ \bibnamefont {Zha}}, \bibinfo {author} {\bibfnamefont {Cha}\
  \bibnamefont {Zhang}}, \bibinfo {author} {\bibfnamefont {Haibin}\
  \bibnamefont {Zhang}}, \bibinfo {author} {\bibfnamefont {Kaili}\ \bibnamefont
  {Zhang}}, \bibinfo {author} {\bibfnamefont {Yiming}\ \bibnamefont {Zhang}},
  \bibinfo {author} {\bibfnamefont {Han}\ \bibnamefont {Zhao}}, \bibinfo
  {author} {\bibfnamefont {Youwei}\ \bibnamefont {Zhao}}, \bibinfo {author}
  {\bibfnamefont {Liang}\ \bibnamefont {Zhou}}, \bibinfo {author}
  {\bibfnamefont {Qingling}\ \bibnamefont {Zhu}}, \bibinfo {author}
  {\bibfnamefont {Chao-Yang}\ \bibnamefont {Lu}}, \bibinfo {author}
  {\bibfnamefont {Cheng-Zhi}\ \bibnamefont {Peng}}, \bibinfo {author}
  {\bibfnamefont {Xiaobo}\ \bibnamefont {Zhu}}, and\ \bibinfo {author}
  {\bibfnamefont {Jian-Wei}\ \bibnamefont {Pan}}} (\bibinfo {year} {2021}),\
  \bibfield  {title} {\enquote {\bibinfo {title} {Strong {{Quantum
  Computational Advantage Using}} a {{Superconducting Quantum Processor}}},}\
  }\href {https://doi.org/10.1103/PhysRevLett.127.180501} {\bibfield  {journal}
  {\bibinfo  {journal} {Physical Review Letters}\ }\textbf {\bibinfo {volume}
  {127}}~(\bibinfo {number} {18}),\ \bibinfo {pages} {180501}}\BibitemShut
  {NoStop}%
\bibitem [{\citenamefont {Xue}\ \emph {et~al.}(2022)\citenamefont {Xue},
  \citenamefont {Russ}, \citenamefont {Samkharadze}, \citenamefont {Undseth},
  \citenamefont {Sammak}, \citenamefont {Scappucci},\ and\ \citenamefont
  {Vandersypen}}]{xueQuantumLogicSpin2022}%
  \BibitemOpen
  \bibfield  {author} {\bibinfo {author} {\bibnamefont {Xue}, \bibfnamefont
  {Xiao}}, \bibinfo {author} {\bibfnamefont {Maximilian}\ \bibnamefont {Russ}},
  \bibinfo {author} {\bibfnamefont {Nodar}\ \bibnamefont {Samkharadze}},
  \bibinfo {author} {\bibfnamefont {Brennan}\ \bibnamefont {Undseth}}, \bibinfo
  {author} {\bibfnamefont {Amir}\ \bibnamefont {Sammak}}, \bibinfo {author}
  {\bibfnamefont {Giordano}\ \bibnamefont {Scappucci}}, and\ \bibinfo {author}
  {\bibfnamefont {Lieven M.~K.}\ \bibnamefont {Vandersypen}}} (\bibinfo {year}
  {2022}),\ \bibfield  {title} {\enquote {\bibinfo {title} {Quantum logic with
  spin qubits crossing the surface code threshold},}\ }\href
  {https://doi.org/10.1038/s41586-021-04273-w} {\bibfield  {journal} {\bibinfo
  {journal} {Nature}\ }\textbf {\bibinfo {volume} {601}}~(\bibinfo {number}
  {7893}),\ \bibinfo {pages} {343--347}}\BibitemShut {NoStop}%
\bibitem [{\citenamefont {Yamamoto}\ \emph {et~al.}(2022)\citenamefont
  {Yamamoto}, \citenamefont {Endo}, \citenamefont {Hakoshima}, \citenamefont
  {Matsuzaki},\ and\ \citenamefont
  {Tokunaga}}]{yamamotoErrormitigatedQuantumMetrology2022}%
  \BibitemOpen
  \bibfield  {author} {\bibinfo {author} {\bibnamefont {Yamamoto},
  \bibfnamefont {Kaoru}}, \bibinfo {author} {\bibfnamefont {Suguru}\
  \bibnamefont {Endo}}, \bibinfo {author} {\bibfnamefont {Hideaki}\
  \bibnamefont {Hakoshima}}, \bibinfo {author} {\bibfnamefont {Yuichiro}\
  \bibnamefont {Matsuzaki}}, and\ \bibinfo {author} {\bibfnamefont {Yuuki}\
  \bibnamefont {Tokunaga}}} (\bibinfo {year} {2022}),\ \bibfield  {title}
  {\enquote {\bibinfo {title} {Error-{{Mitigated Quantum Metrology}} via
  {{Virtual Purification}}},}\ }\href
  {https://doi.org/10.1103/PhysRevLett.129.250503} {\bibfield  {journal}
  {\bibinfo  {journal} {Physical Review Letters}\ }\textbf {\bibinfo {volume}
  {129}}~(\bibinfo {number} {25}),\ \bibinfo {pages} {250503}}\BibitemShut
  {NoStop}%
\bibitem [{\citenamefont {Yang}\ \emph {et~al.}(2021)\citenamefont {Yang},
  \citenamefont {Lu},\ and\ \citenamefont
  {Li}}]{yangAcceleratedQuantumMonte2021}%
  \BibitemOpen
  \bibfield  {author} {\bibinfo {author} {\bibnamefont {Yang}, \bibfnamefont
  {Yongdan}}, \bibinfo {author} {\bibfnamefont {Bing-Nan}\ \bibnamefont {Lu}},
  and\ \bibinfo {author} {\bibfnamefont {Ying}\ \bibnamefont {Li}}} (\bibinfo
  {year} {2021}),\ \bibfield  {title} {\enquote {\bibinfo {title} {Accelerated
  {{Quantum Monte Carlo}} with {{Mitigated Error}} on {{Noisy Quantum
  Computer}}},}\ }\href {https://doi.org/10.1103/PRXQuantum.2.040361}
  {\bibfield  {journal} {\bibinfo  {journal} {PRX Quantum}\ }\textbf {\bibinfo
  {volume} {2}}~(\bibinfo {number} {4}),\ \bibinfo {pages}
  {040361}}\BibitemShut {NoStop}%
\bibitem [{\citenamefont {Yen}\ \emph {et~al.}(2019)\citenamefont {Yen},
  \citenamefont {Lang},\ and\ \citenamefont
  {Izmaylov}}]{yenExactApproximateSymmetry2019}%
  \BibitemOpen
  \bibfield  {author} {\bibinfo {author} {\bibnamefont {Yen}, \bibfnamefont
  {Tzu-Ching}}, \bibinfo {author} {\bibfnamefont {Robert~A.}\ \bibnamefont
  {Lang}}, and\ \bibinfo {author} {\bibfnamefont {Artur~F.}\ \bibnamefont
  {Izmaylov}}} (\bibinfo {year} {2019}),\ \bibfield  {title} {\enquote
  {\bibinfo {title} {Exact and approximate symmetry projectors for the
  electronic structure problem on a quantum computer},}\ }\href
  {https://doi.org/10.1063/1.5110682} {\bibfield  {journal} {\bibinfo
  {journal} {The Journal of Chemical Physics}\ }\textbf {\bibinfo {volume}
  {151}}~(\bibinfo {number} {16}),\ \bibinfo {pages} {164111}}\BibitemShut
  {NoStop}%
\bibitem [{\citenamefont {{Yeter-Aydeniz}}\ \emph {et~al.}(2020)\citenamefont
  {{Yeter-Aydeniz}}, \citenamefont {Pooser},\ and\ \citenamefont
  {Siopsis}}]{yeter-aydenizPracticalQuantumComputation2020}%
  \BibitemOpen
  \bibfield  {author} {\bibinfo {author} {\bibnamefont {{Yeter-Aydeniz}},
  \bibfnamefont {K{\"u}bra}}, \bibinfo {author} {\bibfnamefont {Raphael~C.}\
  \bibnamefont {Pooser}}, and\ \bibinfo {author} {\bibfnamefont {George}\
  \bibnamefont {Siopsis}}} (\bibinfo {year} {2020}),\ \bibfield  {title}
  {\enquote {\bibinfo {title} {Practical quantum computation of chemical and
  nuclear energy levels using quantum imaginary time evolution and {{Lanczos}}
  algorithms},}\ }\href {https://doi.org/10.1038/s41534-020-00290-1} {\bibfield
   {journal} {\bibinfo  {journal} {npj Quantum Information}\ }\textbf {\bibinfo
  {volume} {6}},\ \bibinfo {pages} {63}}\BibitemShut {NoStop}%
\bibitem [{\citenamefont {Yoshioka}\ \emph {et~al.}(2022)\citenamefont
  {Yoshioka}, \citenamefont {Hakoshima}, \citenamefont {Matsuzaki},
  \citenamefont {Tokunaga}, \citenamefont {Suzuki},\ and\ \citenamefont
  {Endo}}]{yoshiokaGeneralizedQuantumSubspace2022}%
  \BibitemOpen
  \bibfield  {author} {\bibinfo {author} {\bibnamefont {Yoshioka},
  \bibfnamefont {Nobuyuki}}, \bibinfo {author} {\bibfnamefont {Hideaki}\
  \bibnamefont {Hakoshima}}, \bibinfo {author} {\bibfnamefont {Yuichiro}\
  \bibnamefont {Matsuzaki}}, \bibinfo {author} {\bibfnamefont {Yuuki}\
  \bibnamefont {Tokunaga}}, \bibinfo {author} {\bibfnamefont {Yasunari}\
  \bibnamefont {Suzuki}}, and\ \bibinfo {author} {\bibfnamefont {Suguru}\
  \bibnamefont {Endo}}} (\bibinfo {year} {2022}),\ \bibfield  {title} {\enquote
  {\bibinfo {title} {Generalized {{Quantum Subspace Expansion}}},}\ }\href
  {https://doi.org/10.1103/PhysRevLett.129.020502} {\bibfield  {journal}
  {\bibinfo  {journal} {Physical Review Letters}\ }\textbf {\bibinfo {volume}
  {129}}~(\bibinfo {number} {2}),\ \bibinfo {pages} {020502}}\BibitemShut
  {NoStop}%
\bibitem [{\citenamefont {Zhang}\ \emph {et~al.}(2020)\citenamefont {Zhang},
  \citenamefont {Lu}, \citenamefont {Zhang}, \citenamefont {Chen},
  \citenamefont {Li}, \citenamefont {Zhang},\ and\ \citenamefont
  {Kim}}]{zhangErrormitigatedQuantumGates2020}%
  \BibitemOpen
  \bibfield  {author} {\bibinfo {author} {\bibnamefont {Zhang}, \bibfnamefont
  {Shuaining}}, \bibinfo {author} {\bibfnamefont {Yao}\ \bibnamefont {Lu}},
  \bibinfo {author} {\bibfnamefont {Kuan}\ \bibnamefont {Zhang}}, \bibinfo
  {author} {\bibfnamefont {Wentao}\ \bibnamefont {Chen}}, \bibinfo {author}
  {\bibfnamefont {Ying}\ \bibnamefont {Li}}, \bibinfo {author} {\bibfnamefont
  {Jing-Ning}\ \bibnamefont {Zhang}}, and\ \bibinfo {author} {\bibfnamefont
  {Kihwan}\ \bibnamefont {Kim}}} (\bibinfo {year} {2020}),\ \bibfield  {title}
  {\enquote {\bibinfo {title} {Error-mitigated quantum gates exceeding physical
  fidelities in a trapped-ion system},}\ }\href
  {https://doi.org/10.1038/s41467-020-14376-z} {\bibfield  {journal} {\bibinfo
  {journal} {Nature Communications}\ }\textbf {\bibinfo {volume} {11}},\
  \bibinfo {pages} {587}}\BibitemShut {NoStop}%
\bibitem [{\citenamefont {Zhu}\ \emph {et~al.}(2020)\citenamefont {Zhu},
  \citenamefont {Johri}, \citenamefont {Linke}, \citenamefont {Landsman},
  \citenamefont {Huerta~Alderete}, \citenamefont {Nguyen}, \citenamefont
  {Matsuura}, \citenamefont {Hsieh},\ and\ \citenamefont
  {Monroe}}]{zhuGenerationThermofieldDouble2020}%
  \BibitemOpen
  \bibfield  {author} {\bibinfo {author} {\bibnamefont {Zhu}, \bibfnamefont
  {D}}, \bibinfo {author} {\bibfnamefont {S.}~\bibnamefont {Johri}}, \bibinfo
  {author} {\bibfnamefont {N.~M.}\ \bibnamefont {Linke}}, \bibinfo {author}
  {\bibfnamefont {K.~A.}\ \bibnamefont {Landsman}}, \bibinfo {author}
  {\bibfnamefont {C.}~\bibnamefont {Huerta~Alderete}}, \bibinfo {author}
  {\bibfnamefont {N.~H.}\ \bibnamefont {Nguyen}}, \bibinfo {author}
  {\bibfnamefont {A.~Y.}\ \bibnamefont {Matsuura}}, \bibinfo {author}
  {\bibfnamefont {T.~H.}\ \bibnamefont {Hsieh}}, and\ \bibinfo {author}
  {\bibfnamefont {C.}~\bibnamefont {Monroe}}} (\bibinfo {year} {2020}),\
  \bibfield  {title} {\enquote {\bibinfo {title} {Generation of thermofield
  double states and critical ground states with a quantum computer},}\ }\href
  {https://doi.org/10.1073/pnas.2006337117} {\bibfield  {journal} {\bibinfo
  {journal} {Proceedings of the National Academy of Sciences}\ }\textbf
  {\bibinfo {volume} {117}}~(\bibinfo {number} {41}),\ \bibinfo {pages}
  {25402--25406}}\BibitemShut {NoStop}%
\end{thebibliography}
%


\clearpage
\pagebreak[4]
\global\pdfpageattr\expandafter{\the\pdfpageattr/Rotate 90}

\begin{turnpage}
\begin{table*}[htb]
    \centering
\resizebox{1.25\textwidth}{!}{
    \begin{threeparttable}
        \begingroup
        \renewcommand{\arraystretch}{2.5}
        \setlength{\tabcolsep}{1em}
        \aboverulesep=0ex
        \belowrulesep=0ex
\begin{ruledtabular}
\begin{tabular}{@{}p{6.7em}  p{8.7em} p{9em}  p{10.3em} p{12em}  p{10.1em}@{}}
Methods & Probabilistic Error Cancellation  & Richardson Extrapolation (Equal-gap\tnote{$\#$} \ ) &Symmetry Verification & Virtual Distillation & Echo Verification \\
\cmidrule(lr{.75em}){2-6}

Main assumptions &Full knowledge of the noise. & Ability to scale the noise. Small $\lambda$\tnote{$\mathparagraph$}. &The ideal state contains inherent symmetry.&\multicolumn{2}{p{24.6em}}{The ideal state $\rho_0$ is pure. The noise is stochastic such that $\rho_0$ is the dominant eigenvector of the noisy state $\rho$. \tnote{$\mathsection$}} \\

Hyper-parameters & Circuit fault rate \emph{after} mitigation: $\lambda_{\mathrm{em}}$ & Number of data points: $M$ &  The projector of the symmetry subspace: $\Pi$& Degree of purification: $M$ & \cell{c}{Nil} \\

Qubit overhead & \cell{c}{$1$} & \cell{c}{$1$} & \cell{c}{$1$\tnote{$\ddagger$}} & \cell{c}{$M$} & \cell{c}{$1$} \\

Circuit runtime overhead & \cell{c}{up to $\sim 2$} & \cell{c}{up to $\sim M$}  & \cell{c}{$1$} & \cell{c}{$\sim 1$\tnote{$\|$}} & \cell{c}{$2$} \\

Sampling overhead ($C_{\mathrm{em}}$) & \cell{c}{$e^{4\left(\lambda - \lambda_{\mathrm{em}}\right)}$} & \cell{c}{$\left(2^M-1\right)^2$} & \raisebox{0pt}{\makecell{Post-selection: $\Tr[\Pi \rho]^{-1}$ \\ Post-processing: $\Tr[\Pi \rho]^{-2}$}} & \cell{c}{$\Tr[ \rho^M]^{-2} \gtrsim \frac{e^{2M\lambda}}{\left[1+\left(e^{\lambda}-1\right)^M\right]^2}$}& \cell{c}{$\Tr[\rho^2]^{-1} \gtrsim \frac{e^{2\lambda}}{1+\left(e^{\lambda}-1\right)^2}$}\\

Fidelity boost\tnote{$*$} & \cell{c}{$e^{\lambda - \lambda_{\mathrm{em}}}$} & \cell{c}{$e^\lambda + \order{\lambda^M}$} & \cell{c}{$\Tr[\Pi \rho]^{-1}$} & \cell{c}{$\frac{\Tr[\rho_0 \rho]^{M-1}}{\Tr[\rho^M]}\gtrsim \frac{e^{\lambda}}{ 1 + \left(e^{\lambda}-1\right)^M} $} &  Same as VD with $M=2$\\

Bias & Can reach $0$ when $\lambda_{\mathrm{em}} = 0$. & \cell{c}{$\order{\lambda^{M}}$} & \multicolumn{3}{p{37em}}{Can be upper-bounded using the error-mitigated fidelity using \cref{eqn:bias_fid_bound}, which in turns is related to the fidelity boost achieved.} 
\end{tabular}
\end{ruledtabular}
\endgroup
\begin{tablenotes}\footnotesize
\item[$*$] Fidelity boost is the factor of increase in the fidelity against the ideal state $\rho_0$ after applying error mitigation (\cref{sec:state_extract_perf}).
\item[$\mathparagraph$] There are successful experiments operating beyond the small-$\lambda$ assumption. The small-$\lambda$ assumption is not needed for some other extrapolation methods. (\cref{sec:zne})
\item[$\#$] There are other variants of Richardson extrapolation that may provide better scalings for the overheads~\cite{sidiPracticalExtrapolationMethods2003}.
\item[$\mathsection$] If $\rho_0$ is not the dominant eigenvector of the noisy state, the ultimate fidelity achieved will be limited by coherent mismatch (\cref{sec:pur}). 
\item[$\ddagger$] Assuming only the inherent symmetries of the physical problems are used. Additional qubit overhead might be needed if we want to introduce additional symmetries.
\item[$\|$] Additional circuit components are needed to swap in between the noisy copies and the corresponding additional runtime required will depend on the connectivity of the hardware.
\end{tablenotes}
\caption{A table summarising the assumptions, costs and performances associated with some of the QEM methods mentioned in this review. Some of the expressions of the sampling overhead and the fidelity boost are derived under the assumption that the occurrence of faults in the circuit follows a Poisson distribution with the circuit fault rate being $\lambda$ as discussed in \cref{sec:circuit_faults}. We use $\rho$ and $\rho_0$ to denote the unmitigated noisy state and the ideal noiseless state, respectively. Only a specific instance of error extrapolation is included here. Measurement error mitigation is not included here since it can be viewed as the special case of probabilistic error cancellation focusing on the measurement noise. Quantum subspace expansion and $N$-representability are also not included in here since their implementation costs and performance are highly problem-specific.}
\label{tab:qem_summary}
\end{threeparttable}
}
\end{table*}
\end{turnpage}

\end{document}